\documentclass[manuscript]{aastex}

\usepackage{subfigure}
\usepackage{natbib}
\usepackage{graphicx}
\usepackage{multirow}
\usepackage{epstopdf}
\usepackage{amsmath}
\usepackage{url}
\usepackage{hyperref}
\usepackage{amssymb}

\shorttitle{BATSE 5B Catalog}
\shortauthors{Goldstein, et al.}

\begin{document}

\title{The BATSE 5B Gamma-Ray Burst Spectral Catalog}

\author{Adam Goldstein\altaffilmark{1}, 
Robert~D.~Preece\altaffilmark{1},
Robert~S.~Mallozzi\altaffilmark{*}, 
Michael~S.~Briggs\altaffilmark{1},
Gerald~J.~Fishman\altaffilmark{2},
Chryssa~Kouveliotou\altaffilmark{2},
William~S.~Paciesas\altaffilmark{3},
J.~Michael~Burgess\altaffilmark{1}}
\altaffiltext{1}{University of Alabama in Huntsville, 320 Sparkman Drive, Huntsville, AL 35899, USA}
\altaffiltext{2}{Space Science Office, VP62, NASA/Marshall Space Flight Center, Huntsville, AL 35812, USA}
\altaffiltext{3}{Universities Space Research Association, 320 Sparkman Drive, Huntsville, AL 35805, USA}
\altaffiltext{*}{deceased}

\begin{abstract}
We present systematic spectral analyses of GRBs detected with the Burst and Transient Source Experiment (BATSE) onboard 
the Compton Gamma-Ray Observatory (CGRO) during its entire nine years of operation.  This catalog contains two types of 
spectra extracted from 2145 GRBs, and fitted with five different spectral models resulting in a compendium of over 19000 
spectra.  The models were selected based on their empirical importance to the spectral shape of many GRBs, and the analysis 
performed was devised to be as thorough and objective as possible.  We describe in detail our procedures and criteria for the 
analyses, and present the bulk results in the form of parameter distributions.  This catalog should be considered an official 
product from the BATSE Science Team, and the data files containing the complete results are available from the High-Energy 
Astrophysics Science Archive Research Center (HEASARC, \url{http://heasarc.gsfc.nasa.gov/W3Browse/cgro/bat5bgrbsp.html}).

\end{abstract}

\keywords{gamma rays: bursts --- methods: data analysis}

\section{Introduction}
Gamma--ray bursts (GRBs) are intense flashes of radiation that appear at unpredictable times from random locations on the 
celestial sphere. They have been studied since their serendipitous discovery by the {\it Vela} Satellite Network \citep
{Klebesadel}.  Despite nearly 40 years of research, the progenitors of these astrophysical phenomena remain largely unknown.  
Several thousand of these events have been detected by many different instruments.  The bulk of the energy emission of GRBs 
occurs in the X--ray  and gamma--ray regime of the electromagnetic spectrum.  Their intensities and durations span many orders 
of magnitude, where the intensity of a burst is usually defined by its peak flux at a specified time resolution and in a specified 
energy range.  The strong bursts are distributed homogeneously in space as the angular distribution is consistent with an 
isotropic distribution \citep{Meegan92, Briggs94, Briggs96}.  Observations of burst afterglow at other wavelengths may provide 
substantial evidence of the burst mechanism.  Research in this area involves attempting to determine the celestial locations of bright 
bursts while they are in progress and to distribute these to observers for rapid follow-up observations at other wavelengths \citep
{Barthelmy94}.  This afterglow search has yielded several observations of GRB afterglows \citep[e.g.;][]{DePasquale06, Roming09}.

The count rate time history of each GRBs is unique, rendering classifications based on morphology unsuccessful [see \citet
{Hrabovsky} for a review].   The shapes of GRB light curves vary widely; they include single-peak and multi-peak events, long 
duration emission at low intensity, and rapidly rising time profiles with an  exponential-like decay phase. Temporal variability on 
time scales of the order of milliseconds has been observed, although smooth light curves with little variability are also 
observed~\citep{Bhat, Lloyd02}.  The study of this variability has lead to theoretical development of the internal emission processes of 
GRBs~\citep{Sari97, Dermer98} The durations of bursts span a wide range, with the shortest bursts of the order of milliseconds, and 
the longest duration bursts lasting for hundreds of seconds.   The logarithmic duration distribution of GRBs observed by BATSE 
appears to exhibit two clusters centered at $\sim$0.3 s and $\sim$40 s, with a deficit of bursts of duration $\sim$2 s dividing the two 
groups.  There is evidence for a duration--spectral hardness correlation in which the short duration bursts are harder than the long 
bursts \citep{Dezalay, Kouveliotou}.  Investigations using other instruments, however, have shown that the division of GRBs based on 
duration is typically energy-dependent~\citep{Bromberg12} and the 2 s division found in BATSE is a result of the detector bandpass.

The prompt spectra of gamma-ray bursts have been measured (e.g., \citet{Cline73, Mazets, Matz, Norris86, Band93, Dingus, 
Pelangeon08, Frontera09, Sakamoto11})  over a wide energy range ($<1$ keV to several GeV). Many burst spectra are consistent 
with a power law at high energies, showing little or no attenuation of high energy photons, while others exhibit an exponential cutoff 
at high energies.  GRB spectra observed by BATSE at lower energies usually exhibit increasing energy flux at lower energies, with a 
break between the low and high energy portions of the spectra usually occurring at approximately 100--300 keV, although the break 
has been found below a few tens of keV in X-Ray Flashes (XRFs) studied by the HETE-2 experiment~\citep{Atteia11}, as well as 
several examples of breaks existing in the MeV range~\citep[e.g.;][]{Briggs99, GBMSpecCat}. Most, but not all, gamma--ray bursts 
exhibit significant spectral evolution, usually evolving from hard to soft.

It is because of the many aforementioned reasons that it is appropriate and enlightening to systematically study the spectral 
properties of GRBs.  Earlier spatial studies focused on the time-resolved properties of GRB spectra, which were only possible for 
very bright bursts \citep{Preece98, Kaneko06}. However, there is also a need for analyses that derive properties for all 
possible bursts.  Indeed, several studies made use of a partially complete set of spectral analyses of BATSE GRBs \citep
{Mallozzi,BandPreece} and one published result has been based upon the complete set \citep{Goldstein10}.  The BATSE dataset of 
GRBs is currently the largest compendium of GRB observations from a single instrument, and as an increasing number of 
observations are made within the lower-energy bandpass $\it Swift$ and the higher-energy and broader bandpass of $\it Fermi$, the 
BATSE dataset will be an important bridge between the sub-keV, MeV, and GeV observations of the prompt emission of GRBs.  
Furthermore, although the energy band of the currently operating {\it Fermi}/GBM completely covers the BATSE energy range, the 
GBM cannot match the sensitivity and effective area that BATSE detectors possessed.  Here, we present the complete set of spectral 
analyses from which these works were derived, corresponding to the burst selection of the 5B BATSE Burst Catalog (Briggs, et al., in 
preparation), covering the entire BATSE mission.  All catalog files are available from a public archive (HEASARC, \url{http://heasarc.gsfc.nasa.gov/W3Browse/cgro/bat5bgrbsp.html}).  

We start with a description of the BATSE detectors and calibration in Section 2.  This is followed in Section 3 by a description of the 
methodology used in the production of this catalog, including  detector selection, data types  used, energy selection and background 
fitting, and the source selection We then offer a description of the spectral models used in this catalog in Section 4.   Finally, in 
Section 5 we present the spectral analysis methods and results.

\section{Detectors and Calibration}
The Compton Gamma--Ray Observatory (CGRO) was placed into low Earth orbit ($\sim$400 km) by the space shuttle Atlantis 
on April 5, 1991.   BATSE was one of four experiments on-board the  17 ton satellite.  It was an eight-module all-sky detector system 
designed to study gamma--rays in the energy band of $\sim$10 keV--20 MeV.  Each of the eight detector modules were mounted 
on the corners of CGRO.  They consisted of two NaI(T$\ell$) scintillation detectors: a Large Area Detector (LAD), optimized for 
temporal resolution,  and a Spectroscopy Detector (SD), optimized for energy resolution.  The  configuration of the experiment 
allowed the maximum unobstructed field of view (approximately $2.6\pi$ steradians) for a low Earth orbit satellite. 

Each LAD contained a NaI crystal $\sim$51 cm in diameter and $\sim$1.3 cm  thick that was uncollimated 
in the forward hemisphere and passively shielded in the aft hemisphere.  The large surface area enabled the collection of a 
large number of gamma--ray photons as compared to previous orbiting scintillation detectors, thus providing a superior 
combination of temporal and energy resolution of observed events. The LAD was mounted on the upper part of the module.  The 
planes of the eight LAD faces formed a canonical octahedron when the modules were in their flight configuration on the 
CGRO.  This ensured that each GRB usually illuminated three or four detectors (with a special case where only two detectors were 
illuminated).  The approximate location of a burst could then be determined by comparing the relative count rates in those 
detectors that observed the burst \citep{Horack}. 

The LAD angular energy response is, to first order, a cosine function in $\theta$ at low energies, where $\theta$ is the angle of the 
GRB  from the normal of the LAD crystal.  The response is flatter than a cosine function for energies greater than $\sim$300 keV 
due to decreasing detector efficiency at higher energies.  The circular configuration and lack of spacecraft interference in the 
forward hemisphere of each LAD essentially removes any azimuthal dependence of the response function.  Hence, the LAD detector 
response matrices (DRMs) used in this study do not incorporate any azimuthal dependence in the response function.  The DRMs are 
mathematical matrix representations of the detector  energy response used to map the observed counts into photons of known 
energy.  Each detector's response is dependent on incident photon energy, the measured detector output energy, and the detector--
source angle and the earth--source--spacecraft geometry \citep{Pendleton}.  For details on the LAD effective area and response, 
see~\citet{McNamara95} and~\citet{Laird06}.

Although most scintillation detectors typically have photomultiplier tubes (PMTs) coupled directly to the crystal, the large surface 
area of  the LADs made this impractical for uniform light collection.   The crystal was instead attached to a collection cone that 
was lined with a highly reflective barium sulfate-based (BaSO$_4$) coating.  Three PMTs collected the scintillation photons and 
these signals were summed at each detector.  The tubes had a minimum quantum efficiency of 26\% at 410 nm \citep{Horack}.  
The light collection cone was lined with lead and tin layers providing passive shielding in the rear hemisphere up to $\sim
$300--400 keV.  The outermost tin layer was designed to absorb the K-shell X--rays produced by gamma interactions in the lead. 
The front of the crystal was covered by a plastic scintillator whose light was collected by two 5 cm PMTs, the signals of which 
were also summed at each detector. The purpose of this plastic scintillator, called the Charged Particle Detector (CPD), was to 
detect particles  that entered the LADs.  The instrument could be prevented from triggering due to  radiation produced by 
interactions of these particles with the NaI crystal by the detectors' optional coincidence/anti-coincidence circuitry  (the instrument 
operated in anti-coincidence mode for the entire mission).  No energy information was available for incident particles; the plastic 
scintillator was used solely to aid in identification of charged particle events.  The threshold energy deposition for the charged 
particle detector  was $\sim$500 keV.

\begin{figure}
	\begin{center}
		\subfigure[Near Start of Mission]{\label{beg511Calib}\includegraphics[scale=0.45]{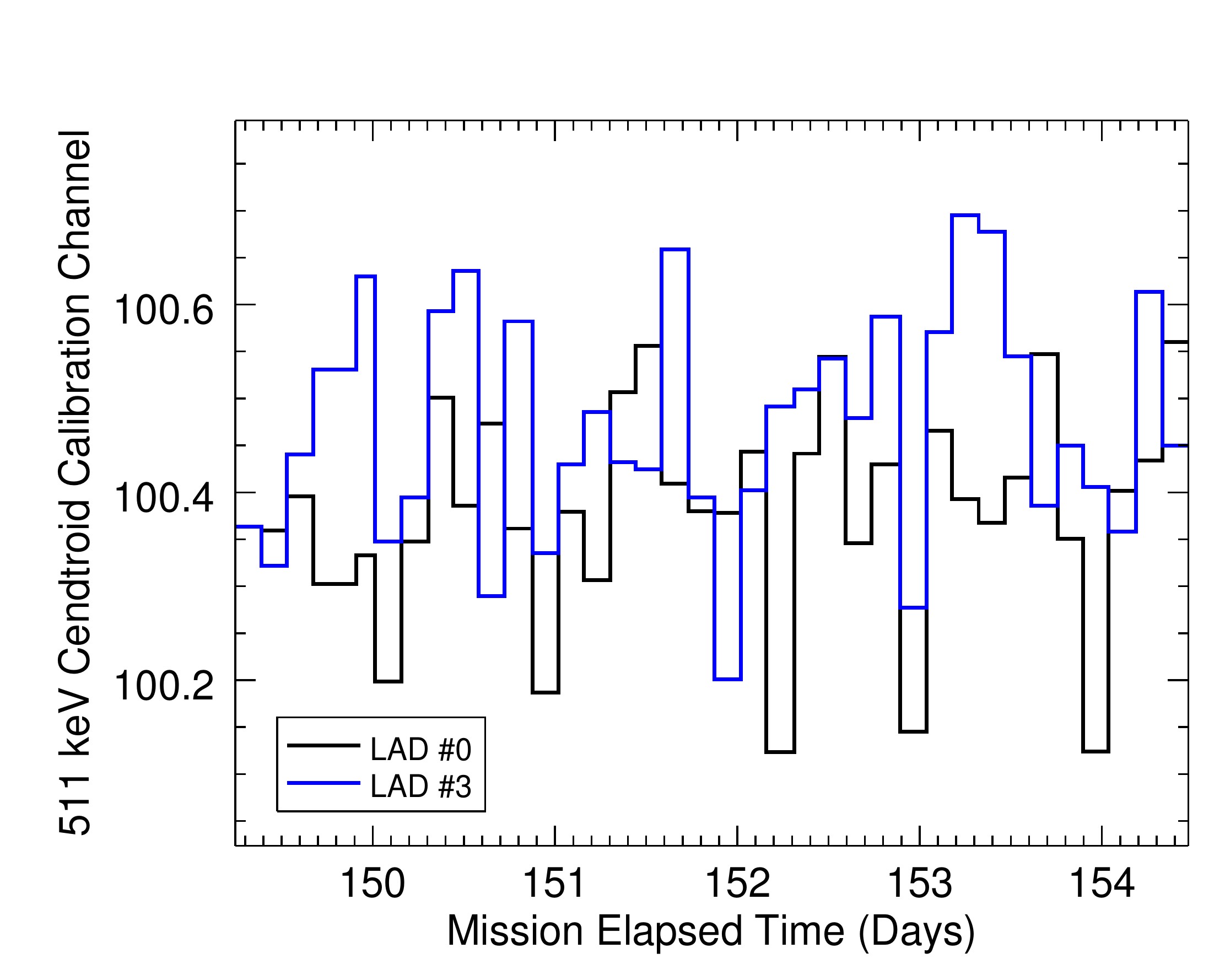}}\\
		\subfigure[Near End of Mission]{\label{end511Calib}\includegraphics[scale=0.45]{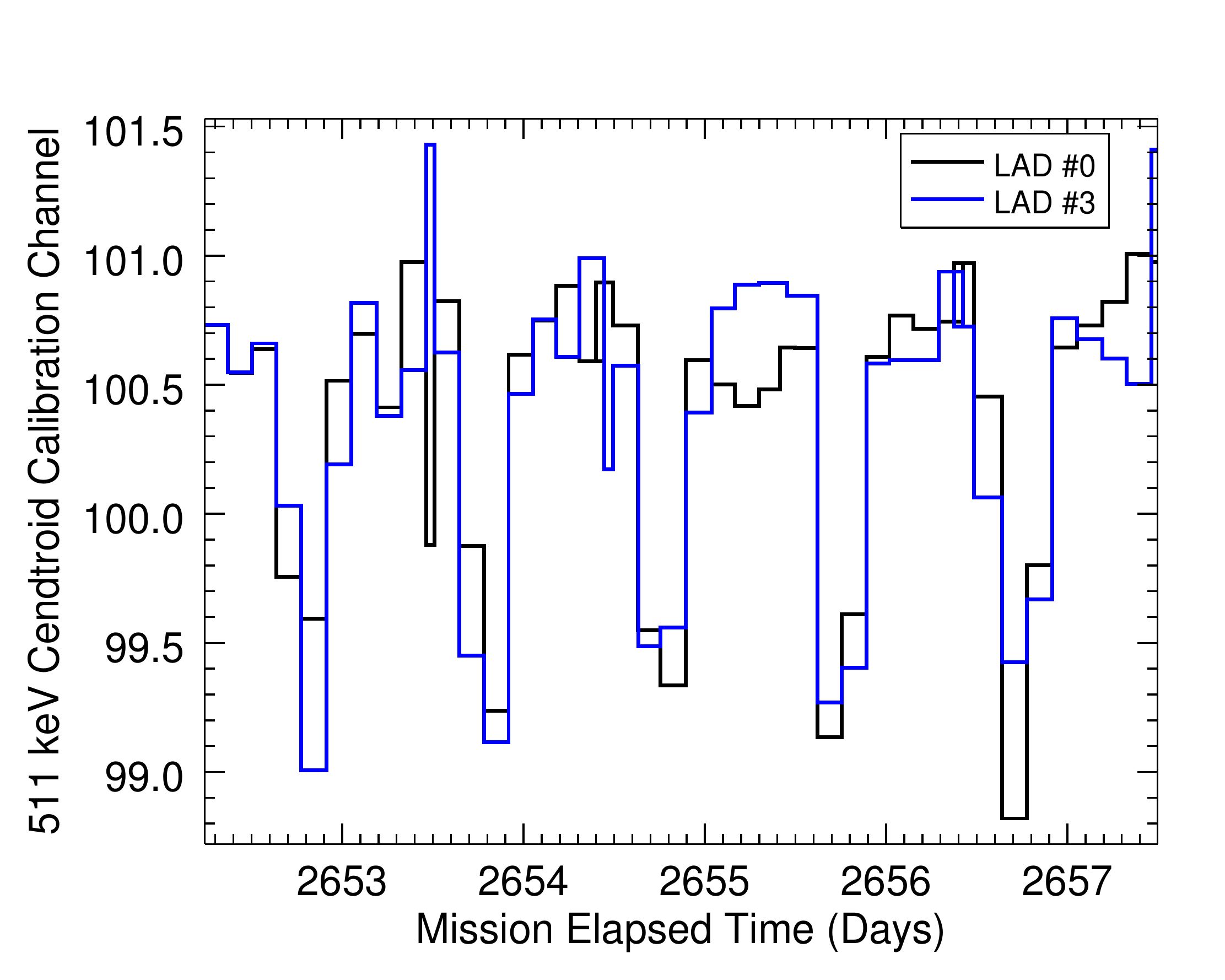}}
	\end{center}
\caption{The variation of the centroid of the 511 keV calibration line in channel space for two detectors from a period of time near the start and end of the mission.  \label{511Calib}}
\end{figure}

\begin{figure}
	\begin{center}
		\subfigure[Near Start of Mission]{\label{begBackSpec}\includegraphics[scale=0.45]{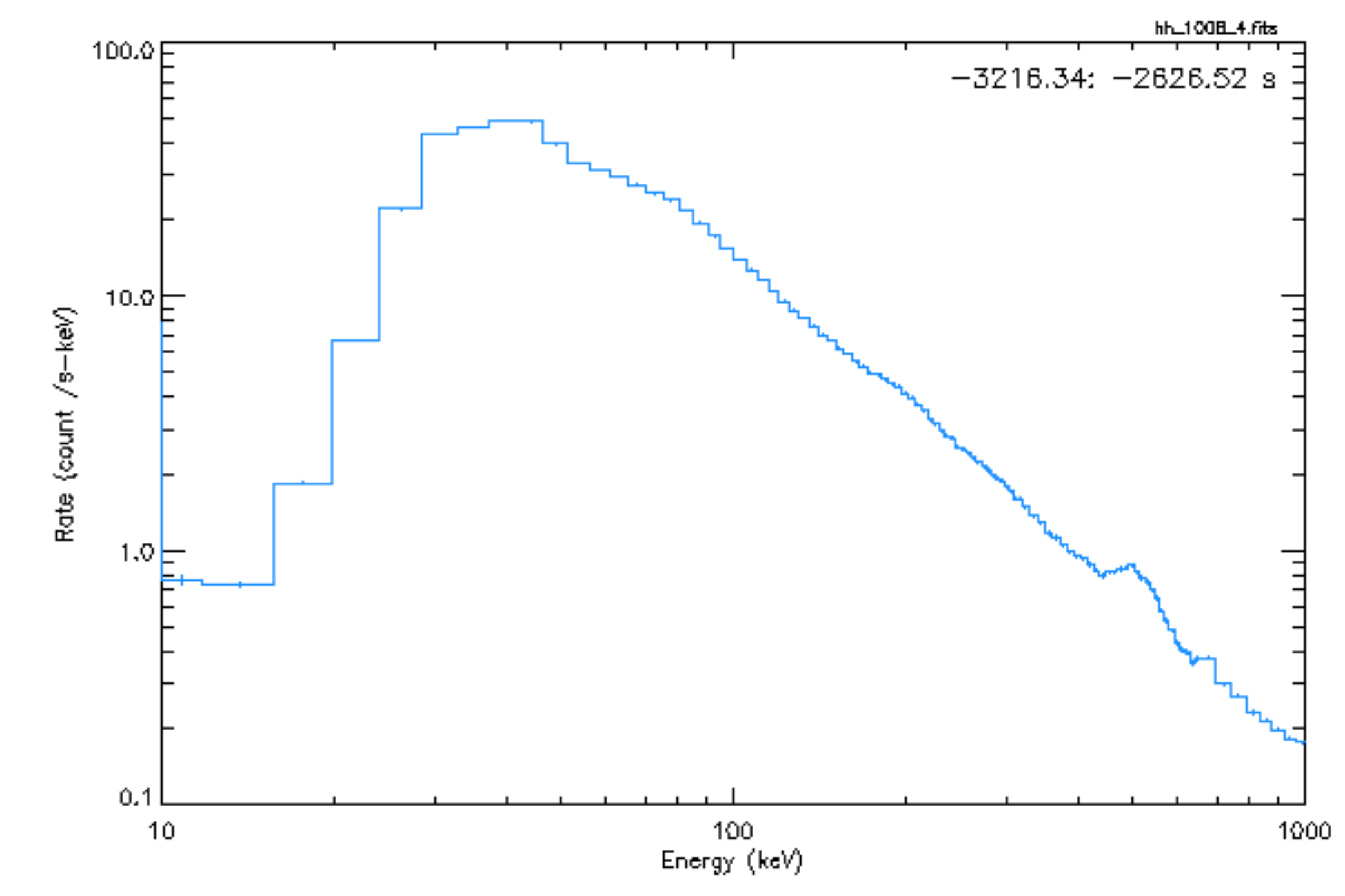}}\\
		\subfigure[Near End of Mission]{\label{endBackSpec}\includegraphics[scale=0.45]{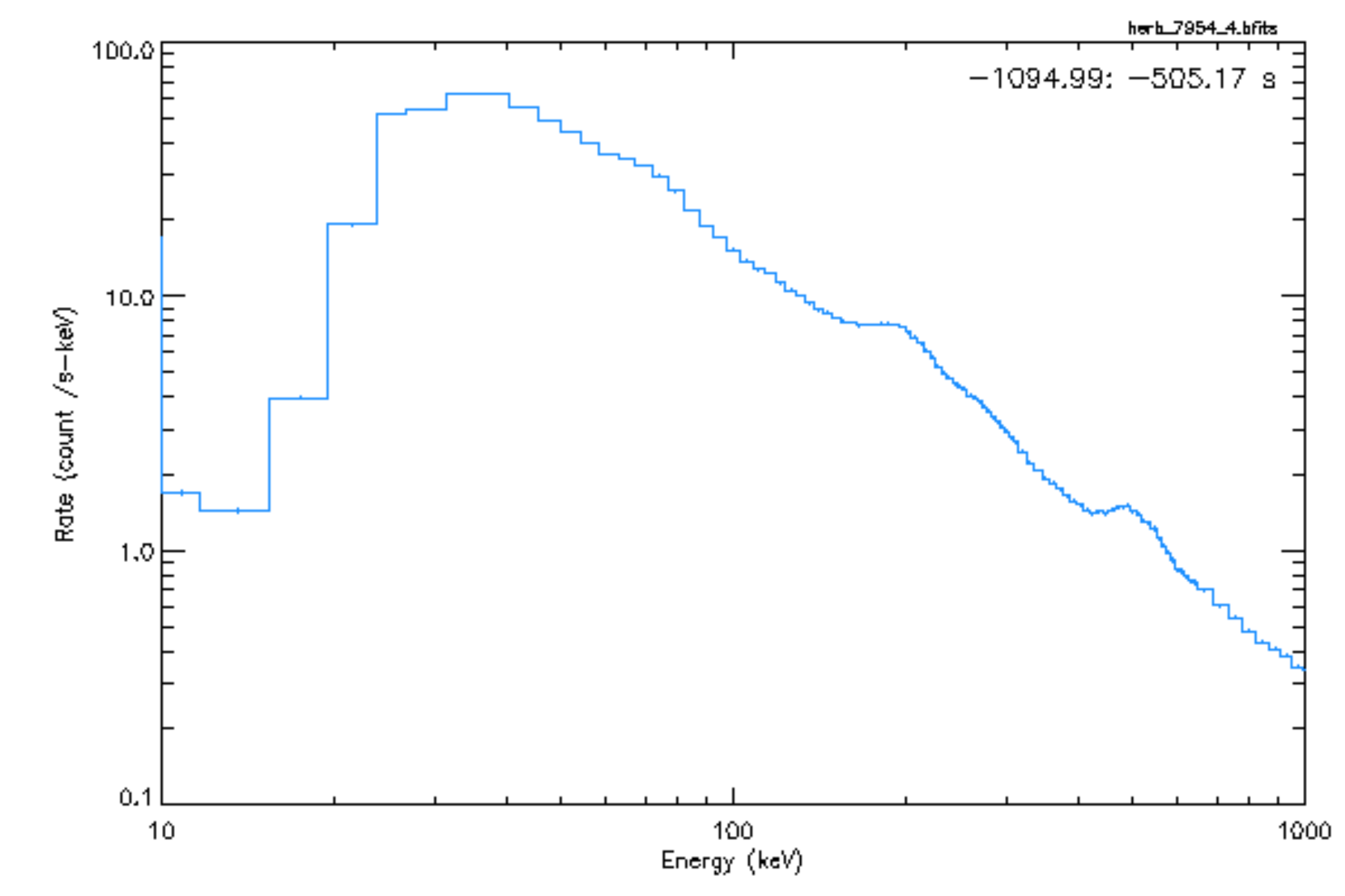}}
	\end{center}
\caption{Spectra of the background from LAD \#4 from near the beginning and end of the BATSE mission. \label{backSpec}}
\end{figure}

Each LAD was equipped with a system that controlled the high voltages (gain) of the photomultiplier tubes with minimal 
intervention from controllers on the ground.   The Automatic Gain Control (AGC) algorithm computed and executed
adjustments to the PMT high voltages so that a feature in the count spectrum remained in a specified energy  channel.  The 
background feature nominally monitored by the AGC was the 511 keV electron--positron annihilation line that is present in the 
gamma--ray background.  The background was calculated as a straight line over a specified range of energy channels and  was 
used to produce a background-subtracted count spectrum. The channel centroid of the 511 keV line was was computed, and if this 
centroid was different from a specified channel, the high voltage of a single PMT was adjusted to correct the computed value.  If the 
background line drifted too far from the programmed energy channel, the gains of the PMTs were automatically adjusted $\sim$4 
volts higher or lower to move the annihilation line back to the correct energy channel.  The PMTs were adjusted cyclically, and 
the range of voltage adjustment was clamped.  If the AGC attempted to change the high voltage to values outside of the specified 
range, an error was issued.  This procedure, which occurred approximately every 5 minutes, ensured that the eight LADs had 
nearly equal energies in a given pulse-height channel.  A sample of the variation in the 511 keV calibration line near the 
beginning and end of the mission is shown in Figure~\ref{511Calib}, where it is shown that generally the gain varied by only a 
fraction of a percent.  Furthermore, examples of the observed background spectrum near the start and end of the mission can be 
found in Figure \ref{backSpec}, which displays an enhancement of an activation line at $\sim$191 keV near the end of the mission.  
Additional details of the energy calibration of the LADs such as the Crab Nebula spectral analysis and energy--PHA  relation can be 
found elsewhere~\citep{Band92, Pendleton94, Preece98}.

\section{Method}
During its entire 3323 days of operation (an effective exposure of $\sim$2390 days in the GRB triggering energy band), BATSE 
triggered on 2704 GRBs, 2145 of which are presented in this catalog.  Bursts that were excluded include those with a low 
accumulation of count rates or a lack of spectral/temporal coverage.  In some cases the data collected onboard the spacecraft would 
be corrupted during storage or during transmission to ground.  For this reason, many bursts do not have contiguous data and so were 
not included in the catalog.  In a few cases data were available but contained incomplete time history, either through data corruption 
or count rate truncation from an extreme number of detected counts ($\sim 5\times10^4$ counts/energy channel/s above 
background, where `energy channel' refers to the channels in the triggering energy band).  These bursts were also omitted, as were 
extremely bright detections where the count rates were so large in each energy channel that caused detector saturation, and in some 
cases pulse pile-up.  In order to provide the most useful analysis to the community, we have attempted to make the method as 
objective, systematic and uniform as possible. When we deviate from uniformity we indicate the circumstances clearly.

\subsection{Detector Selection}
BATSE employed a total of 16 detectors in pairs of two on each the eight corners of the spacecraft.  Each pair comprised a LAD 
and SD detector.  For the purposes of this catalog, we have chosen to use only the LADs because of their much larger effective 
area, and the fact that the DRMs for the SDs exhibit poor photopeak efficiencies and large off-diagonal DRM elements at 
energies $>$3 MeV \citep{Kaneko05}.  The area of each LAD was 2025 $\rm cm^2$ with a spectral coverage from $\sim$20 keV 
to $\sim$2 MeV and a peak spectral response at $\sim$50--200 keV.  To ensure sufficient detector response, only detectors with 
viewing angles to the burst less than $65^\circ$ with respect to the LAD normal were included in the analysis.  These were 
selected from the subset of the four brightest detectors for each GRB, since the count rate in the detector is highly dependent on 
the incident angle.  In most cases this resulted in multiple detectors per burst.  Rather than performing joint spectral fits of all 
relevant detectors for a burst, we integrated the count rates over all relevant detectors, while preserving the energy edges, as well 
as integrating over the detector responses.  This method boosts the signal-to-noise and allows the inclusion of many weaker 
bursts into the catalog.  Another benefit of this method is that it helps to minimize the dilution of noise in the signal selection, 
especially for bursts less than 2 seconds.  A trade-off, however, is that there is an increase in systematic uncertainties that may not 
be completely accounted for in the error propagation of our results.  In Appendix D we show the impact of this is minimal when 
compared to the benefit of improved statistics.

\subsection{Data Types}
The primary data type used in this catalog was the 2.048 second resolution CONT data, which provided semi-continuous count 
rate and 16-channel spectral coverage during the entire BATSE lifetime.  Other BATSE datatypes such as HERB or SHERB data 
provide a much higher energy resolution (128 and 256 channels respectively) at the cost of time resolution.  These datatypes 
provided pre-burst time resolution on the order of $\sim$300 s, which complicates accurate background subtraction.  
Additionally, the trigger time resolution can vary depending on the count rate, and in many cases the high temporal resolution 
and trigger data end before the end of the GRB, further compromising the spectral analysis.  Therefore, CONT data are a prime 
choice for time-integrated spectral fitting since they allow the analysis of any precursor or late-time prompt emission that is 
not covered by other data types.  Unfortunately, this datatype places all of the emission of classically defined short GRBs into a single 
time bin and at times includes background in the signal selection.  The background contamination in this catalog, however, is 
marginal since most of the short GRBs in this catalog have a high signal-to-noise ratio, and they are only slightly affected by the 
inclusion of a relative small amount of noise.  An example of this is trigger \#206, which is representative of a lower than average 
intensity short GRB observed by BATSE.  The signal-to-noise ratio (SNR) for this GRB is 5.7 using the single 2.048 s bin in the CONT 
data, compared to a SNR of 7.6 when selecting the region in the 16 ms resolution MER data.  

The spectral coverage of the CONT data is split into 16 bins, including the noise-prone low-energy channel and high-energy overflow 
channel, so we used the remaining 14 energy channels for spectral fitting.  While this is a coarser energy resolution than some data 
types such as HERB or SHERB, 14 channels are more than twice the number of free parameters of the models being fit, so they are 
statistically sufficient for model fitting and comparison of GRBs (see Appendix A for simulations confirming the accuracy of CONT 
data as compared to HERB data).  When CONT data were not available (usually due to data corruption), the associated MER data 
were used if possible.  The native time resolution of MER data was 16 ms but only started at trigger time and extended to less than 
200 s after trigger.  The energy resolution is the same as CONT data, so the count rates were binned to 2.048 s in order to be 
compatible with CONT data. MER data were used only if a background model could be fit to the post-burst background and 
extrapolated through the duration of the burst.  MER data was used for only 15 GRBs in this catalog.

\subsection{Energy Selection and Background Fitting}
\begin{figure}
\begin{center}
	\includegraphics[scale=0.7, angle=90]{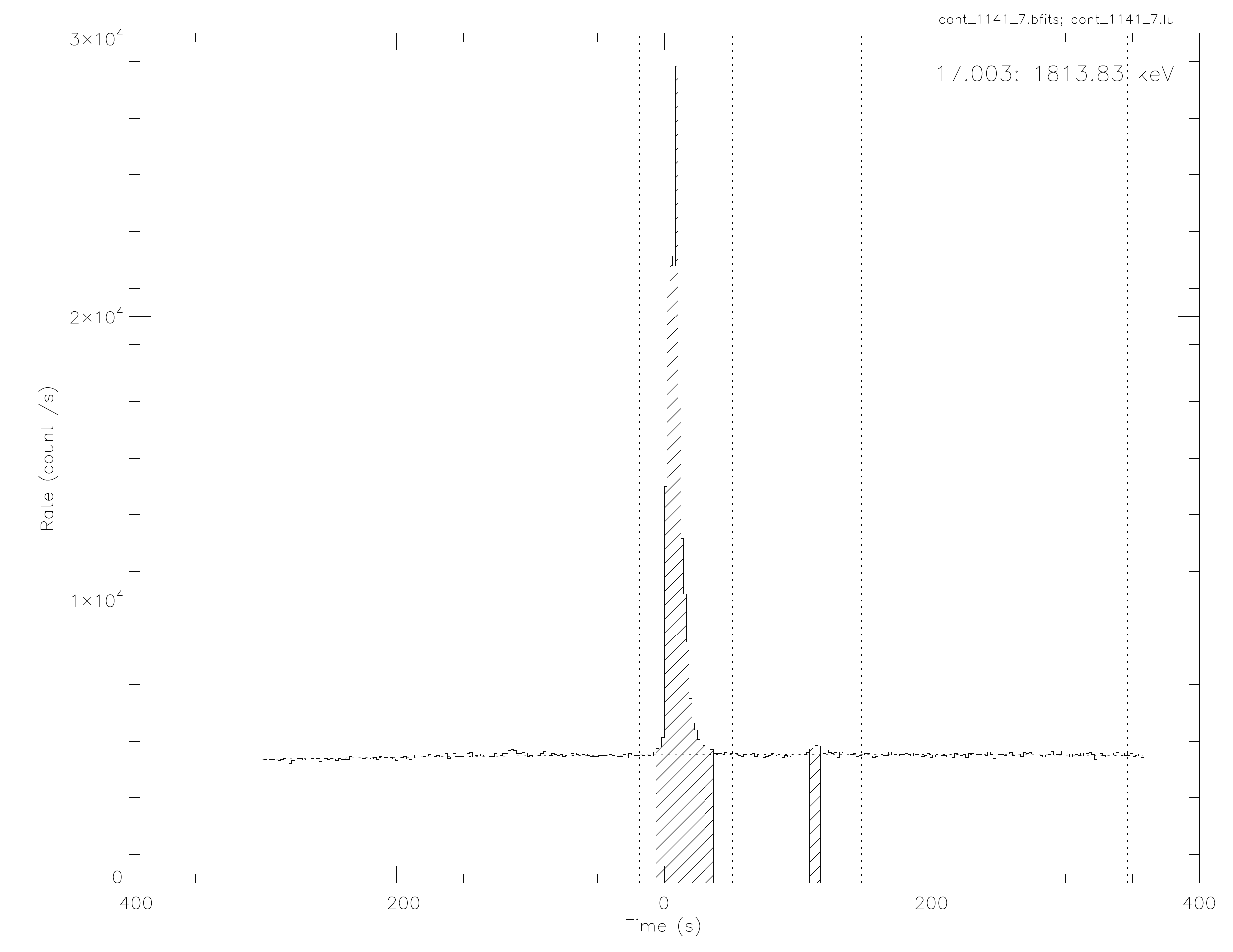}
	\caption{An example of the background selections from trigger \#1141.  The vertical dotted lines demarcate the three regions in 
	which the background is fit.  In this particular example, a second order polynomial was fit to the background regions. \label
	{background}}
\end{center}
\end{figure}

With the optimum subset of detectors selected, the best time and energy ranges are then chosen to fit the data.  
From the available energy channels in the LADs we select channels 1-14, corresponding to energies between $\sim$30 keV 
and $\sim$1.8 MeV. This selection is performed to exclude the high-energy overflow channel and the low-energy channel where 
the instrument response is poor and the background is high.  With the resulting time series, we select long pre- and 
post-burst background intervals to sufficiently model the background and fit a single polynomial (up to $4^{\rm th}$ order) to each 
energy channel in the background selection. For each detector the time selection and polynomial order are varied until the $
\chi^2$ statistic map over all energy channels is minimized resulting in an adequate background fit.  Typically a first- or second-order 
polynomial was fit to the background, since BATSE generally had a fairly stable background due to its constant inertial-pointing.  
Shown in Figure \ref{background} is an example of the regions selected as background for a GRB.  Only the bursts using MER 
data  that had a background that could be sufficiently fit with a first order polynomial were included in the catalog, since only post-
burst background was available for those bursts with only MER data.

\subsection{Source Selection}
\begin{figure}
	\begin{center}
		\subfigure[]{\label{acumtime}\includegraphics[scale=0.35]{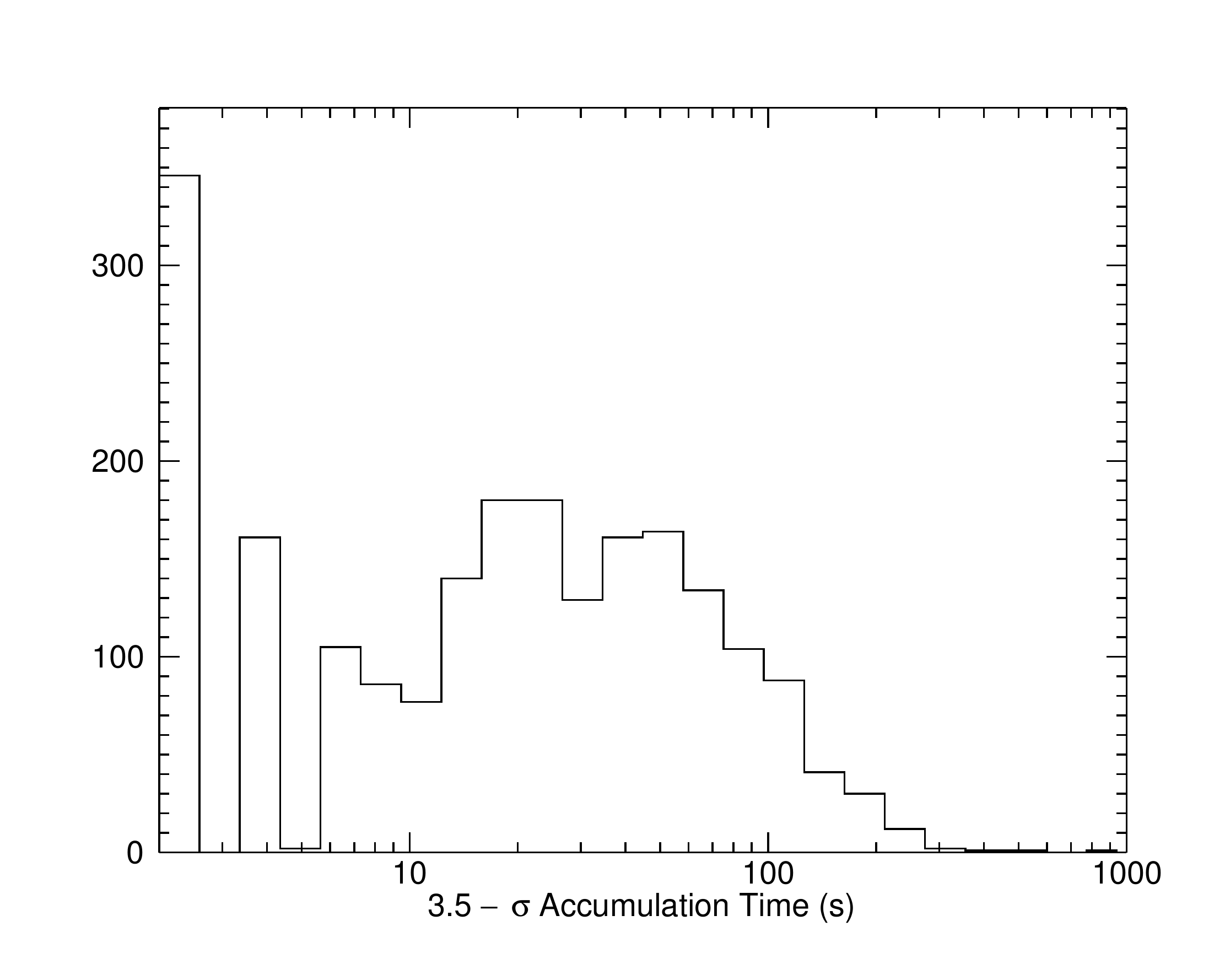}}\\
		\subfigure[]{\label{timepfluence}\includegraphics[scale=0.35]{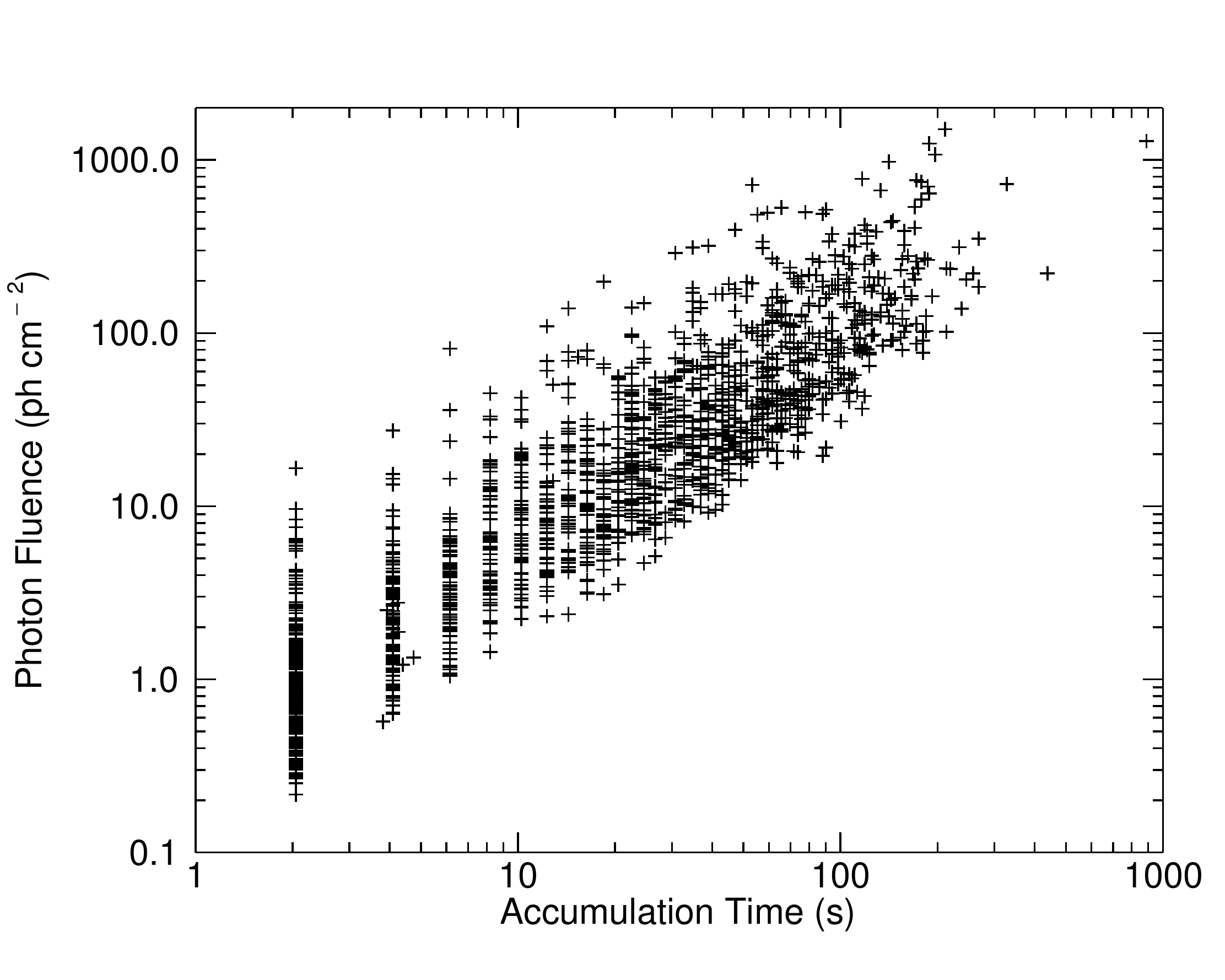}}
		\subfigure[]{\label{timepflux}\includegraphics[scale=0.35]{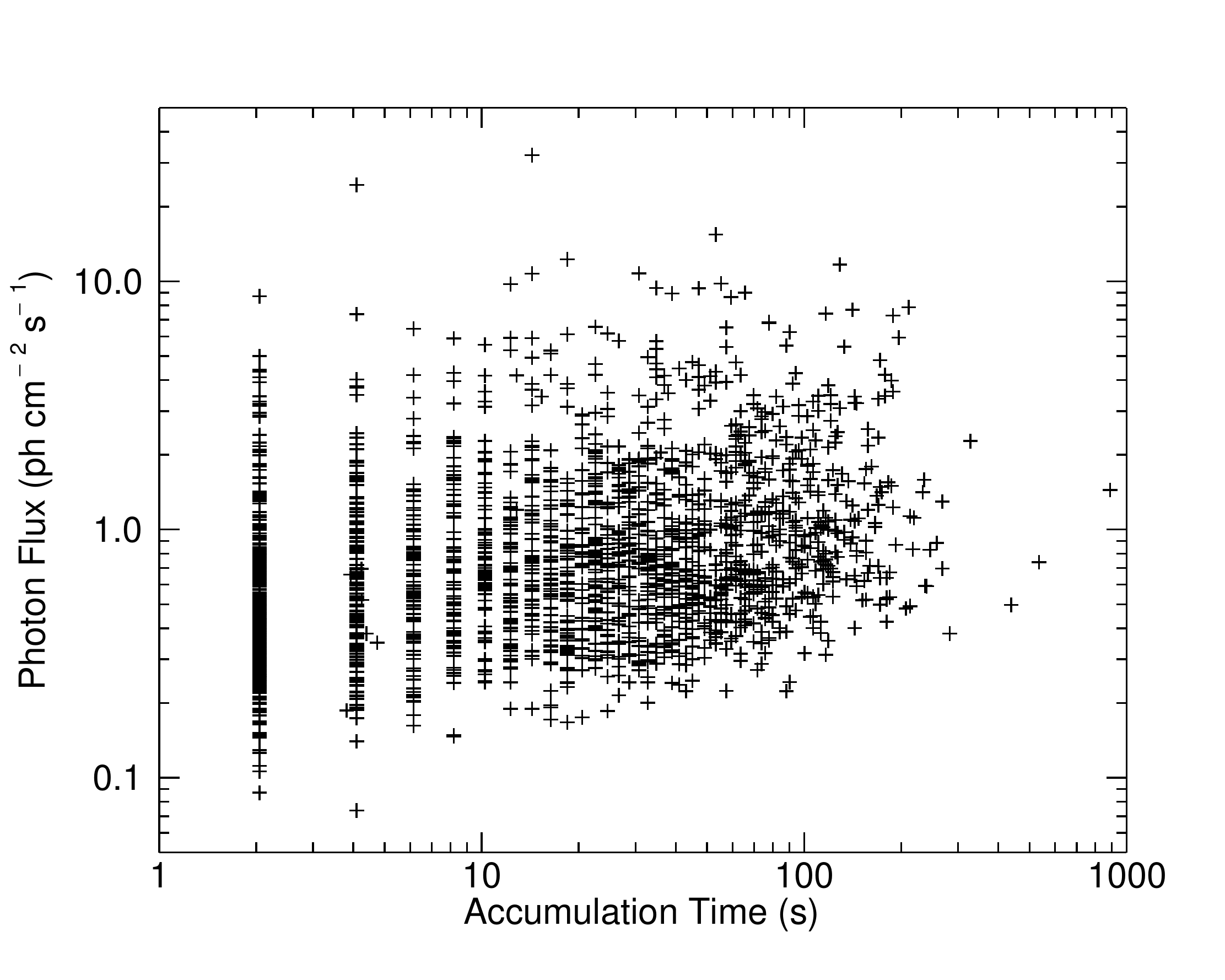}}
	\end{center}
\caption{\ref{acumtime} is the distribution of the accumulation time based on the 3.5$\sigma$ signal-to-noise selections.  Note the 
similarity to the traditional $T_{90}$ distribution, with the minimum near 1 second.  No other estimation of the duration was 
factored into the production of the accumulation time.  \ref{timepfluence} and \ref{timepflux} show the comparison of the model 
photon fluence and photon flux to the accumulation time respectively.  The fluxes and fluences shown in these figures are from 
the estimated BEST model fits. \label{allTime}}
\end{figure}

Once the background count rates are determined, we subtract the background, convert to counts, and calculate the 
signal-to-noise ratio (SNR) in the 20 keV - 2 MeV band for each time bin.  Only the bins that had a SNR greater or equal to 3.5 sigma 
were selected as signal.  This criterion ensures that there is adequate signal to successfully perform a spectral fit and constrain the 
parameters of the fit. This does however eliminate some faint bursts from the catalog sample (i.e., those with no time bins with signal 
above 3.5 sigma).  In addition, this strict cut was performed out of the need to provide an objective catalog, so we note that it is 
possible that not all signal from a burst was selected.  However, most of the signal below 3.5 sigma is likely indiscernible from the 
background fluctuations, so a spectral analysis including those bins would likely only increase the uncertainty in the 
measurements without improving the measurement.  This selection is referred to as the ``fluence'' selection, since it is a 
time-integrated selection, and is representative of the fluence over the duration of the burst as defined by the count rate bins that 
are above the SNR threshold.  The other selection performed is based on a 2.048 s peak photon flux, namely selecting the single 
time-history bin of signal with the highest background-subtracted count rate, (i.e., the most intense part of the burst).  We define the 
accumulation time as the total amount of signal selected using the 3.5 sigma SNR criterion, which ignores quiescent periods during 
the burst.  Figure \ref{allTime} shows the distribution of accumulation time based on the signal-to-noise selection criteria.  The 
accumulation time reported is similar to the observed emission time of the burst, excluding quiescent periods, similar to previous 
studies by \citet{Mitrofanov}.  As a result of the datatype used in this catalog, the accumulation time is quantized by multiples of 2.048 
s, thereby eliminating evidence of bimodality of the accumulation time.  Figure \ref{allTime} also includes the comparisons of the 
model photon fluence and photon flux compared to the accumulation time.  Note that both comparisons contain two distinct regions 
associated with short and long GRBs.  While there appears to be a very clear correlation between the photon fluence and the 
accumulation time, there is little correlation between the burst-averaged photon flux and the accumulation time.

\section{Models}
We chose five spectral models to fit the spectra of GRBs in our selection sample. These models include a 
single power law (PL), Band's GRB function (BAND), an exponential cut-off power-law (COMP), a smoothly-connected 
broken power law (SBPL), and a $\rm Log_{10}$ Gaussian (GLOGE). All models are formulated in units of photon flux with 
energy (E) in keV and multiplied  by a normalization constant A ($ \rm ph \ s^{-1} \ cm^{-2} \ keV^{-1}$). Below we detail each 
model and its features.

\subsection{Power-Law Model}
An obvious first model choice ubiquitous in astrophysical spectra, is the simple single power law with two free parameters,
\begin{equation}
f_{ PL } ( E ) = A \left( \frac{ E }{ E_{piv} } \right )^{ \lambda }
\end{equation}
where \emph{A} is the amplitude and $\lambda$ is the spectral index. The pivot energy ($E_{piv}$) normalizes the model to the 
energy range under inspection and helps reduce cross-correlation of other parameters.  In all cases in this catalog, $E_{piv}$ is 
held fixed at 100 keV.  While most GRBs exhibit a spectral break in the BATSE passband, some weak GRBs are too weak to 
adequately constrain this break in the fits and therefore we chose to fit these with the PL model.

\subsection{Band's GRB function}
Band's GRB function \citep{Band93} has become a standard spectral form for fitting GRB spectra, and therefore we include it in 
our analysis::
\begin{equation}
f_{ BAND } ( E ) = A \begin{cases} 
\biggl(\frac{E}{100 \ \rm keV }\biggr)^{\alpha} \exp \biggl[- \frac{ (\alpha +2) E}{ E_{peak} } \biggr] & E \geq \frac{ (\alpha - \beta) \ 
E_{peak} } { \alpha +2} \\
\biggl( \frac{E}{ 100 \ \rm keV } \biggr)^{ \beta } \exp (\beta -\alpha) \biggl[ \frac{(\alpha-\beta ) E_{peak}}{100 \ \rm keV \ (\alpha 
+2)} \biggr]^{\alpha-\beta } & E < \frac{(\alpha -\beta ) \ E_{peak}}{\alpha +2}
\end{cases}
\end{equation}
The four free parameters are the amplitude, \emph{A}, the low and high energy spectral indices, $\alpha$ and $\beta$, 
respectively, and the $\nu F_{\nu}$ peak energy, $E_{peak}$. This function is essentially a smoothly broken power law with a 
curvature defined by its spectral indices. The low-energy index spectrum asymptotically becomes a power law. 

\subsection{Comptonized Model}
This model is an exponentially cutoff power-law which is a subset of the Band function in the limit 
that $\beta \to -\infty$:
\begin{equation}
f_{COMP}(E) = A \ \Bigl(\frac{E}{E_{piv}}\Bigr) ^{\alpha} \exp \Biggl[ -\frac{(\alpha+2) \ E}{E_{peak}}  \Biggr]
\end{equation}	  
The three free parameters are the amplitude \emph{A}, the low energy spectral index $\alpha$ and $E_{peak}$. $E_{piv}$ is 
again fixed to 100 keV, as for the power law model.

\subsection{Smoothly Broken Power-Law}
Another model that we consider in this catalog is a broken power-law characterized by one break with flexible curvature able 
to fit spectra with both sharp and smooth transitions between the low and high energy spectra.  This model, first published in 
\citet{Ryde} where the logarithmic derivative of the photon flux is a continuous hyperbolic tangent, has been re-parametrized 
\citep{Kaneko06} as:
\begin{equation}
f_{SBPL}(E)=A \biggl(\frac{E}{E_{piv}} \biggr)^b  \ 10^{(a - a_{piv})}
\end{equation}
where
\begin{equation}
	\begin{split}
&a=m\Delta \ln \biggl(\frac{e^q+e^{-q}}{2}\biggr), \quad a_{piv}=m\Delta \ln \biggl(\frac{e^{q_{piv}}+e^{-q_{piv}}}{2} \bigg), \\
&\\
&q=\frac{\log (E/E_b)}{\Delta}, \quad q_{piv}=\frac{\log(E_{piv}/E)}{\Delta},\\
&\\
&m=\frac{\lambda_2-\lambda_1}{2}, \quad b=\frac{\lambda_1+\lambda_2}{2}.
	\end{split}
\end{equation}
In the above relations, the low- and high-energy power law indices are $\lambda_1$ and $\lambda_2$ respectively, $E_b$ is 
the break energy in keV, and $\Delta$ is the break scale given in decades of energy.  The break scale is independent and not 
coupled to the power law indices as it is with the Band function, and as such represents an additional degree of freedom.  
However, \citet{Kaneko06} found that an appropriate value for $\Delta$ for GRB spectra is 0.3, therefore we fix $\Delta$ at this 
value.

\subsection{$\bf Log_{10}$ Gaussian}
The final model that we consider in this catalog is a Gaussian parametrized with logarithmic energies, or GLOGE.  The photon 
model is represented by
\begin{equation}
f_{GLOGE}(E) = \frac{A} { \sqrt{2 \pi} s} \ \exp \biggl[ - \frac{1} {2} \biggl( \frac{ \log_{10} E - \log_{10} E_{cen} } {s} \biggr)
^2\biggr],
\end{equation}
where $E_{cen}$ is the centroid energy in keV and $s$ is the standard deviation at $E_{cen}$ in decades of keV.  This 
model is identical to the log-parabolic function that is common in investigating BL Lac spectra, and has recently been used to 
study GRB spectra \citep{Massaro}.

\section{Data Analysis \& Results}
To study the spectra resulting from the BATSE detectors, a method must be established to associate the energy deposited in the 
detectors to the energy of the detected photons.  This association is dependent on effective area and the angle of the detector to 
the incoming photons.  To do this, detector response matrices (DRMs) are used to convert the photon energies into detector 
channel energies.  The DRM is a mathematical model of the deposition of photon energy in the crystal---a photon that interacts by 
the photoelectric effect will deposit 100\% of its energy, subject to resolution broadening, while a photon that interacts by a single 
Compton scatter may deposit only a portion of its energy.  The exception to this is when a photon carrying the energy of the iodine 
K--shell escapes the crystal.  The energy calibration determines the energy boundaries of the energy deposition channels.  The 
response matrices for all GRBs in the catalog were made using DRM\_LAD\_GEN v2.07 of the response generator for BATSE~\cite{Pendleton}.  

The spectral analysis of all bursts was performed using RMfit, version 3.4rc1.  RMfit employs a modified, forward-folding 
Levenberg-Marquardt algorithm for spectral fitting.  The Castor C-Statistic, which is a modified log-likelihood statistic 
based on the Cash parametrization \citep{Cash} is used in the model-fitting process as a figure of merit to be minimized.  This 
statistic is preferable over the more traditional $\chi^2$ statistic minimization because of the non-Gaussian counting statistics 
present when studying dim GRBs.  Although it is advantageous to perform the spectral fitting using C-Stat, the statistic provides no 
estimation of the goodness-of-fit, since there exists no standard probability distribution for likelihood statistics.  For this reason, we 
also calculate $\chi^2$ for each spectral fit performed by minimizing C-Stat.  This allows an estimation of the goodness-of-fit of a 
function to the data even though $\chi^2$ was not minimized (see Appendix C for simulations).  This also allows for easy comparison 
between nested models.

\begin{figure}
	\begin{center}
		\includegraphics[scale=0.7]{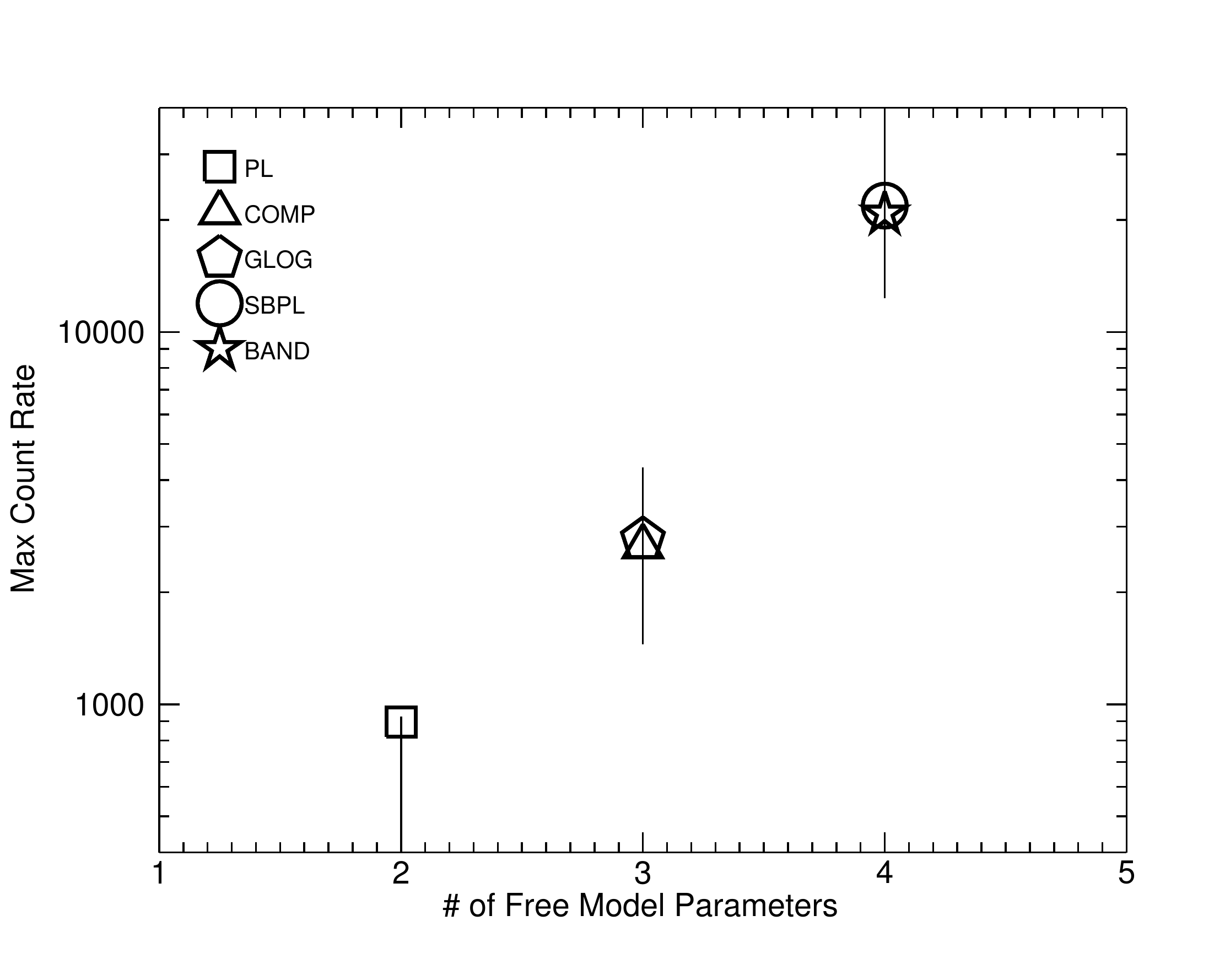}
	\end{center}
\caption{Plot of the average maximum background-subtracted count rates (as a proxy for observed intensity) versus the number 
of degrees of freedom of the best fit model. The maximum count rate is defined as the single 2048 ms bin that contains the 
highest background-subtracted count rate.  The best fit model was determined for each burst, and a geometric average was 
calculated for the maximum count rates of the bursts for each best fit model.  The error bars shown are the 1$\sigma$ standard 
deviations of the distributions of maximum count rates for each best fit model.  \label{countrate}}
\end{figure}

We fit the five functions described in Section 4 to the spectra of each burst.  The BAND and COMP functions are parametrized 
with $E_{peak}$, the peak in the power density spectrum, while the SBPL is parametrized with the break energy, $E_{break}$, 
and the GLOGE model is parametrized with the centroid energy $E_{cen}$.  We choose to fit these five different functions 
because the measurable spectrum of GRBs is dependent on intensity, as is shown in Figure \ref{countrate}.  Observably less 
intense bursts provide less data to support a large number of parameters.  This allows us to determine why, in many situations, a 
particular empirical function provides a poor fit, while in other cases it provides an accurate fit.    For example, the energy spectra of 
GRBs are normally well fit by two smoothly joined power laws.  For particularly bright GRBs, the BAND and SBPL functions are 
typically an accurate description of the spectrum, while for weaker bursts the COMP or GLOGE function is most acceptable.  Bursts 
that have signal significance on the order of the background fluctuations do not have a detectable distinctive break in their spectrum 
and so the power law is the most acceptable function.  Although weaker GRBs do not statistically prefer a model with more 
parameters, it is instructive to study the parameters of even the weaker bursts.  In addition, the actual physical GRB processes can 
have an effect on the spectra and different empirical models may fit certain bursts better than others.  The spectral results, including 
the best fit spectral parameters and the photon model, are stored in files following the FITS standard similar to those described in the 
Appendix of \citet{GBMSpecCat} and are hosted as a public data archive on HEASARC (\url{http://heasarc.gsfc.nasa.gov/W3Browse/cgro/bat5bgrbsp.html}).  Many of the important spectral 
quantities are also available in tables available in the electronic version of this catalog following the formats listed in Tables~\ref
{BANDparms}-\ref{SBPLparms}.

When inspecting the distribution of the parameters for the fitted models, we first define a data cut based on the goodness-of-fit.  
We require that the $\chi^2$ statistic for the fit to be within the  3$\sigma$ expected region for the $\chi^2$ distribution of the given 
degrees of freedom, and we define a subset of each parameter distribution of this data cut as GOOD if the error is within certain 
limits.  Following \citet{Kaneko06}, for the low-energy power law indices, we consider GOOD values to have errors less than 0.4, 
and for high-energy power law indices we consider GOOD values to have errors less than 1.0.  For all other parameters we 
consider a GOOD value to have a relative error of 0.4 or better.  The motivation for this is to show well-constrained parameter 
values, rather than basing possible interpretations on parameters that are poorly constrained.

In addition, we define a BEST sample where we compare the goodness-of-fit of all spectral models for each burst and select 
the most preferred model based on the difference in $\chi^2$ per degree of freedom.  The criterion for accepting a model with 
a single additional parameter is a change in $\chi^2$ of at least 6 since the probability for achieving this difference is $\sim
$0.01.  The parameter distributions are then populated with the spectral parameters from the BEST spectral fits.

\subsection{Fluence Spectra}
\begin{figure}
	\begin{center}
		\subfigure[]{\label{indexlof}\includegraphics[scale=0.35]{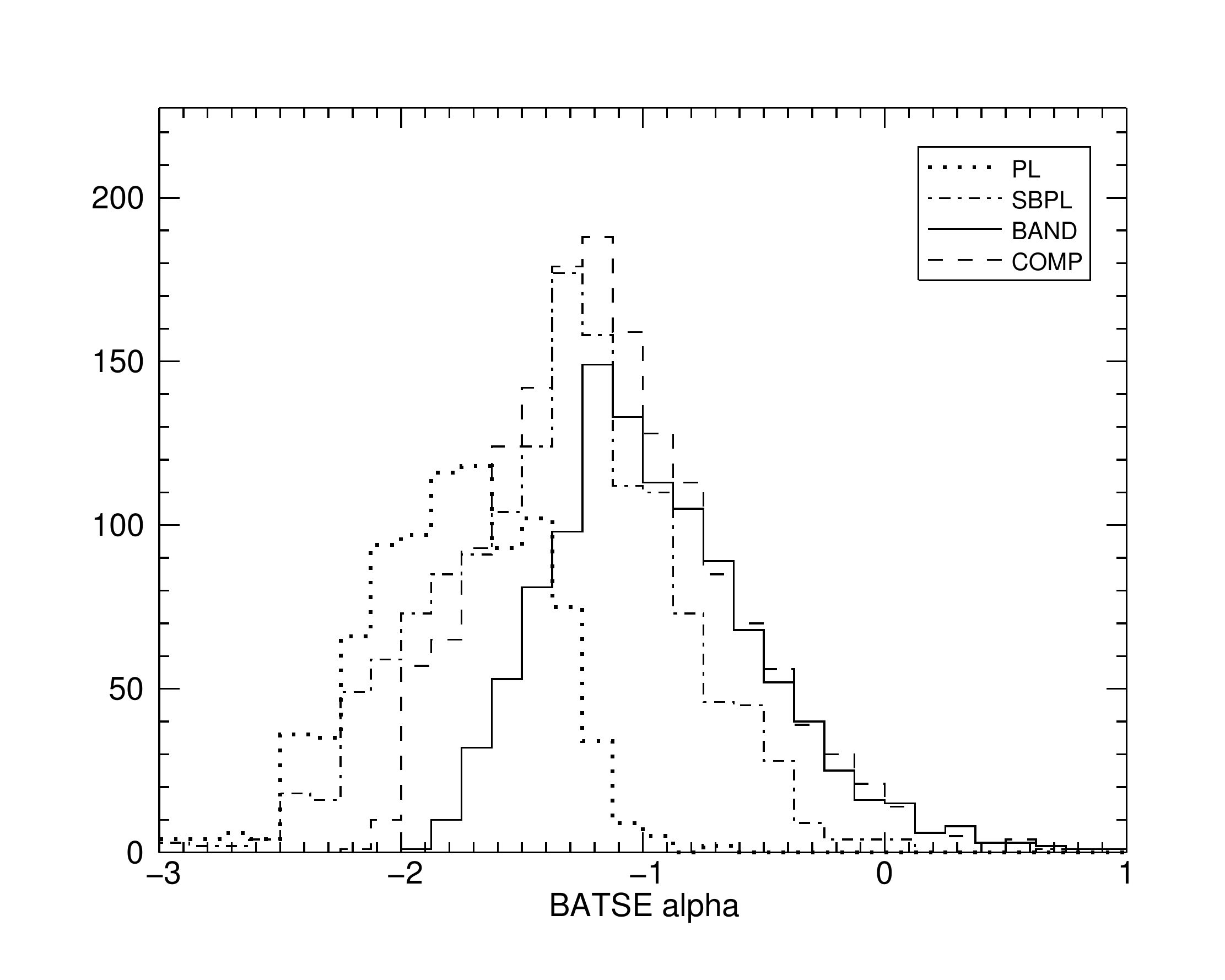}}
		\subfigure[]{\label{alphasbplf}\includegraphics[scale=0.35]{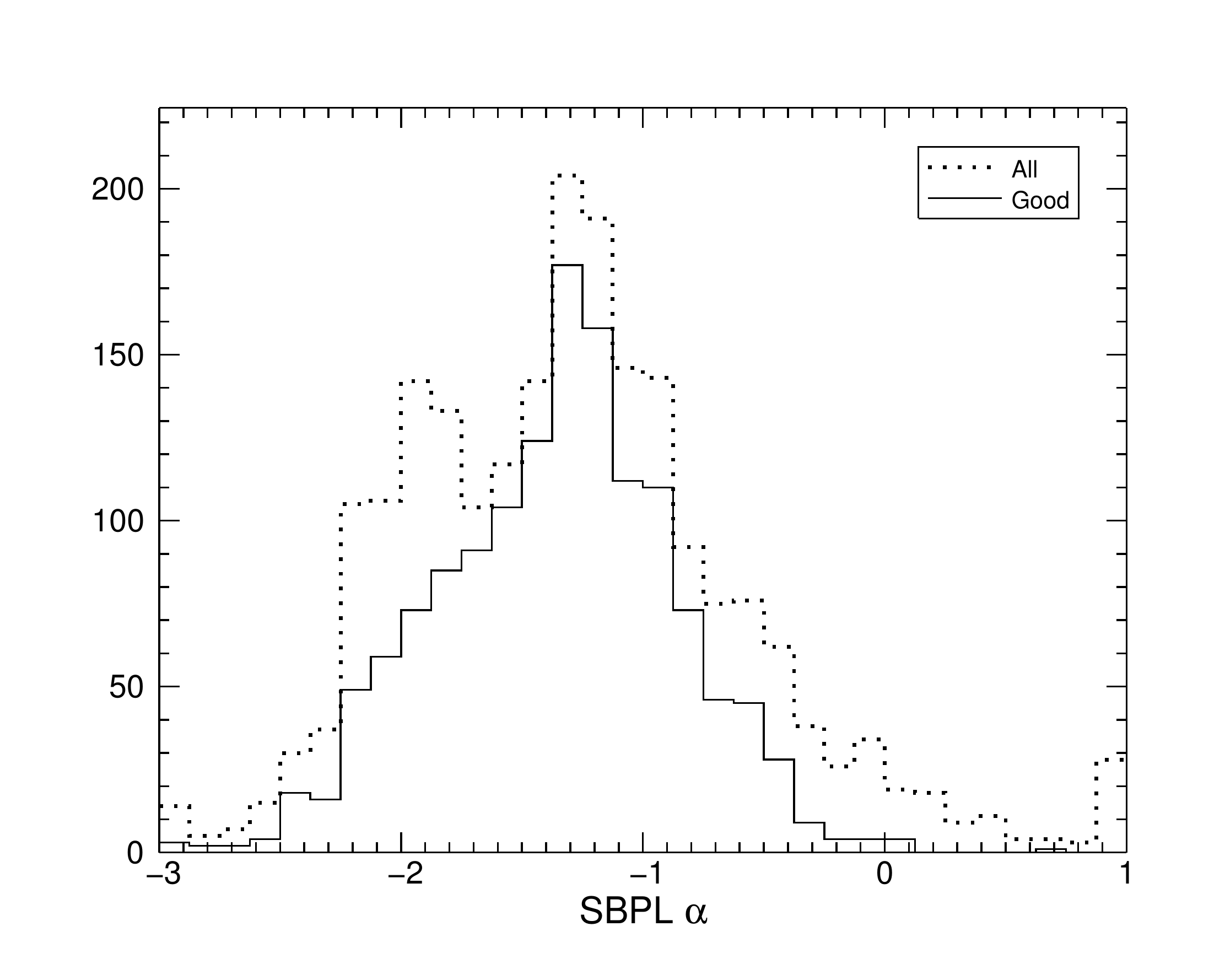}}\\
		\subfigure[]{\label{alphabandf}\includegraphics[scale=0.35]{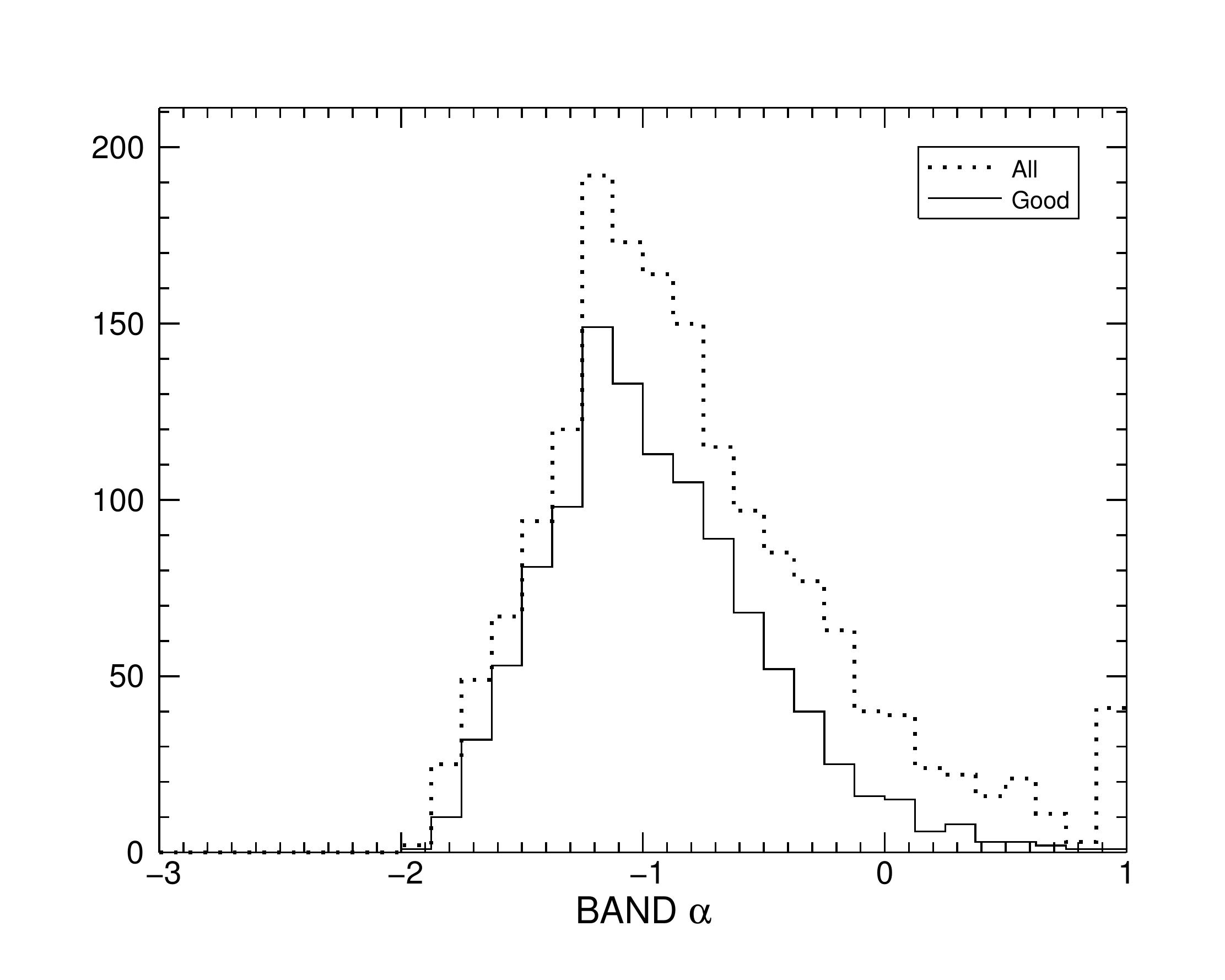}}
		\subfigure[]{\label{alphacompf}\includegraphics[scale=0.35]{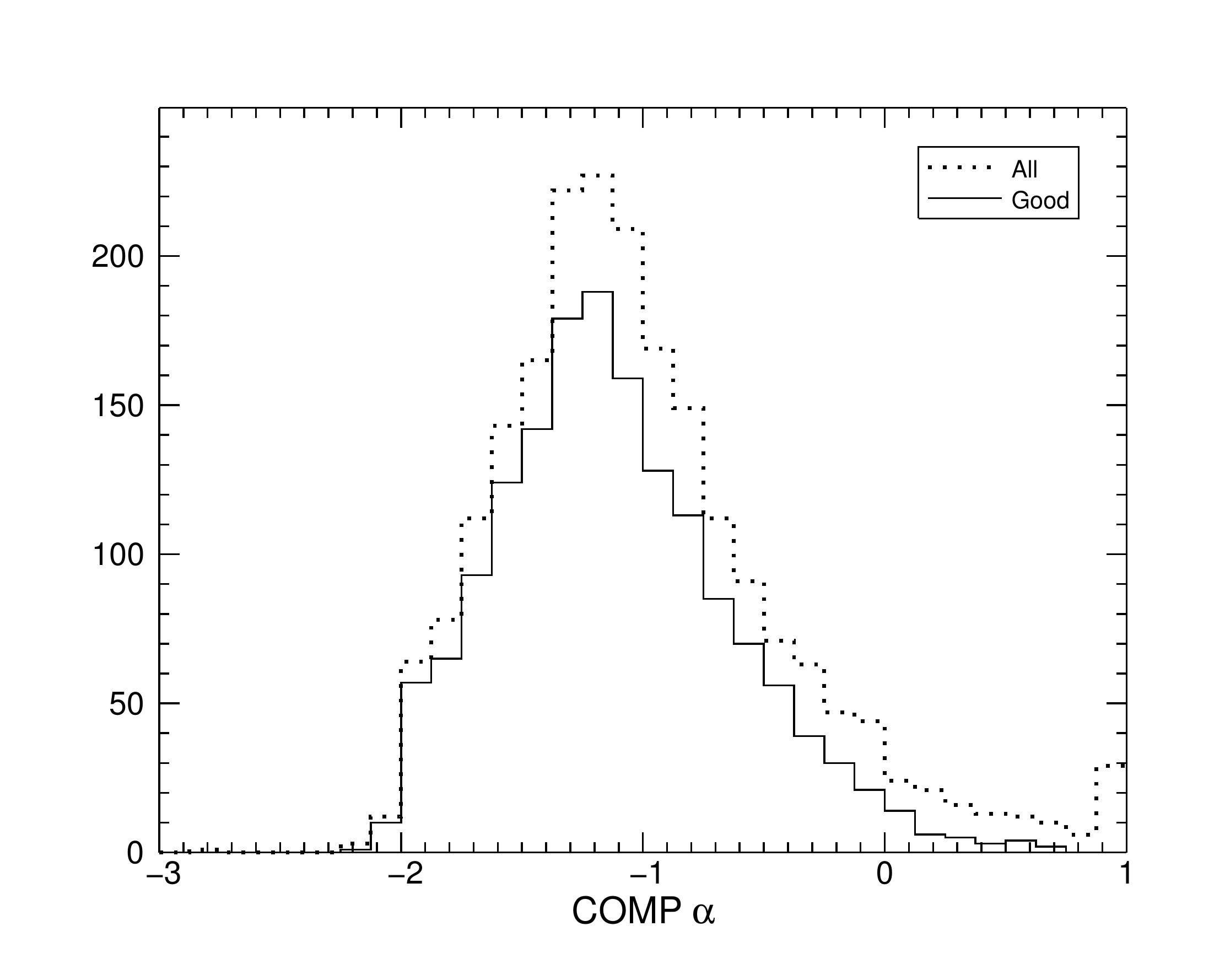}}
	\end{center}
\caption{Distributions of the low-energy spectral indices from fluence spectral fits.  \ref{indexlof} shows the distributions of 
GOOD parameters and compares to the distribution of PL indices.  \ref{alphasbplf}--\ref{alphacompf} display the comparison 
between the distribution of GOOD parameters and all parameters with no data cuts.  The last bin includes values greater than 
1. \label{loindexf}}
\end{figure}

\begin{figure}
	\begin{center}
		\subfigure[]{\label{indexhif}\includegraphics[scale=0.35]{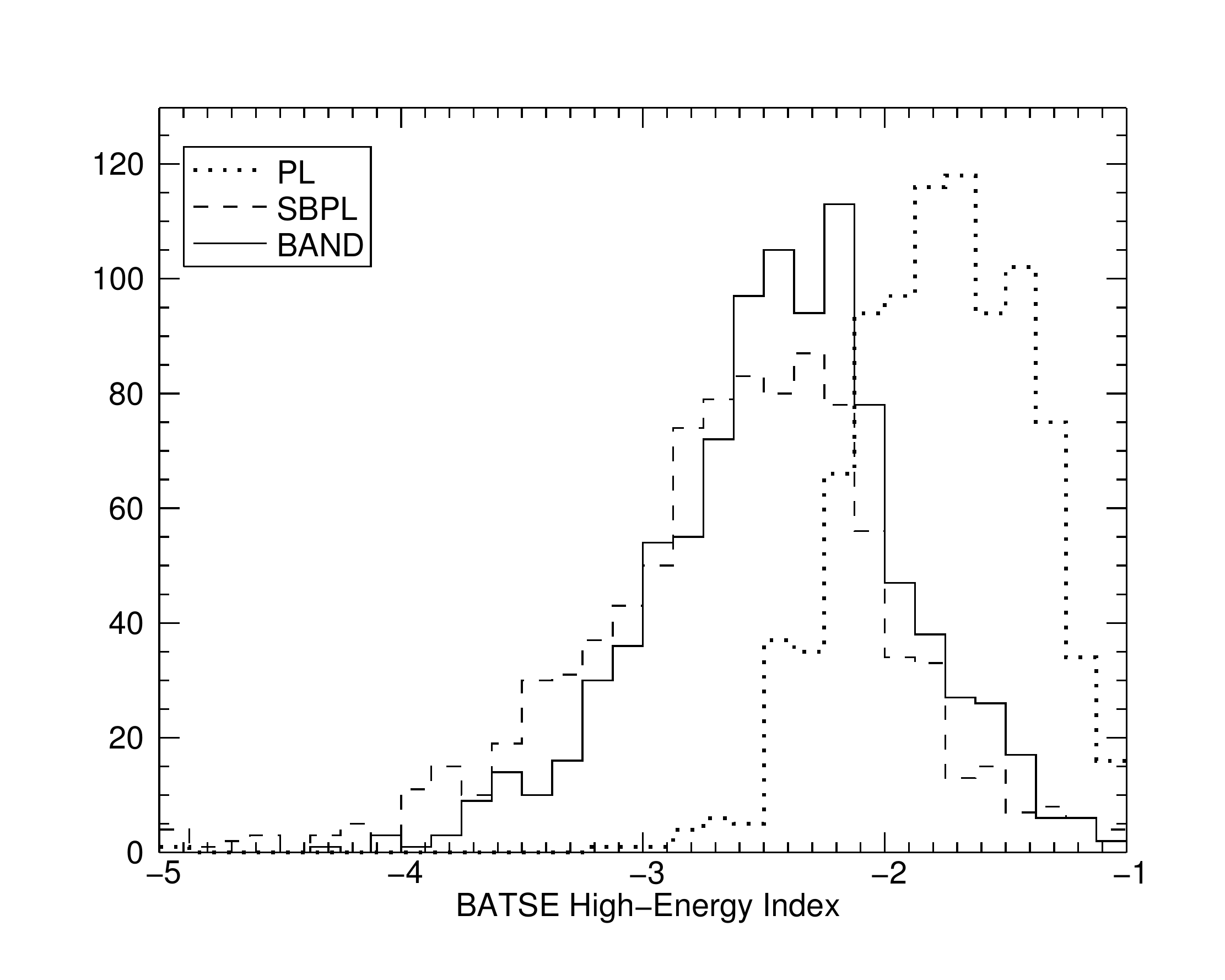}}
		\subfigure[]{\label{betasbplf}\includegraphics[scale=0.35]{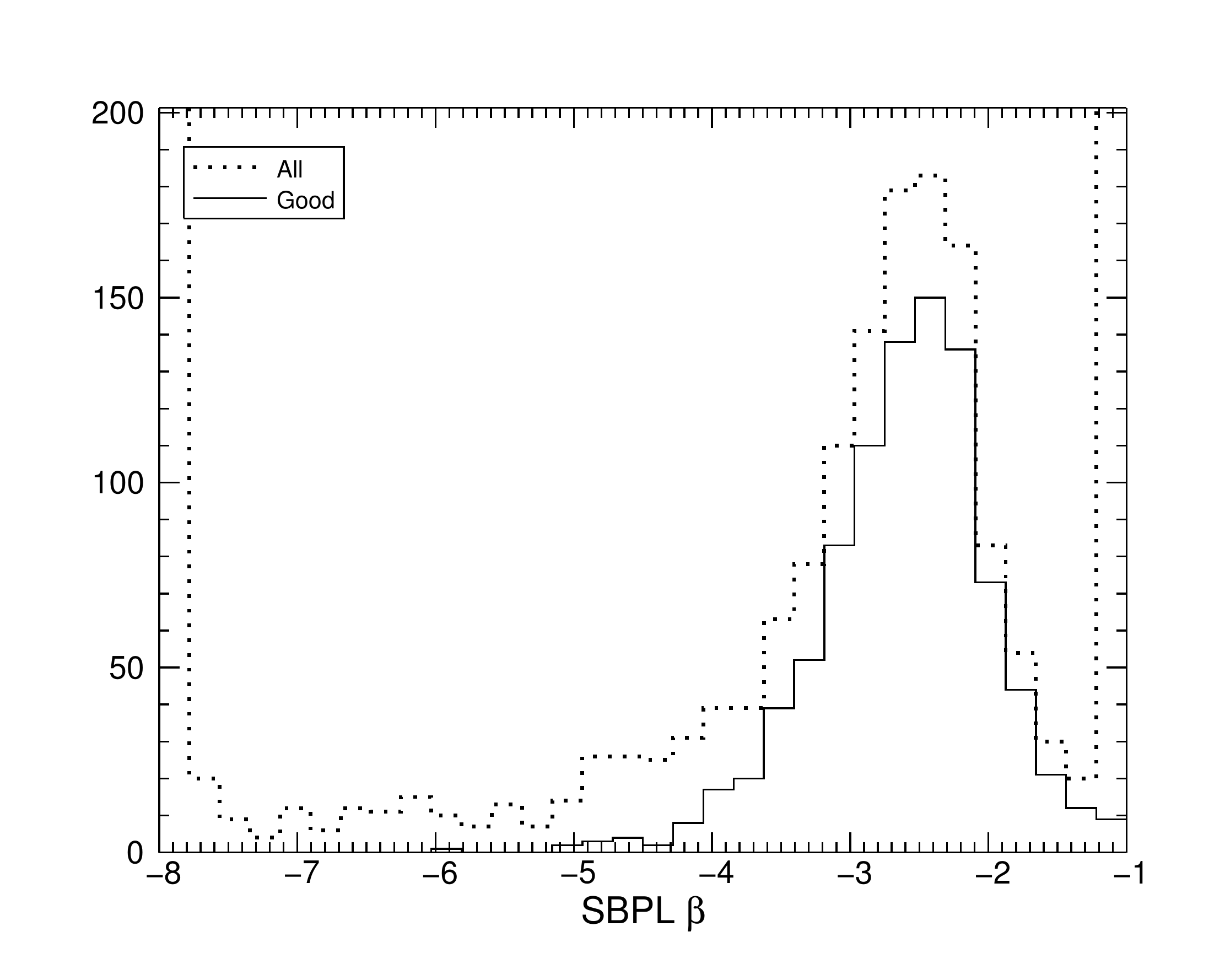}}\\
		\subfigure[]{\label{betabandf}\includegraphics[scale=0.35]{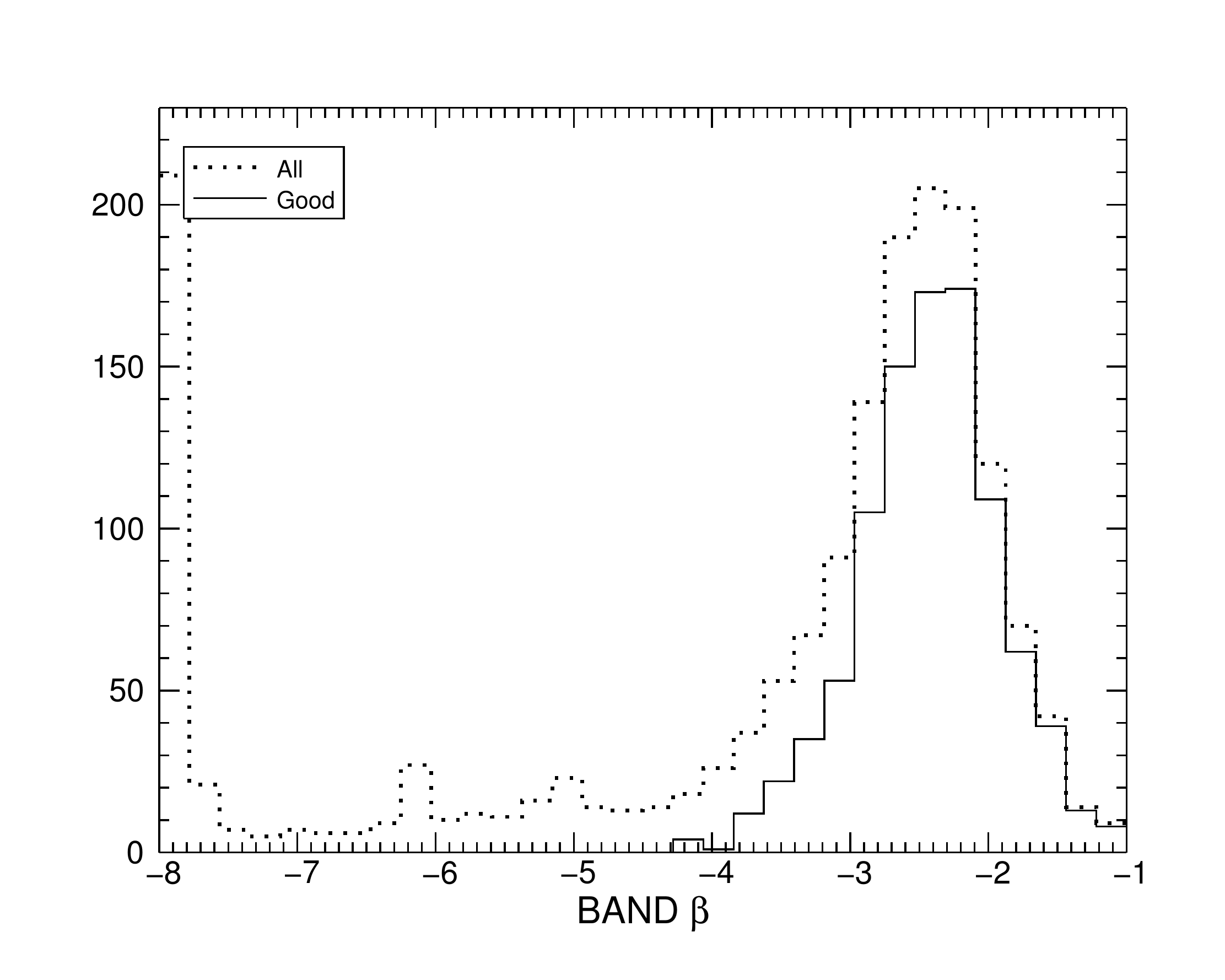}}
		\subfigure[]{\label{deltasf}\includegraphics[scale=0.35]{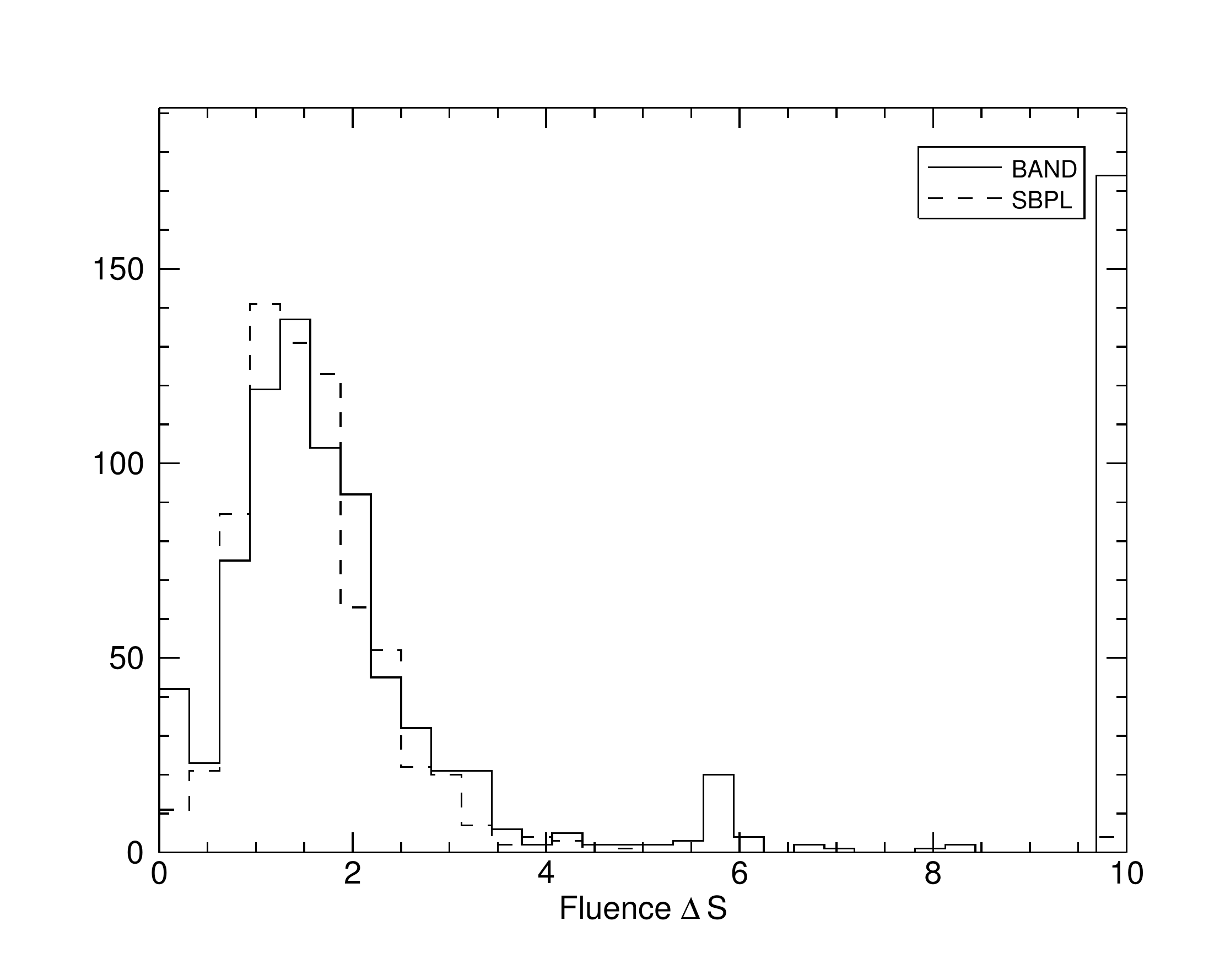}}
	\end{center}
\caption{\ref{indexhif} - \ref{betabandf} are distributions of the high-energy spectral indices from fluence spectral fits.  \ref
{indexhif} shows the distributions of GOOD parameters and compares to the distribution of PL indices.  \ref{betasbplf} and \ref
{betabandf}  display the comparison between the distribution of GOOD parameters and all parameters with no data cuts.  The 
first bins include values less than -8 and the last bin include values greater than -1.  \ref{deltasf} shows the distribution of the 
difference between the low- and high-energy indices. The first bin contains values less than 0, indicating that the centroid value 
of alpha is steeper than the centroid value of beta. \label{hiindexf}}
\end{figure}

The time-integrated fluence distributions are estimated over the duration of the observed emission, where the observed emission is 
defined as 3.5$\sigma$ over the estimated background in the 20-2000 keV energy range.  It should be noted that the following 
distributions do not take into account any spectral evolution that may exist within bursts.  The low-energy indices, as shown in Figure 
\ref{loindexf}, distribute about a $-1$ power law typical of most GRBs.  Accounting for parameter uncertainty, up to 14\% of the GOOD 
low-energy indices violate the $-2/3$ synchrotron ``line-of-death'' \citep{Preece}, while an additional 57\% of the indices violate the 
$-3/2$ synchrotron cooling limit.  The high-energy indices in Figure \ref{hiindexf} peak at a slope slightly steeper than $-2$ and have 
a long tail toward steeper indices.  Note that the large number of unconstrained (or very steep) high-energy indices in the distribution 
of all high-energy index values indicates that a large number of GRBs are  better fit by the COMP model, which is equivalent to a 
BAND function with a high-energy index of $- \infty$.  The comparison of the simple power law index to the low- and high-energy 
indices makes evident that the simple power law index is averaged over the break energy, resulting in a index that is on average 
steeper than the low-energy index yet shallower than the high-energy index.  We also show in Figure \ref{deltasf} the difference 
between the time-integrated low- and high-energy spectral indices, $\Delta S =(\alpha-\beta)$.  This quantity is useful since the 
synchrotron shock model makes predictions of this value in a number of cases \citep{Preece}.  The peak of this distribution is at $
\sim$1.5, which indicates a peak electron energy spectral index of $\sim$4 assuming the traditional synchrotron shock model~\citep
{Sari96}.  Note that the large overflow of $\Delta S$ at $\sim$10 for the BAND model is due to the extremely steep slope of $\beta$, 
where the high-energy spectrum approximates an exponential cut-off. 

\begin{figure}
	\begin{center}
		\subfigure[]{\label{ecentf}\includegraphics[scale=0.35]{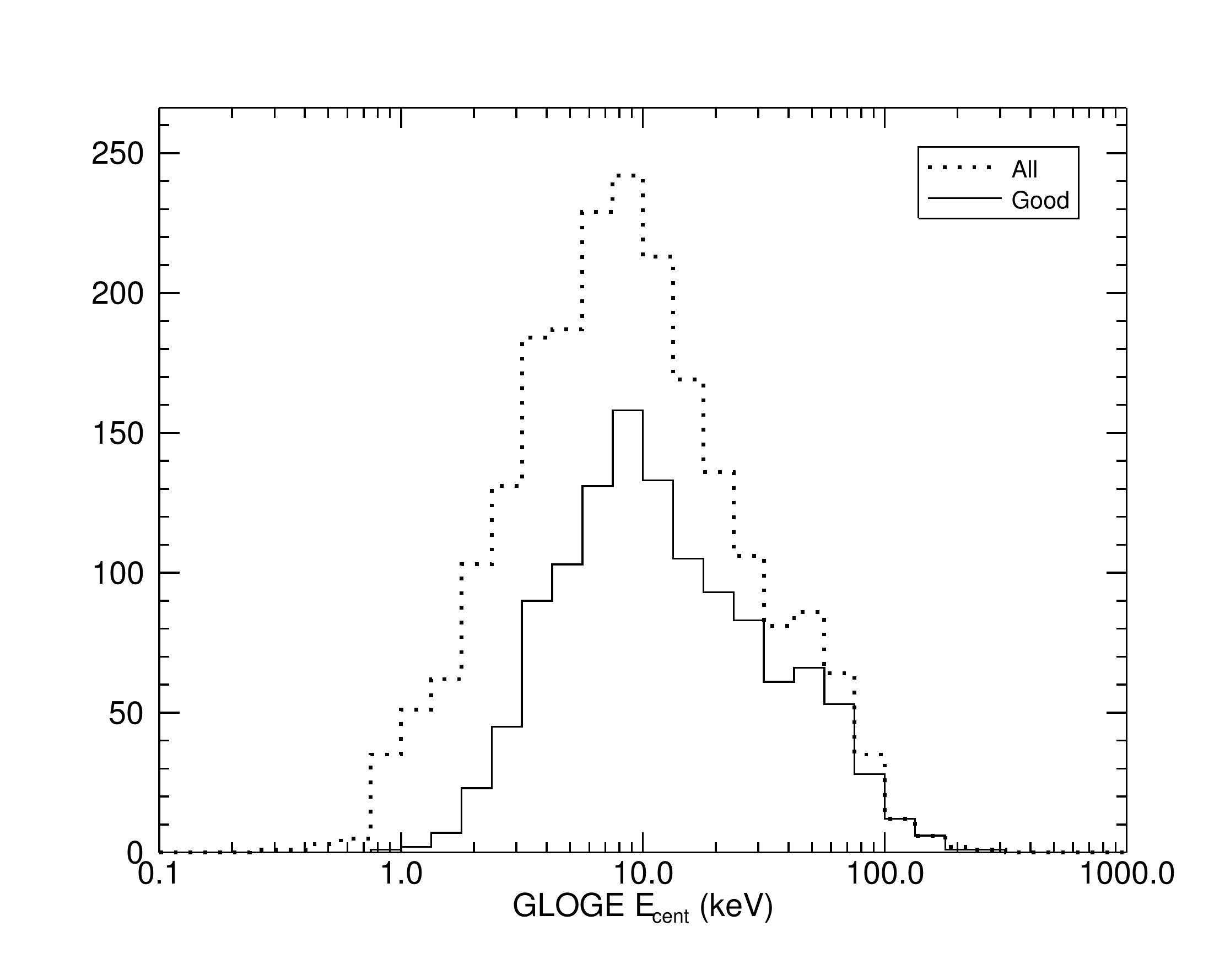}}
		\subfigure[]{\label{fwhmf}\includegraphics[scale=0.35]{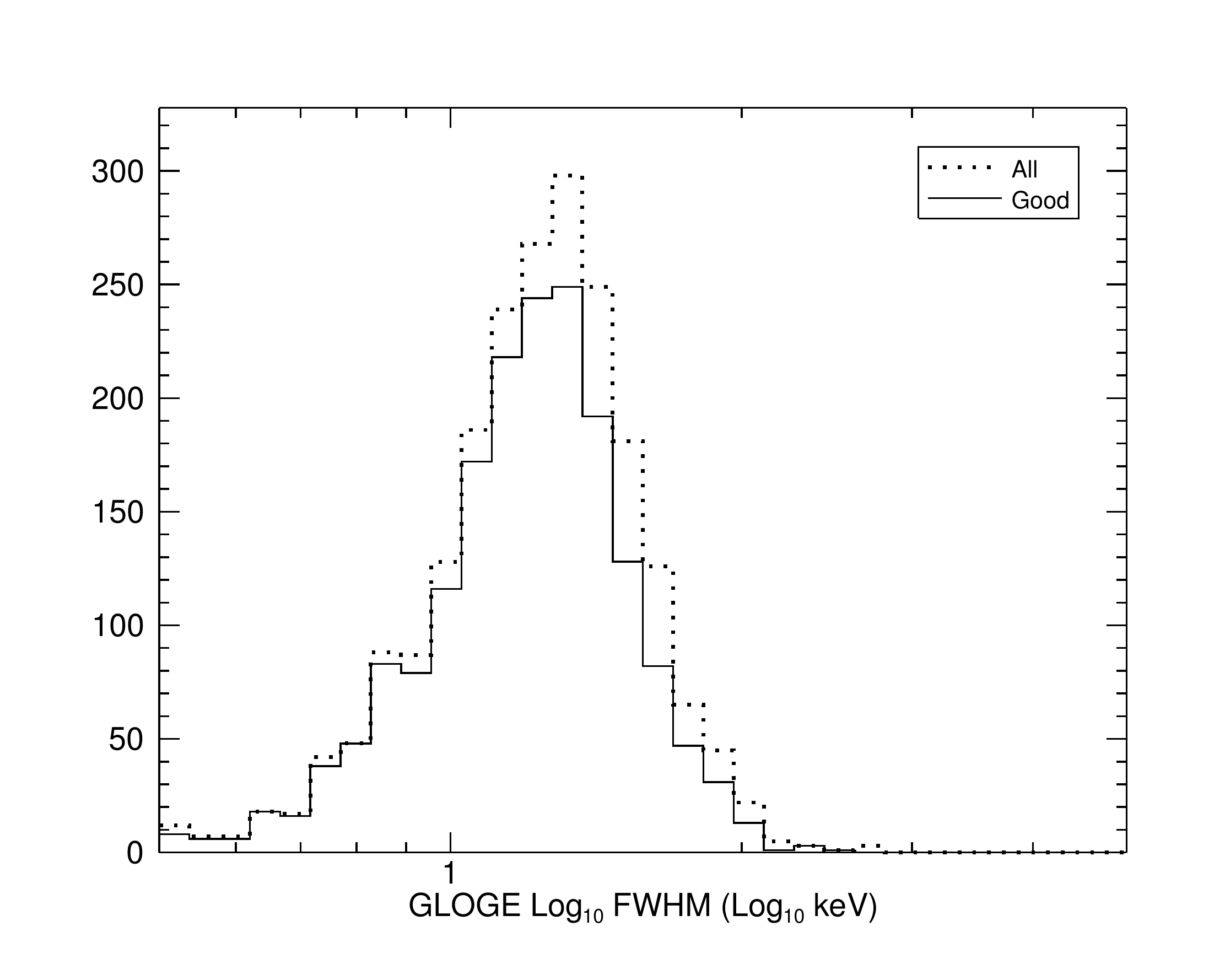}}
	\end{center}
\caption{Distributions of the GLOGE $E_{cent}$ and FWHM parameters from the fluence spectral fits.  \label{GLOGEf}}
\end{figure}

\begin{figure}
	\begin{center}
		\includegraphics[scale=0.7]{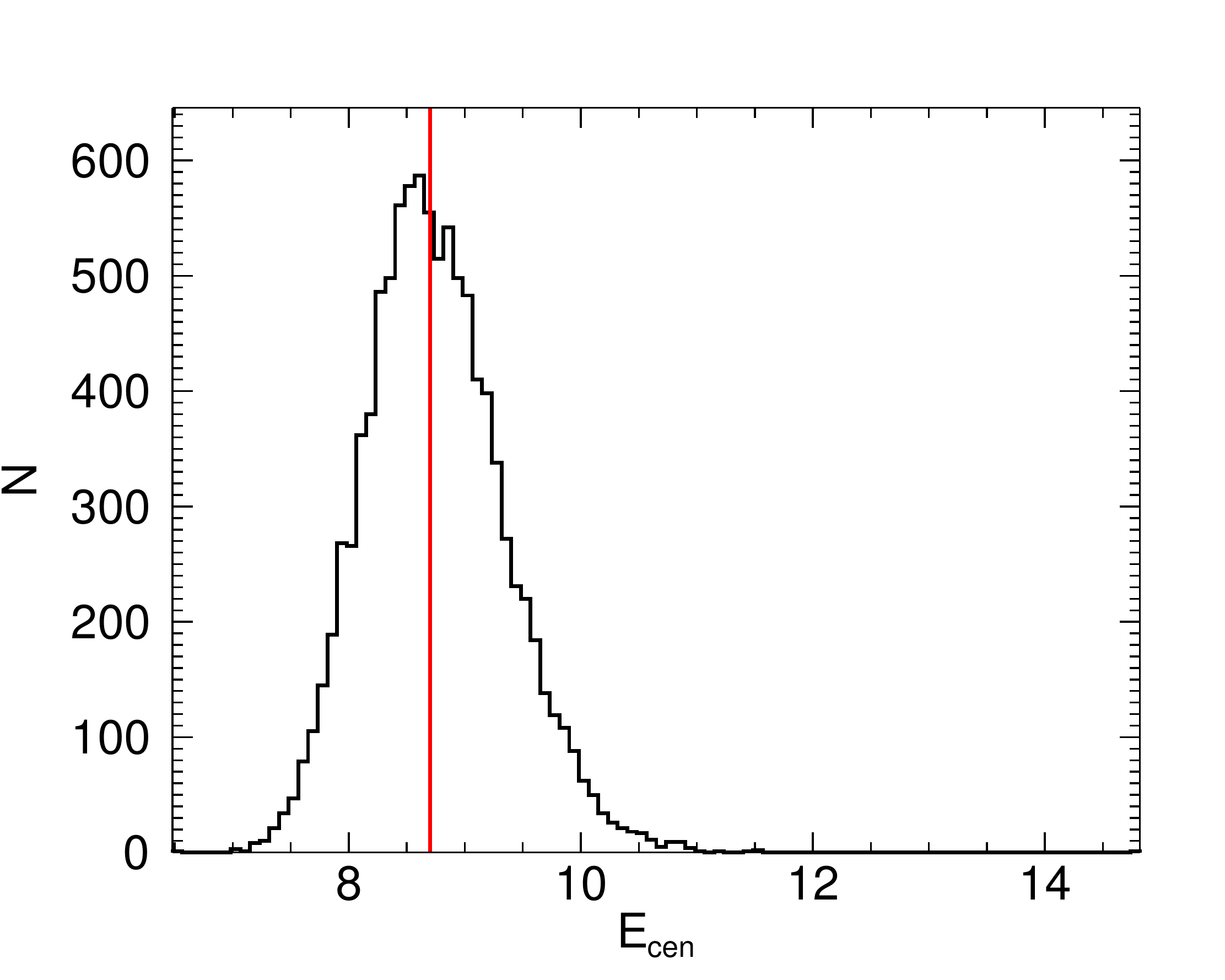}
	\end{center}
\caption{Distribution of GLOGE $E_{cen}$ from fitting synthetic spectra with input $E_{cen} = 8.7$ keV.  The red vertical line indicates 
the input $E_{cen}$ value. \label{EcenSim}}
\end{figure}

Shown in  Figure \ref{GLOGEf} are the distributions for the centroid energy, $E_{cen}$ in keV, and the full-width at half maximum, 
FWHM in log keV, of the GLOGE function.  The $E_{cen}$ distribution peaks at $\sim$10 keV and covers two orders of 
magnitude while the FWHM peaks at $\sim$1.22 and covers less than one order of magnitude.  Note that the curvature of the 
GLOGE function can be well constrained by the data even when the centroid of the GLOGE fit is below the BATSE bandpass.  As an 
example of this behavior, $10^4$ synthetic GLOGE spectra were generated with $E_{cen} = 8.7$ keV, the spectra were folded 
through a CONT DRM, and Poisson fluctuations were added to the total observed counts as well as the background counts. Figure~
\ref{EcenSim} shows the resulting distribution of $E_{cen}$ from performing spectral fits on the synthetic spectra.  As can be seen, the 
data has no problem in constraining a centroid value below 20 keV.

\begin{figure}
	\begin{center}
		\subfigure[]{\label{ebreaksbplf}\includegraphics[scale=0.35]{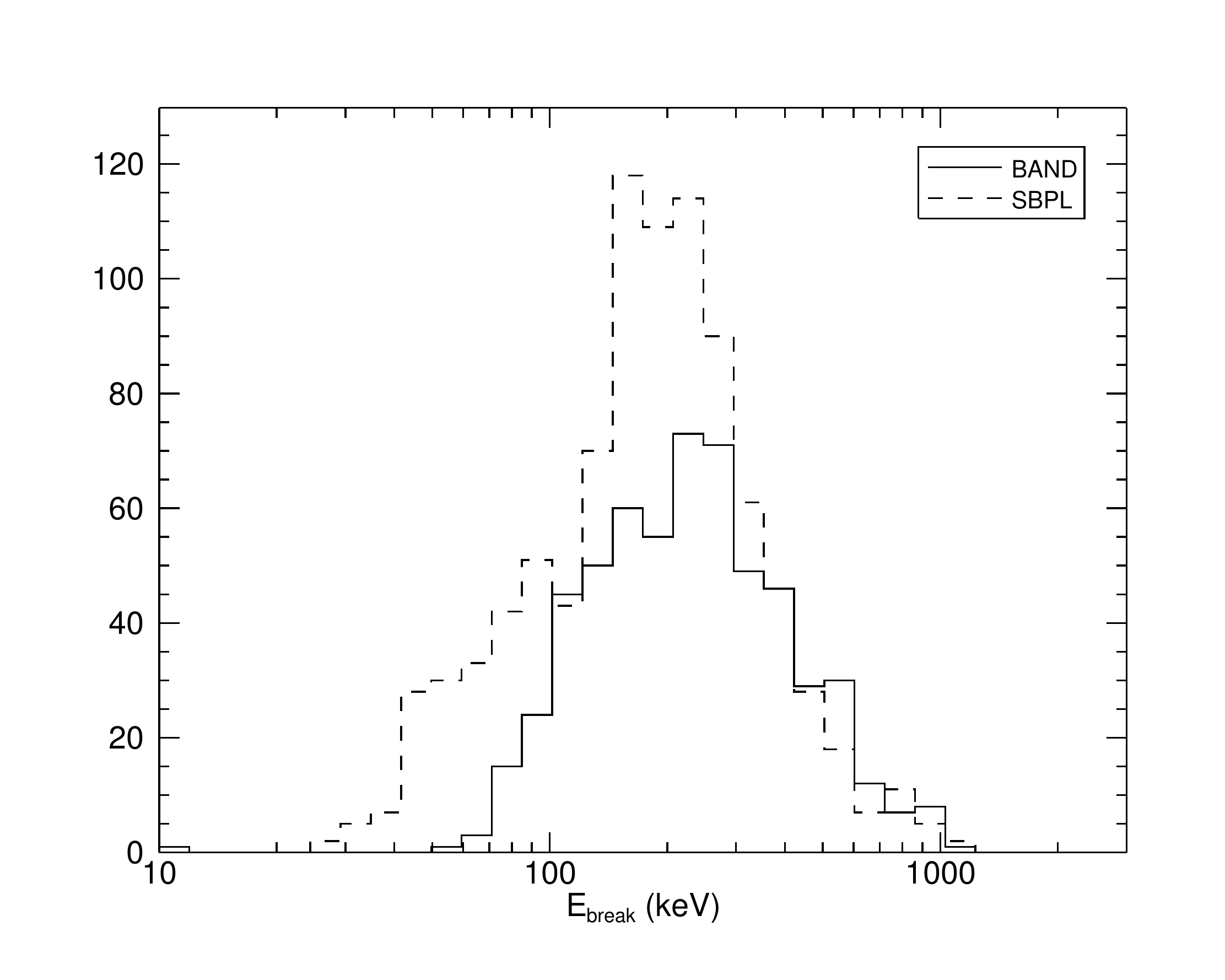}}
		\subfigure[]{\label{epeakf}\includegraphics[scale=0.35]{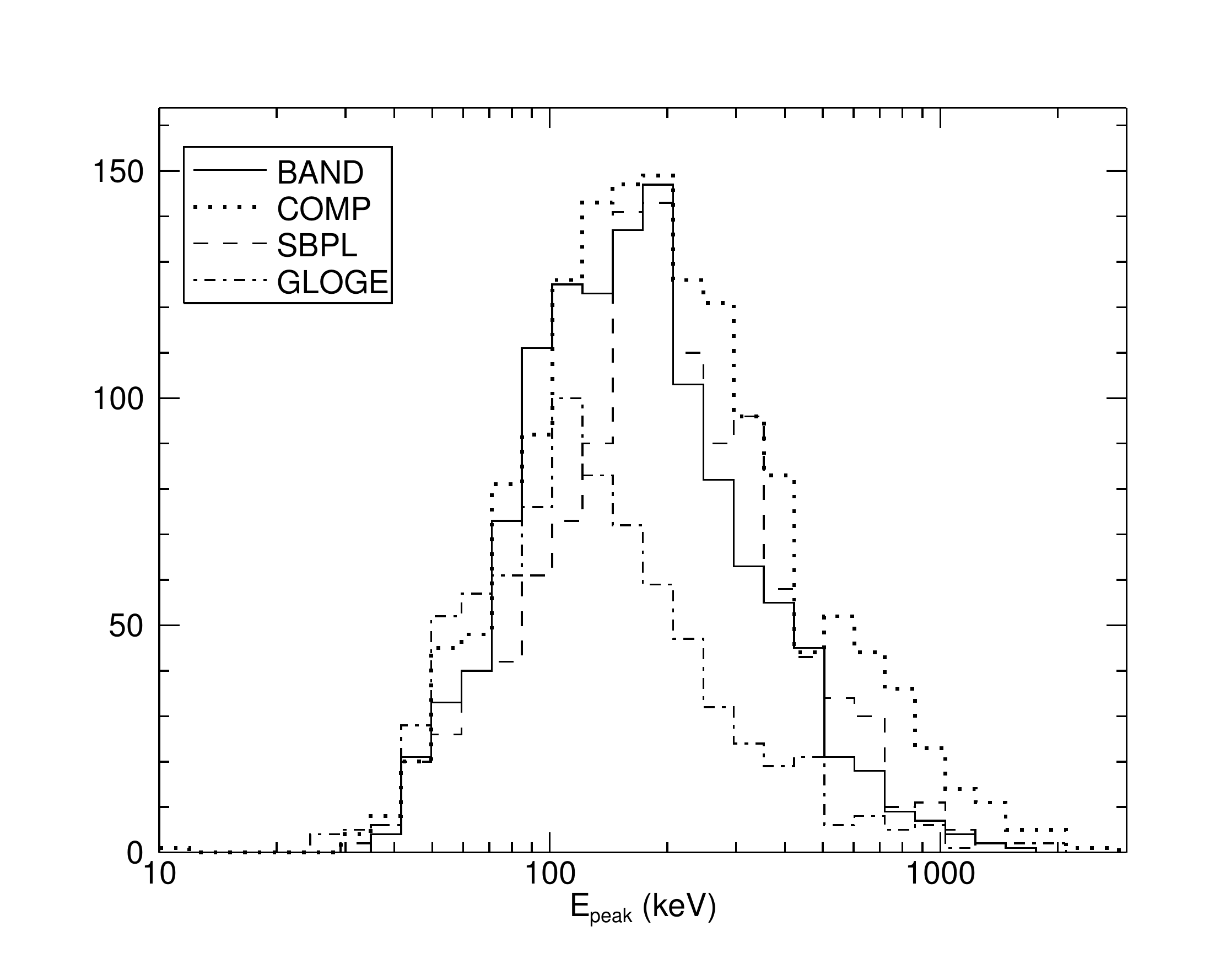}}\\
		\subfigure[]{\label{epeakbandf}\includegraphics[scale=0.35]{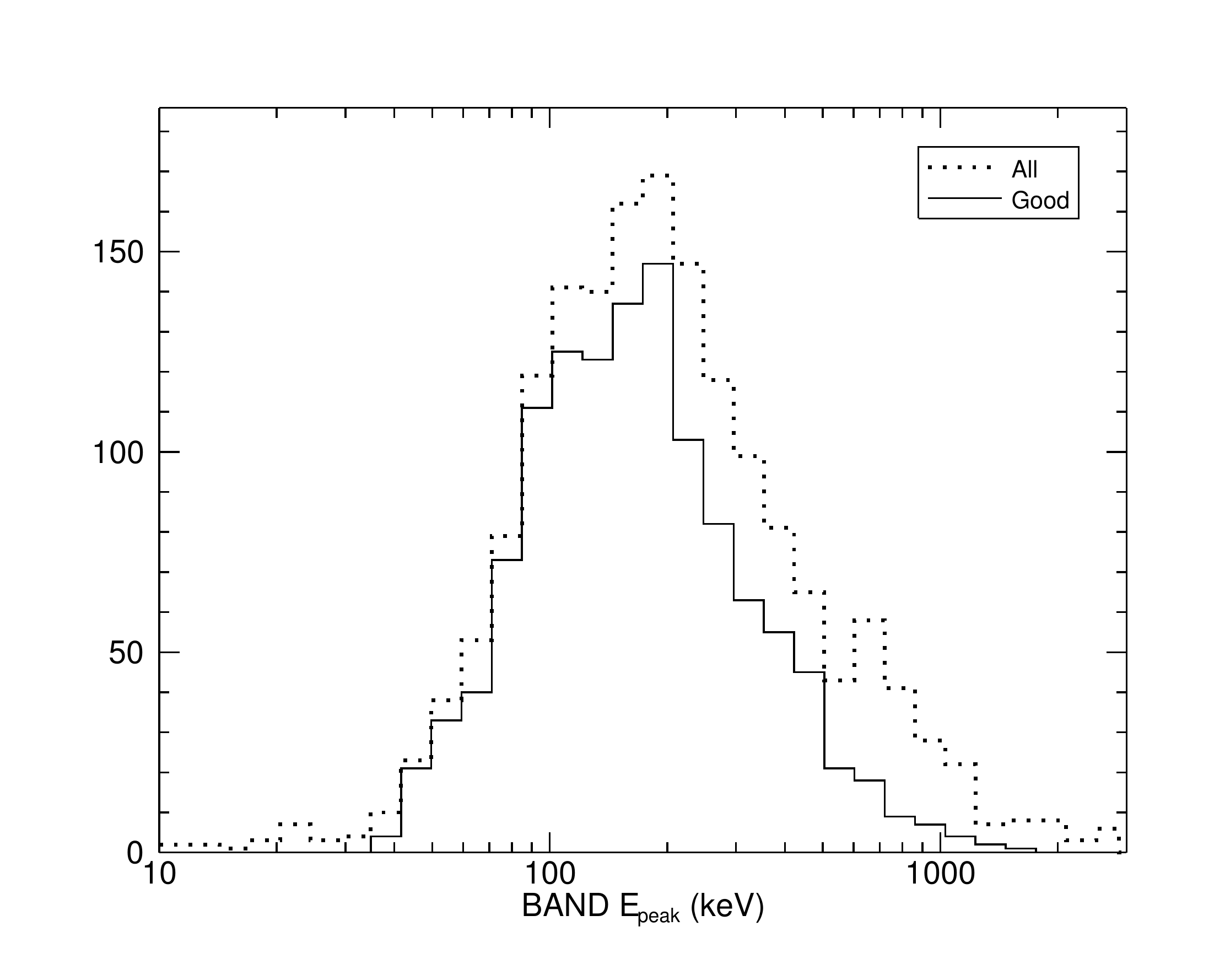}}
		\subfigure[]{\label{epeakcompf}\includegraphics[scale=0.35]{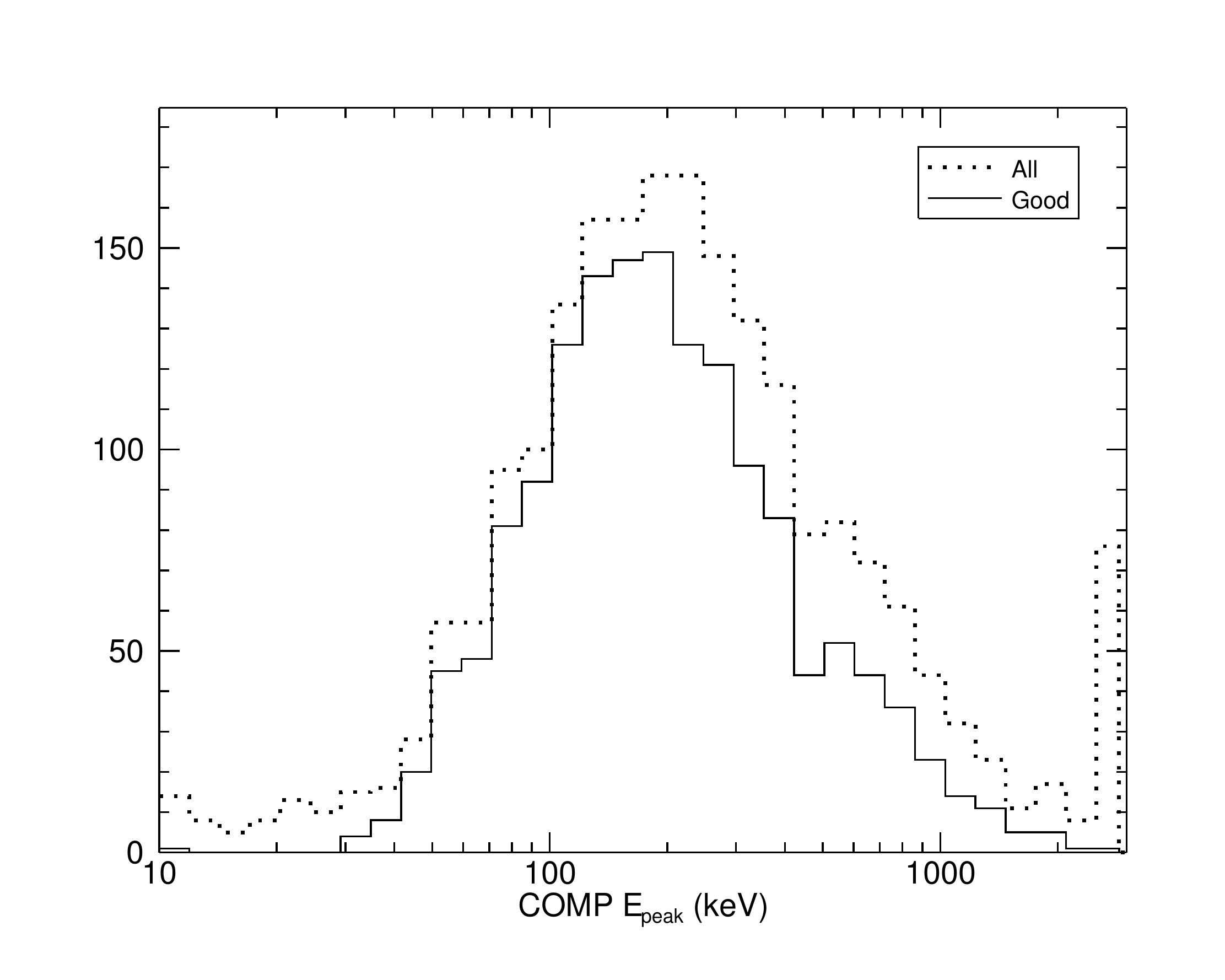}}
	\end{center}
\caption{Distributions of $E_{break}$ and $E_{peak}$ from fluence spectral fits.  \ref{ebreaksbplf} displays the comparison 
between the distribution of GOOD $E_{break}$ and $E_{break}$ with no data cuts.  \ref{epeakf} shows the distributions of 
GOOD $E_{peak}$ for BAND, SBPL, COMP, and GLOGE.  \ref{epeakbandf} and \ref{epeakcompf} display the comparison between the distribution of GOOD paramters and all parameters with no data cuts. \label{epeakebreakf}}
\end{figure}

\begin{figure}
	\begin{center}
		\includegraphics[scale=0.45]{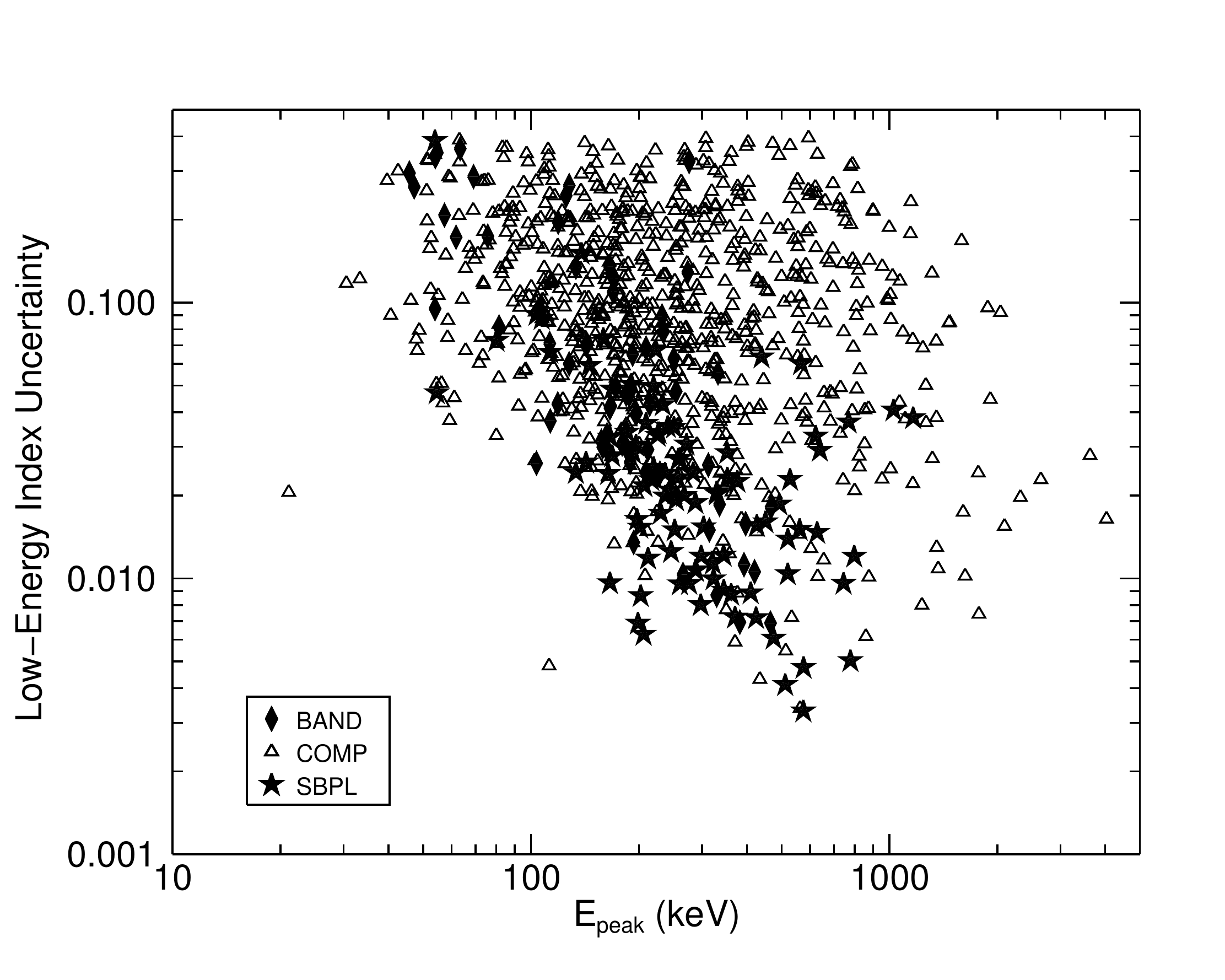}
	\end{center}
\caption{Comparison of the low-energy index uncertainty and $E_{peak}$ for three models from the fluence spectral fits.  This 
comparison reveals a correlation between the $E_{peak}$ energy and the uncertainty on the low-energy index: generally a  lower 
energy $E_{peak}$ tends to result in a less constrained low-energy index. \label{alphaepeak}}
\end{figure}

In Figure \ref{epeakebreakf}, we show the distributions for the break energy, $E_{break}$ and the peak of the power density 
spectrum, $E_{peak}$.  $E_{break}$ is the energy at which the low- and high-energy power laws are joined, which is not necessarily 
representative of the $E_{peak}$.  As discussed in \citet{Kaneko06}, although the SBPL is parametrized with $E_{break}$, the $E_
{peak}$ can be derived from the functional form.  Note that we have calculated the $E_{peak}$ for all bursts with low-energy index 
shallower than -2 and high-energy index steeper than -2, and we have used the covariance matrix to formally propagate and 
calculate the errors on the derived $E_{peak}$. Similarly, we show in Appendix B, the method to derive $E_{peak}$ from the GLOGE 
model.  The $E_{break}$ for the SBPL peaks near 200 keV, but has a significant tail towards lower energies.  Comparatively, the 
peak for the BAND distribution is also near 200 keV, but less broad and there is no apparent evidence for a low-energy tail.  The $E_
{peak}$ distributions for all functions generally peak around 200 keV, except the derived $E_{peak}$ for GLOGE, which peaks 
closer to 100 keV.  Note that the lower $E_{peak}$ resulting from GLOGE spectral fits are likely due to the $E_{cen}$ peaking 
below the BATSE bandpass.  All $E_{peak}$ distribution cover about two orders of magnitude, which is consistent with previous 
findings \citep{Mallozzi, Lloyd00}.  The peak and overall distribution of $E_{peak}$ is similar to that found by bursts observed by the 
GBM NaI and BGO detectors \citep{GBMSpecCat}, which had a much smaller collecting area but larger bandwidth.  This would seem 
to indicate that it is unlikely for there to be a hidden population undiscovered by either instrument within the $\sim$20 keV -- 2 MeV 
range.  Additionally, the value of $E_{peak}$ can strongly affect the measurement of the low-energy index of the spectrum, as 
shown in Figure \ref{alphaepeak}.  A general trend appears to show that lower $E_{peak}$ values tend to increase the uncertainty 
in the measurement of the low-energy index, mostly due to the fact that a spectrum with a low $E_{peak}$ will exhibit most of its 
curvature near the lower end of the instrument bandpass.  In many cases, if the low-energy index is found to be reasonably steep 
($\lesssim -1$), the uncertainty of the index is minimized even if $E_{peak}$ is low.

\begin{figure}
	\begin{center}
		\includegraphics[scale=0.45]{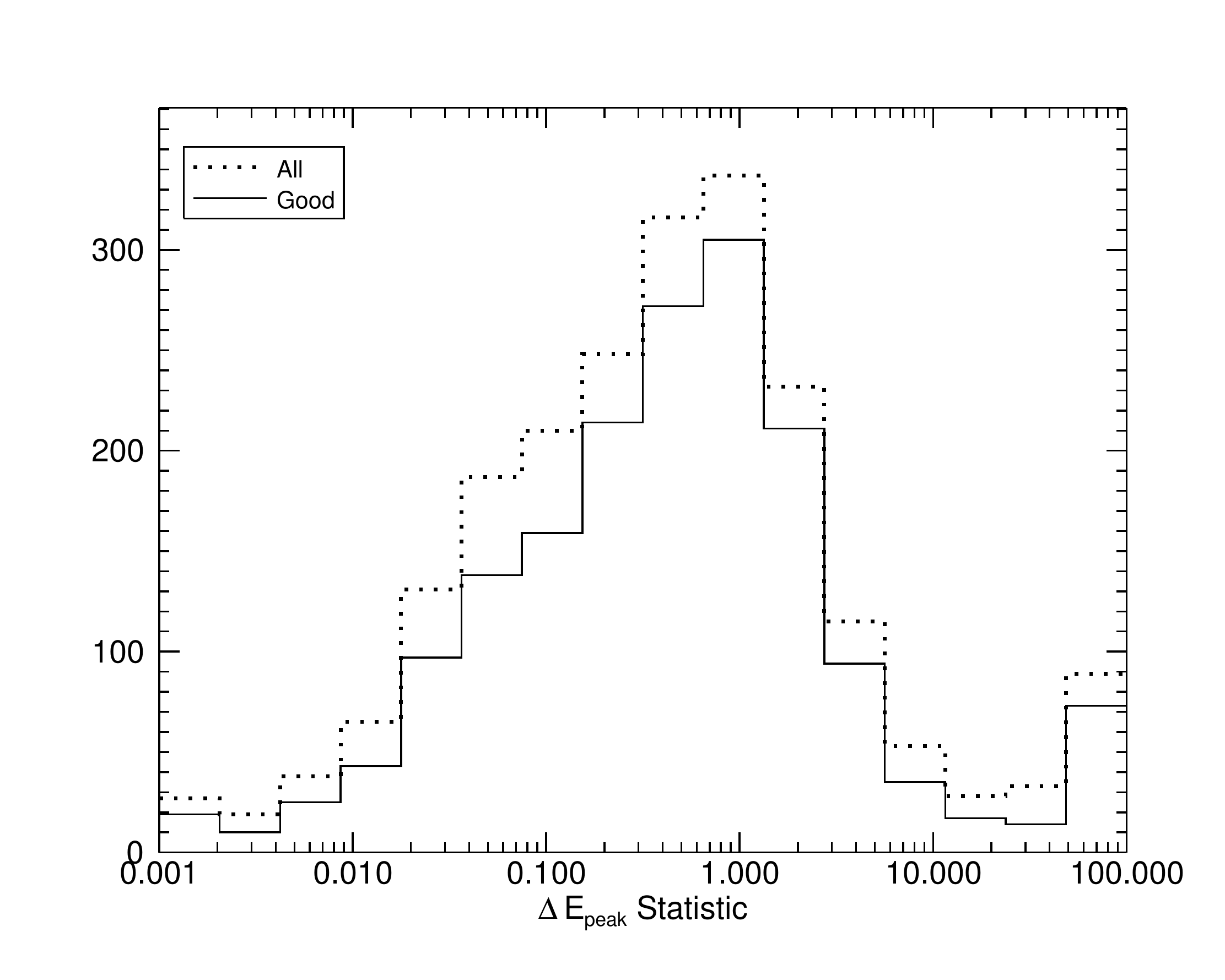}
	\end{center}
\caption{Distribution of the $\Delta E_{peak}$ statistic for the COMP and BAND models from fluence spectral fits.  A value less 
than 1 indicates the $E_{peak}$ values are similar within errors, while a value larger than 1 indicates the $E_{peak}$ values are
 not within errors. \label{deltaepeakf}}
\end{figure}

It is of interest to study the difference in the value of $E_{peak}$ between the BAND and 
COMP functions.  The relative deviation between the two values can be calculated from a statistic based on the difference 
between the values, taking into account their errors.  This statistic can be calculated by
\begin{equation}
	\Delta S=\frac{|E_{peak}^{C}-E_{peak}^{B}|}{\sigma_{E_{peak}}^{C}+\sigma_{E_{peak}}^{B}}
\end{equation}
where C and B indicate the COMP and BAND values respectively.  This statistic has a value of unity when the deviation 
between the $E_{peak}$ values exactly matches the sum of the $1\sigma$ errors.  A value less than one indicates the $E_
{peak}$ values are similar within errors, and a value greater than one indicates that the $E_{peak}$ values are not within errors of 
each other.  Figure \ref{deltaepeakf} depicts the distribution of the statistic and roughly 33\% of the BAND and COMP $E_{peak}
$ values are found to be outside of the combined errors.  This indicates that, although COMP is a special case of BAND, a 
significant fraction of the $E_{peak}$ values can vary by more than 1$\sigma$ based on which model is chosen.

\begin{figure}
	\begin{center}
		\subfigure[]{\label{pflux1mevf}\includegraphics[scale=0.35]{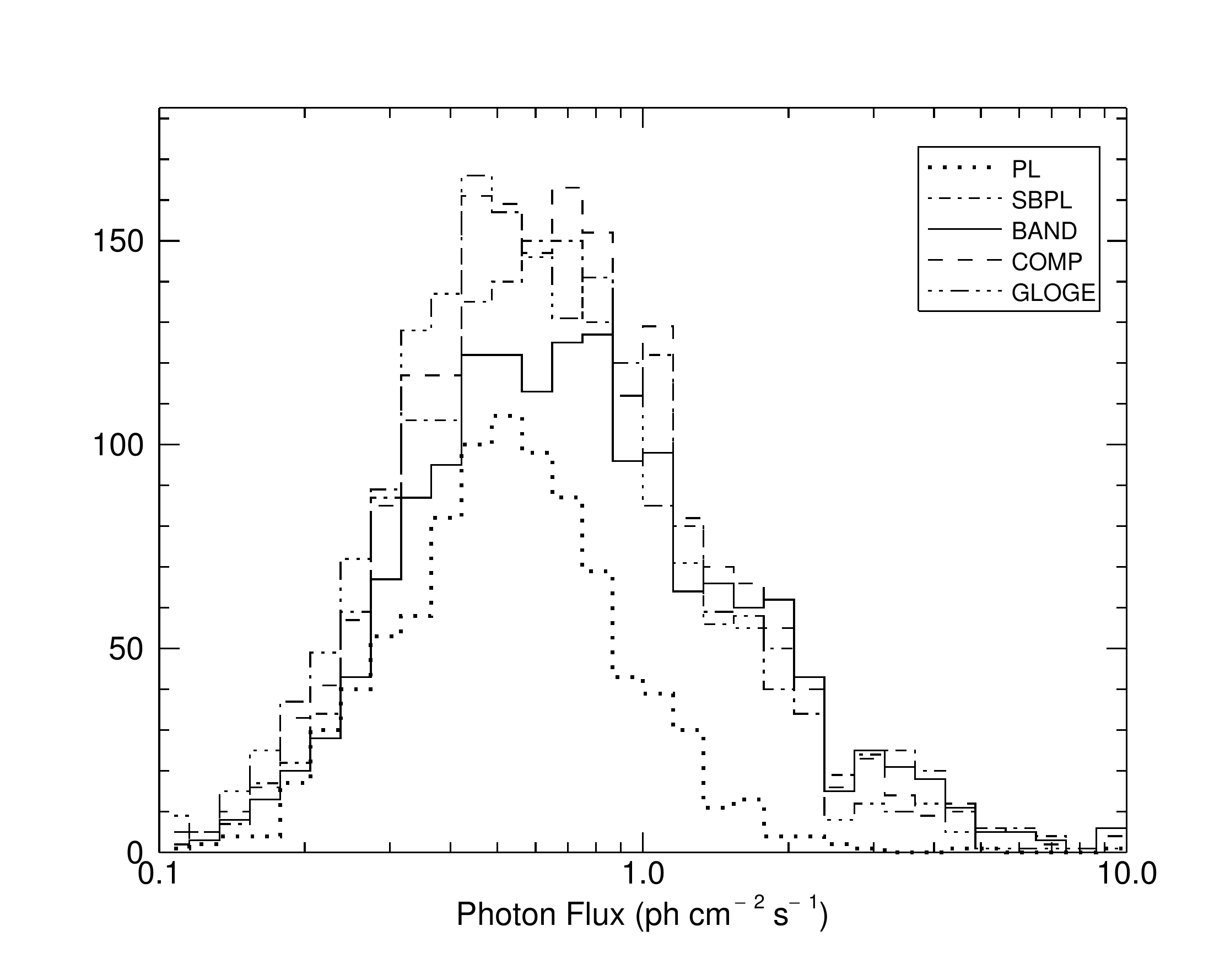}}
		\subfigure[]{\label{eflux1mevf}\includegraphics[scale=0.35]{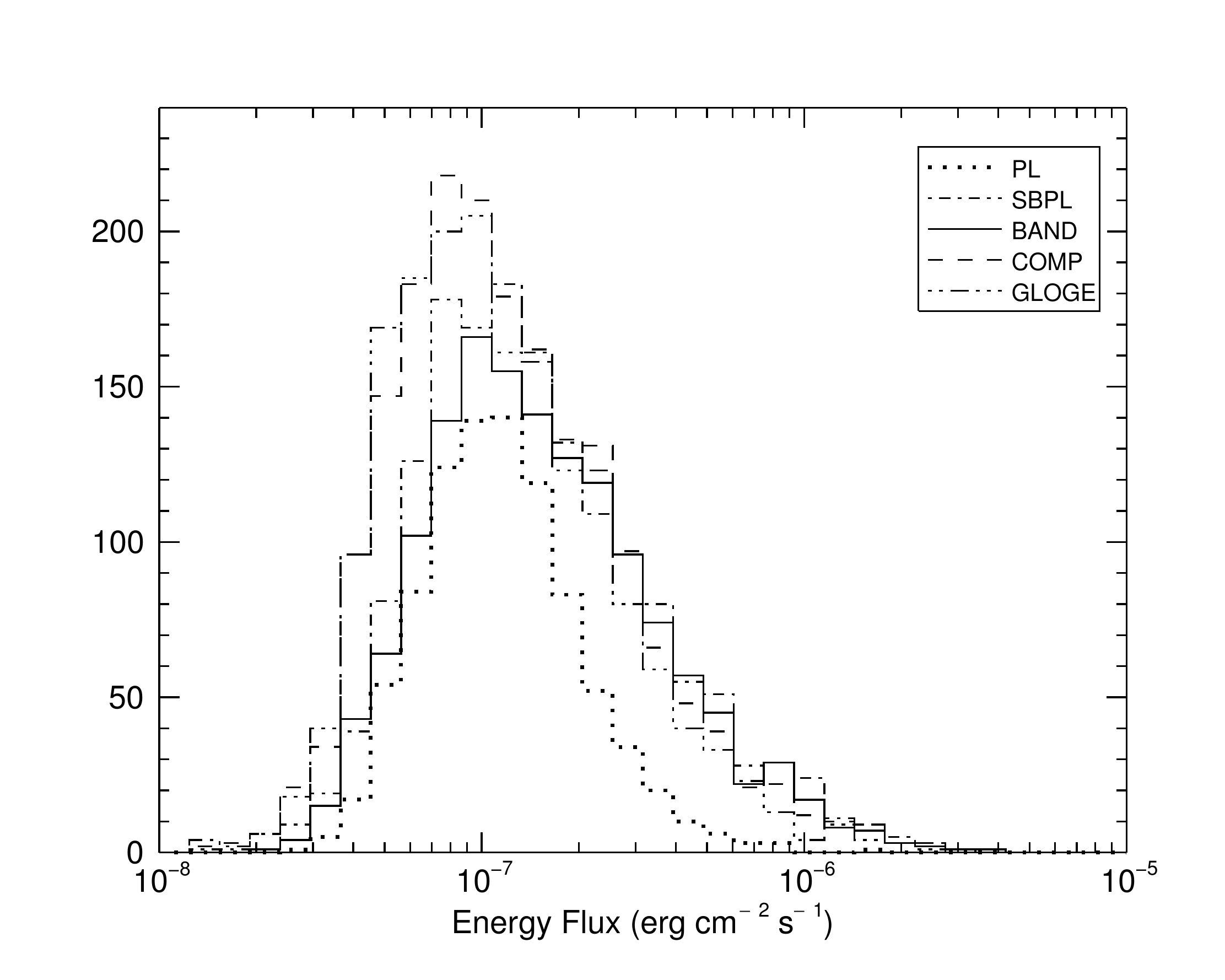}}
	\end{center}
\caption{Distributions of photon and energy flux from fluence spectral fits.  \ref{pflux1mevf} and \ref{eflux1mevf} display the flux 
distributions for the 20 keV--2 MeV band. \label{fluxf}}
\end{figure}

\begin{figure}
	\begin{center}
		\subfigure[]{\label{pfluence1mev}\includegraphics[scale=0.35]{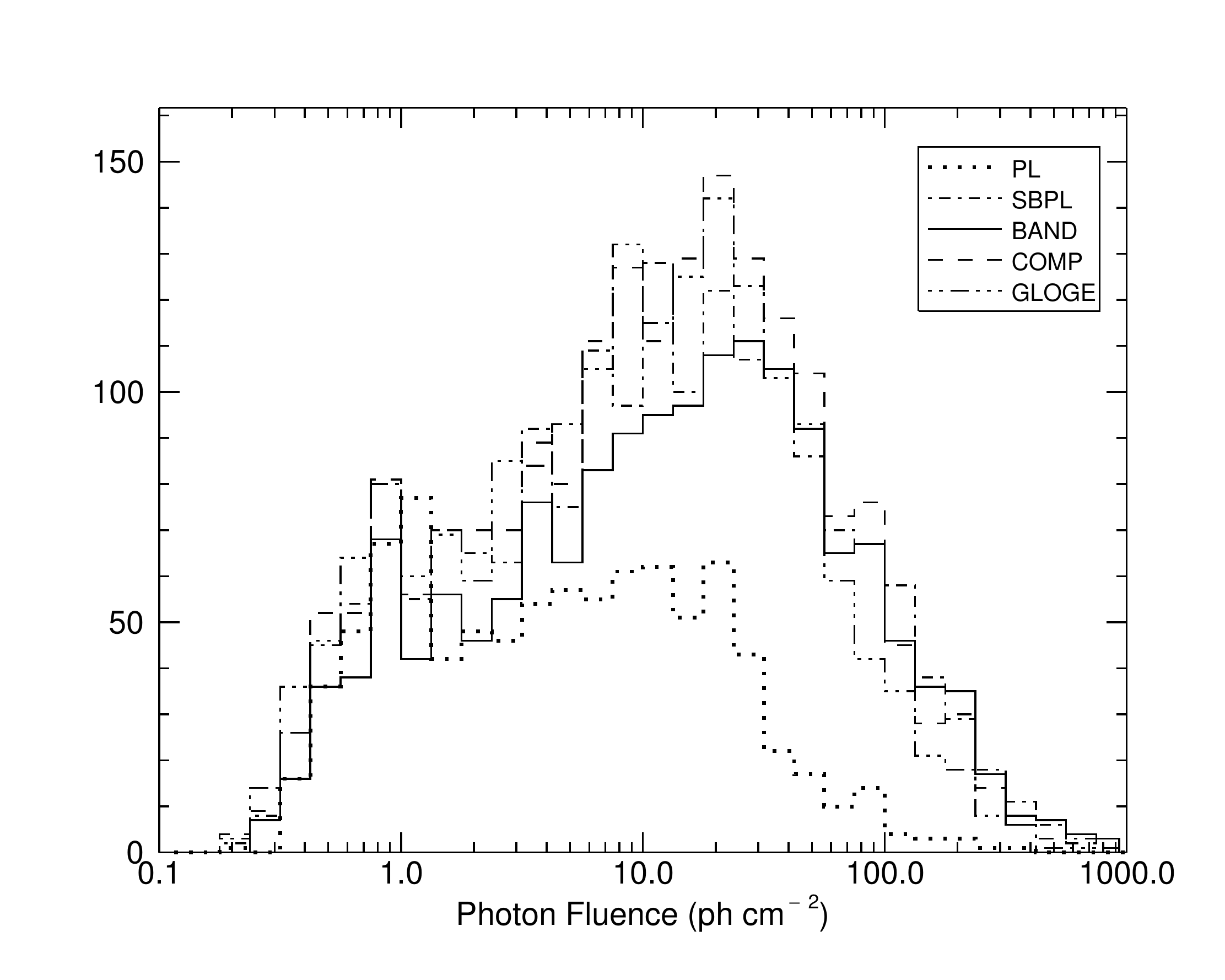}}
		\subfigure[]{\label{efluence1mev}\includegraphics[scale=0.35]{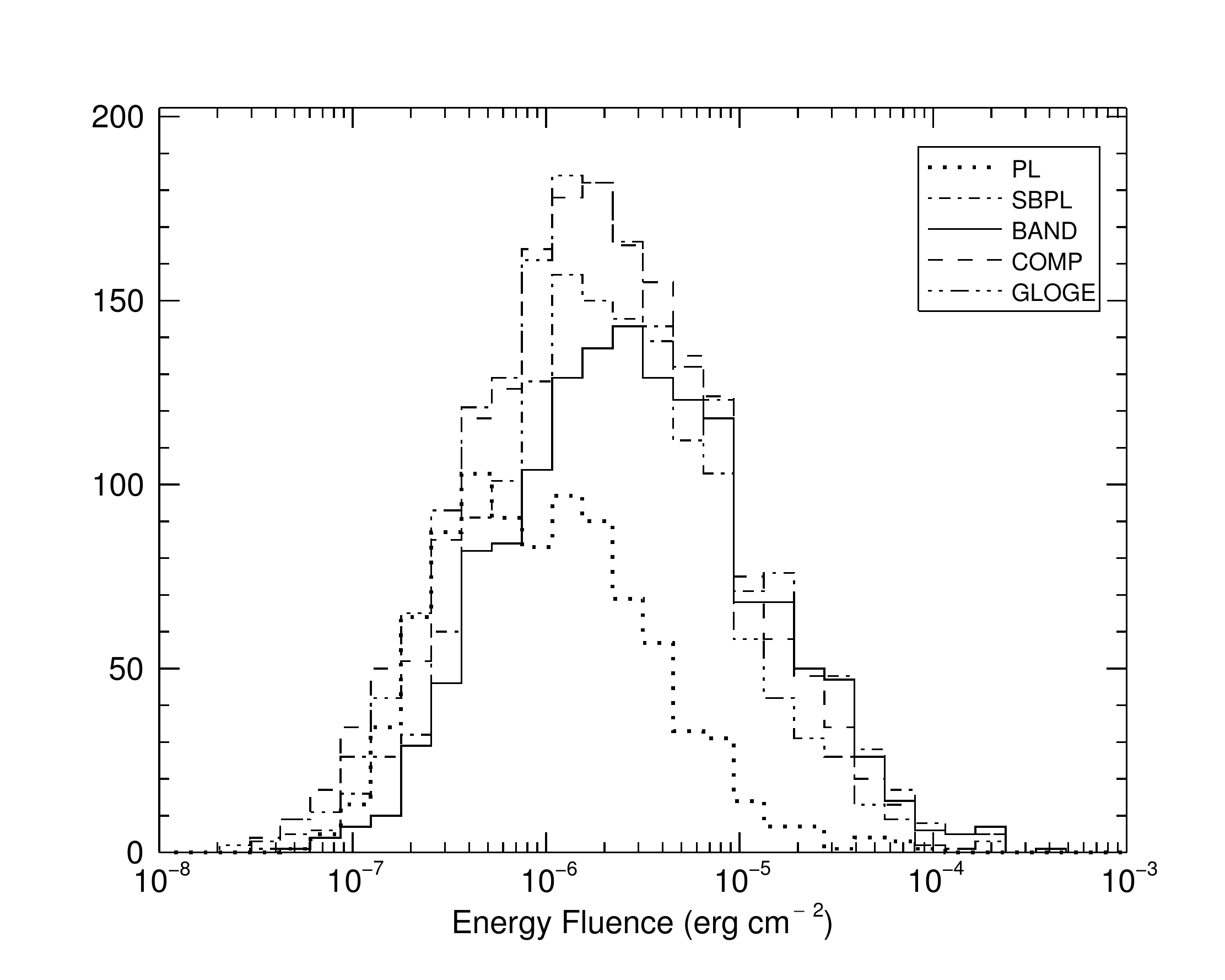}}
	\end{center}
\caption{Distributions of photon and energy fluence.  \ref{pfluence1mev} and \ref{efluence1mev} display the fluence 
distributions from the 20 keV--2 MeV band.  \label{fluence}}
\end{figure}

The distributions for the time-integrated photon flux and energy flux are shown in Figure \ref{fluxf}.  The photon flux peaks at $\sim
$0.4--0.7 photons cm$^{-2}$  s$^{-1}$, and the energy flux peaks at $8 \times 10^{-8}$ ergs cm$^{-2}$ s$^{-1}$ in the 
20-2000 keV band.  Similarly in Figure \ref{fluence}, the distributions for the photon fluence and energy fluence are depicted.  
The plots for the photon fluence appear to contain evidence of the duration bimodality of GRBs, and has discriminant peaks at $
\sim$1 and 20 photons cm$^{-2}$.  The energy fluence in the full BATSE bandpass peaks at $\sim2 \times 10^{-6}$ erg cm$^{-2}$.  
The brightest GRB contained in this catalog based on time-averaged photon flux is GRB 931206 (trigger \#2680) with a flux of $\rm 
>32\ ph\ s^{-1}\ cm^{-2}$ and the burst with the largest average energy flux is GRB 991216 (trigger \#7906) with an energy flux of $
\rm \sim3.8 \times 10^{-6}\ erg\ s^{-1} cm^{-2}$.  The most fluent burst (although not the longest duration) in the catalog is GRB 
990104B (trigger \#7301) with a photon fluence of $\rm >1500\ ph\ cm^{-2}$ and an energy fluence of $\rm >3.8 \times 10^{-4}\ erg\ 
cm^{-2}$.

\subsection{Peak Flux Spectra}
\begin{figure}
	\begin{center}
		\subfigure[]{\label{indexlop}\includegraphics[scale=0.35]{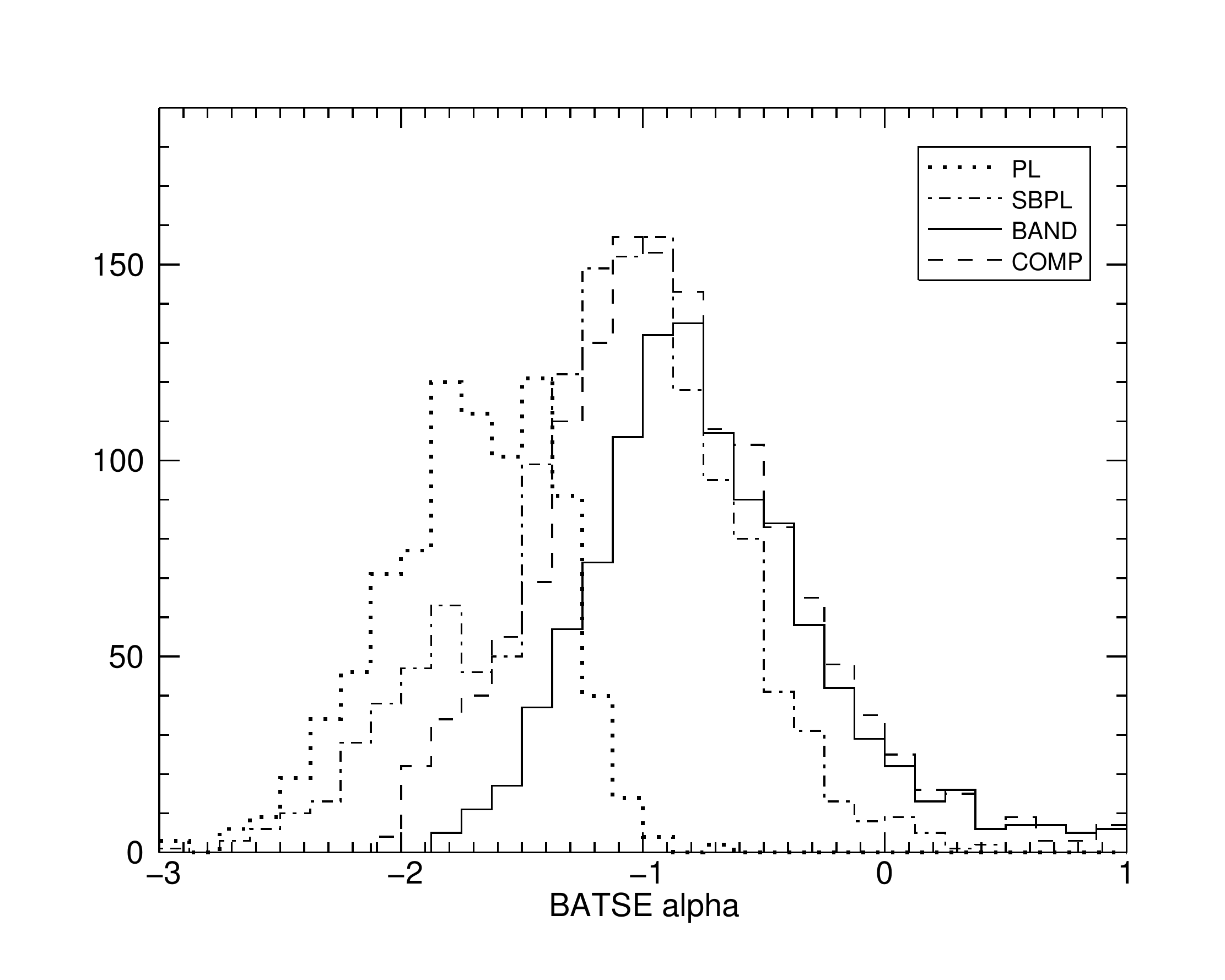}}
		\subfigure[]{\label{alphasbplp}\includegraphics[scale=0.35]{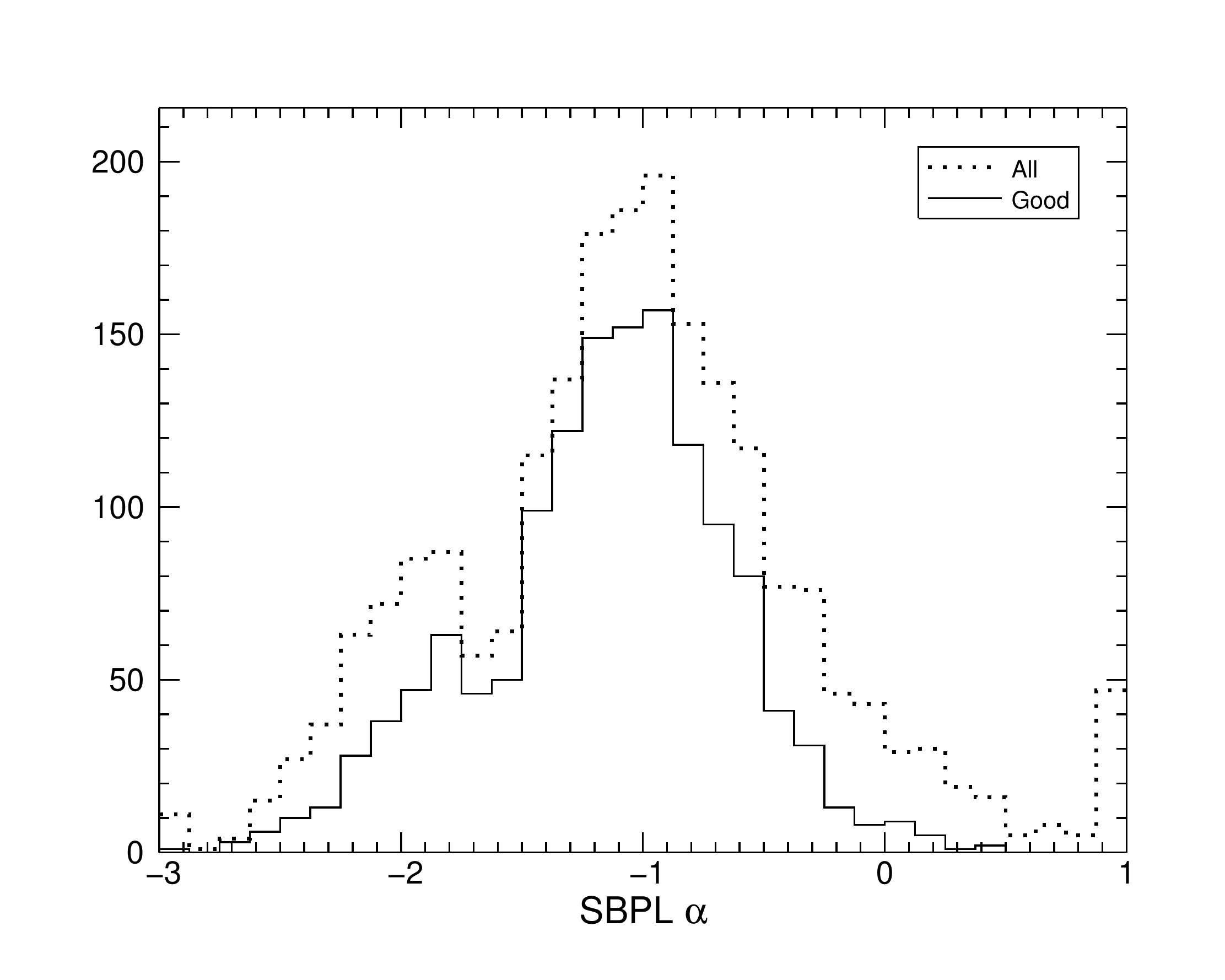}}\\
		\subfigure[]{\label{alphabandp}\includegraphics[scale=0.35]{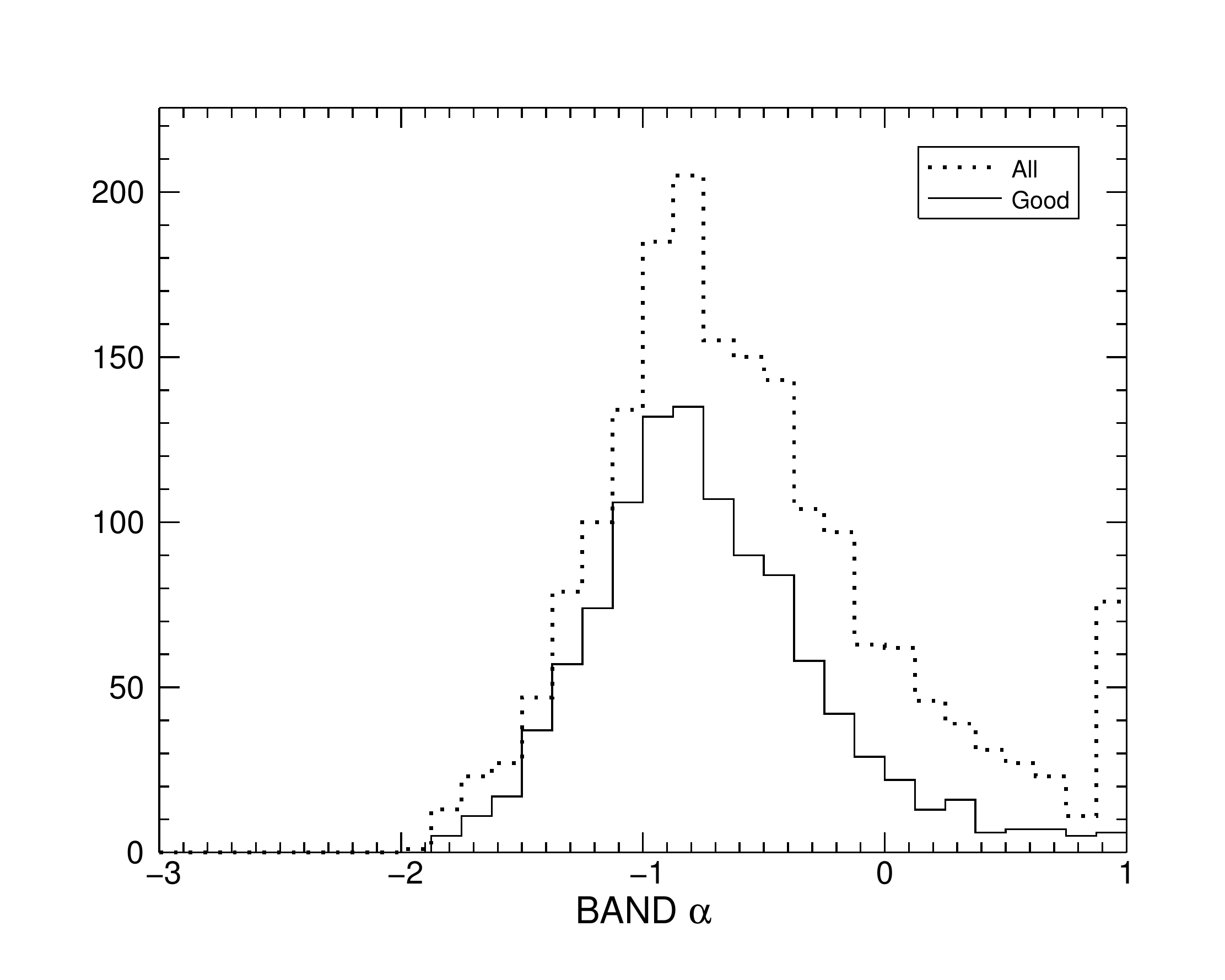}}
		\subfigure[]{\label{alphacompp}\includegraphics[scale=0.35]{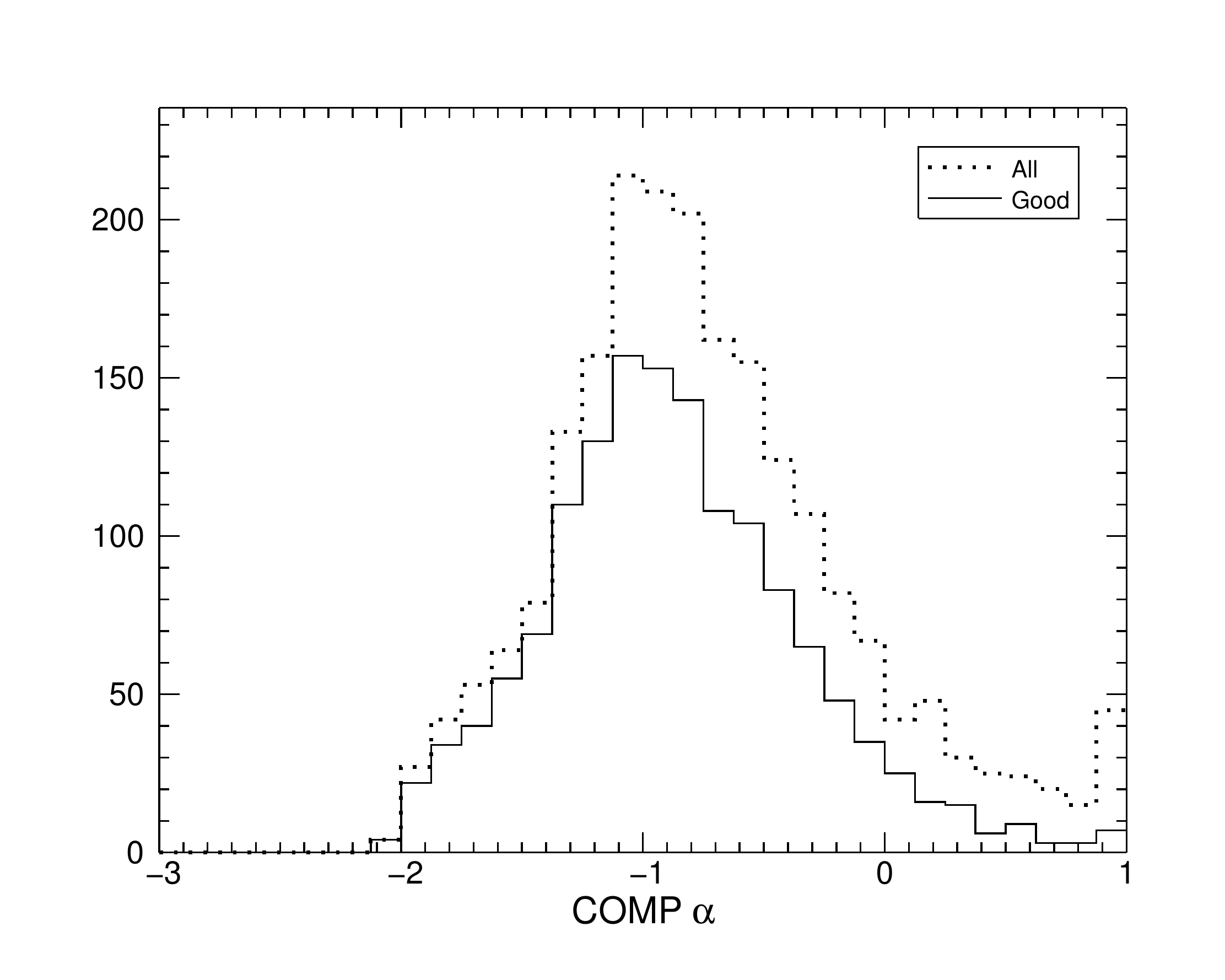}}
	\end{center}
\caption{Distributions of the low-energy spectral indices from peak flux spectral fits.  \ref{indexlop} shows the distributions of 
GOOD parameters and compares to the distribution of PL indices.  \ref{alphasbplp}--\ref{alphacompp} display the comparison 
between the distribution of GOOD parameters and all parameters with no data cuts.  The last bin includes values greater than 
1. \label{loindexp}}
\end{figure}

\begin{figure}
	\begin{center}
		\subfigure[]{\label{indexhip}\includegraphics[scale=0.35]{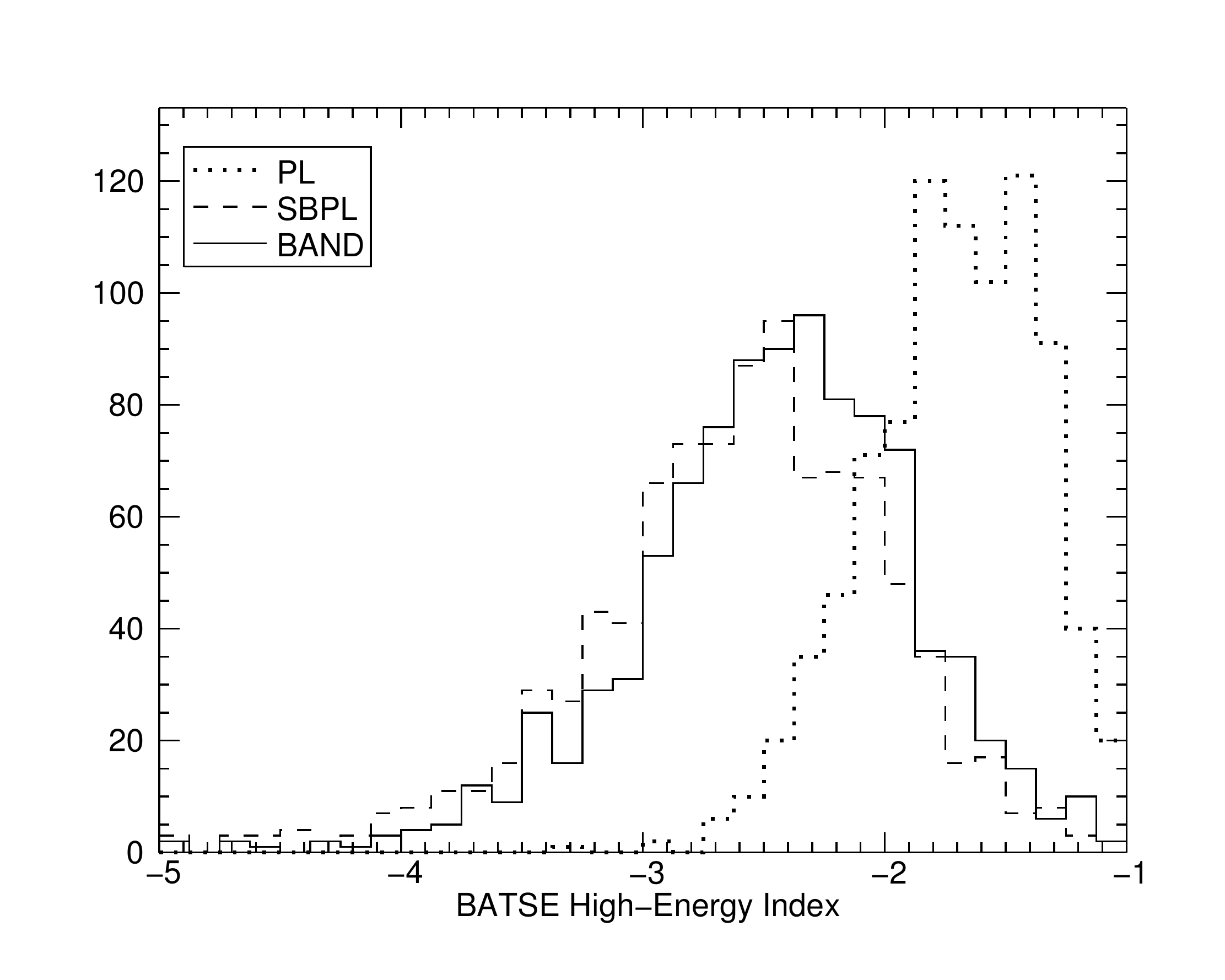}}
		\subfigure[]{\label{betasbplp}\includegraphics[scale=0.35]{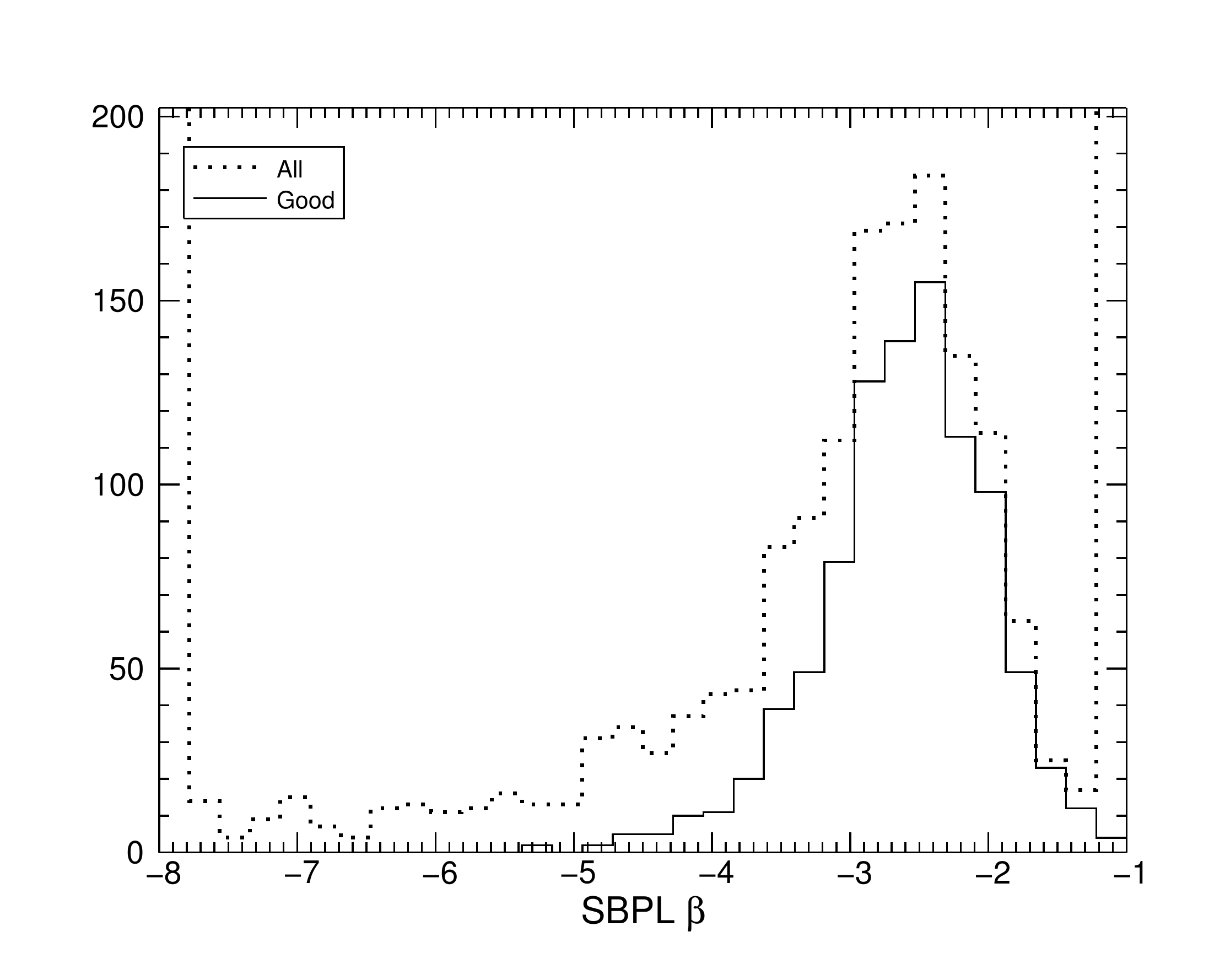}}\\
		\subfigure[]{\label{betabandp}\includegraphics[scale=0.35]{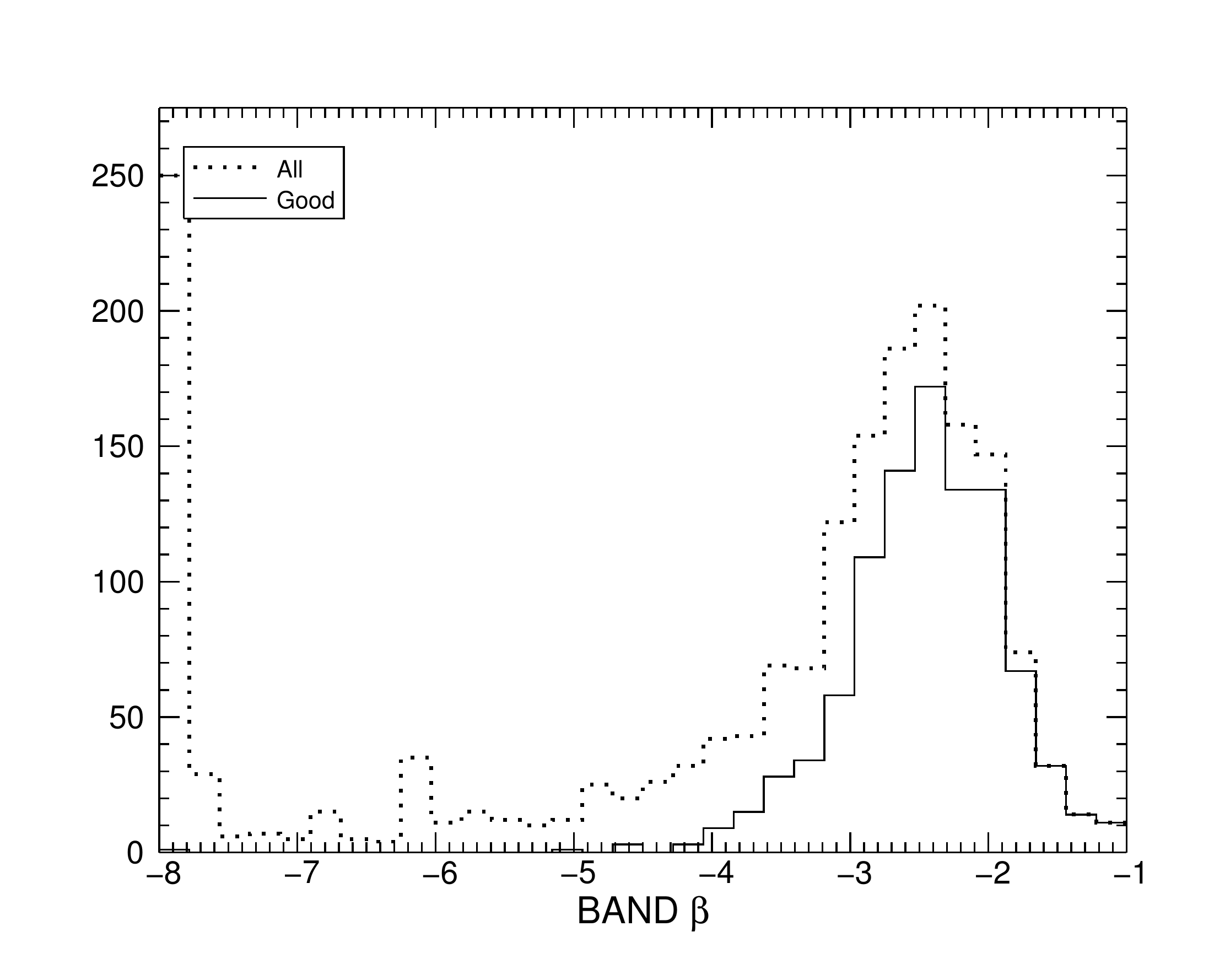}}
		\subfigure[]{\label{deltasp}\includegraphics[scale=0.35]{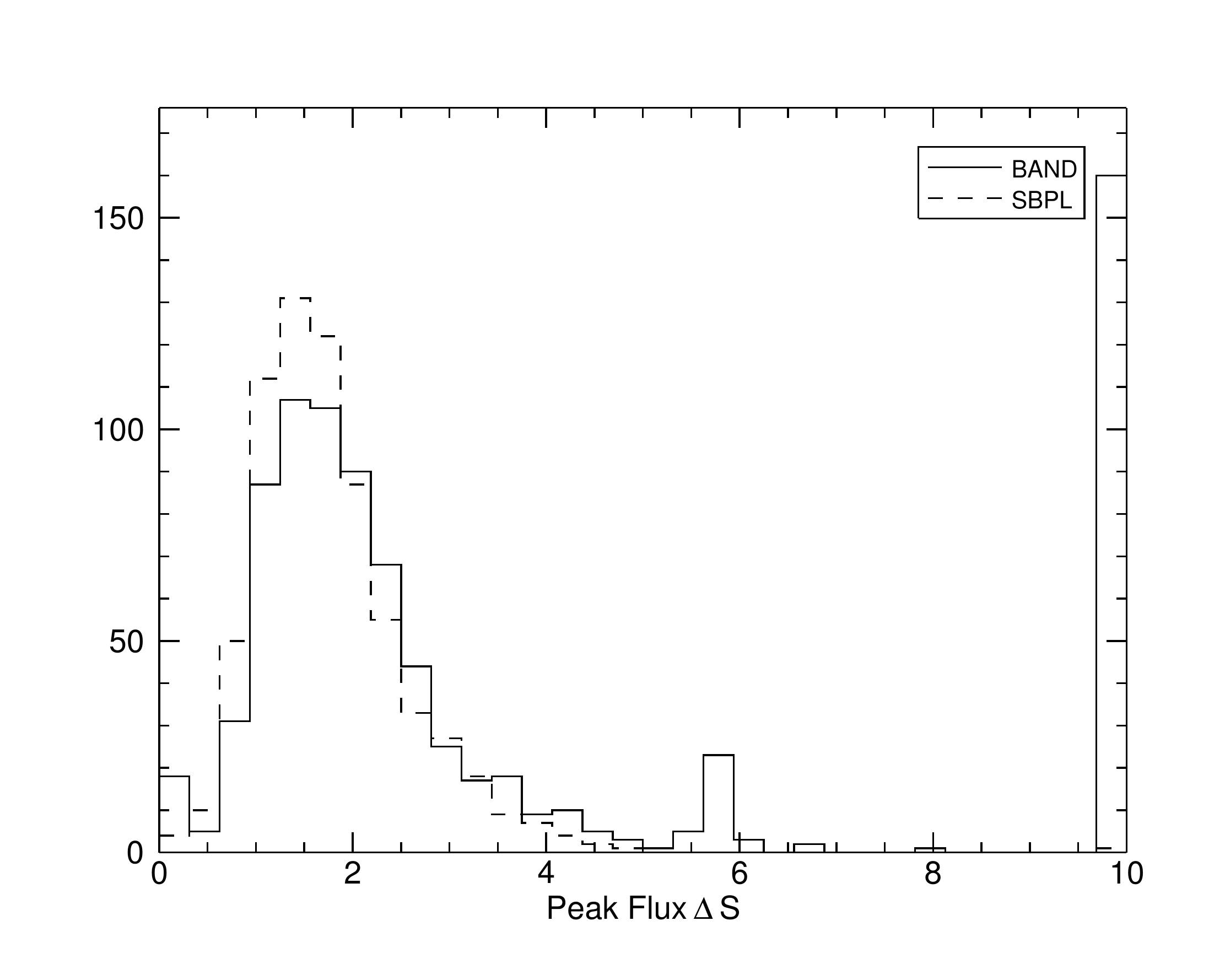}}
	\end{center}
\caption{\ref{indexhip} - \ref{betabandp} are distributions of the high-energy spectral indices from peak flux spectral fits.  \ref
{indexhip} shows the distributions of GOOD parameters and compares to the distribution of PL indices.  \ref{betasbplp} and \ref
{betabandp}  display the comparison between the distribution of GOOD parameters and all parameters with no data cuts.  The 
first bins include values less than -8 and the last bin include values greater than -1. \ref{deltasp} shows the difference between 
the low- and high-energy indices.  The first bin contains values less than 0, indicating that the centroid value of alpha is 
steeper than the centroid value of beta.\label{hiindexp}}
\end{figure}

The following peak flux spectral distributions have been produced by fitting the GRB spectra over the 2048 ms peak count flux.  
Note that short bursts with durations less than 2048 ms will constitute a single bin in the lightcurve and it is that single bin 
that is selected for analysis.  The low-energy indices, as shown in Figure \ref{loindexp} distribute about a $-1$ power law typical 
of most GRBs.  Accounting for the parameter uncertainty, up to 28\% of the GOOD low-energy indices violate the $-2/3$ synchrotron 
``line-of-death'', while an additional 55\% of the indices violate the $-3/2$ synchrotron cooling limit.  The high-energy indices in Figure 
\ref{hiindexp} peak at a slope of $\sim -2$ and have a long tail toward steeper indices.  As shown with the fluence spectra, the PL 
index serves as an average between low- and high-energy indices for the BAND and SBPL functions.  Shown in Figure \ref{deltasp} 
is the $\Delta S$ distribution for the peak flux spectra.  Similar to the fluence spectra, this distribution peaks at $\sim$1.5, which 
implies a median electron energy spectral index of $\sim$4 assuming the synchrotron shock model.  As was the case with the 
fluence spectra, the overflow of $\Delta S$ at $\sim$10 for the BAND model is due to the extremely steep slope of $\beta$, where the 
high-energy spectrum approximates an exponential cut-off. 

\begin{figure}
	\begin{center}
		\subfigure[]{\label{ecentp}\includegraphics[scale=0.35]{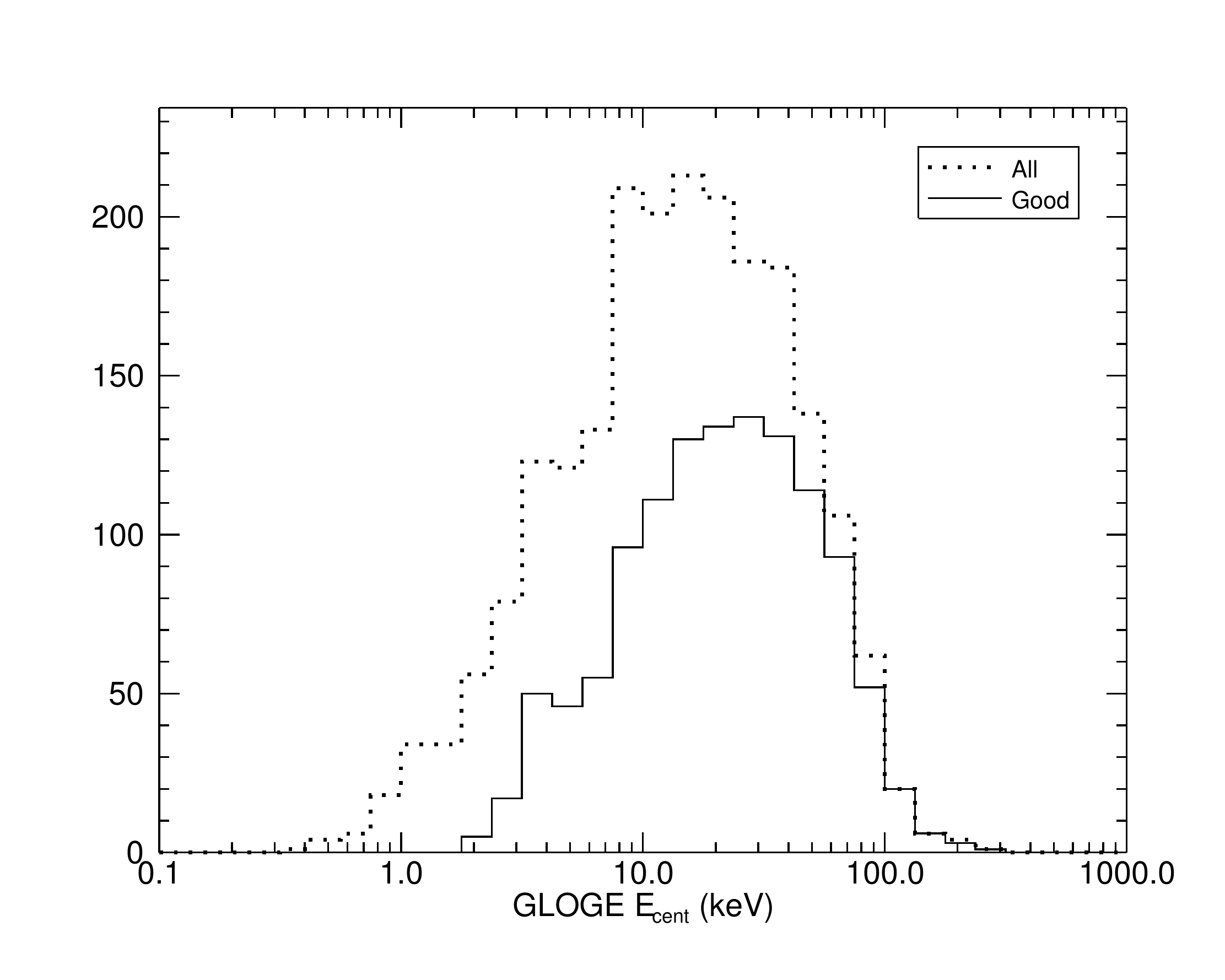}}
		\subfigure[]{\label{fwhmp}\includegraphics[scale=0.35]{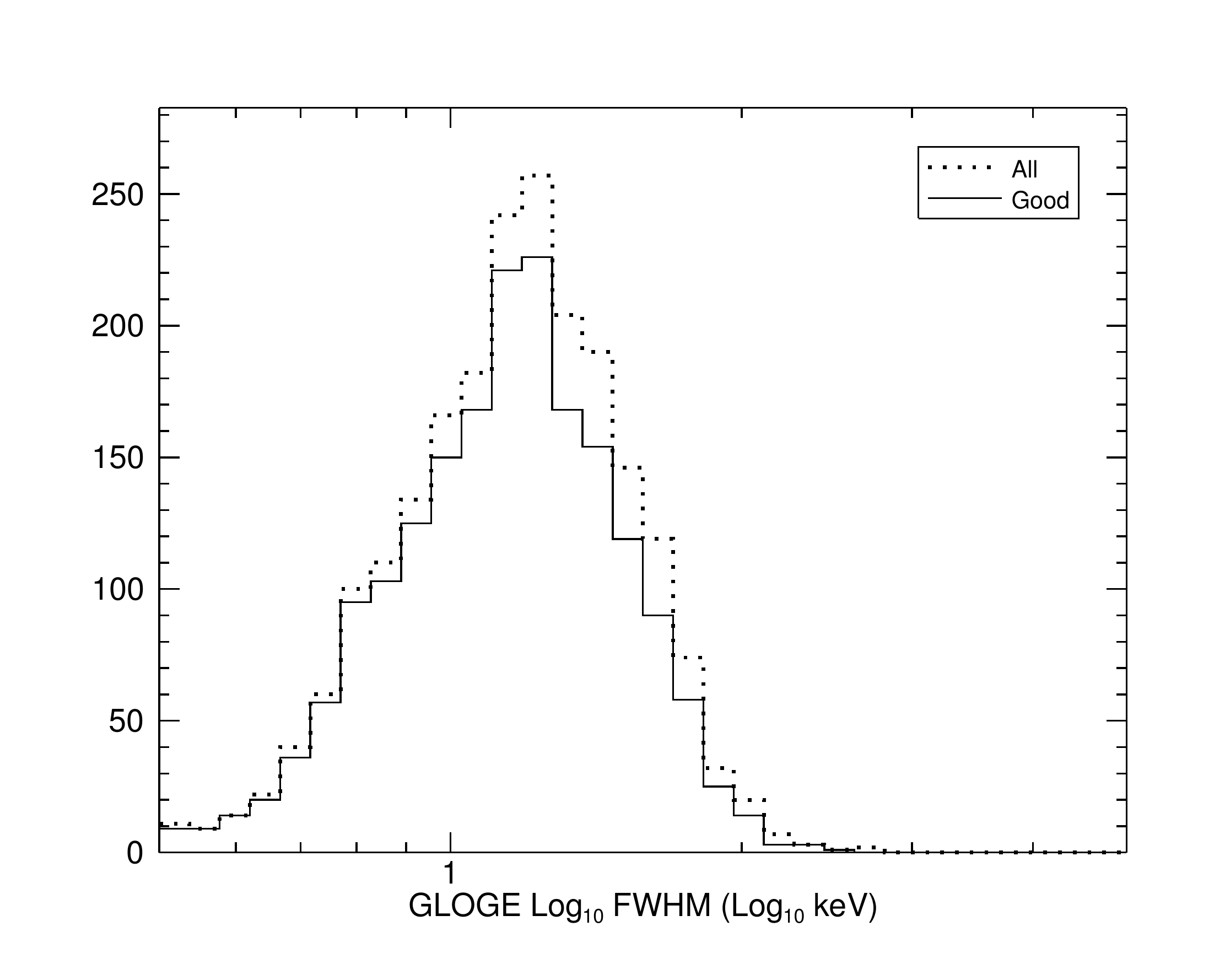}}
	\end{center}
\caption{Distributions of the GLOGE $E_{cent}$ and FWHM parameters from the peak flux spectral fits.  \label{GLOGEp}}
\end{figure}

\begin{figure}
	\begin{center}
		\subfigure[]{\label{ebreaksbplp}\includegraphics[scale=0.35]{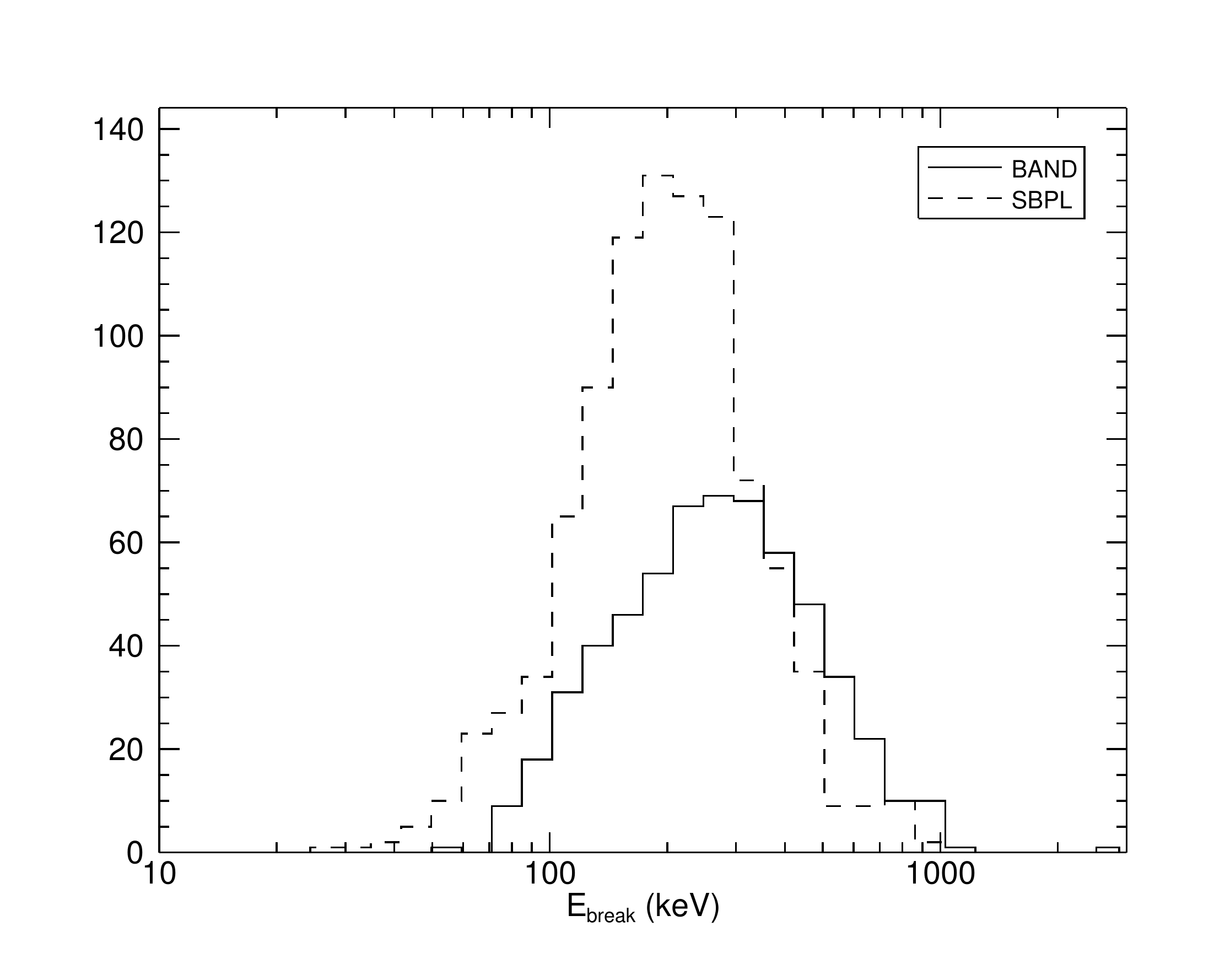}}
		\subfigure[]{\label{epeakp}\includegraphics[scale=0.35]{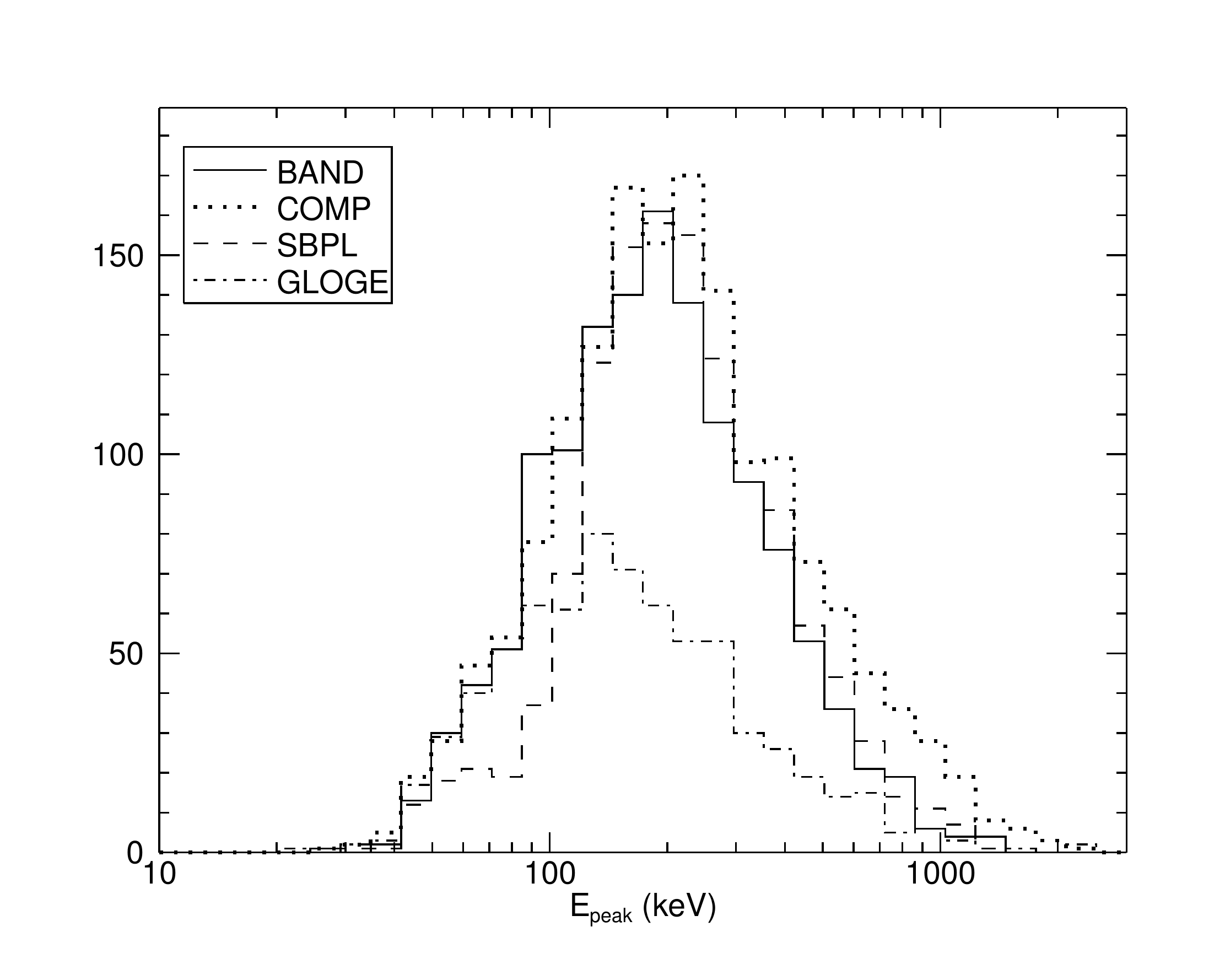}}\\
		\subfigure[]{\label{epeakbandp}\includegraphics[scale=0.35]{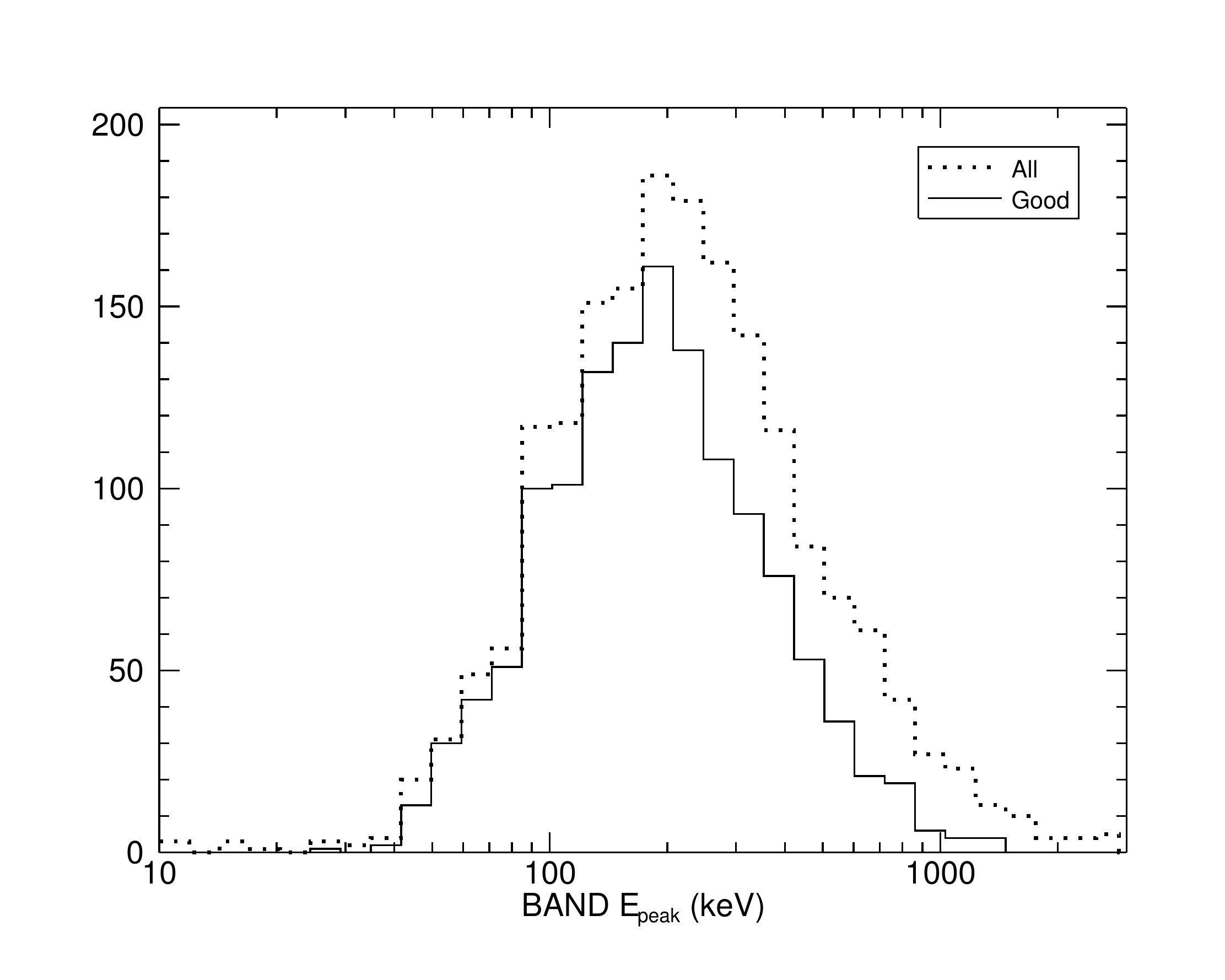}}
		\subfigure[]{\label{epeakcompp}\includegraphics[scale=0.35]{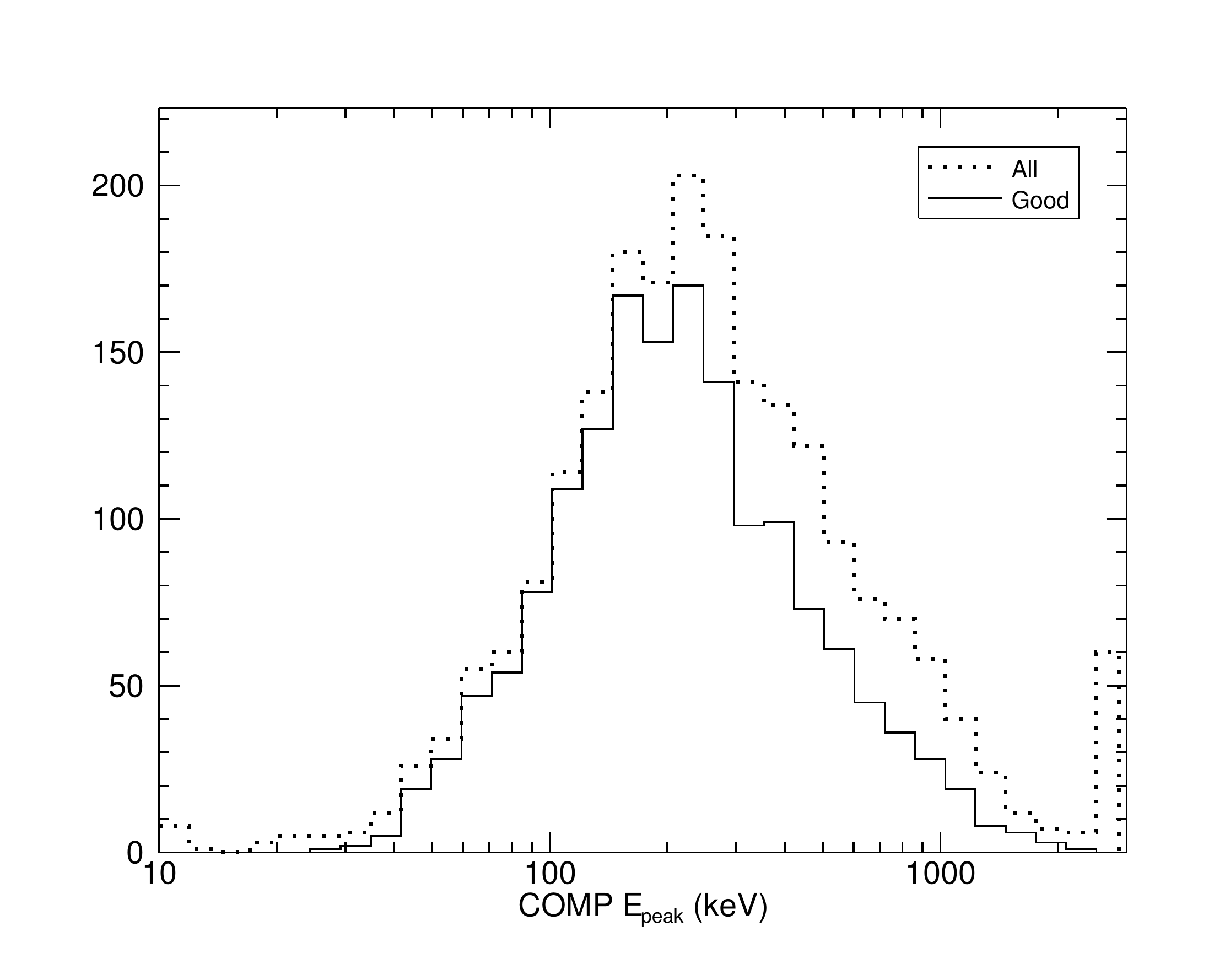}}
	\end{center}
\caption{Distributions of $E_{break}$ and $E_{peak}$ from peak flux spectral fits.  \ref{ebreaksbplp} displays the comparison 
between the distribution of GOOD $E_{break}$ and $E_{break}$ with no data cuts.  \ref{epeakp} shows the distributions of 
GOOD $E_{peak}$ for BAND, SBPL,  and COMP.  \ref{epeakbandp} and \ref{epeakcompp} display the comparison between the 
distribution of GOOD parameters and all parameters with no data cuts. \label{epeakebreakp}}
\end{figure}

\begin{figure}
	\begin{center}
		\includegraphics[scale=0.7]{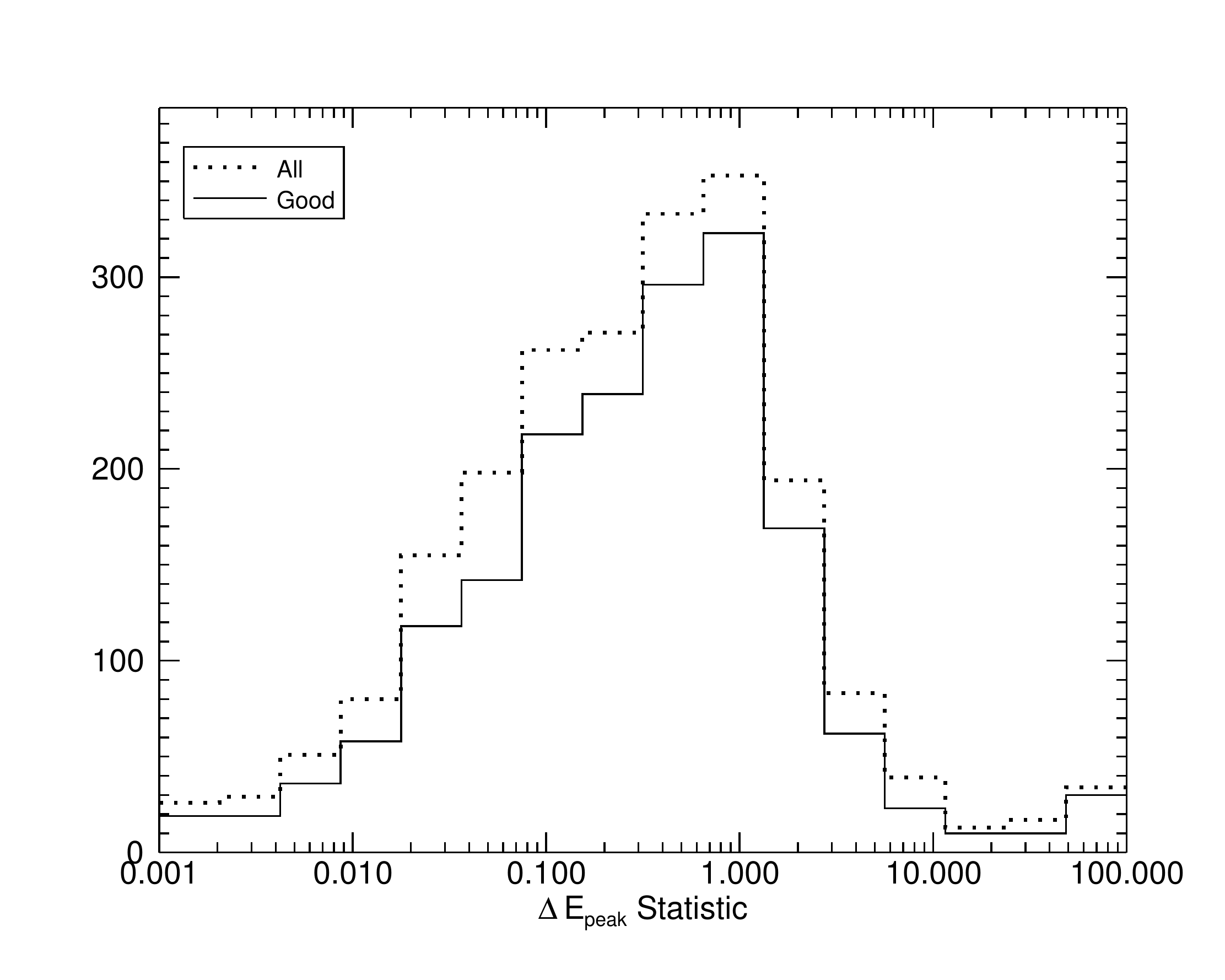}
	\end{center}
\caption{Distribution of the $\Delta E_{peak}$ statistic for the COMP and BAND models from peak flux spectral fits.  A value less 
than 1 indicates the $E_{peak}$ values are similar within errors, while a value larger than 1 indicates the $E_{peak}$ values are 
not within errors. \label{deltaepeakp}}
\end{figure}

As shown in Figure \ref{GLOGEp}, the GLOGE $E_{cen}$, covering two orders of magnitude, is shifted to peak at $\sim$30 keV, while 
the FWHM distribution peaks at $\sim$1.22 in contrast to the fluence spectral results.  In Figure \ref{epeakebreakp}, we show the 
distributions for $E_{break}$ and  $E_{peak}$.  As was evident from the fluence spectra, the $E_{break}$ from the SBPL fits 
appears to peak at 200 keV with a low-energy tail, meanwhile the $E_{break}$ from BAND is harder and peaks at about 300 keV.  
Consistent with previous findings \citep{Preece98, Kaneko06}, the $E_{peak}$ distributions for all models peak around 200 keV, 
except for the derived $E_{peak}$ from GLOGE which peaks at slightly more than 100 keV,  and covers over two orders of 
magnitude.  We calculate and show the $\Delta E_{peak}$ statistic in Figure \ref{deltaepeakp} and roughly 24\% of the BAND and 
COMP $E_{peak}$ values are outside the combined errors.  

\begin{figure}
	\begin{center}
		\subfigure[]{\label{pflux1mevp}\includegraphics[scale=0.35]{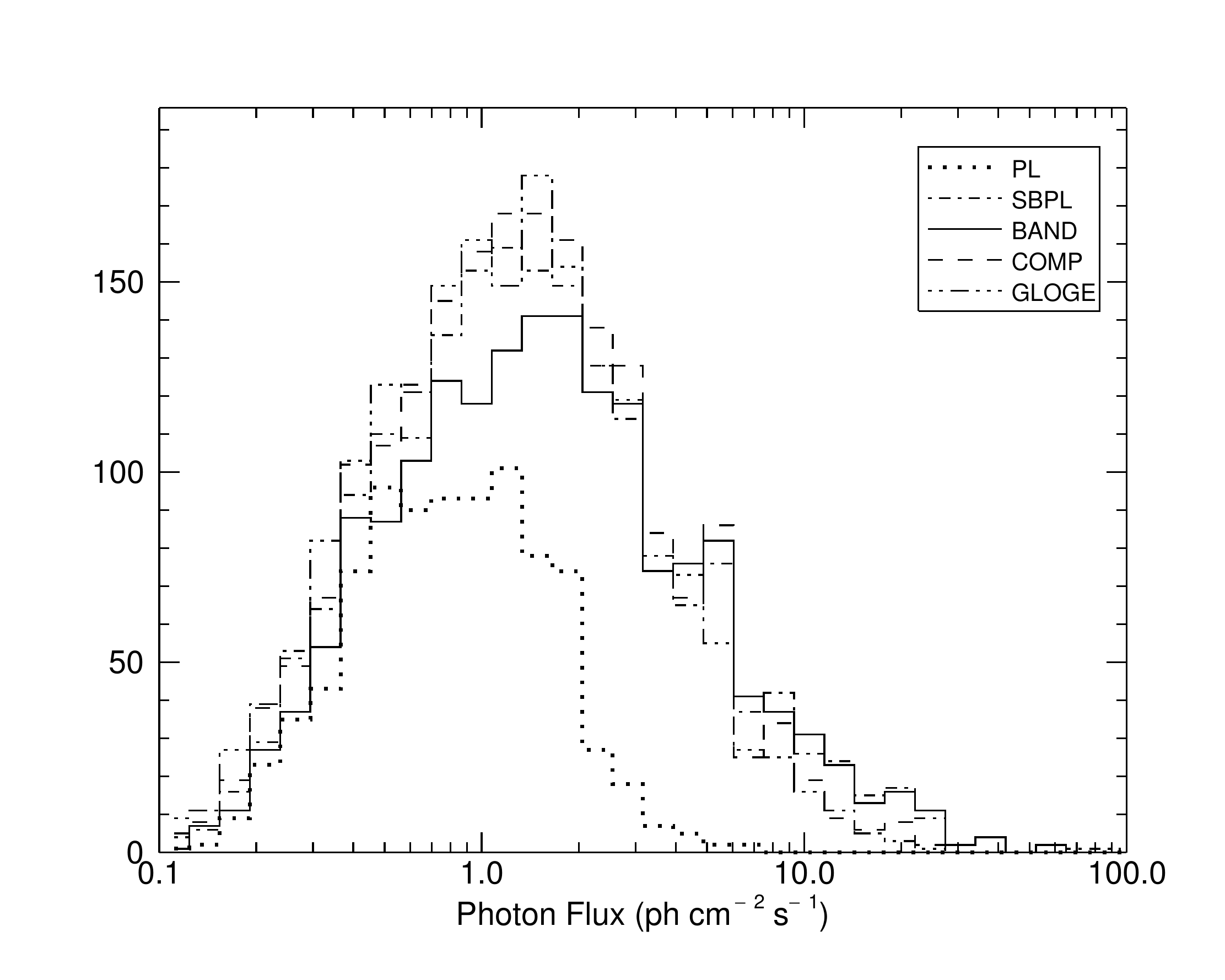}}
		\subfigure[]{\label{eflux1mevp}\includegraphics[scale=0.35]{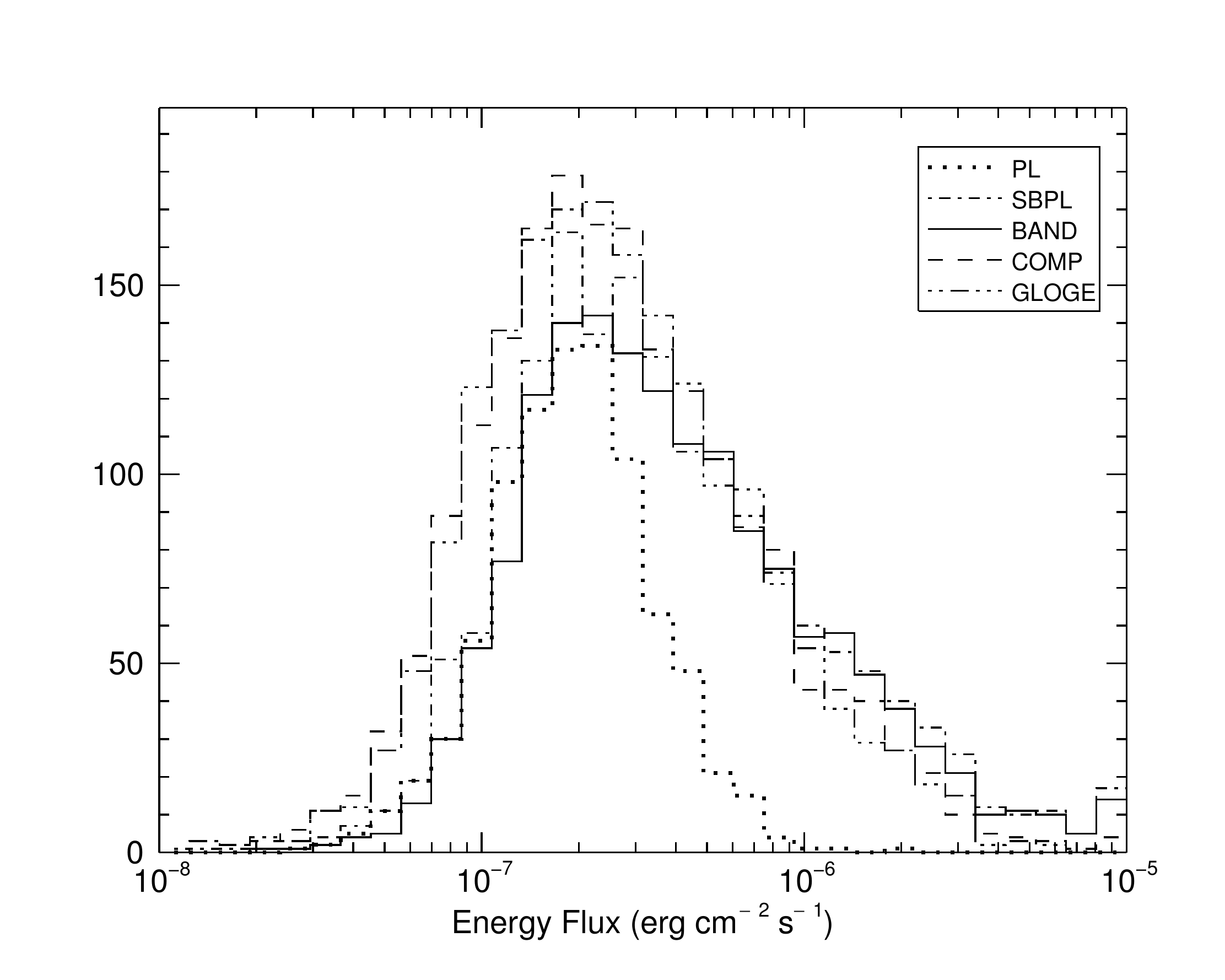}}
	\end{center}
\caption{Distributions of photon and energy flux from peak flux spectral fits.  \ref{pflux1mevp} and \ref{eflux1mevp} display the 
flux distributions for the 20 keV--2 MeV band. \label{fluxp}}
\end{figure}

The distributions for the peak photon flux and energy flux are shown in Figure \ref{fluxp}.  The photon flux peaks around 1.5 
photons cm$^{-2}$  s$^{-1}$, and the energy flux peaks at $2 \times 10^{-7}$ ergs cm$^{-2}$ s$^{-1}$ in the 20--2000 keV band.  
The GRB with the brightest peak photon flux is GRB 931206 (trigger \#2680) at $\rm >350\ ph\ s^{-1}\ cm^{-2}$.  This GRB is also 
the brightest GRB in terms of time-averaged photon flux. Similar to the time-averaged energy flux, the burst with highest peak energy 
flux is GRB 991216 (trigger \#7906) at $\rm >5.7 \times 10^{-5}\ ergs\ s^{-1}\ cm^{-2}$.

\begin{figure}
	\begin{center}
		\subfigure[]{\label{pffalpha}\includegraphics[scale=0.35]{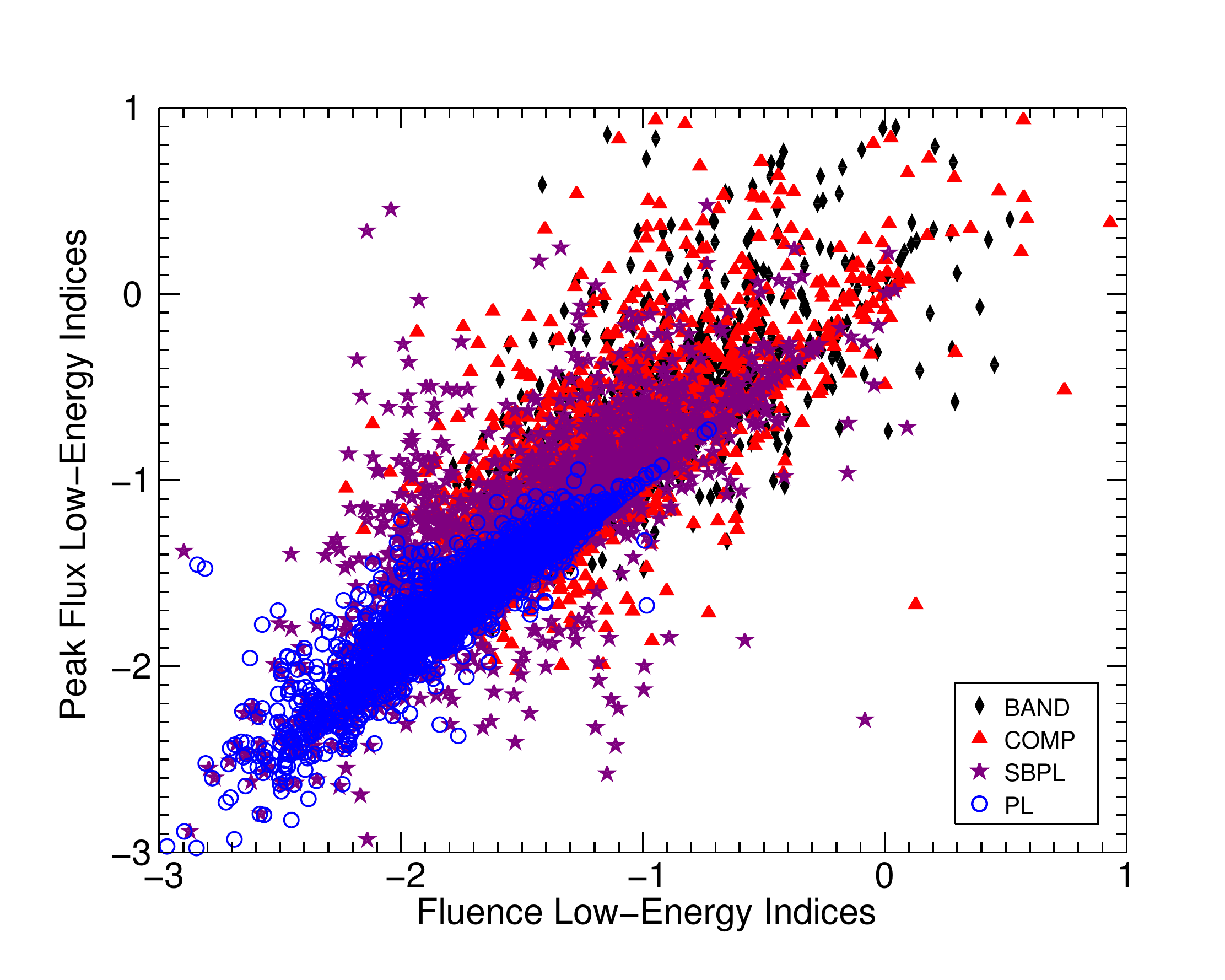}}
		\subfigure[]{\label{pffbeta}\includegraphics[scale=0.35]{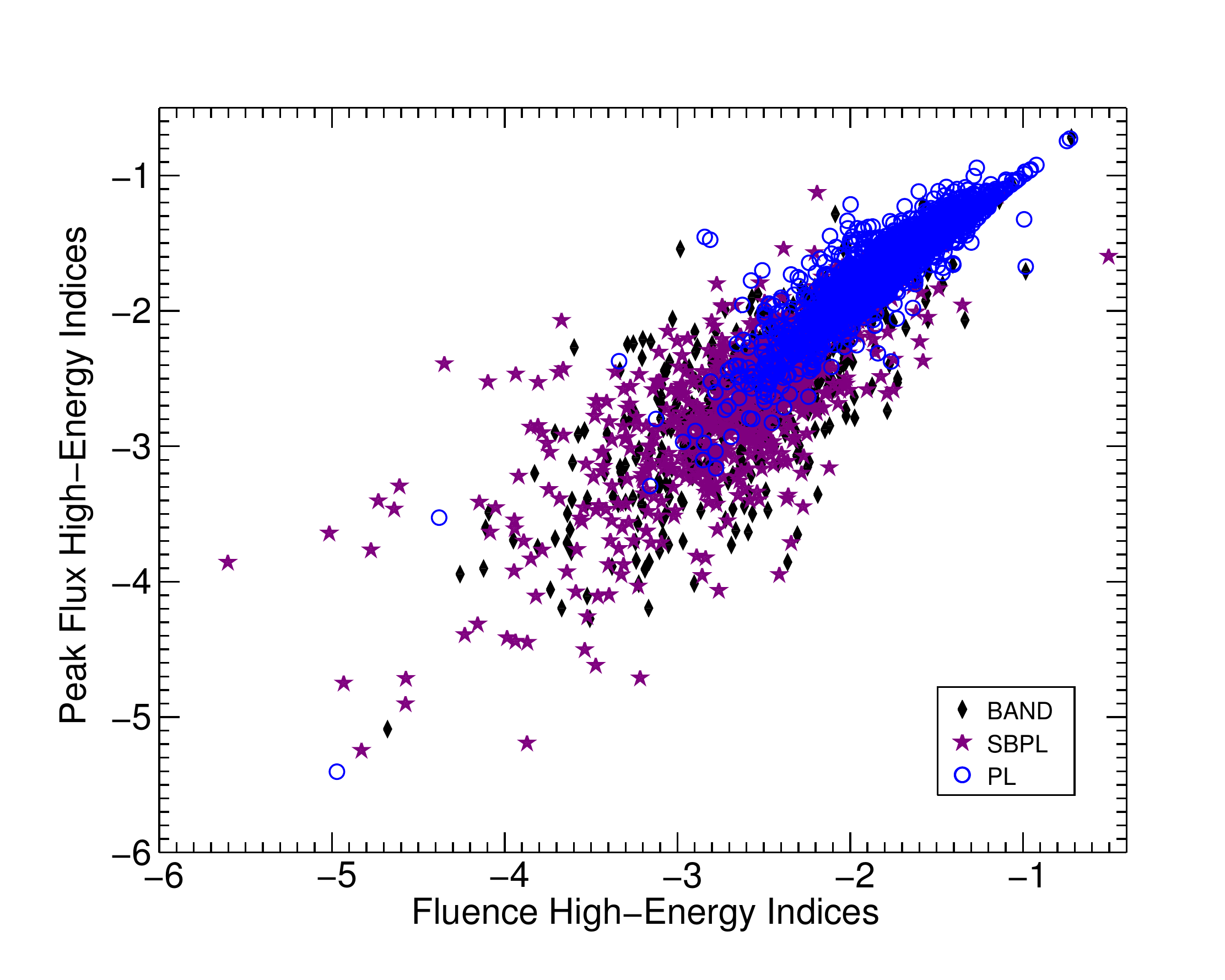}}\\
		\subfigure[]{\label{pffepeak}\includegraphics[scale=0.35]{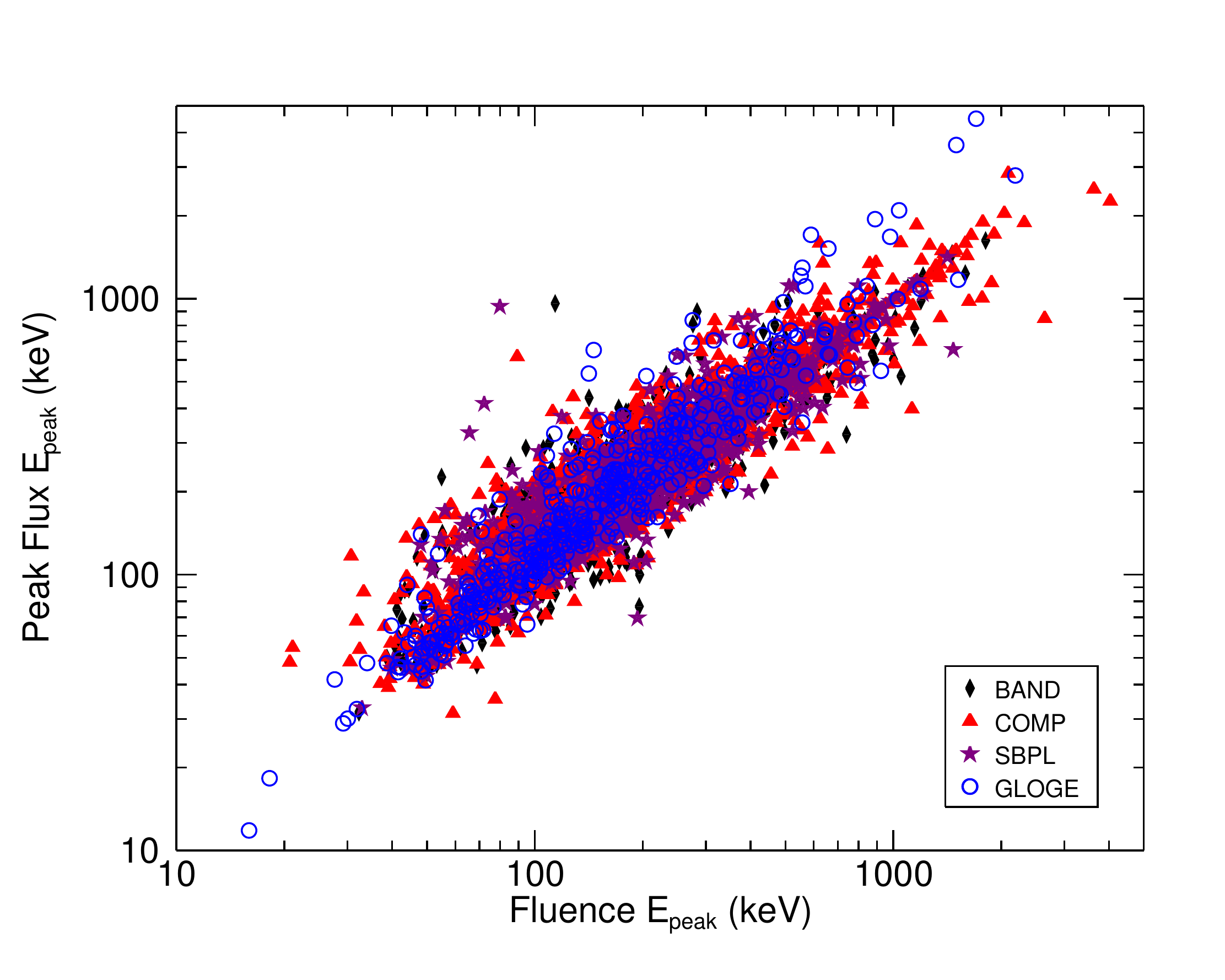}}
	\end{center}
\caption{Peak flux spectral parameters as a function of the fluence spectral parameters. For all three parameters there is 
evidence for a strong correlation between the parameters found for the fluence spectra and those for the peak flux spectra.  Note that the PL index is shown in both \ref{pffalpha} and \ref{pffbeta} for comparison. \label
{pff}}
\end{figure}

When studying the two types of spectra in this catalog, it is instructive to study the similarities and differences between the 
resulting parameters.  Plotted in Figure \ref{pff} are the low-energy indices, high-energy indices, and $E_{peak}$ energies of the 
peak flux spectra as a function of the corresponding parameters from the fluence spectra.  Most of the peak flux spectral 
parameters correlate with the fluence spectral parameters on the order of unity.  There are particular regions in each plot where 
outliers exist, and these areas are indications that either the GRB spectrum is poorly sampled or there exists significant spectral 
evolution in the fluence measurement of the spectrum that skews the fluence spectral values.  Examples of the former case are 
when the low-energy index is atypically shallow ($\gtrsim-0.5$) or the high-energy index is steeper than average ($\lesssim-3$).  
An example where spectral evolution may skew the correlation between the two types of spectra is apparent in the comparison 
of $E_{peak}$.  Here, it is likely that a fluence spectrum covering significant spectral evolution will produce a lower energy $E_
{peak}$ than is measured when inspecting the peak flux spectrum. Additionally, a Kolmogorov-Smirnov test was performed 
between each fluence and peak flux parameter, and the results are listed in Table~\ref{FluPFluxCompare}.  From these tests, it 
appears that only the distributions of the high-energy power law index for both the BAND and SBPL are significantly similar between 
the fluence and peak flux spectra.

\subsection{The BEST Sample}
The BEST parameter sample produces the best estimate of the observed properties of GRBs.  By using model comparison, the 
most preferred model is selected, and the parameters are inspected for that model.  The models contained herein and in most 
GRB spectral analyses are empirical models, based only on the data received; therefore the data from different GRBs tend to 
support different models.  Perhaps it will be possible to determine the physics of the emission process by investigating the 
tendencies of the data to support a particular model over others.  It is this motivation, as well as the motivation to provide a 
sample that contains the best picture of the global properties of the data, that prompted our investigation of the BEST sample.

\begin{figure}
	\begin{center}
		\subfigure[]{\label{indexbestf}\includegraphics[scale=0.35]{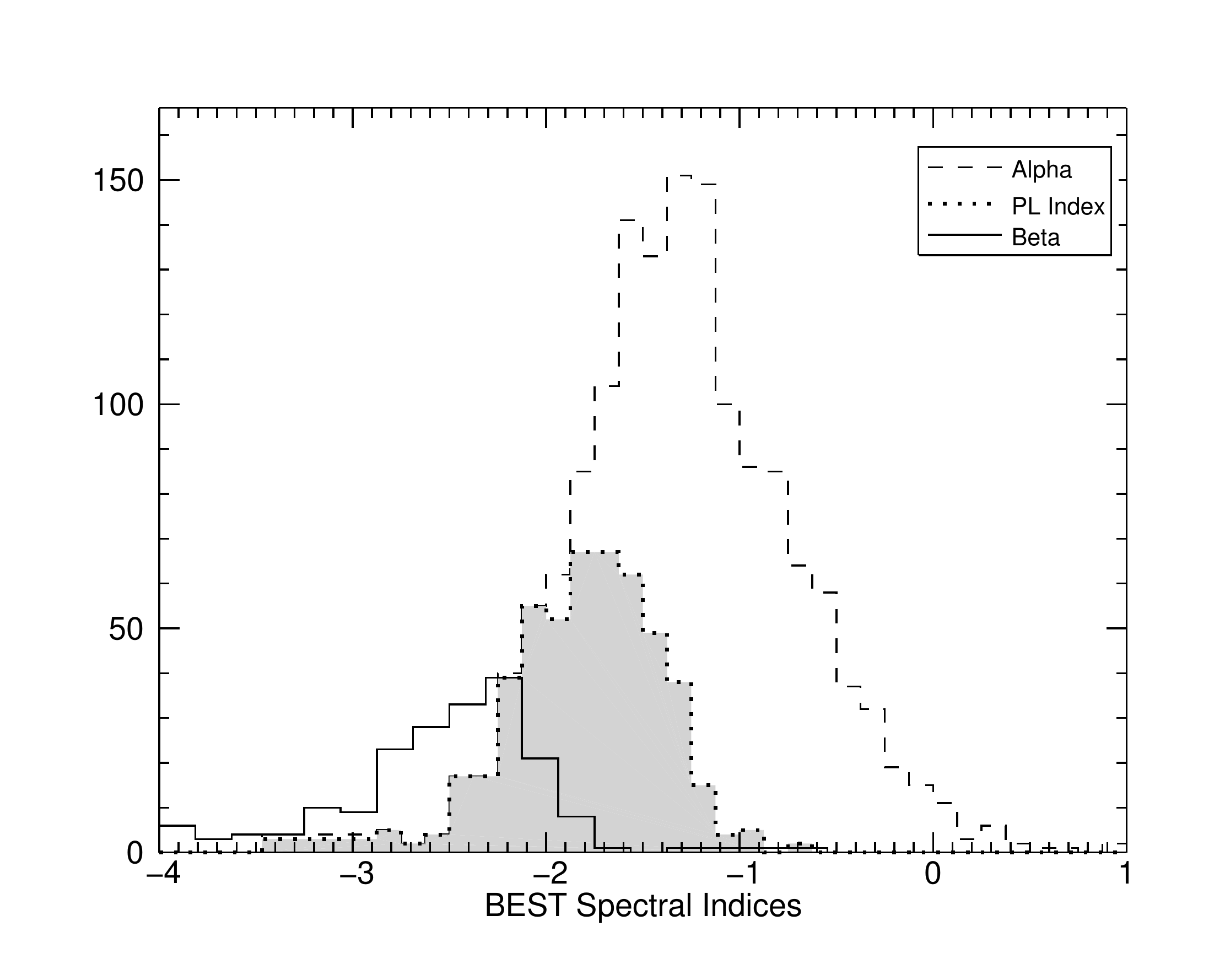}}
		\subfigure[]{\label{epeakbestf}\includegraphics[scale=0.35]{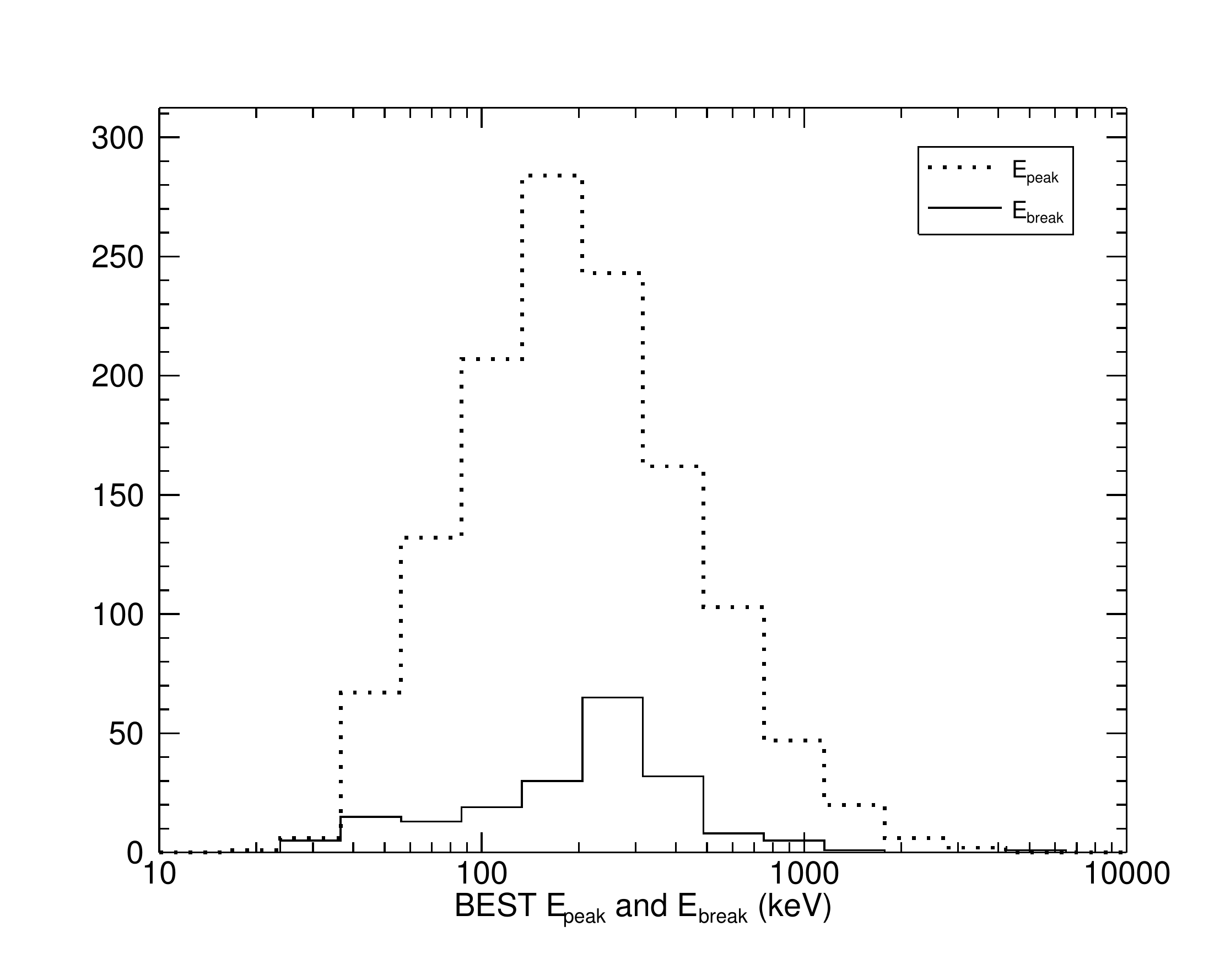}}\\
		\subfigure[]{\label{pfluxbestf}\includegraphics[scale=0.35]{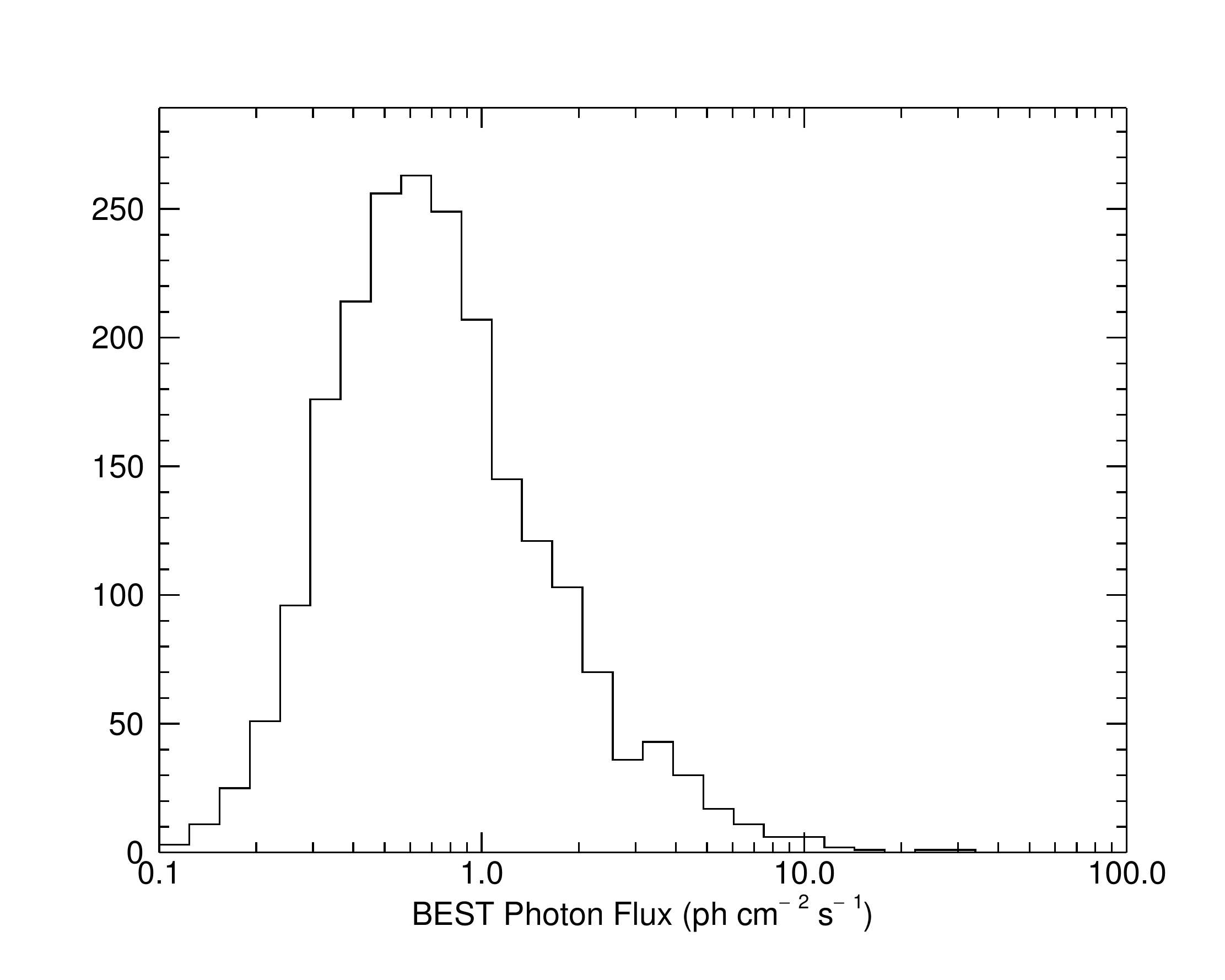}}
		\subfigure[]{\label{efluxbestf}\includegraphics[scale=0.35]{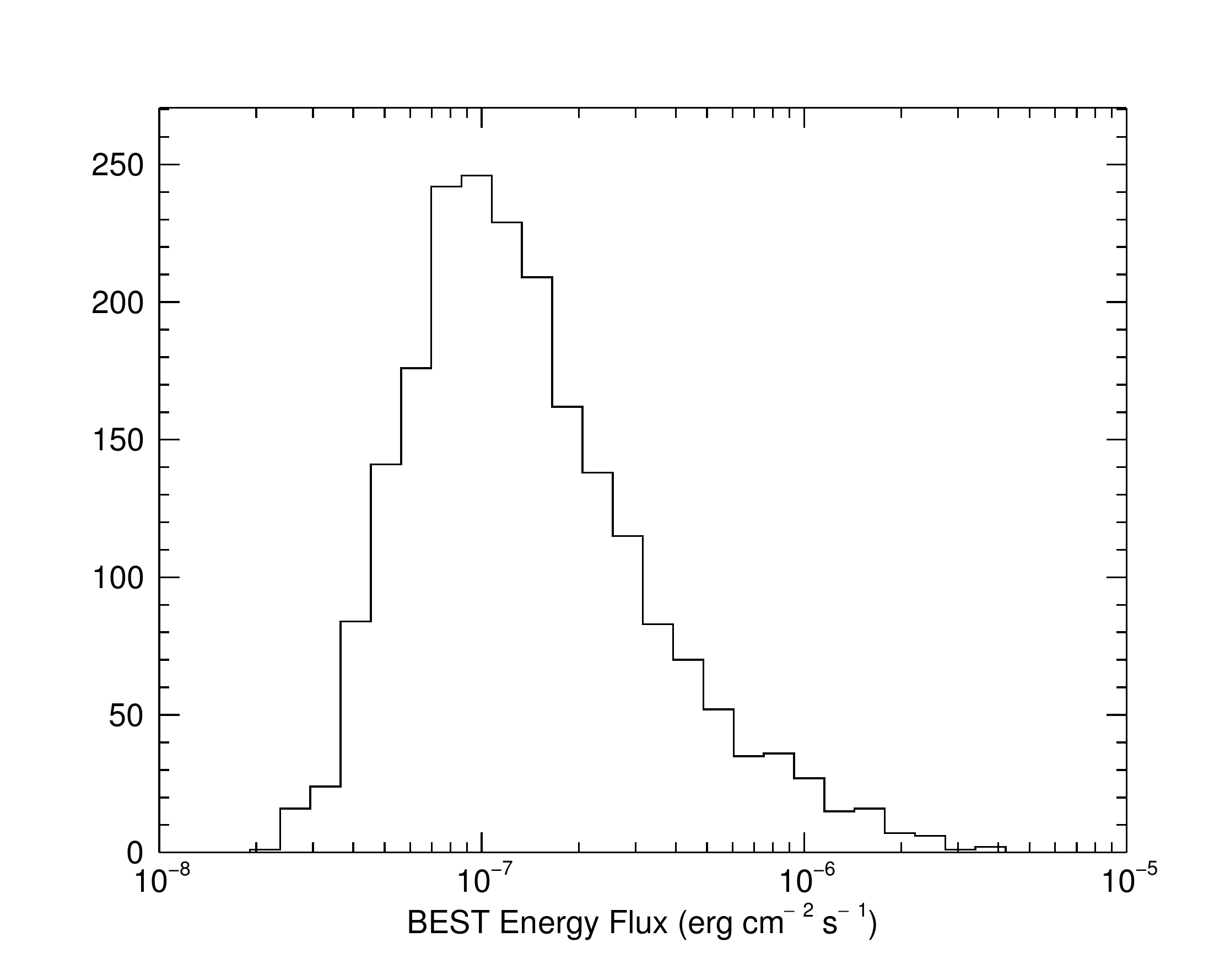}}
	\end{center}
\caption{Distributions of the BEST spectral parameters for the fluence spectra.  \ref{indexbestf} displays the selection of best 
low-energy and high-energy spectral indices.  The shaded distribution depicts the location of the distribution of the PL index.  
\ref{epeakbestf} shows the selection of the best $E_{peak}$ and $E_{break}$.  \ref{pfluxbestf} and \ref{efluxbestf} show the 
selection of the best photon flux and energy flux respectively.  \label{bestf}}
\end{figure}

\begin{figure}
	\begin{center}
		\subfigure[]{\label{indexbestp}\includegraphics[scale=0.35]{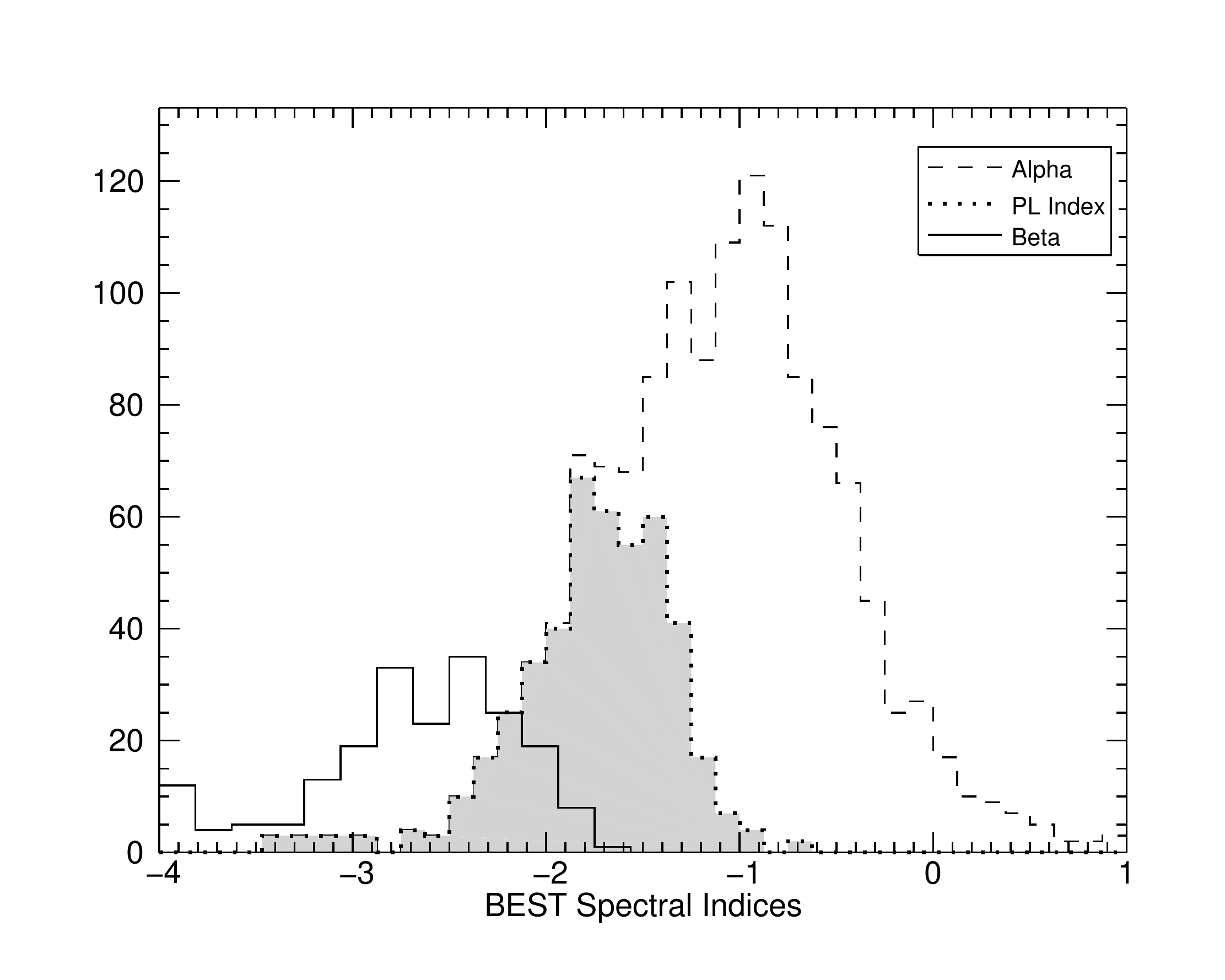}}
		\subfigure[]{\label{epeakbestp}\includegraphics[scale=0.35]{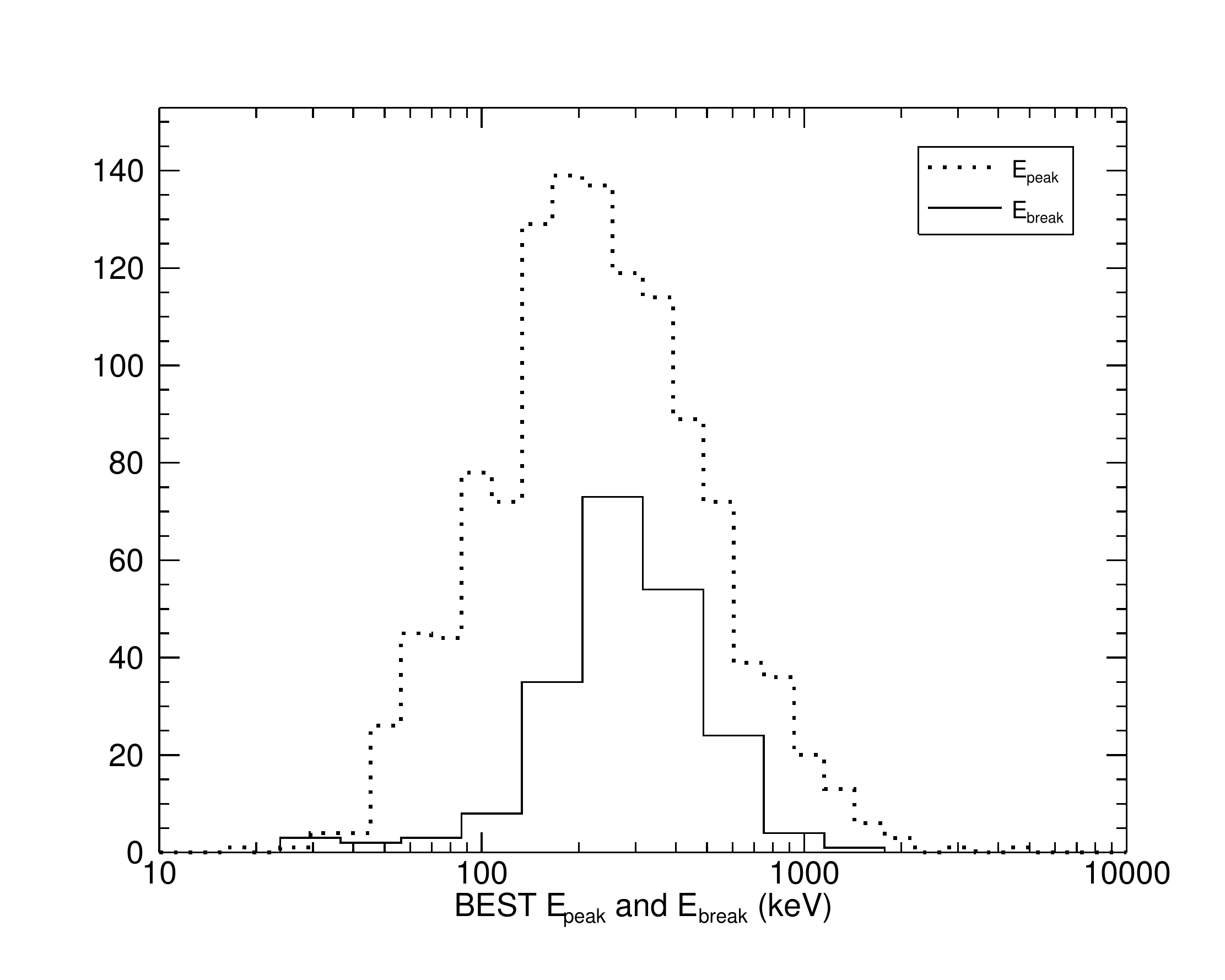}}\\
		\subfigure[]{\label{pfluxbestp}\includegraphics[scale=0.35]{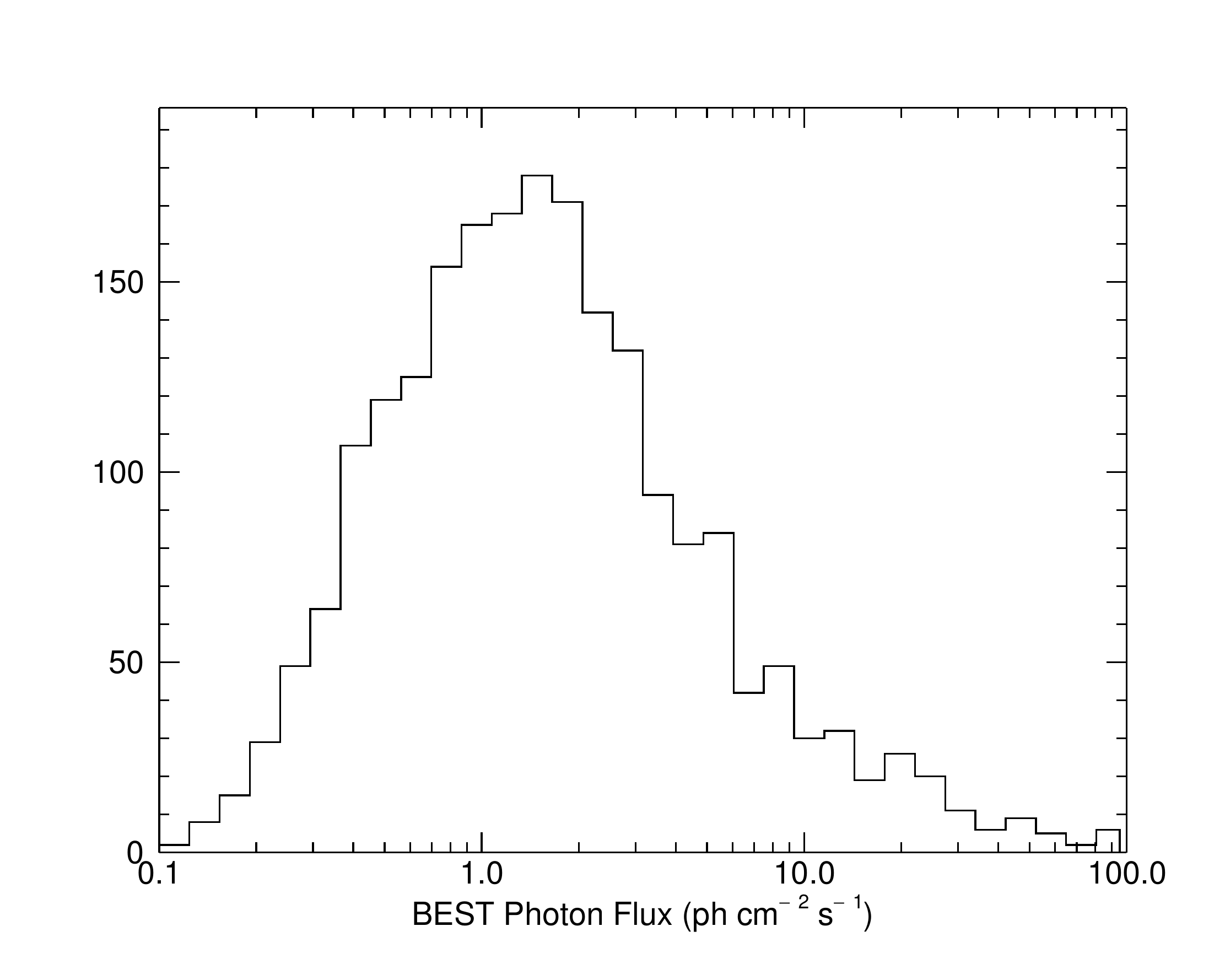}}
		\subfigure[]{\label{efluxbestp}\includegraphics[scale=0.35]{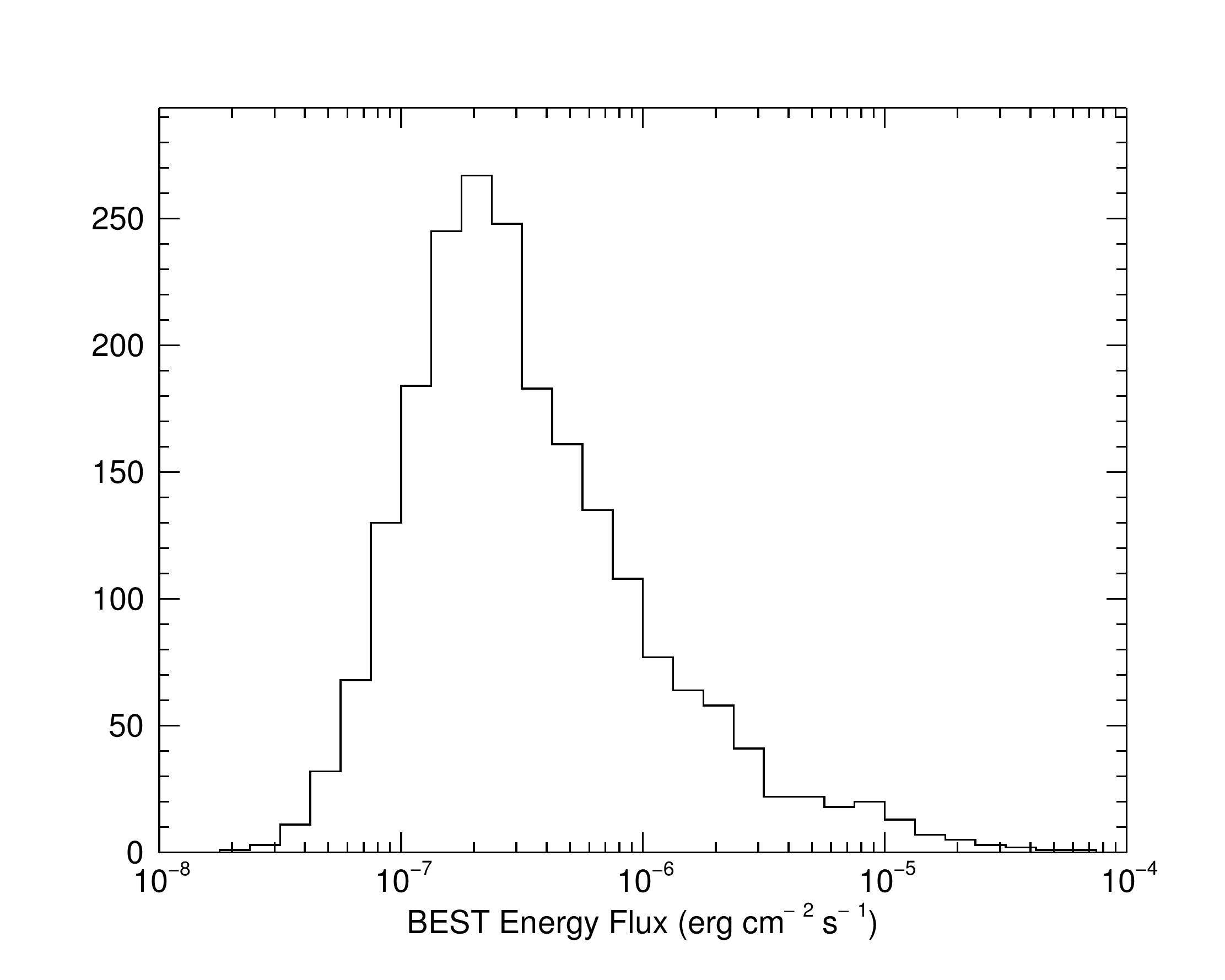}}
	\end{center}
\caption{Distributions of the BEST spectral parameters for the peak flux spectra.  \ref{indexbestp} displays the selection of best 
low-energy and high-energy spectral indices.  The shaded distribution depicts the location of the distribution of the PL index.  
\ref{epeakbestp} shows the selection of the best $E_{peak}$ and $E_{break}$.  \ref{pfluxbestp} and \ref{efluxbestp} show the 
selection of the best photon flux and energy flux respectively.  \label{bestp}}
\end{figure}

In Table \ref{BestTable} we present the composition of models for the BEST samples.  From this table, it is apparent that the 
spectral data from BATSE strongly favors the COMP model over the others in nearly half of all GRBs.  The BAND and SBPL are 
favored by a relative few GRBs in the catalog, possibly because of the limited spectral resolution of the data.  Table \ref{ParamTable} 
lists the sample mean and and standard deviations comparing the GOOD fluence and peak flux spectra as well as the BEST fluence 
and peak flux spectra.  In figures \ref{bestf} and \ref{bestp} the same error cuts used in the GOOD samples were also used for the 
BEST parameters.  Note that the PL index is statistically an averaging of the low- and high-energy power laws; due to the fact that 
BATSE spectral responses have a peak effective area at lower energies ($\sim$ 70 keV), we have included PL indices into the BEST 
low-energy index distribution.  Although in a number of cases the PL model is statistically preferred over the other models in this 
catalog, the spectral shape represented by the PL is inherently different from the shape of the other models.  Therefore, the PL index 
is not necessarily representative of either the low- or high-energy indices from the other models.  In Figures \ref{indexbestf} and \ref
{indexbestp} we show where the distribution of PL indices exists relative to the alpha and beta distributions.  The PL index represents 
23\% of the BEST fluence alpha distribution and 21\% of the BEST peak flux alpha distribution.  The fluence spectra, on the whole, 
have a steeper measured alpha and shallower beta than the peak flux spectra.  The alpha distribution for the fluence spectra 
peaks at slightly steeper than $-1$, while the peak flux low-energy spectral index peaks at slightly shallower than $-1$.  Conversely, 
the beta distribution for the fluence spectra peaks at $-2.1$ and the peak flux high-energy spectral index peaks at $-2.4$.

The $E_{peak}$ and $E_{break}$ distributions are similar between the two spectra, however.  The fluence spectra $E_{peak}$ 
peaks at $\sim$200 keV as does the peak flux spectra $E_{peak}$ and the $E_{break}$ peaks at more than 200 keV.  Note that 
the $E_{break}$ distributions are significantly smaller in comparison to the $E_{peak}$ distributions, especially for the fluence 
spectra, simply because the errors are not as well constrained  for $E_{break}$.  As is shown in Figures \ref{pfluxbestf} and \ref
{pfluxbestp} respectively, the fluence spectra photon flux peaks at $0.75^{+1.09}_{-0.37}$ photons $\rm cm^{-2} \ s^{-1}$ 
and the peak flux photon flux peaks at $1.61^{+4.32}_{-1.02}$ photons $\rm cm^{-2} \ s^{-1}$.  Comparatively, the fluence 
energy flux peaks at $(1.40^{+2.49}_{-0.71}) \times 10^{-7} \rm erg \ cm^{-2} \ s^{-1}$, and the peak flux energy flux peaks at  $(3.44^{+11.5}_{-2.09}) \times 10^{-7} \rm erg \ cm^{-2} \ s^{-1}$, as is shown in Figures \ref{efluxbestf} and \ref{efluxbestp}.

\begin{figure}
	\begin{center}
		\subfigure[]{\label{fluencealpha}\includegraphics[scale=0.35]{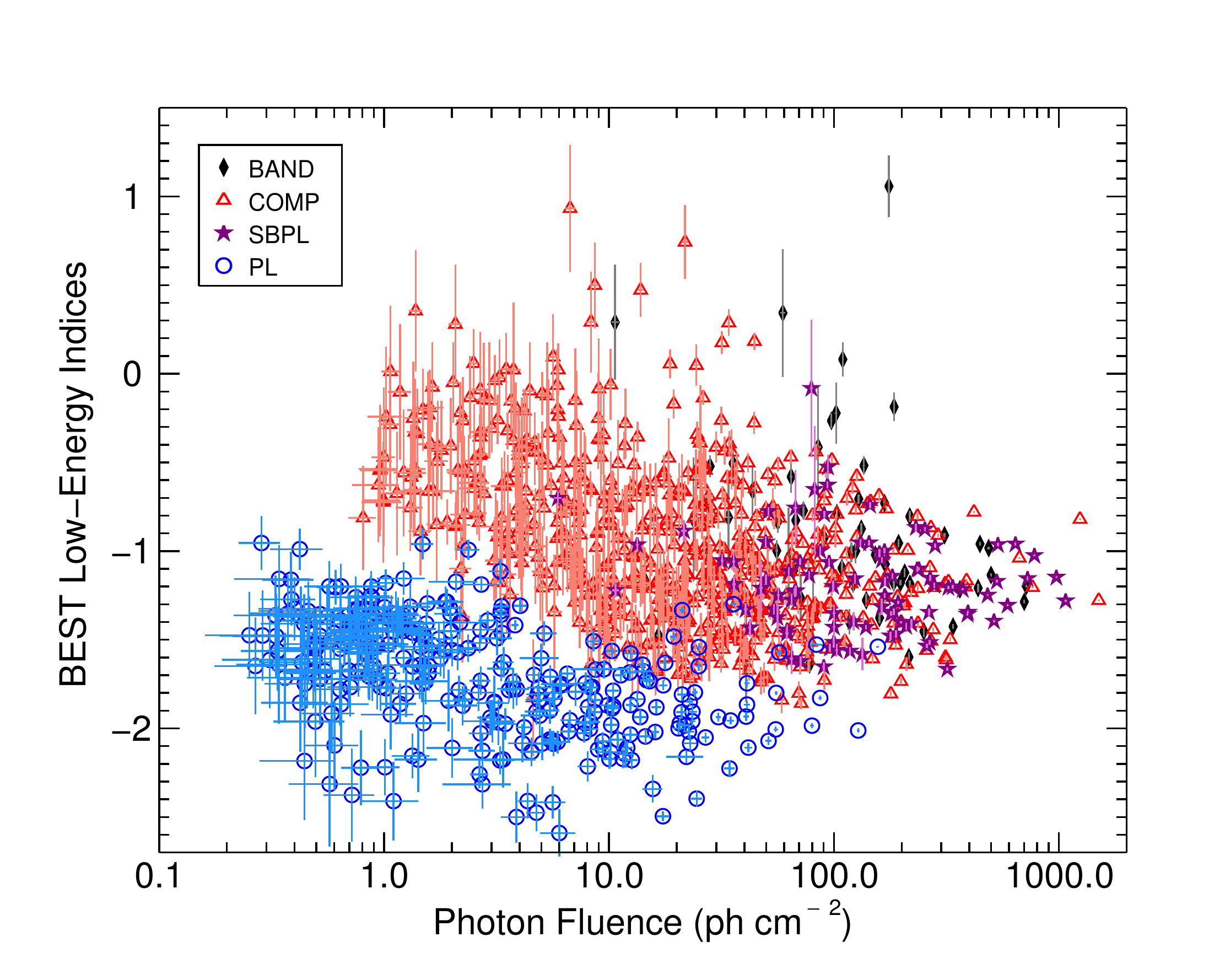}}
		\subfigure[]{\label{fluencebeta}\includegraphics[scale=0.35]{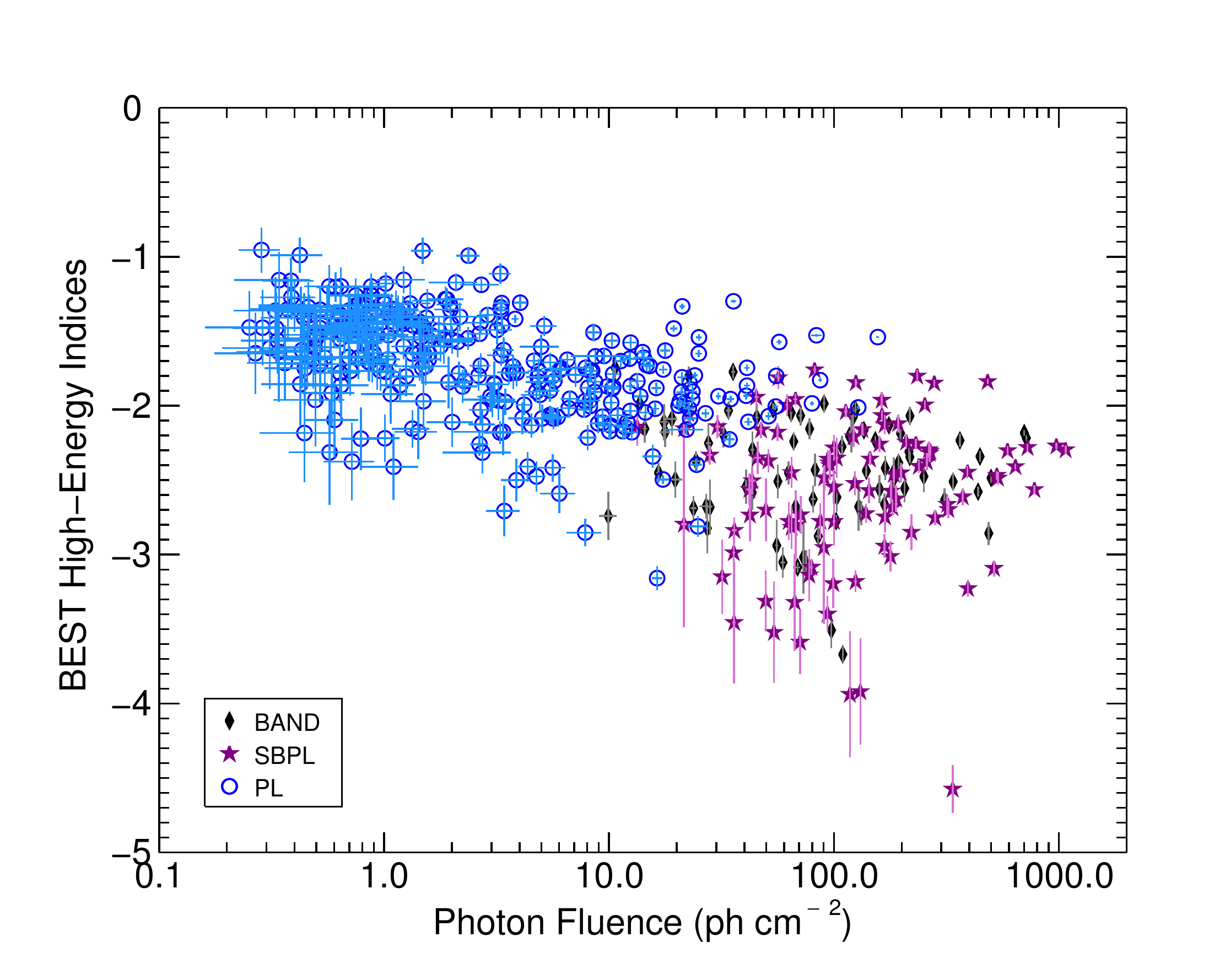}}\\
		\subfigure[]{\label{fluenceepeak}\includegraphics[scale=0.35]{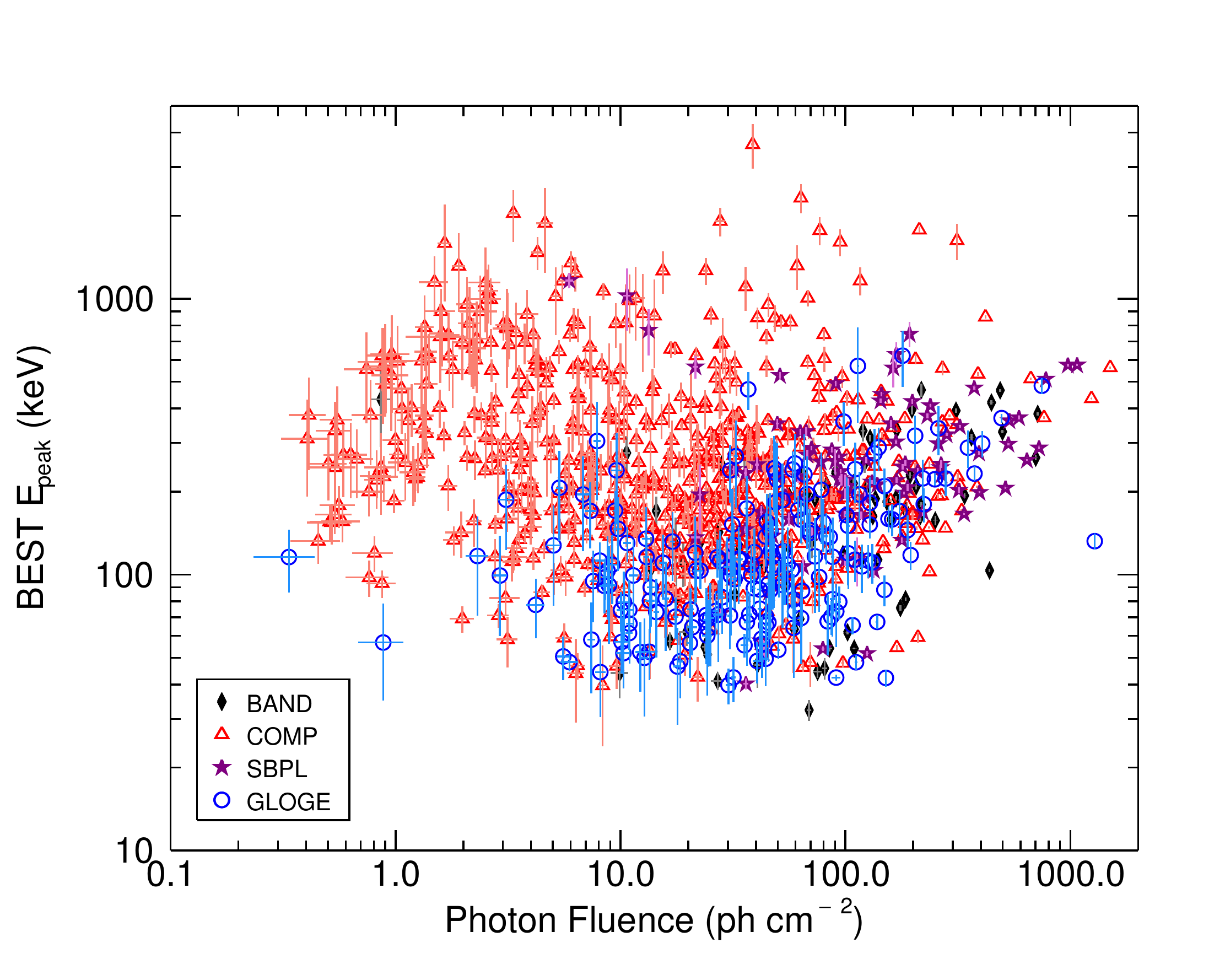}}
	\end{center}
\caption{BEST fluence spectral parameters as a function of the model photon fluence.  Note that the PL index is shown in both 
\ref{fluencealpha} and \ref{fluencebeta} for comparison. \label{fluenceparms}}
\end{figure}

\begin{figure}
	\begin{center}
		\subfigure[]{\label{fluxalpha}\includegraphics[scale=0.35]{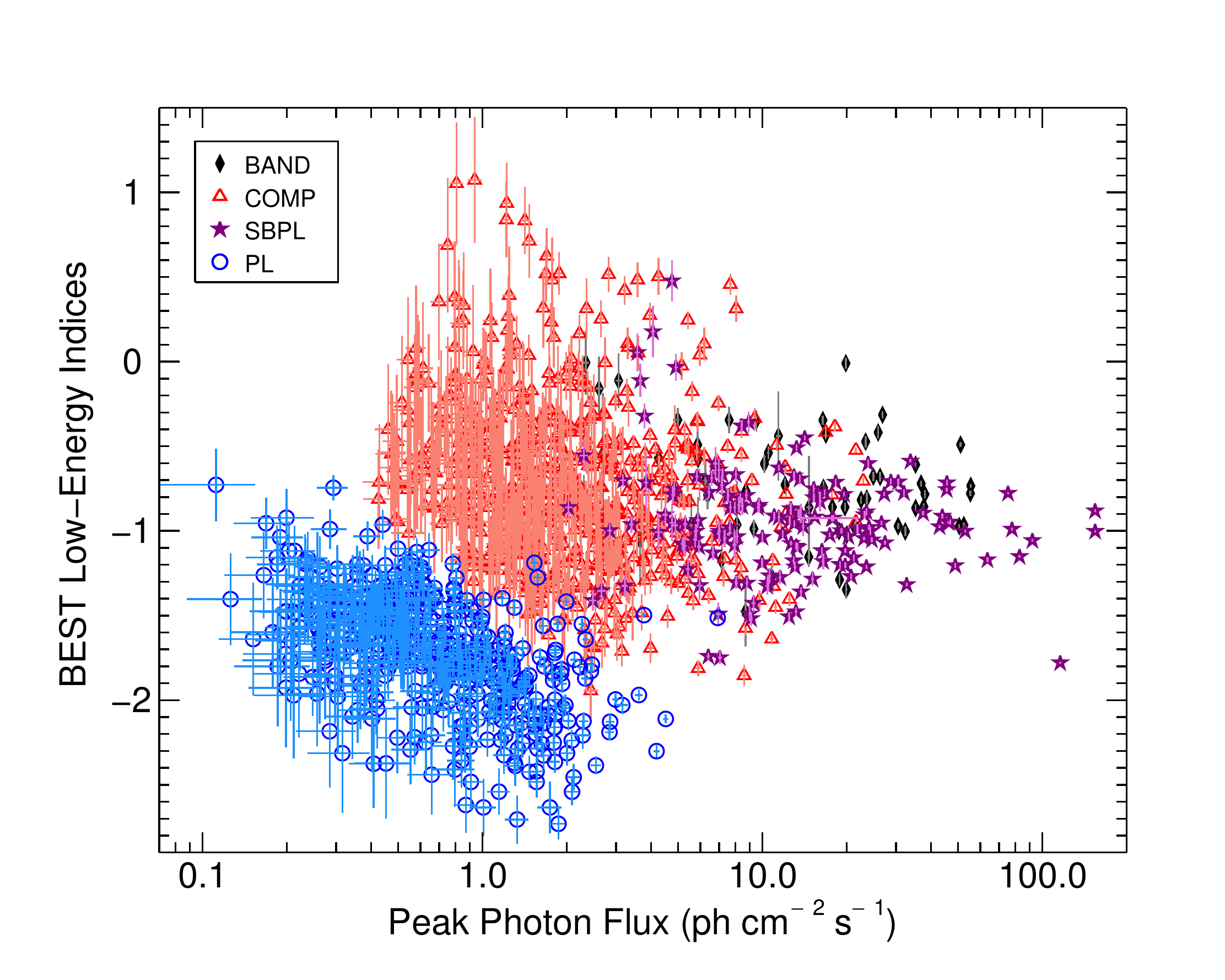}}
		\subfigure[]{\label{fluxbeta}\includegraphics[scale=0.35]{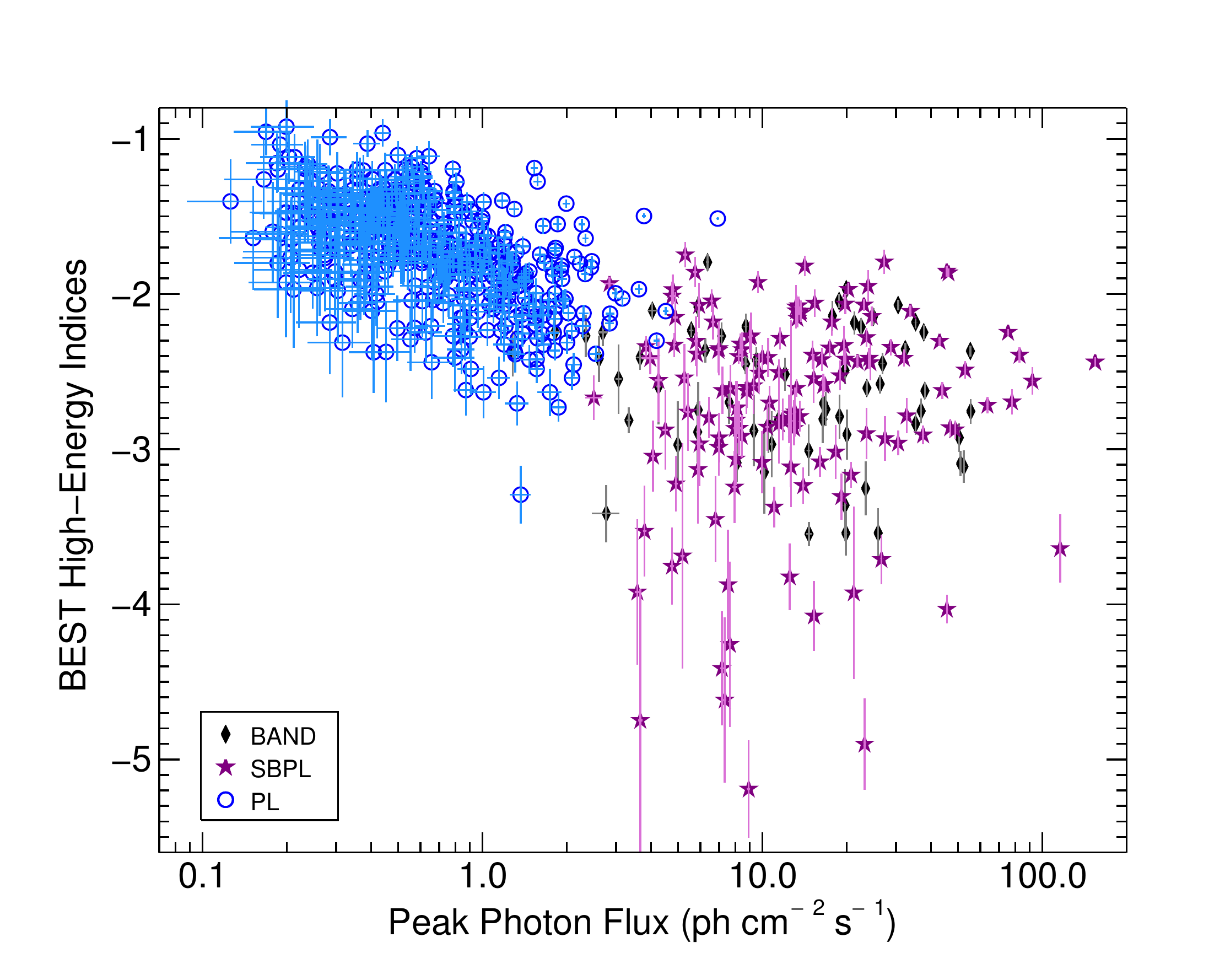}}\\
		\subfigure[]{\label{fluxepeak}\includegraphics[scale=0.35]{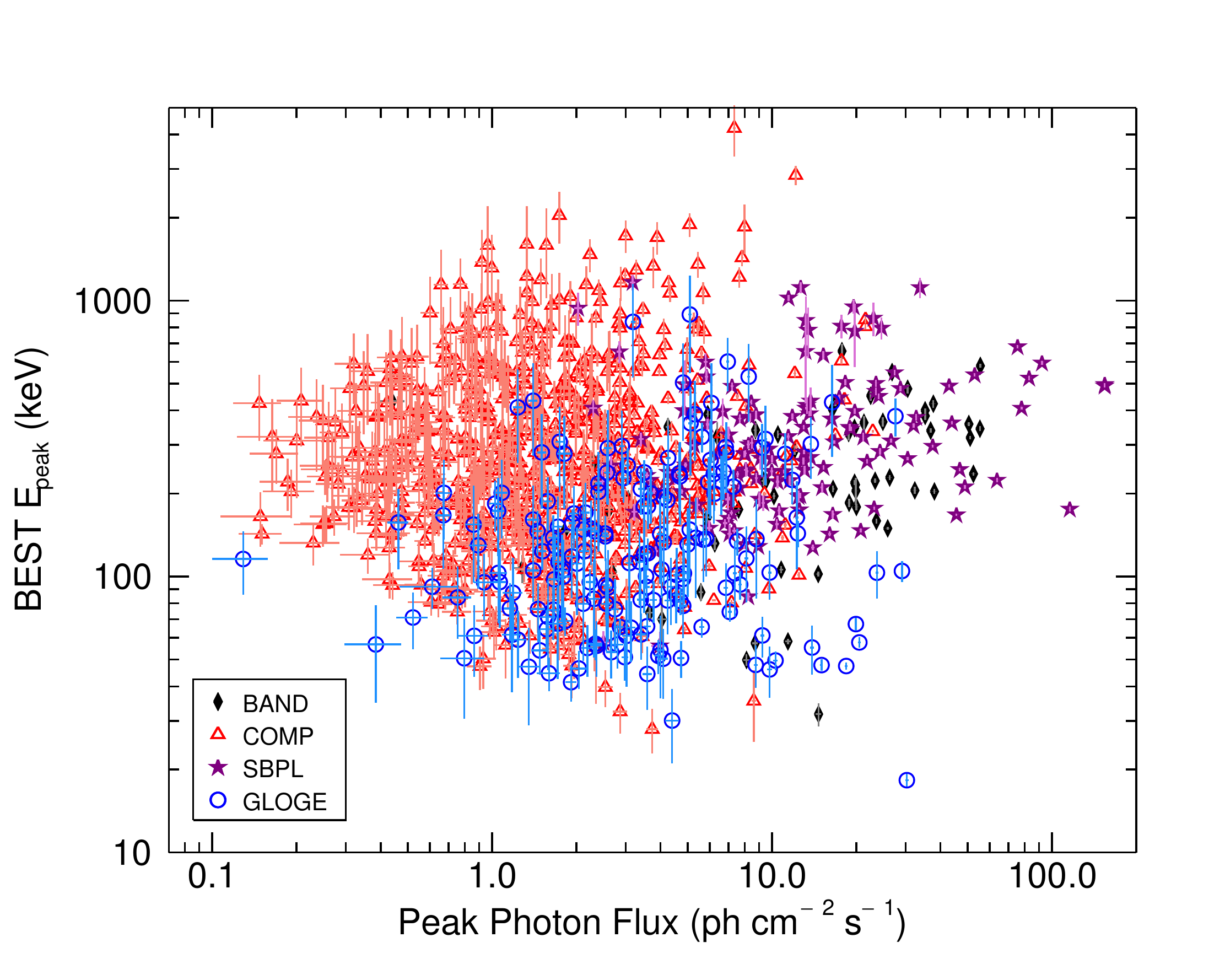}}
	\end{center}
\caption{BEST peak flux spectral parameters as a function of the model peak photon flux.  Note that the PL index is shown in both \ref
{fluxalpha} and \ref{fluxbeta} for comparison. \label{fluxparms}}
\end{figure}

To aid in the study of the systematics of the parameter estimation, as well as to garner the effect statistics has on the fitting 
process, we investigate the behavior of the parameter values as a function of the photon fluence and peak photon flux for the fluence 
and peak flux BEST spectra, respectively.  These distributions are shown in Figures \ref{fluenceparms} and \ref{fluxparms}.  When  
fitting the time-integrated spectrum of a burst, we find the low- and high-energy indices trend toward steeper values for 
exceedingly more fluent spectra.  These figures show that the simple PL index trends from shallow value of $\sim-1.3$ to a steeper 
value of $\sim-2.3$.  The low-energy index for a spectrum with curvature tends to exhibit an unusually shallow value of $\sim-0.4$ for 
extremely low fluence spectra, and steepens to $\sim-1.5$.  Similarly, the high-energy index trends from $\sim-1.6$ at low 
fluence to $\sim-3$ at high fluence, although this is complicated by unusually steep and poorly constrained indices that 
indicate that an exponential cutoff results in a more reliable spectral fit.  There are a number of cases where the COMP spectral 
index is unusually steep ($\sim$--2), which may indicate that the GRB is particularly soft and that $E_{peak}$ exists sufficiently below 
the detector bandpass.  In these cases, BATSE is only observing the high-energy power law of the GRB, which may display an 
exponential cutoff similar to other GRBs at higher energies.  Note that in this situation, the measured $E_{peak}$ will not be the true 
peak of the $\nu F_\nu$ spectrum, since the true $E_{peak}$ will exist below 20 keV.  In other cases, the high-energy power law may 
not be cutoff and is well-modeled by a PL model, which could be indicated by high flux GRBs with index steeper than --2.
 
\begin{figure}
	\begin{center}
		\includegraphics[scale= 0.7]{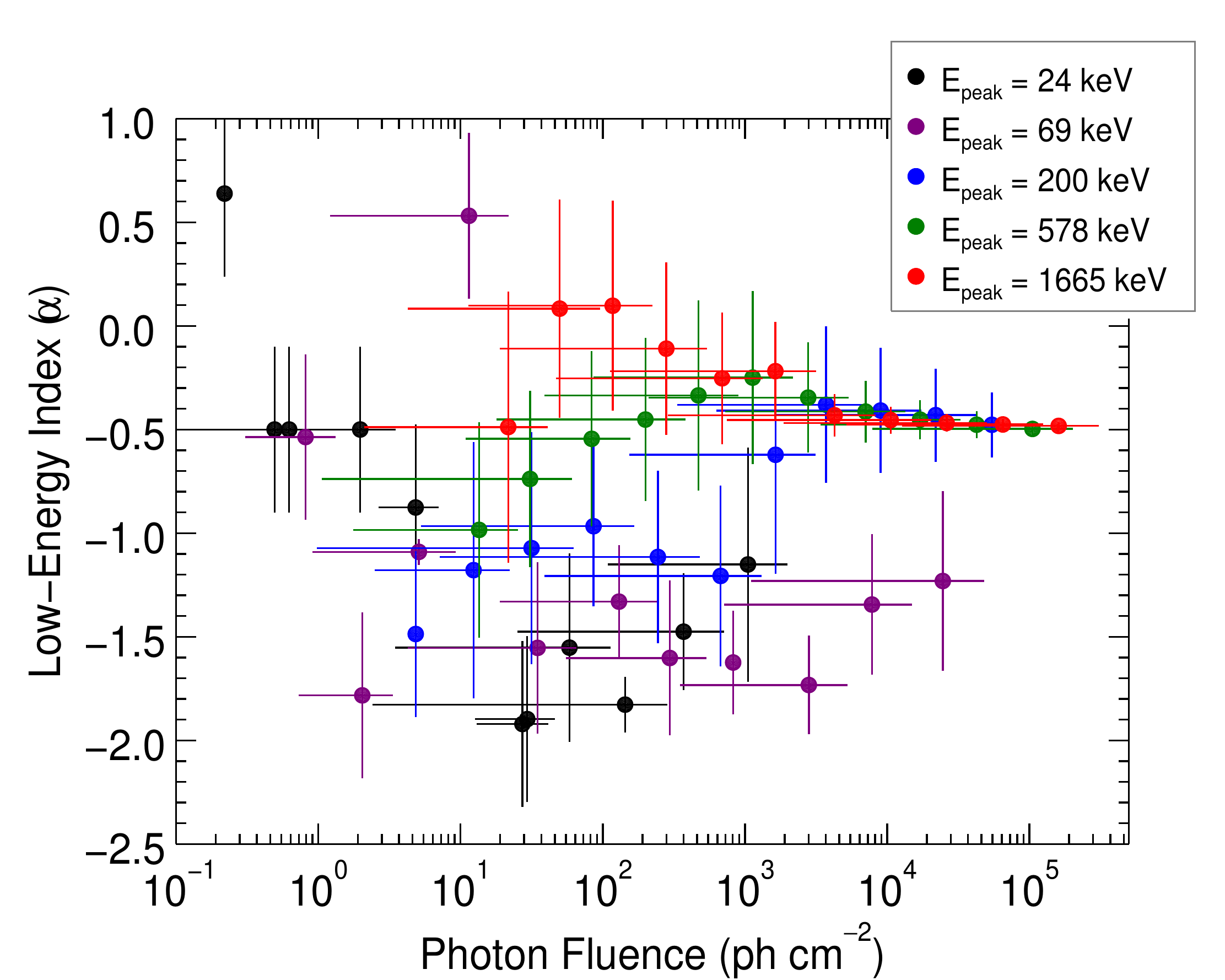}
	\end{center}
\caption{Simulations of fitting the BAND low-energy index (alpha) as a function of the spectral photon fluence for five different values of $E_{peak}$. \label{AlphaPFluenceSim}}
\end{figure}

Additionally, to test if the trend seen in the low-energy index versus photon fluence is a product purely of unknown systematics 
issues from the spectral fitting of weak GRBs, we performed five sets of 10000 simulations of the BAND model by using five different 
$E_{peak}$ values and a low energy index of --0.5 and a high energy index of --2.6.  Different photon fluences were also simulated 
by changing the simulated live time of the spectra using 11 equally spaced values in logarithmic space spanning from 0.128 -- 1049 
s.  The simulations were then fit and the mean and uncertainty of the resulting parameters from each set of simulations was 
calculated, and the BAND alpha-photon fluence figure is shown in Figure \ref{AlphaPFluenceSim}.  From these simulations we 
deduce that the trend seen in the catalog does not appear to be from systematic effects due to the low fluence of GRBs. In fact, these 
simulations show that on the average, a GRB with a fluence  $<$10 photons $\rm cm^{-2}$ will tend to under-predict the true value of 
the BAND alpha, which would show a positive trend if there was only this one systematic effect.  The other finding from these 
simulations is that only for $E_{peak} \gtrsim 100$ keV can the BAND alpha value be reliably found even out to $\sim$1000 photons 
$\rm cm^{-2}$.

When inspecting the $E_{peak}$ as a function of photon fluence, a trend is much less apparent.  If a burst is assumed to have 
significant spectral evolution, then obviously the $E_{peak}$ will change values through the time history of the burst, typically 
following the traditional hard-to-soft energy evolution.  For this reason, spectra that are integrated over increasingly longer time 
intervals will tend to suppress the highest energy of $E_{peak}$ within the burst, so a general decrease in $E_{peak}$ is expected 
with longer integration times.  However, the photon fluence convolves the integration time with the photon flux so that an intense 
burst with a short duration may have approximately the same fluence as a much longer but less intense burst, albeit in a higher 
$E_{peak}$.  This causes significant broadening to the decreasing trend as shown in Figure \ref{fluenceepeak}.  The distribution of 
parameters as a function of the peak photon flux show similar trends, although due to the smaller statistics of the shorter 
integration time, the trends are much more dispersed.  The distributions shown in Figure \ref{fluxparms} are more susceptible to 
uncertainty because of the generally smaller statistics involved in the study of the peak flux of the GRB, except in some cases 
where the peak photon flux is similar to the photon fluence.  Ignoring the regions where the parameters are poorly constrained, 
another trend emerges from the low-energy indices; they appear to become slightly more shallow as the photon flux increases.  
The high-energy indices, however, appear to be unaffected by the photon flux, except those that are unusually steep and indicate 
that an exponential cutoff may be preferred.  Additionally, an obvious trend emerges from all Figures in \ref{fluenceparms} and \ref
{fluxparms}: GRBs with lower photon flux and photon fluence are more likely to be best fit by a simple power law, and brighter GRBs 
with higher photon flux or fluence will typically be best fit by the more complex BAND or SBPL models.  This trend displays a model-
dependent analog to Figure \ref{countrate}, which shows that the preferred model complexity is correlated to the amount of data 
present from the burst.

\subsection{Comparisons to Previous Results}
A comparison between different catalogs can be instructive, especially when noting the difference in detector geometry, 
sensitivity, and bandpass.  For example, the {\it Fermi}/GBM Spectral Catalog \citep{GBMSpecCat} contains spectral information 
of many GRBs that were observed by detectors with a total collecting area that is $\sim$1/8 of BATSE, but with a larger bandpass 
($\sim$8 keV -- 40 MeV) and increased high-energy sensitivity.  In general, there are no large discrepancies between the two 
catalogs, although there are subtle differences between the parameter distributions.  Table \ref{BestComparison} displays a 
comparison of the sample mean and standard deviations of the GBM spectral catalog and this catalog. The GBM spectral catalog 
contains, on average, shallower low- and high-energy power law indices when compared to this catalog.  The differences are a 
few tenths of an index, and the distributions in the GBM catalog are not as wide, most likely due to a much smaller sample of 
GRBs.  The average $E_{peak}$ in this catalog is slightly softer and the average $E_{break}$ is slightly harder than was found 
from the GBM bursts.  As expected, the photon flux distribution for BATSE is shifted to smaller fluxes due to the much larger 
collecting area of BATSE, enabling the detection and observation of weaker bursts.  Similarly, the energy flux distribution for GBM 
is shifted to larger fluxes due to its increased high-energy sensitivity from the BGO detectors.

Additionally, when comparing our results to the BATSE catalog of bright GRBs \citep{Kaneko06}, there are some marked 
differences as well, although many of these arise because the \citet{Kaneko06} sample studied only the 350 brightest BATSE 
GRBs.  We investigated the sample mean of the time-integrated spectra for comparison (see Table \ref{BestComparison}) and 
found that the average low-energy index and high-energy index on average was slightly steeper when compared to \citet
{Kaneko06}.  The time-averaged $E_{peak}$ in this catalog, however, is on average softer, most likely due to the fact that included all 
bursts, not just the brightest bursts detected.  This also affects the $E_{break}$, which is also found to be slightly 
softer in this catalog; as expected the \citet{Kaneko06} catalog reports larger average photon and energy fluxes.

\section{Summary}
BATSE provided a uniquely large, homogenous sample of GRB spectra over the 20 keV -- 2 MeV energy band.  The broad energy 
range and long operational life of BATSE translates to currently the largest and most comprehensive collection of GRB temporal 
and spectral properties.  The distributions contained here are similar to those shown by previous studies, yet contain differences 
that display the usefulness of studying GRBs over different energy bands and sensitivities.  We have shown in many cases that 
the fitted spectrum of a GRB depends in large part on its intensity as well as the detector sensitivity.  This observation implies that 
weak GRBs may have the same inherent spectrum as their more intense counterparts, yet we are unable to accurately determine 
their spectrum, reinforcing the importance of comparing the spectral parameter distributions from different acceptable models.  

In conclusion, we have presented a systematic analysis of 2145 GRBs detected with BATSE during its entire period of operation. 
This catalog contains five basic photon model fits to each burst, using two different selection criteria to produce over 19000 
spectra and facilitate an accurate estimate of the spectral properties of these GRBs.  We have described subsets of the full results 
in the form of data cuts based on parameter uncertainty, as well as employing model comparison techniques to select the most 
statistically preferred model for each GRB.  The analysis of each GRB was performed as objectively as possible, in an attempt to 
minimize biased systematic errors inherent in subjective analysis.  The methods we have described treat all bursts equally, and 
we have presented the ensemble of observed spectral properties of BATSE GRBs. This catalog represents the largest sample of 
GRBs to date, and constitutes a wide array of GRB spectral properties. Certainly there are avenues of investigation that require 
more detailed work and analysis or perhaps a different methodology.  This catalog should be treated as a starting point for future 
research on interesting bursts and ideas.  As has been the case in previous GRB spectral catalogs, we hope this catalog will be of 
great assistance and importance to the search for the physical properties of GRBs and other related studies.

\section{Acknowledgments}
AG acknowledges the support of the Graduate Student Researchers Program funded by NASA.

\clearpage

\section{Appendix A \\
Comparison of 16-Channel CONT Spectra to 128-Channel HERB Spectra}
The CONT data used in this catalog contains a total of 16 spectral channels (14 usable) over a broad $\sim$20--2000 keV energy 
range.  To ensure that there is not significant loss of spectral information compared the 128-channel HERB data over the same 
energy range, a set of simulations was performed to ascertain the effectiveness of each data type to accurately find the correct 
parameters of a Band function spectrum.  To perform these simulations, a 4-dimensional grid was constructed in parameter space, 
and each dimension was evenly sampled with 11 input parameter values sufficiently spanning the parameter space.  The values 
were as follows:
\begin{itemize}
\item Amplitude:  0.001, 0.003,  0.01, 0.03,  0.1, 0.3, 1.0, 3.0, 10, 31, 100 
\item $E_{peak}$: 1, 3, 8, 24, 69, 200, 577, 1665, 4804, 13863, 40000
\item Low-Energy Index: -2.0, -1.7, -1.4, -1.1, -0.8, -0.5, -0.2, +0.1, +0.4, +0.7, +1.0
\item High-Energy Index: -5.0, -4.6, -4.2, -3.8, -3.4, -3.0, -2.6, -2.2, -1.8, -1.4, -1.0
\end{itemize}
Each grid point comprised a value for each parameter, such that there were $11^4$ grid points.  At each grid point, 
a set of 1000 simulated spectra were created with input parameters at that grid point, the resulting photon spectra were folded 
through the LAD response matrix, and Poisson noise was added to the total and background count rate for each spectrum.  The 
resulting synthetic spectra were then fit with a Band function to determine the spectral parameters.  This procedure was completed 
using the CONT DRM as well as the HERB DRM from Trigger \# 1815.  Note that in many cases the combination of parameters of the 
simulated spectrum may represent an extreme parametrization of the Band function, which may not necessarily reflect the typical 
observed GRB spectrum.

\begin{figure}
	\begin{center}
		\subfigure[]{\label{ampsigma}\includegraphics[scale=0.35]{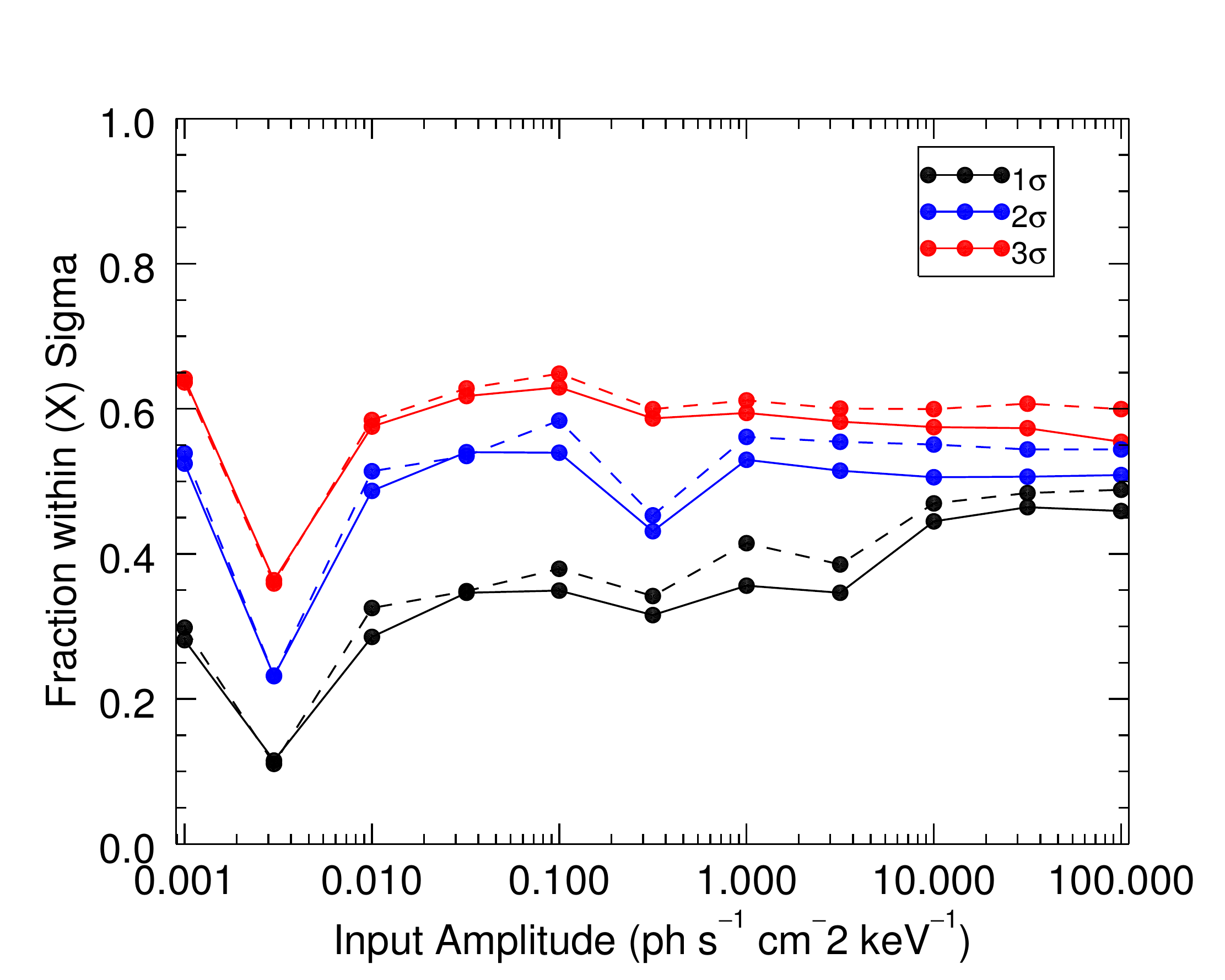}}
		\subfigure[]{\label{epeaksigma}\includegraphics[scale=0.35]{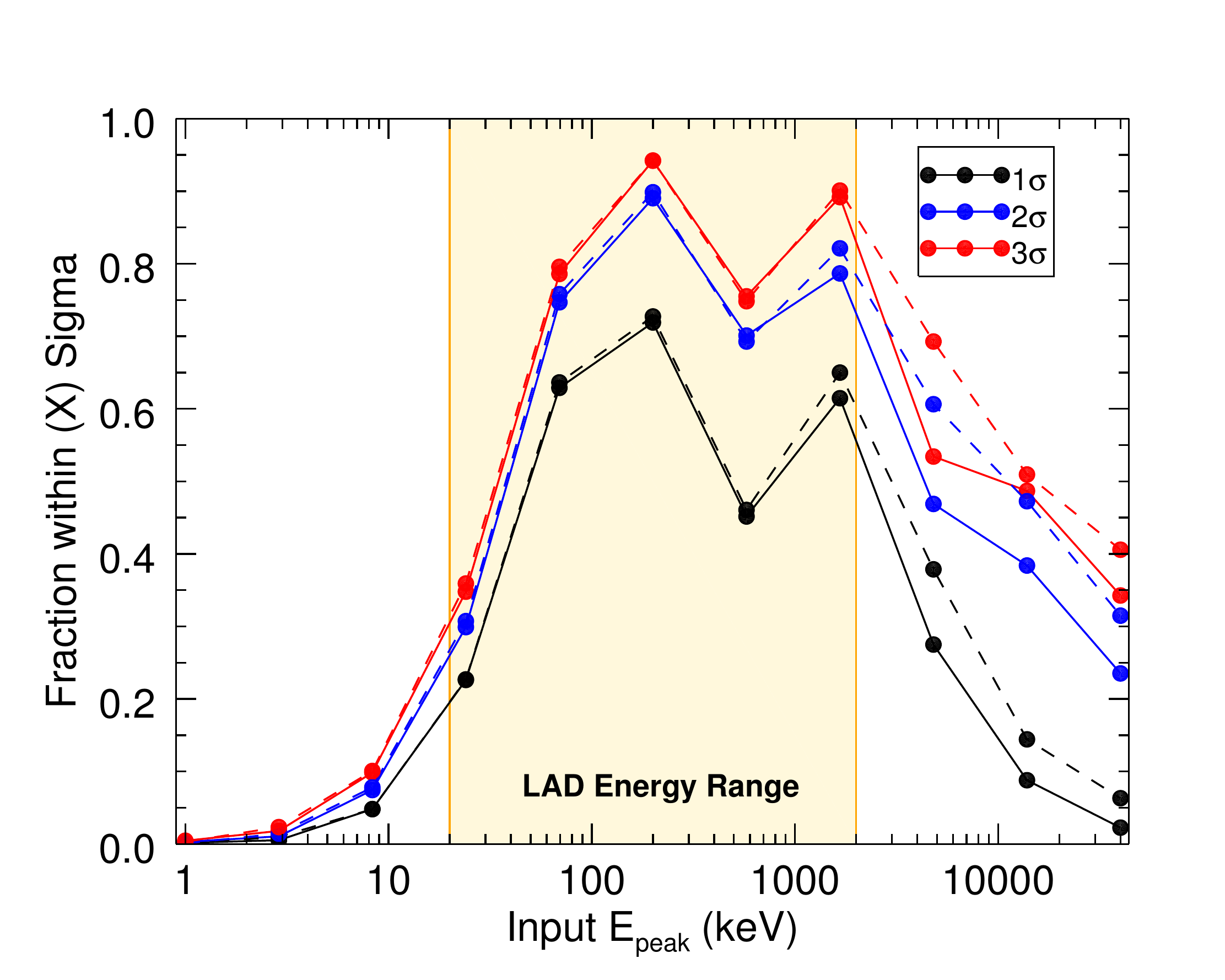}}\\
		\subfigure[]{\label{alphasigma}\includegraphics[scale=0.35]{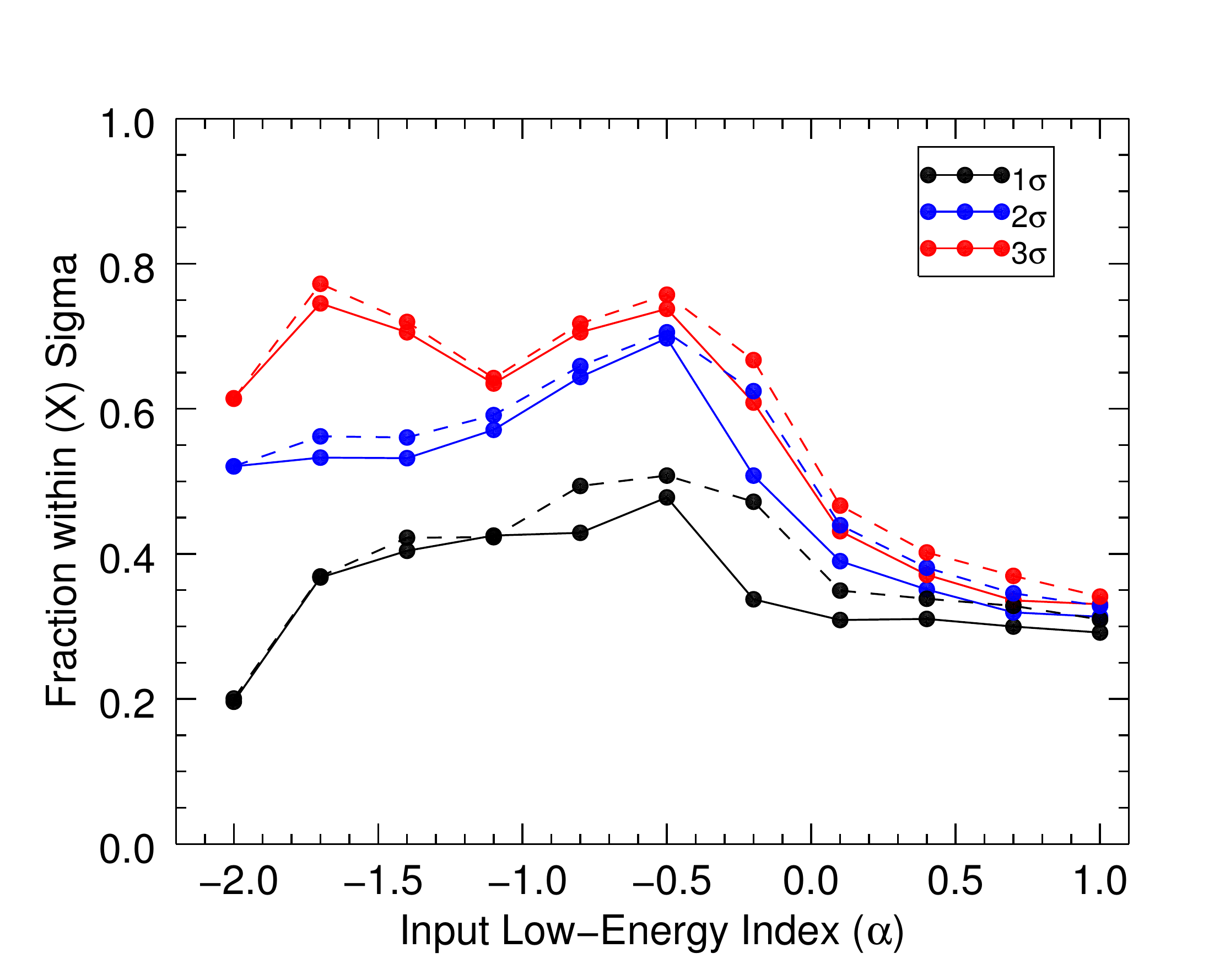}}
		\subfigure[]{\label{betasigma}\includegraphics[scale=0.35]{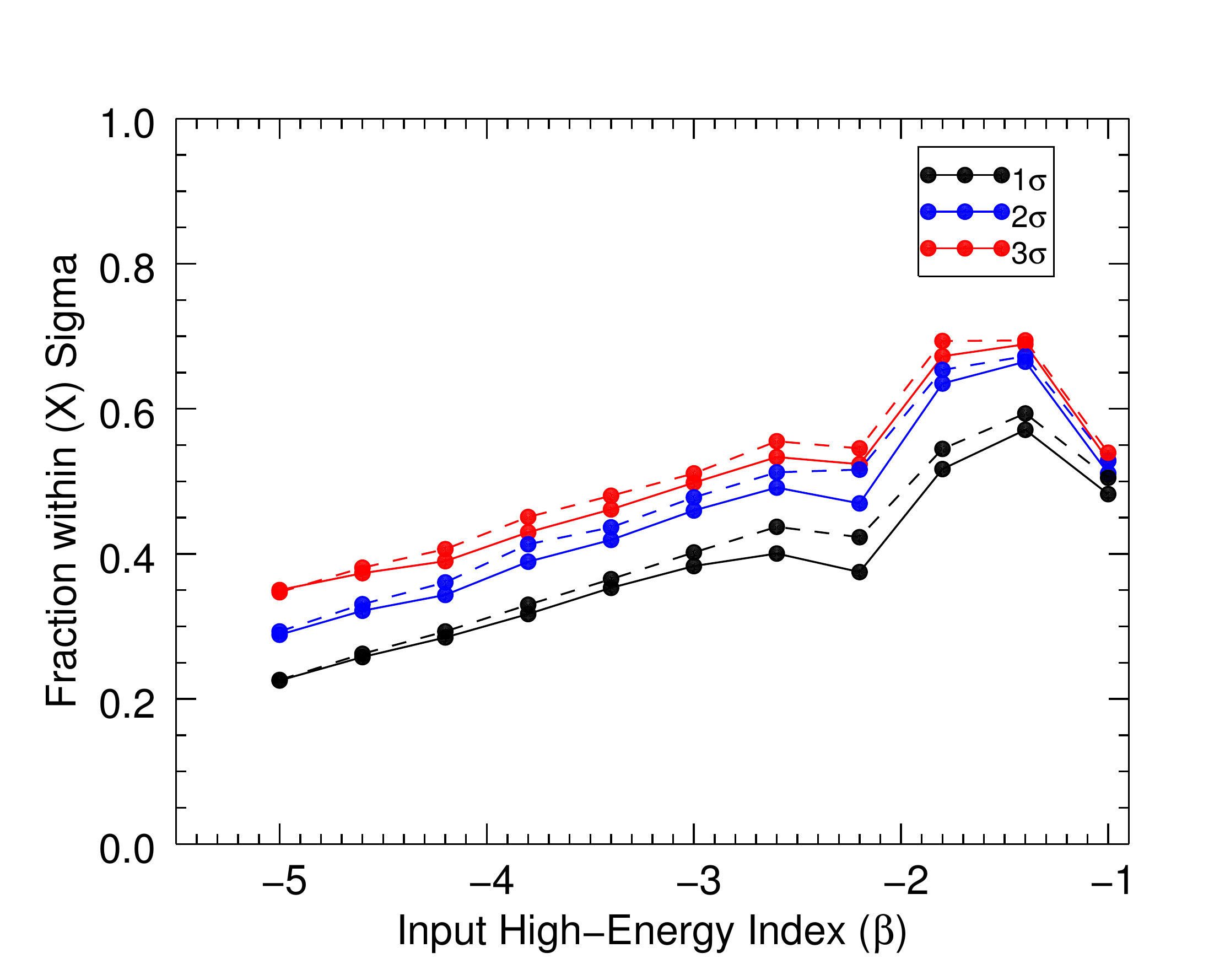}}
	\end{center}
\caption{Plots of the effectiveness of the BATSE LADs when fitting a Band spectrum.  The solid line signifies the 16-channel CONT 
data, while the dashed line represents the 128-Channel HERB data.  The three colors signify the distance in standard deviations 
between the mean value of the parameter from the spectral fits and the input spectral values.  The standard deviation for each 
parameter is defined by the 68\% confidence interval resulting from each set of simulations.\label{SigmaSims}}
\end{figure}

The resulting spectral fits to the simulations were compiled, and the mean and standard deviation of each set of 1000 simulations 
was computed for each grid point.  Finally, for each individual input parameter value, the distance between the parameter value of the 
fit and the input was calculated, integrating over all other dimensions of the sample parameter space.  Figure \ref{SigmaSims} shows 
the fraction of spectral fits that contain the input spectral parameter within the 1$\sigma$, 2$\sigma$, and 3$\sigma$ confidence 
intervals for each parameter while integrating over the other dimensions of the parameter space.  The key result from Figure \ref
{SigmaSims} is that there appears to be minimal loss of spectral information when using the 16-channel CONT data compared to the 
128-Channel HERB data.

\section{Appendix B \\
Derivation of $E_{peak}$ for GLOGE}
The GLOGE model is parametrized with $E_{cen}$, which is the centroid energy in keV, and $s$ which is the standard 
deviation in units of $\rm Log_{10} \ keV$ and is related to the full-width at half-maximum (FWHM) by $s=\rm FWHM/
2.35482$.  Although it is not parametrized with $E_{peak}$, the power density spectrum is a parabolic function in log-log space, 
therefore, there will always exist an $E_{peak}$ although it is not guaranteed to exist within the detector bandpass.  It is 
relatively easy to derive $E_{peak}$ for GLOGE, which can be found by
\begin{equation}
\frac{d}{dE}(\nu F_{\nu})=0
\end{equation}
where $\nu F_{\nu}=E^2 \ f_{GLOGE}(E)$.  Solving this equation for $E_{peak}$ results in
\begin{equation}
E_{peak}=E_{cen} \ e^{2 d s^2}
\end{equation} 
where $d=\ln^2(10)$.  In order to correctly calculate the errors on the derived $E_{peak}$, the errors in $E_{cen}$ and $s$ 
must be propagated properly.  Assuming higher-order terms in the Taylor expansion of the equation are negligible, the error to 
first order is given by
\begin{equation}
\sigma^2_{E_{peak}}=\biggl| \frac{\partial E_{peak} }{\partial E_{cen}} \biggr|^2 \sigma^2_{E_{cen}} + \biggl| \frac{\partial E_
{peak}}{\partial s} \biggr|^2 \sigma^2_{s}  + 2 \frac{\partial E_{peak}}{\partial E_{cen}} \frac{\partial E_{peak}}{\partial s} {\rm Cov}
(E_{cen},s)
\end{equation}
where ${\rm Cov}(E_{cen},s)$ is the covariance between $E_{cen}$ and $s$.  Once the partial derivatives have been 
determined,
\begin{equation}
\frac{\partial E_{peak}}{\partial E_{cen}}=\frac{E_{peak}}{E_{cen}};
\end{equation}
\begin{equation}
\frac{\partial E_{peak}}{\partial \sigma}=4 d s E_{peak},
\end{equation}
and we have the covariance between $E_{cen}$ and $s$ (from the covariance matrix returned after performing the 
spectral fit) we can properly calculate the error on $E_{peak}$.

\clearpage

\section{Appendix C \\
$\bf \chi^2$ Distributions}
\begin{figure}
	\begin{center}
		\subfigure[]{\label{redchisqf}\includegraphics[scale=0.35]{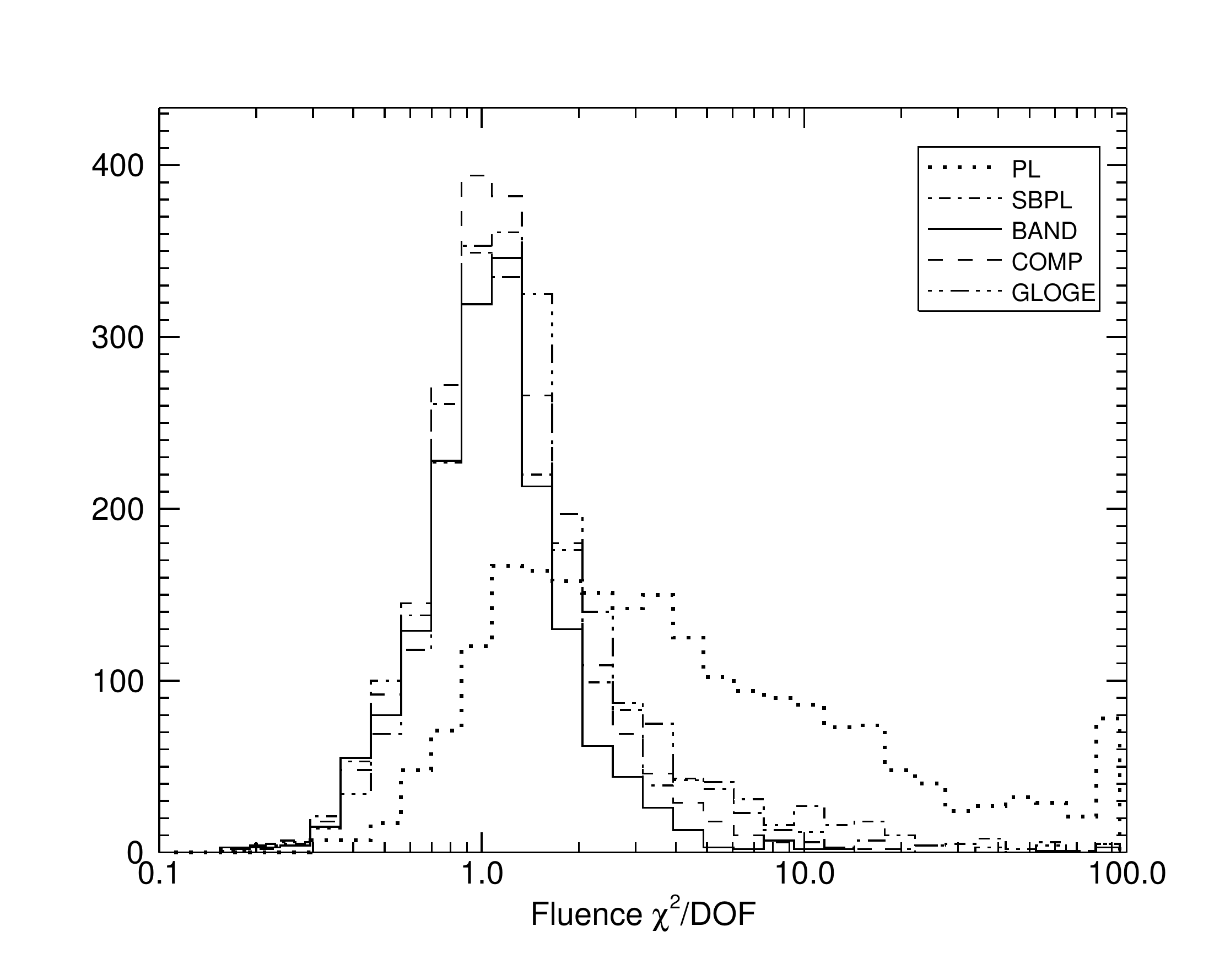}}
		\subfigure[]{\label{redchisqp}\includegraphics[scale=0.35]{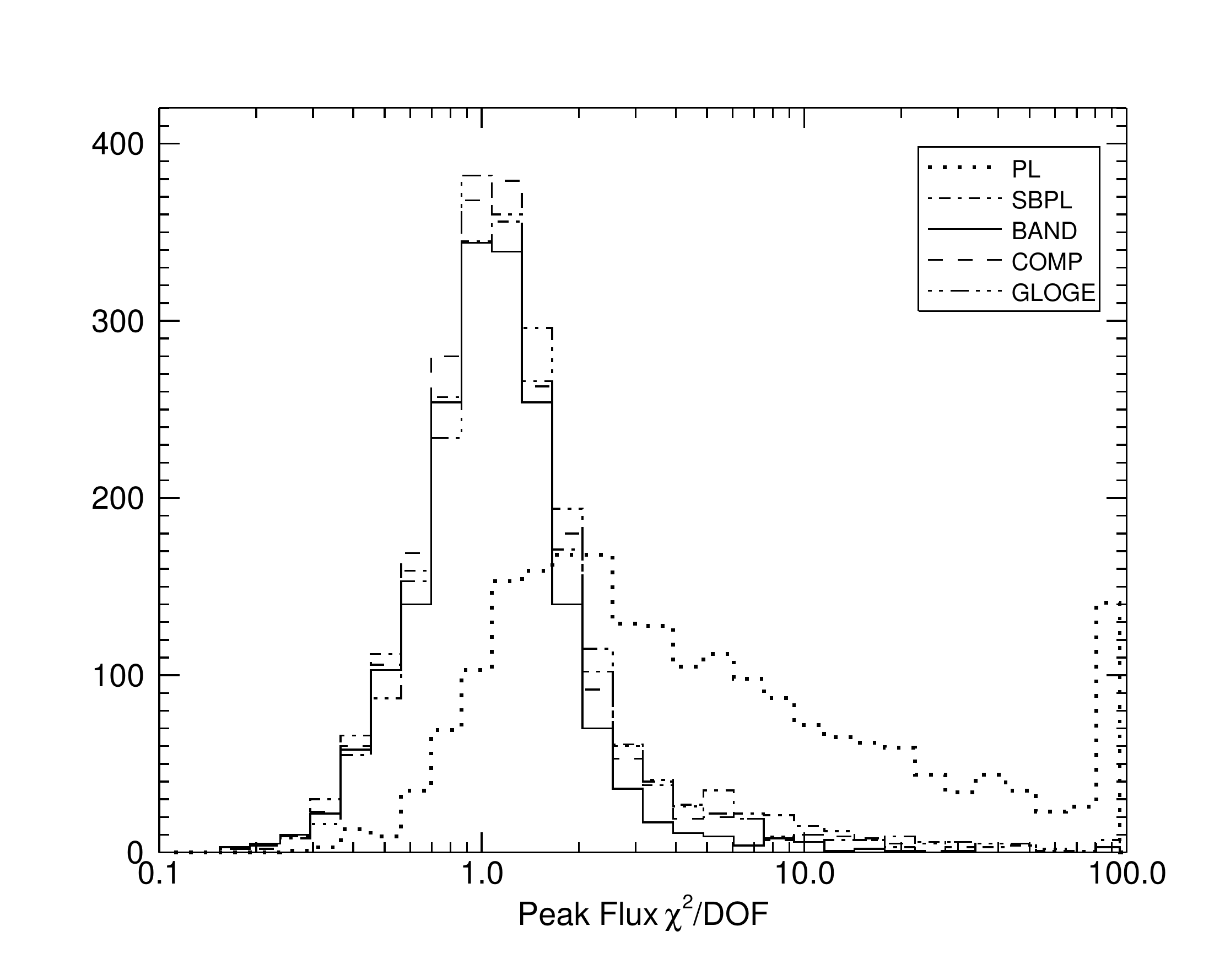}}\\
		\subfigure[]{\label{redchisqbestf}\includegraphics[scale=0.35]{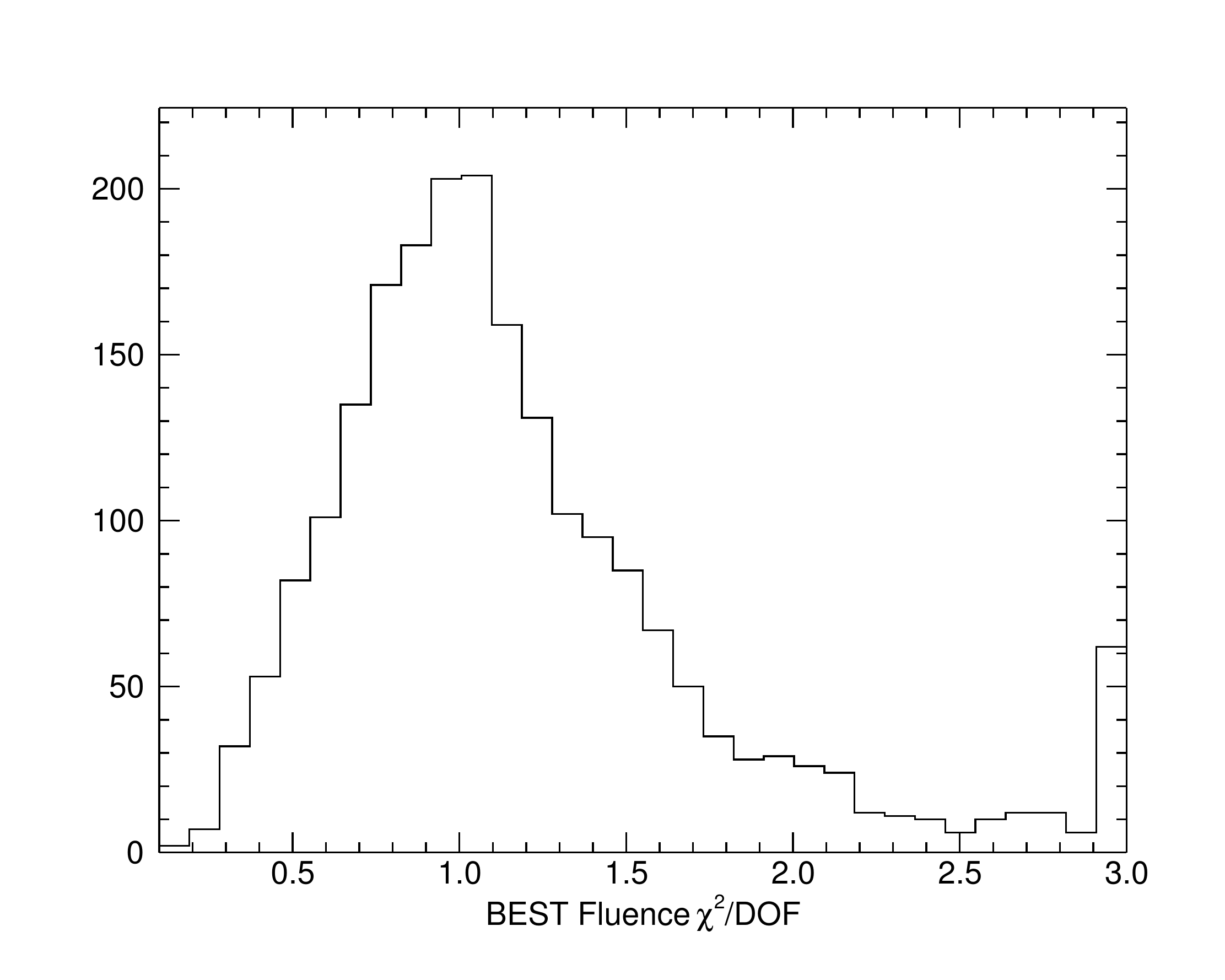}}
		\subfigure[]{\label{redchisqbestp}\includegraphics[scale=0.35]{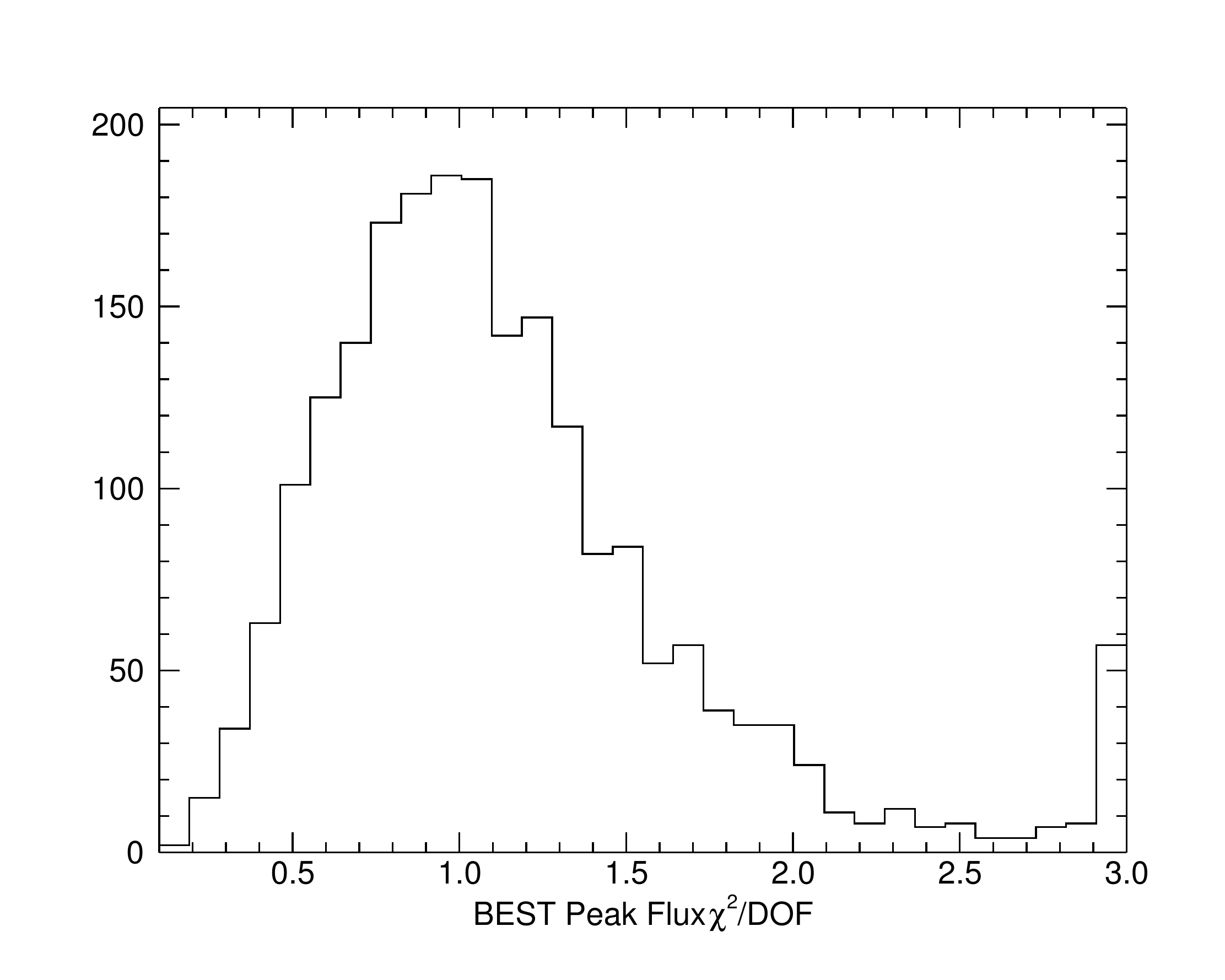}}
	\end{center}
\caption{Distributions of the reduced $\chi^2$ for each model.  \ref{redchisqf} and \ref{redchisqp} are distributions of the 
reduced $\chi^2$ for each model for each burst.  Note that these distributions include all fits, including unconstrained fits.  \ref
{redchisqbestf} and \ref{redchisqbestp} show the distribution of BEST reduced $\chi^2$ values.  The first and last bins in all 
distributions contain overflow values.  \label{redchisq}}
\end{figure}

Although the least-squares fitting process did not minimize $\chi^2$ as a figure of merit, we can calculate the $\chi^2$ 
goodness-of-fit statistic comparing the model to the data.  To do this, we difference the background-subtracted count rates from 
the model rates, summing first over all energy channels in each detector, and then over all detectors.  This is shown by
\begin{equation}
\chi^2= \sum_i \sum_j \Biggl[ \frac{O_{ij} - B_{ij} - M_{ij}}{\sqrt{\sigma_{M_{ij}}^2}} \Biggr]^2,
\end{equation}
where the $O_{ij}$ are the observed count rates, $B_{ij}$ are the background rates, $M_{ij}$ are the model rates, and $
\sigma_{M_{ij}}^2$ are the derived model variances.  In the ideal situation, and assuming acceptable spectral fits (i.e. when 
performing spectral analysis of simulated data), the reduced $\chi^2$ value ($\chi^2$/d.o.f.) will tend to distribute around a 
value of 1.  In this way, $\chi^2$ gives an estimate on the acceptability of the fit.  Figure \ref{redchisq} shows the reduced $
\chi^2$ distributions for both the fluence and peak flux spectra, as well as the corresponding BEST distributions.  It is 
important to note that the distributions peak at slightly larger or smaller values than 1, which is acceptable since even the BEST 
spectral fits represent a rough approximation to the actual spectra of GRBs due to their empirical nature.  The fluence $\chi^2$ 
distributions appear similar to the peak flux $\chi^2$ distributions even though the fluence spectra contain longer time integration 
intervals and many GRBs experience spectral evolution.  The peak flux sample captures the spectra of all bursts in a small slice 
of time at the same stage of the lightcurve, meanwhile the fluence sample integrates over the duration of the emission, in many 
cases over several pulses.  The proximity of most of the BEST reduced $\chi^2$ values to the nominal value is indicative of 
acceptable spectral fits.

\begin{figure}
	\begin{center}
		\subfigure[]{\label{fluencechisq}\includegraphics[scale=0.35]{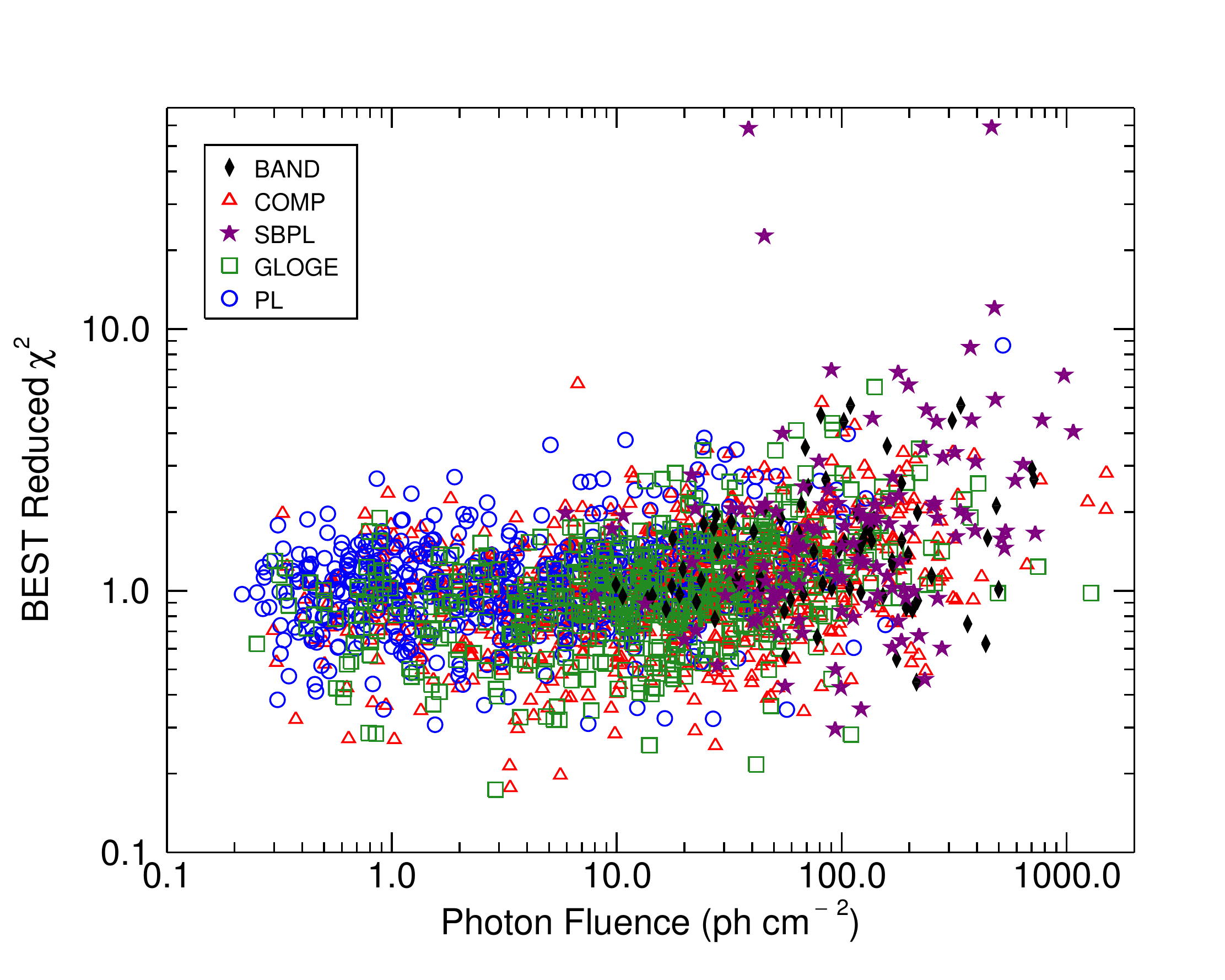}}
		\subfigure[]{\label{fluxchisq}\includegraphics[scale=0.35]{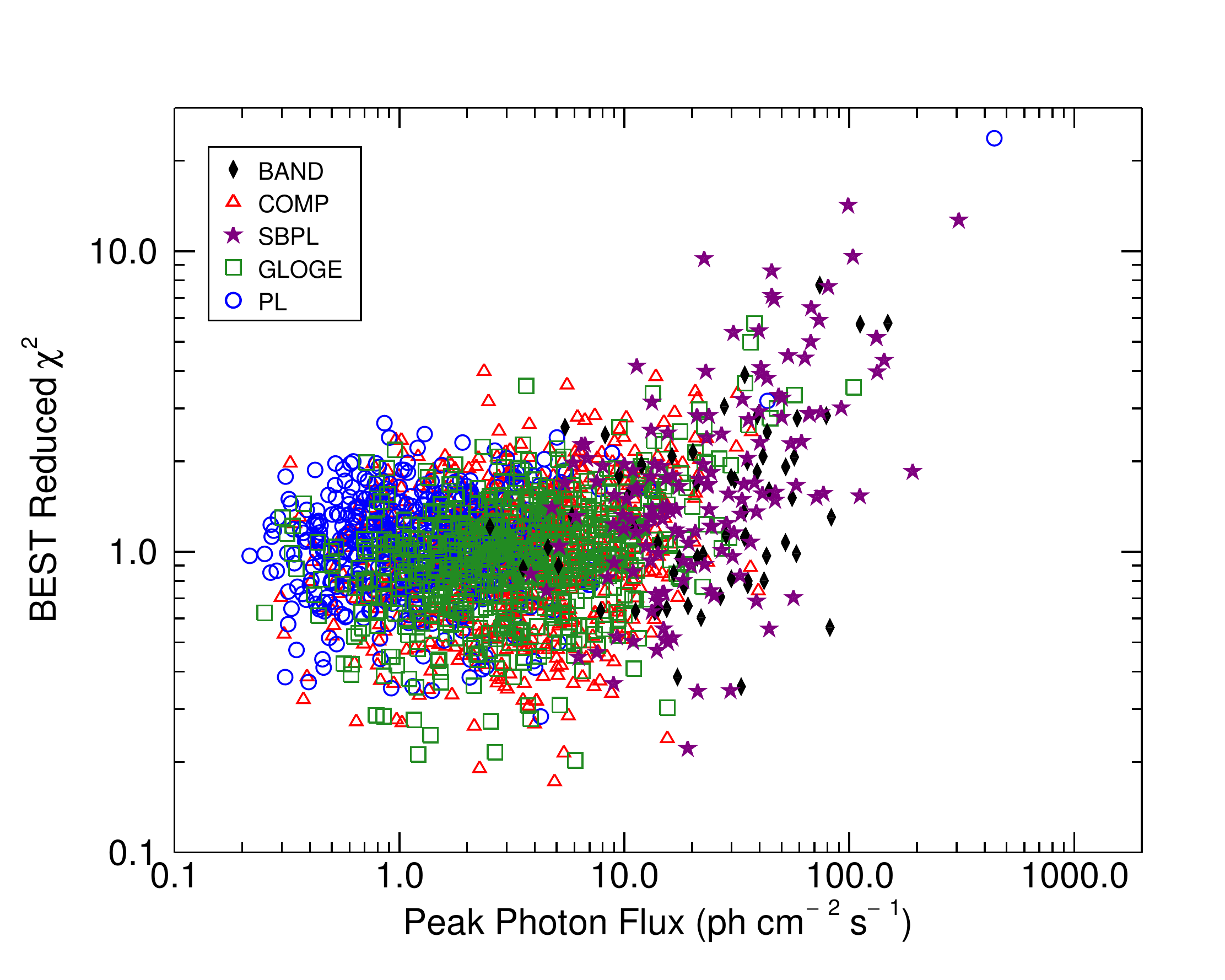}}
	\end{center}
\caption{\ref{fluencechisq} Reduced $\chi^2$ as a function of the model photon fluence.  \ref{fluxchisq} Reduced $\chi^2$ as a 
function of the model peak photon flux.   \label{ffredchisq}}
\end{figure}

In addition, Figure \ref{ffredchisq} plots the BEST reduced $\chi^2$ as a function of photon fluence and peak photon flux for the 
fluence and peak flux spectra, respectively.  The reduced $\chi^2$ for the fluence fits shows a slight upward trend as the 
photon fluence increases.  The average reduced $\chi^2$ starts at approximately a value of unity at a low fluence of $\sim$0.2 $
\rm photons \ cm^{-2}$ and increases to a value of $\sim$2 at $\sim$1000 $\rm photons \ cm^{-2}$.  Additionally there are several 
outliers to the trend that exist at high fluence and exhibit even larger reduced $\chi^2$ values.  This indicates that the goodness-
of-fit is increasingly worse the more fluent the burst.  This follows from the fact that in most cases extremely fluent bursts are long 
and may exhibit significant spectral evolution, therefore the time-integrated spectral fit will average over the evolution and will 
produce a significantly worse fit.  When inspecting the reduced $\chi^2$ as a function of the photon flux, we find that the trend is 
flat until $\sim$20 $\rm photons \ cm^{-2}$.  Most of the large reduced $\chi^2$ values from the peak flux spectra result from 
higher flux bursts where systematic uncertainties tend to dominate the statistical uncertainties.

\clearpage

\section{Appendix D \\
Simulations of Spectral Parameters from Summing Detectors}
\begin{figure}
	\begin{center}
		\includegraphics[scale=0.7]{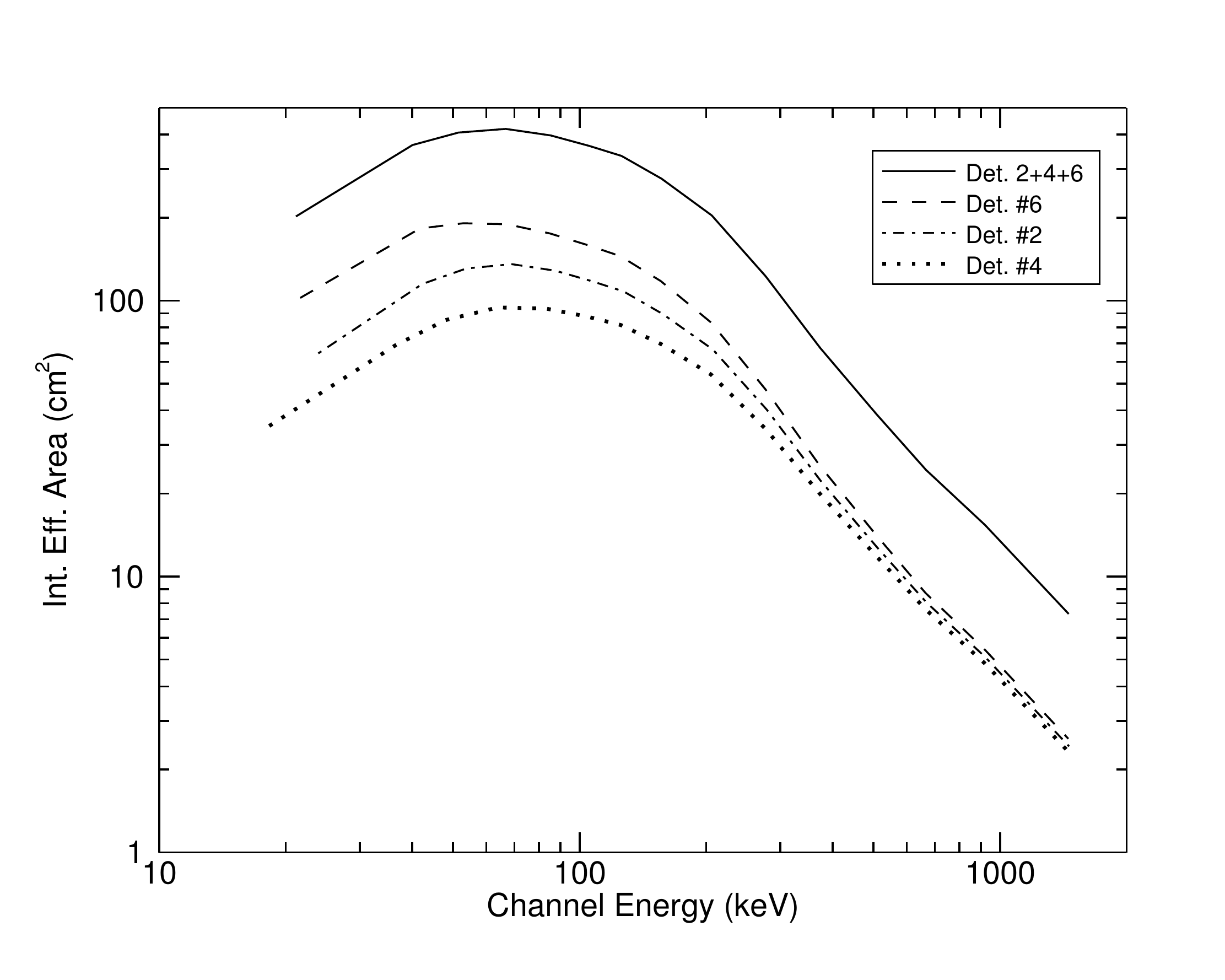}
	\end{center}
\caption{Plot of the effective area integrated over photon energies of BATSE detectors for GRB 910503 (trigger \#0143).  Each 
detector possesses its own response that is dependent on angle and energy.  Adding these responses together effectively 
averages the features of the individual responses and spectral resolution is decreased.  \label{sumdrm}}
\end{figure}

For this catalog, we have chosen to sum appropriate detectors for each burst to perform a spectral analysis.  This increases 
the signal--to--noise ratio of the summed datasets when compared to using single detectors, which increases the amount of signal 
investigated.  This method also provides a simple, objective approach to selecting signal from a GRB on which to do analysis.  
There are, however, a few consequences to summing detector counts prior to spectral analysis.  One of these consequences is that a 
new DRM for the summed detectors must be created.  The DRMs, which contain the effective area of each detector as a function of 
detector--to--source angle and energy, can be summed to produce a new DRM appropriate for the summed detectors.  The photon 
energy and channel energy edges must be averaged accordingly to provide an accurate mapping from channel to 
photon energy.  Because of this averaging, spectral resolution decreases.  As shown in Figure \ref{sumdrm}, the effective area 
integrated over photon energies for the summed DRM can average features found in the individual DRMs.  For example, the 
typical peak response energy for BATSE detector is $\sim$60--70 keV, and while the summed DRM preserves this 
approximately, the \emph{region} around the peak response energy is slightly broadened due to the summation of the DRMs.

\begin{figure}
	\begin{center}
		\subfigure[]{\label{simplindex}\includegraphics[scale=0.35]{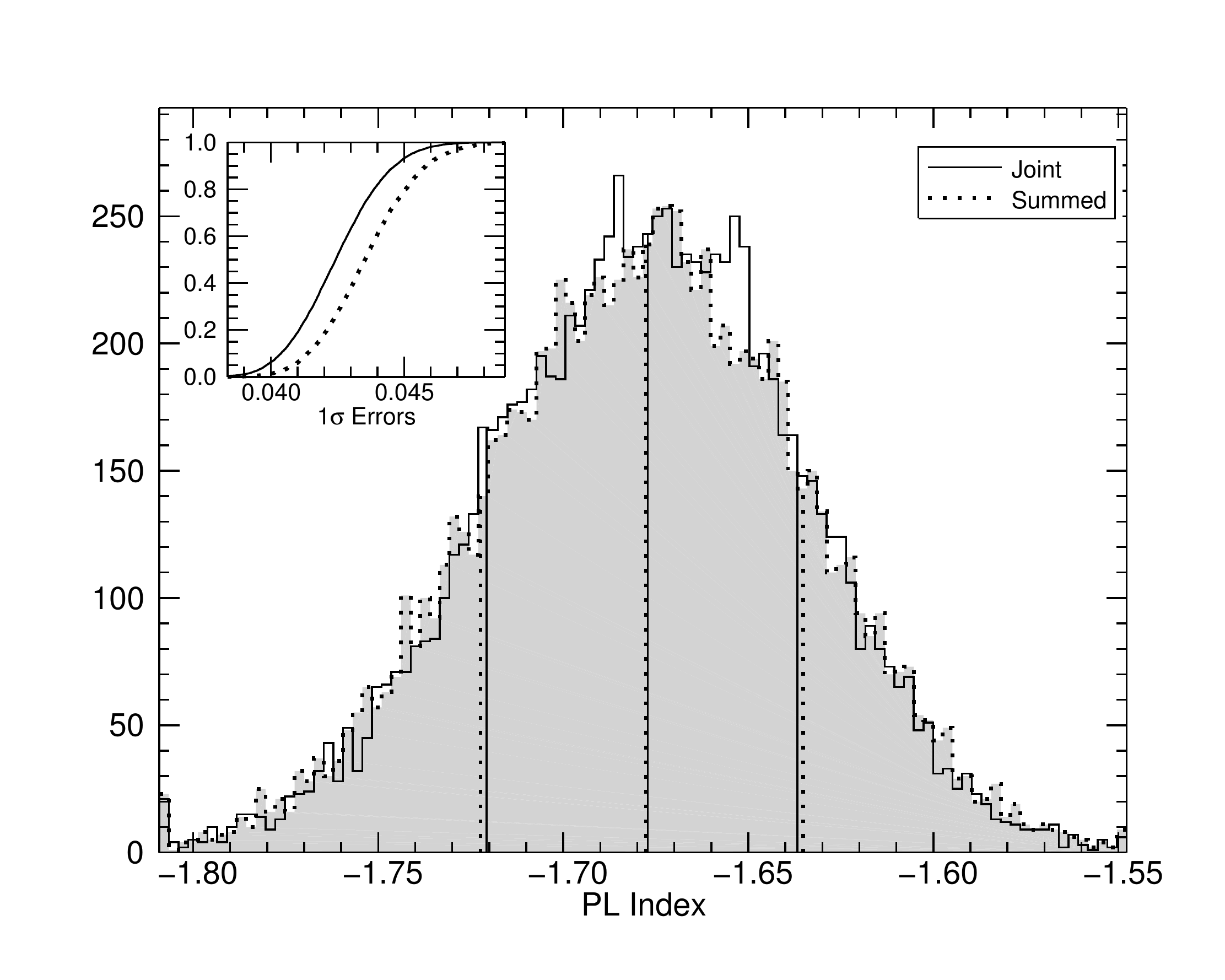}}\\
		\subfigure[]{\label{simglogecen}\includegraphics[scale=0.35]{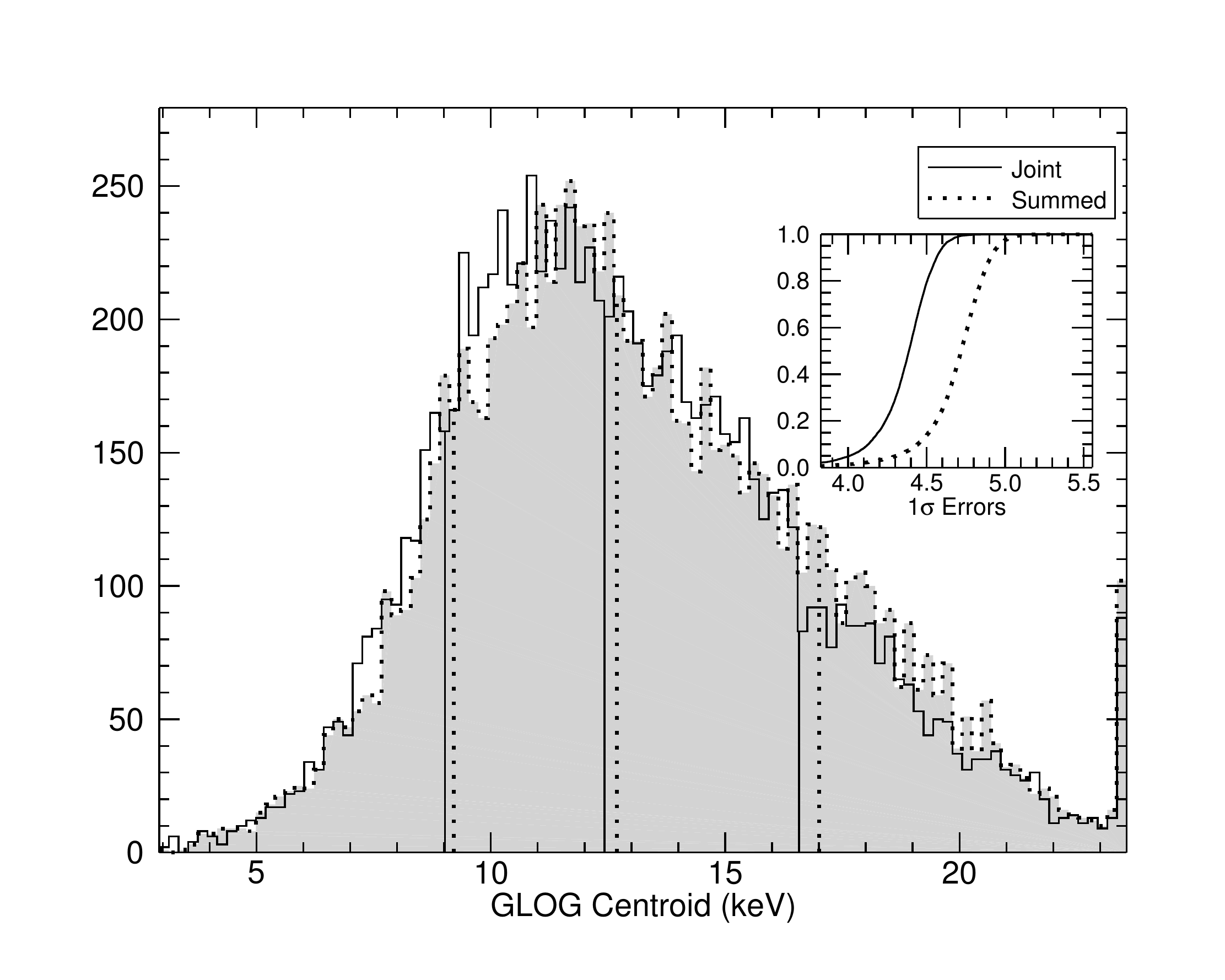}}
		\subfigure[]{\label{simglogfwhm}\includegraphics[scale=0.35]{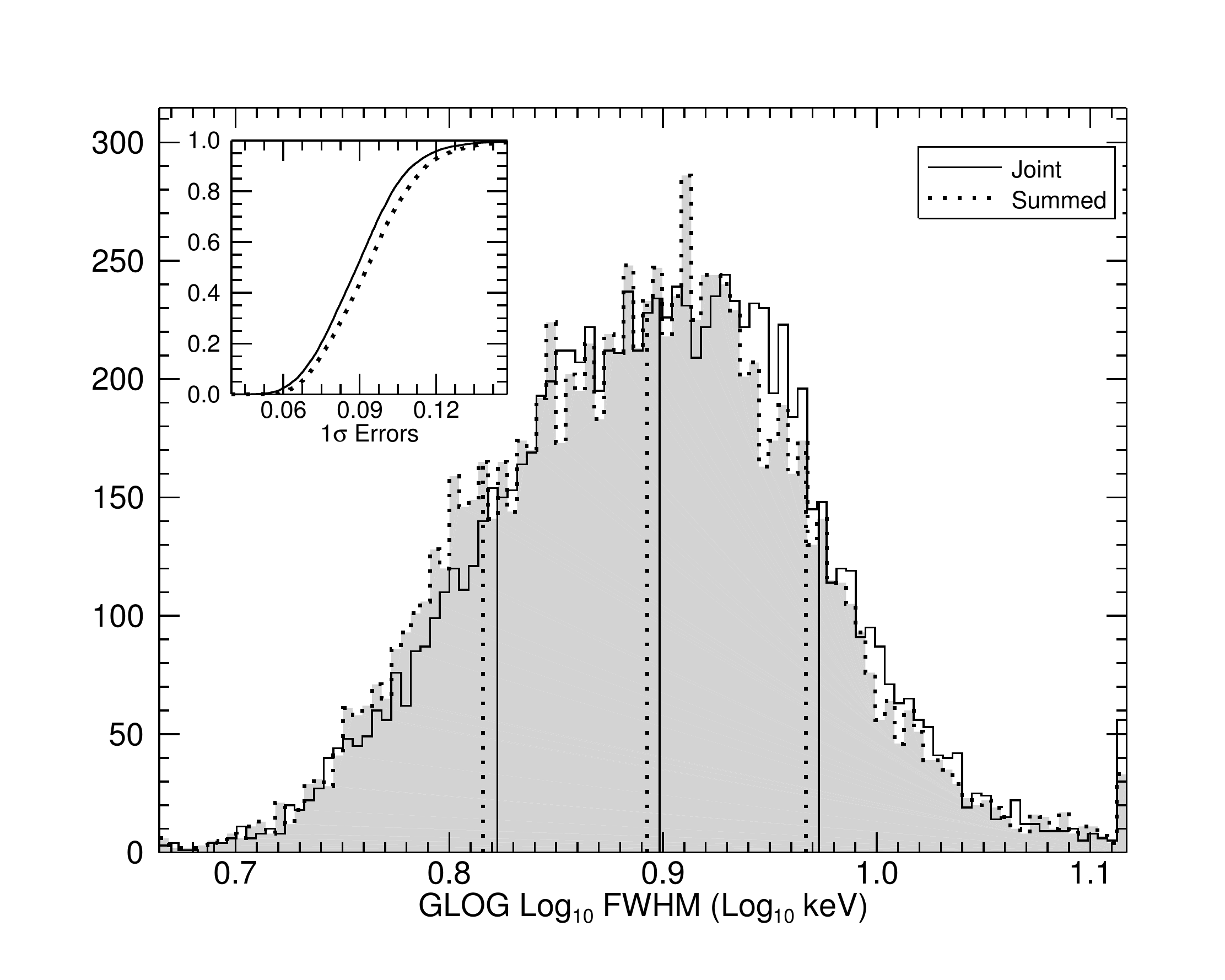}}
	\end{center}
\caption{Simulations of PL (\ref{simplindex}) and GLOGE (\ref{simglogecen} \& \ref{simglogfwhm}) parameters.  \ref{simplindex} 
compares the summed detectors with the joint detectors for GRB 910425 (BATSE Trigger \#0110) and \ref{simglogecen} \& 
\ref{simglogfwhm} shows the same comparison for GRB 910521 (BATSE Trigger \#0214).  The shaded histograms 
represent the parameter distributions for the summed detectors, and the empty histograms represent the joint detectors.  The 
central vertical lines represent the sample mean, while the vertical lines on either side of the mean represent the asymmetric 
sample standard deviation.  The inset plots show the cumulative distributions of the parameter errors.
\label{simplglog}}
\end{figure}

\begin{figure}
	\begin{center}
		\subfigure[]{\label{simcompepeak}\includegraphics[scale=0.35]{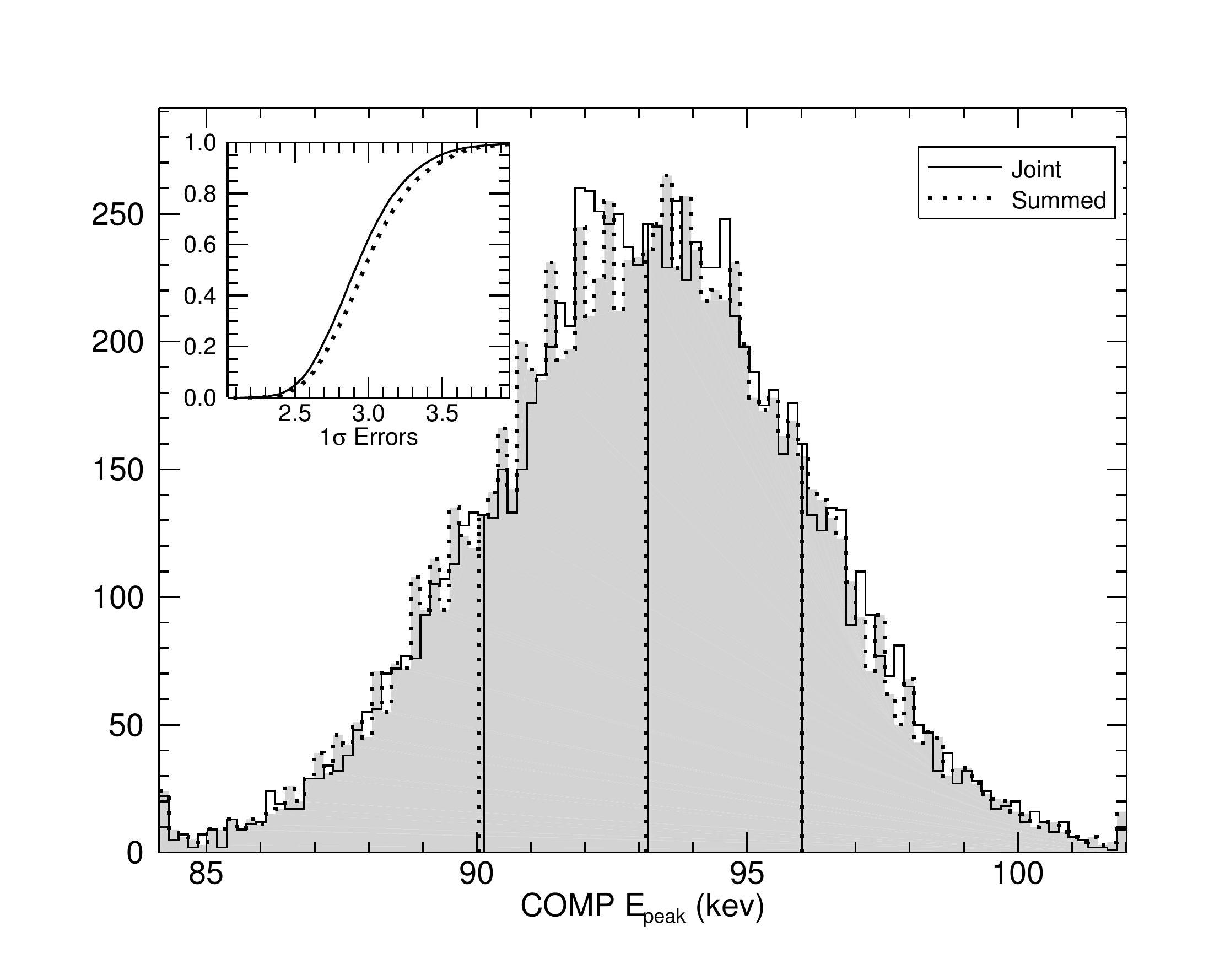}}
		\subfigure[]{\label{simcompindex}\includegraphics[scale=0.35]{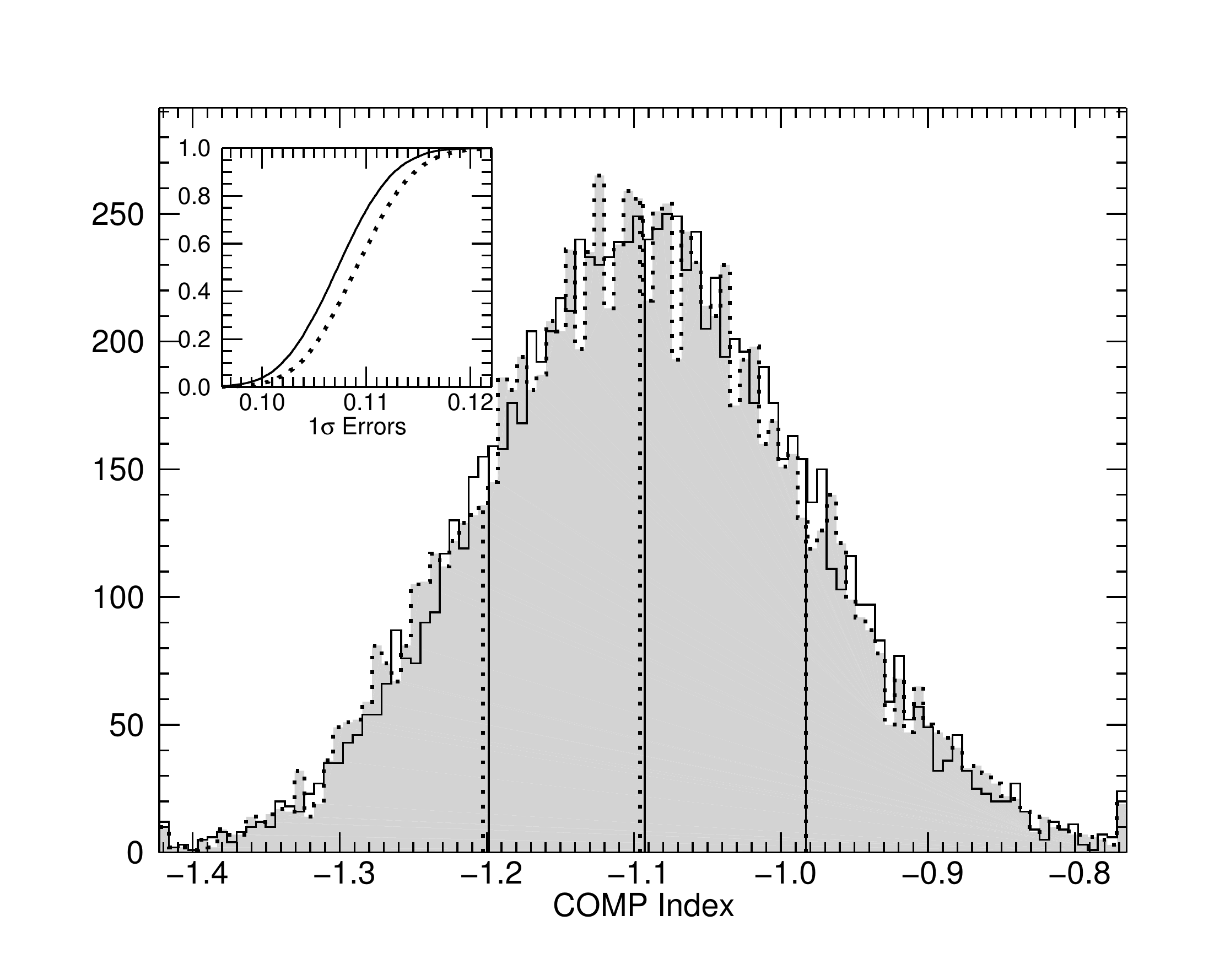}}
	\end{center}
\caption{Simulations of COMP parameters.  The plots compare the summed detectors with the joint detectors for GRB 910426 
(BATSE Trigger \#0111).  The shaded histograms represent the parameter distributions for the summed detectors, and the empty 
histograms represent the joint detectors.  The central vertical lines represent the sample mean, while the vertical lines on either 
side of the mean represent the asymmetric sample standard deviation.  The inset plots show the cumulative distributions of the 
parameter errors.
\label{simcomp}}
\end{figure}

\begin{figure}
	\begin{center}
		\subfigure[]{\label{simbandepeak}\includegraphics[scale=0.35]{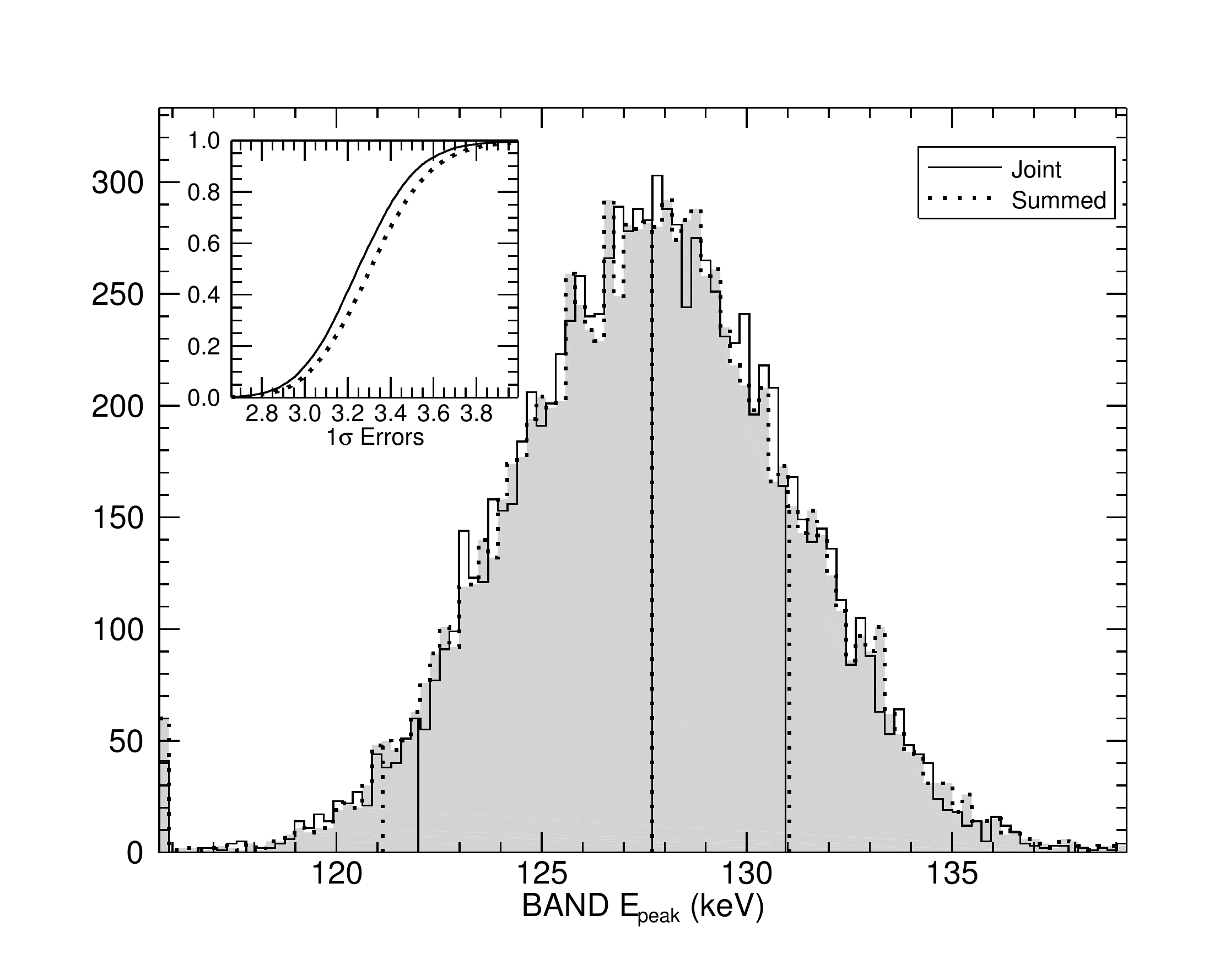}}\\
		\subfigure[]{\label{simbandalpha}\includegraphics[scale=0.35]{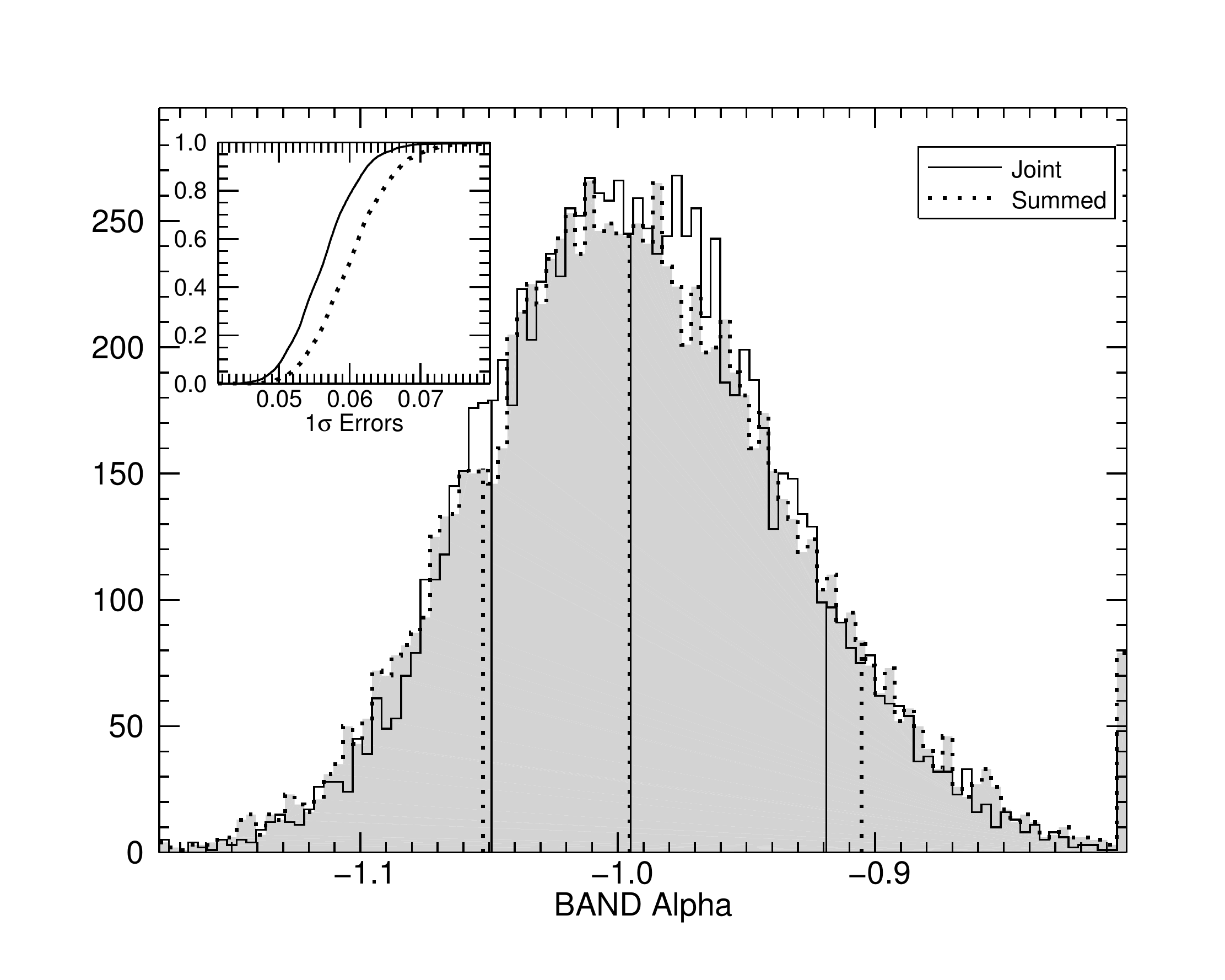}}
		\subfigure[]{\label{simbandbeta}\includegraphics[scale=0.35]{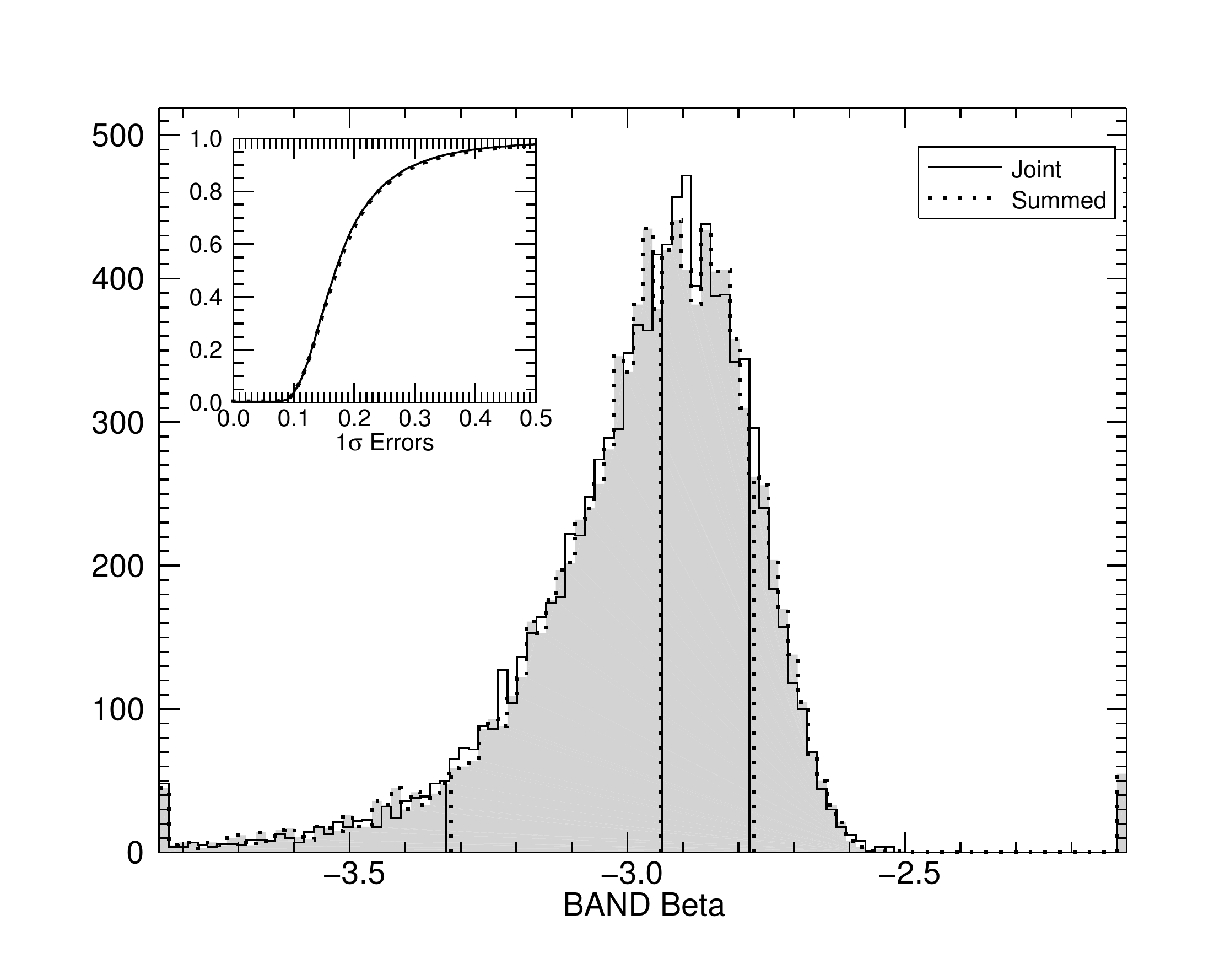}}
	\end{center}
\caption{Simulations of BAND parameters.  The plots compare the summed detectors with the joint detectors for GRB 910421 
(BATSE Trigger \#0105).  The shaded histograms represent the parameter distributions for the summed detectors, and the empty 
histograms represent the joint detectors.  The central vertical lines represent the sample mean, while the vertical lines on either 
side of the mean represent the asymmetric sample standard deviation.  The inset plots show the cumulative distributions of the 
parameter errors.
\label{simband}}
\end{figure}

\begin{figure}
	\begin{center}
		\subfigure[]{\label{simsbplebreak}\includegraphics[scale=0.35]{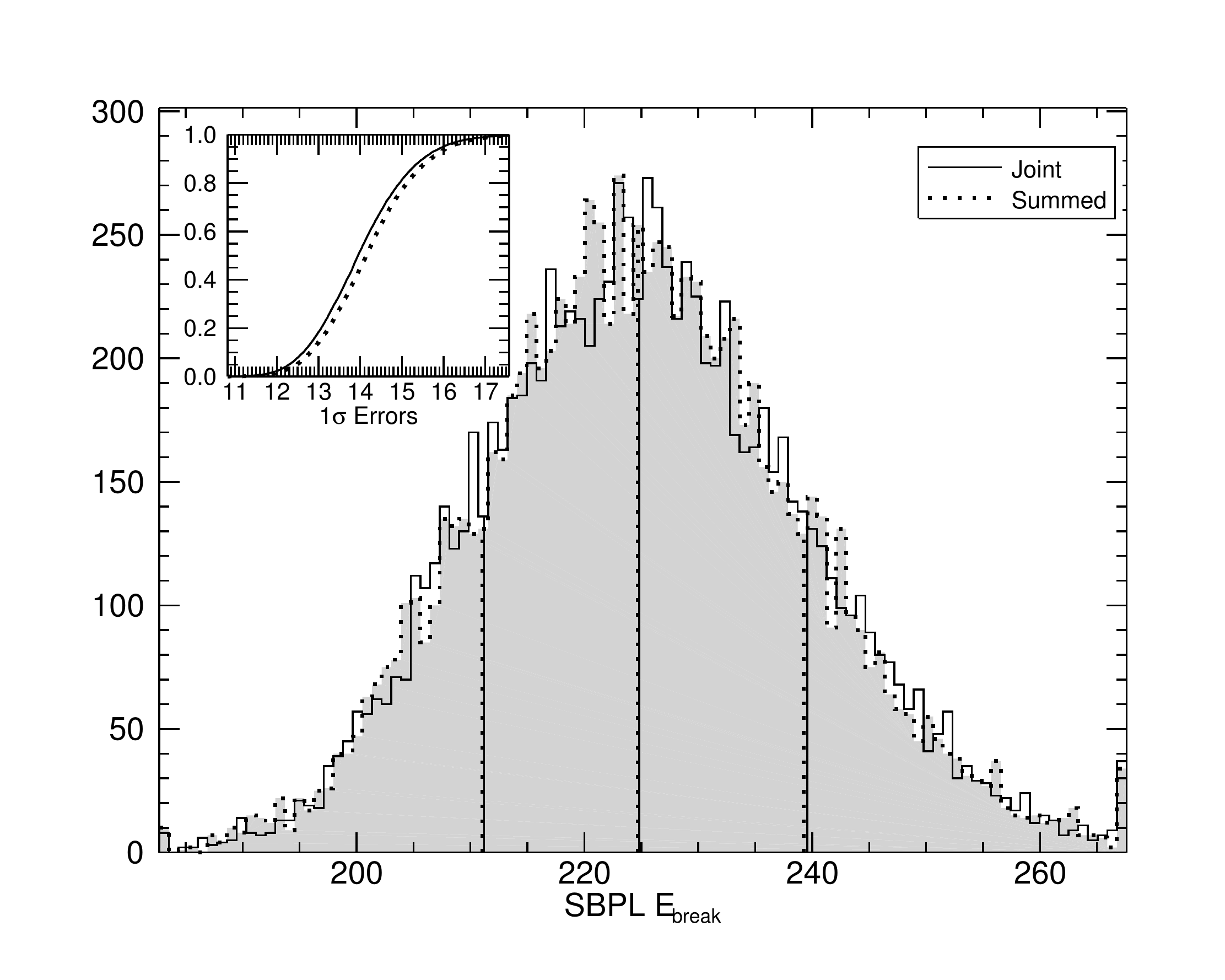}}\\
		\subfigure[]{\label{simsbplalpha}\includegraphics[scale=0.35]{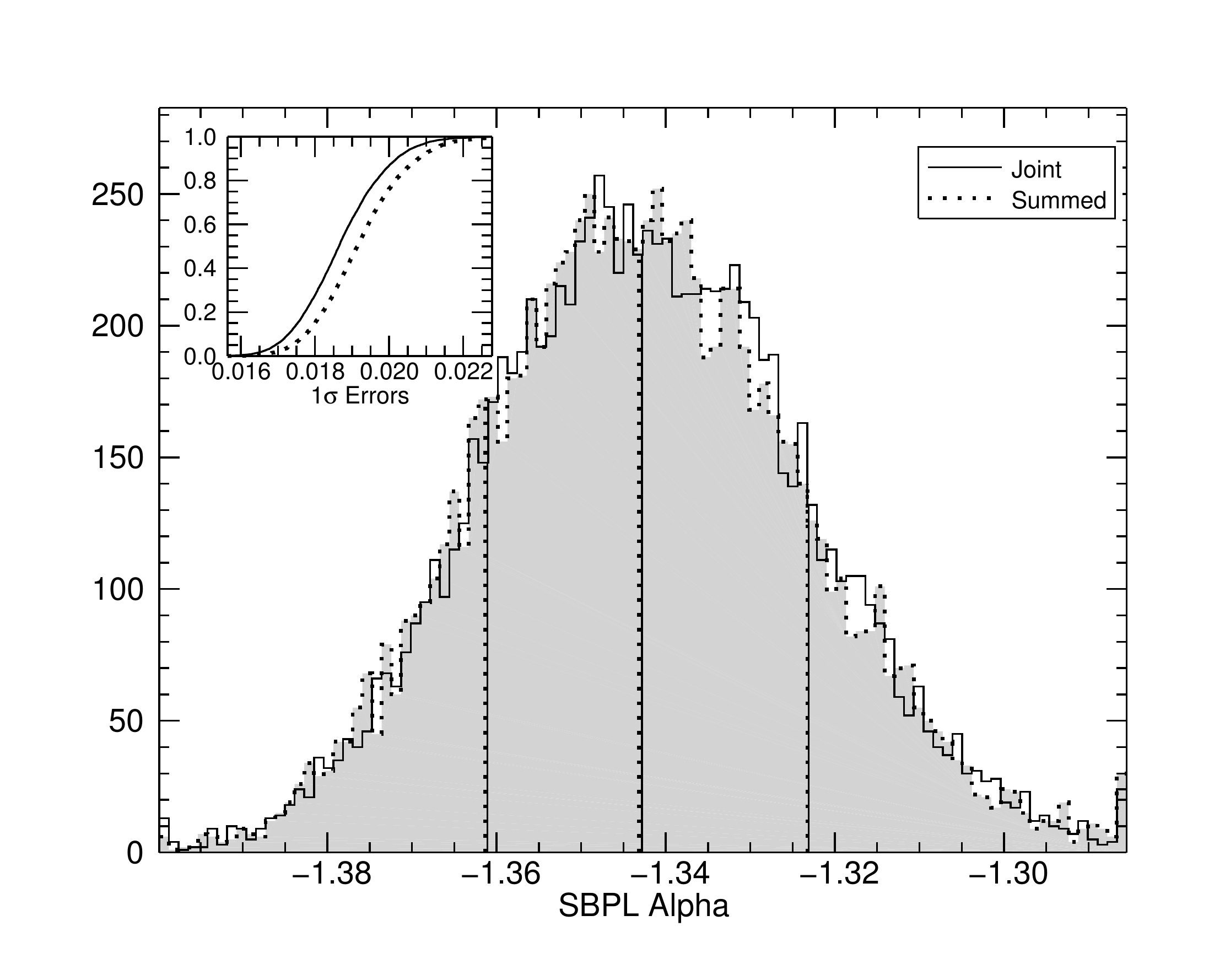}}
		\subfigure[]{\label{simsbplbeta}\includegraphics[scale=0.35]{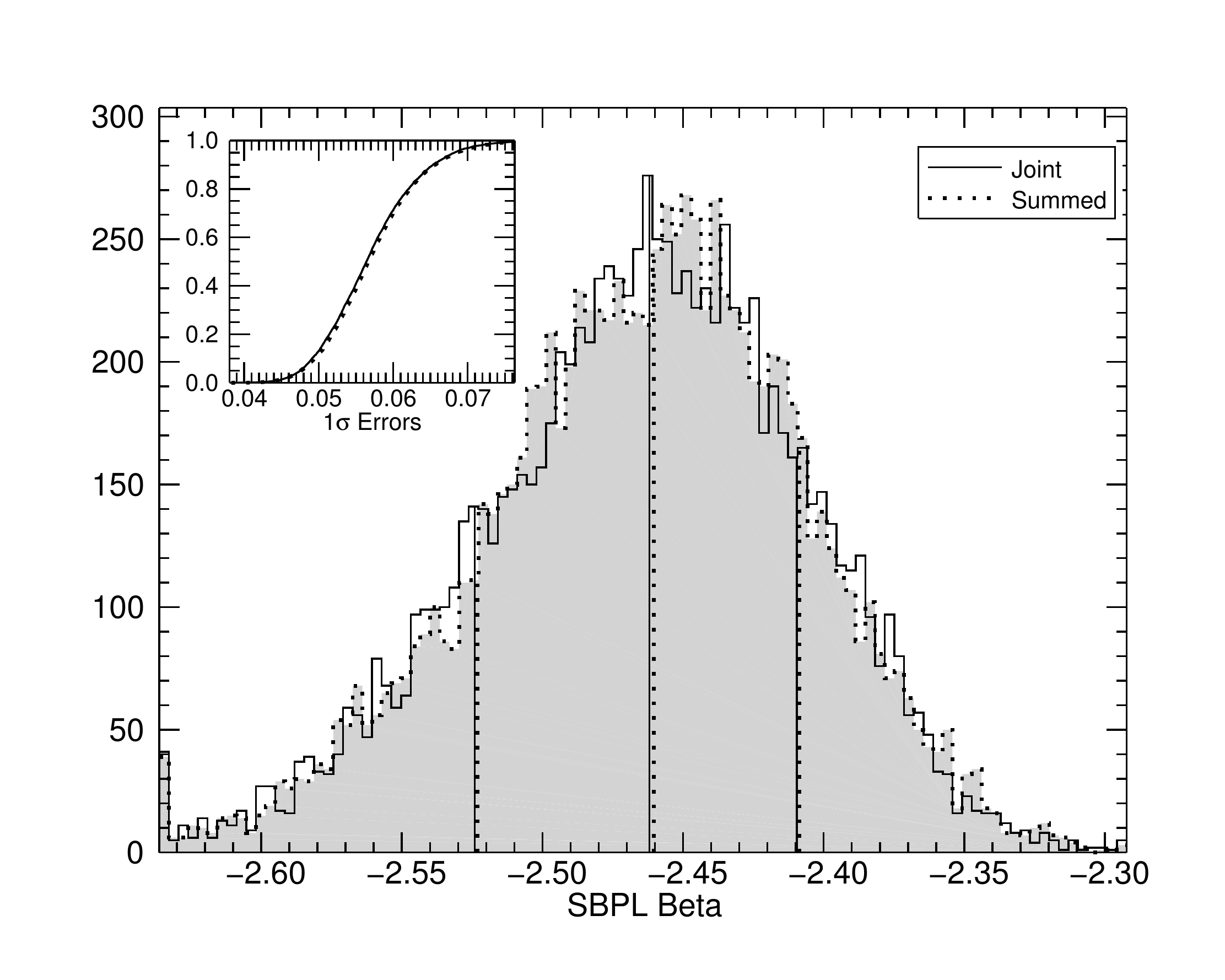}}
	\end{center}
\caption{Simulations of SBPL parameters.  The plots compare the summed detectors with the joint detectors for GRB 910522 
(BATSE Trigger \#0219).  The shaded histograms represent the parameter distributions for the summed detectors, and the empty 
histograms represent the joint detectors.  The central vertical lines represent the sample mean, while the vertical lines on either 
side of the mean represent the asymmetric sample standard deviation.  The inset plots show the cumulative distributions of the 
parameter errors.
\label{simsbpl}}
\end{figure}

In addition to the loss of spectral resolution, there may be additional systematic errors associated with performing summed 
spectral fits.  To investigate these effects, we have performed several Monte Carlo simulations of GRB spectra and compared the 
results from summing detectors to the joint spectral analysis.  For each simulation, we chose a single GRB that was best fit by 
each model used in this catalog.  We created 20,000 simulated spectra for each burst using the best fit fluence spectral 
parameters.  Half of the spectra for each burst were created using the summed detector and response, and the other half were 
created using the individual detectors and responses.  Each simulated spectrum was produced with a Poisson deviate 
background and convolved through the detectors' response to produce an accurate representation of an actual GRB observed 
by that detector.  Each background--subtracted simulated spectrum was then folded through the associated DRM(s) and fit with 
the corresponding best fit function.  The parameter distributions resulting from the simulations are shown in Figures 
\ref{simplglog}--\ref{simsbpl}.  It is evident that the parameter estimates between the two methods are similar.  No single 
parameter has a mean value that deviates more than 2\% when comparing the summed and joint detectors, and in general the 
deviations are much less.  In all cases the 1$\sigma$ standard deviations of the summed spectral parameters are slightly larger 
than those of the joint spectral parameters.  This increased variance is the systematic error that is added to the parameter 
estimations when the detectors are summed.  In all cases the difference in the standard deviation is a small fraction of one 
standard deviation.

\begin{figure}
	\begin{center}
		\subfigure[]{\label{comparepl}\includegraphics[scale=0.25]{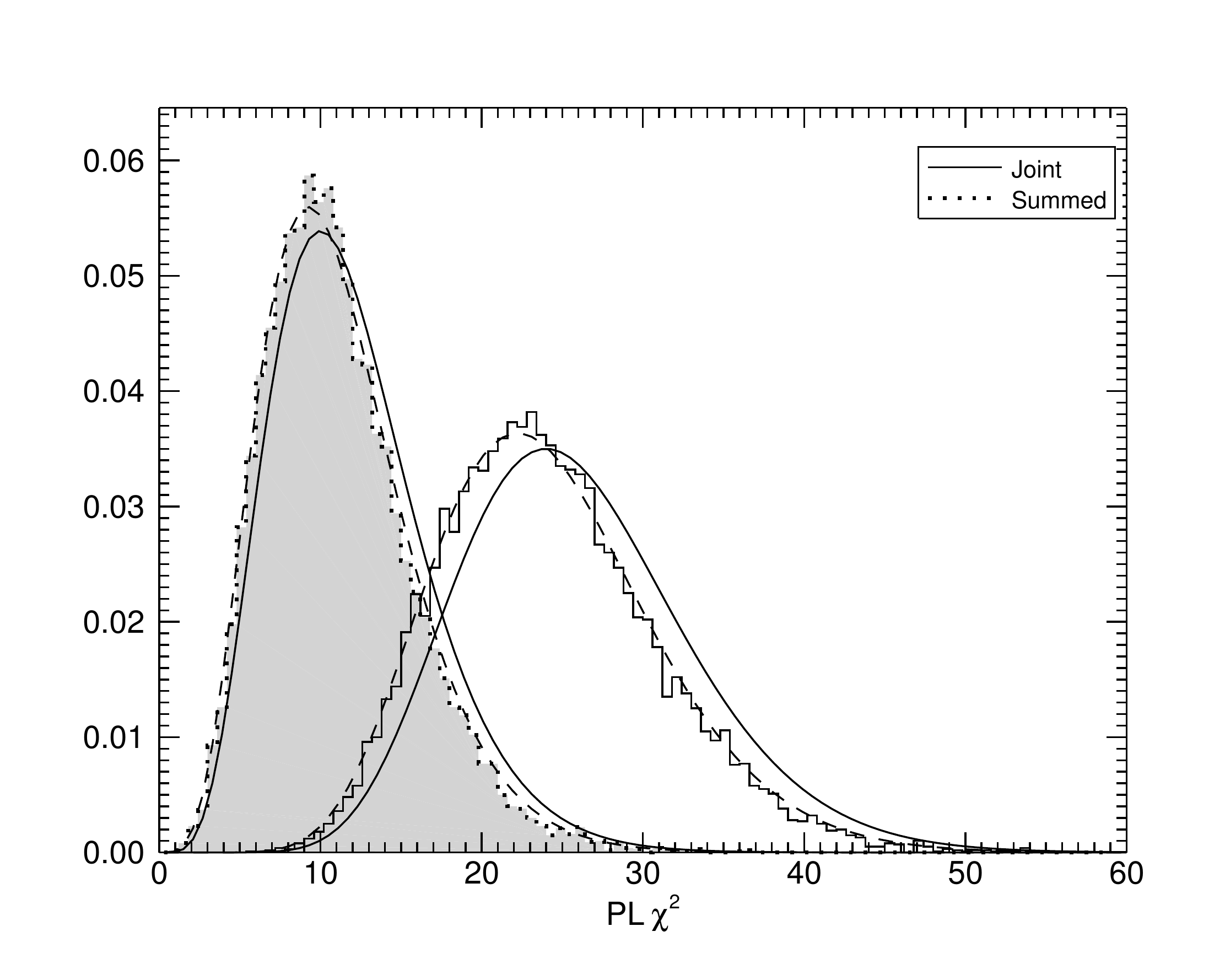}}
		\subfigure[]{\label{compareglog}\includegraphics[scale=0.25]{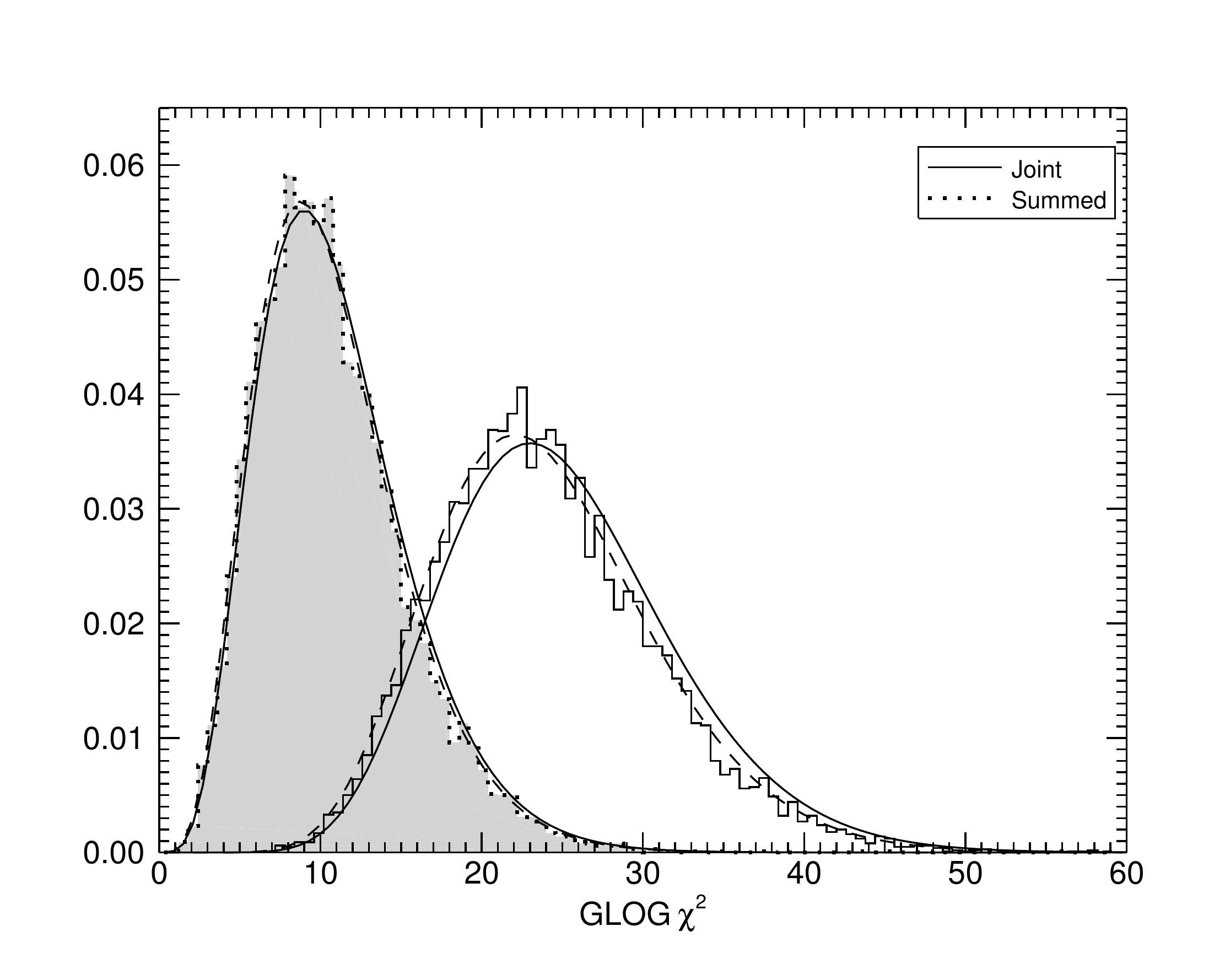}}\\
		\subfigure[]{\label{comparecomp}\includegraphics[scale=0.25]{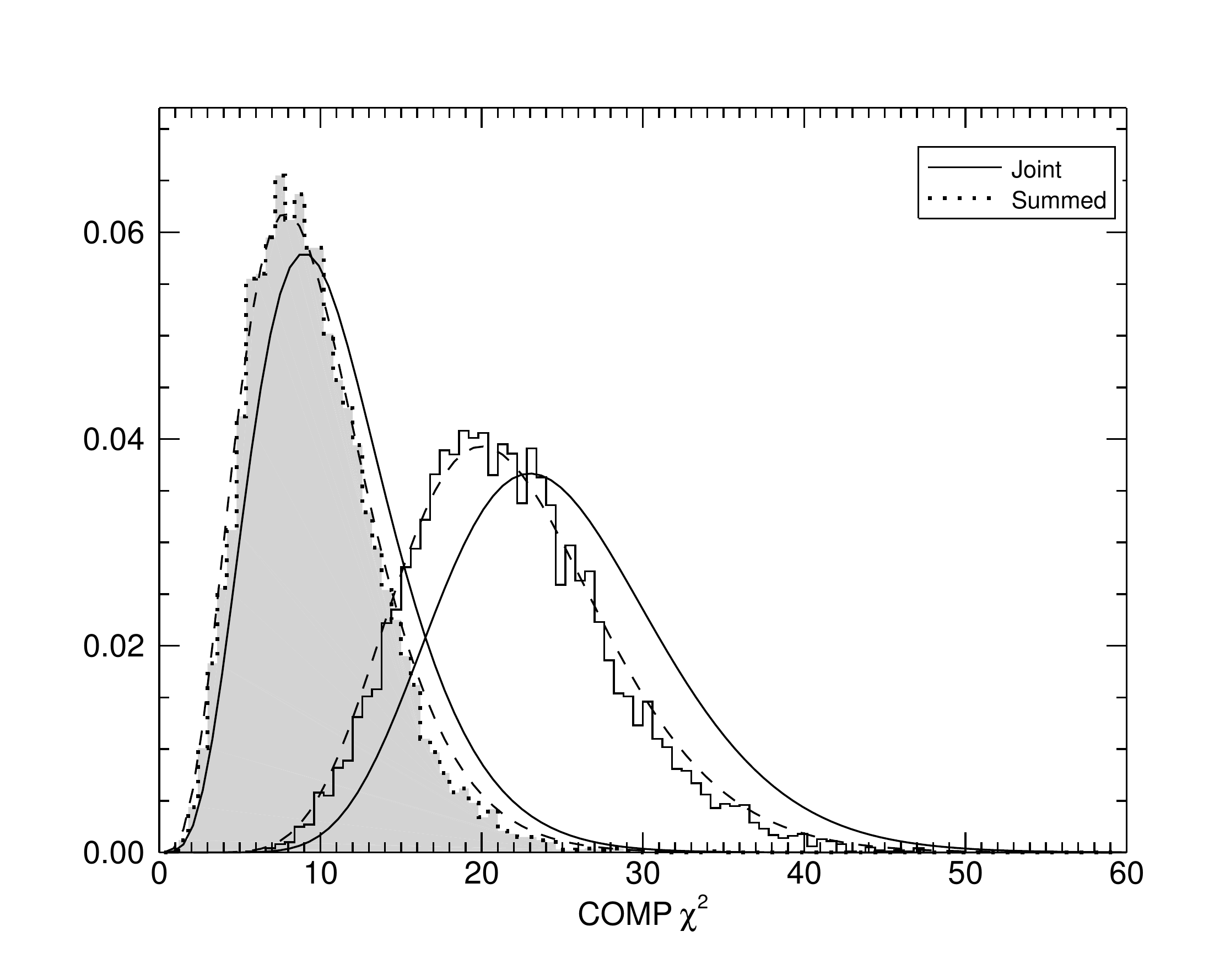}}
		\subfigure[]{\label{compareband}\includegraphics[scale=0.25]{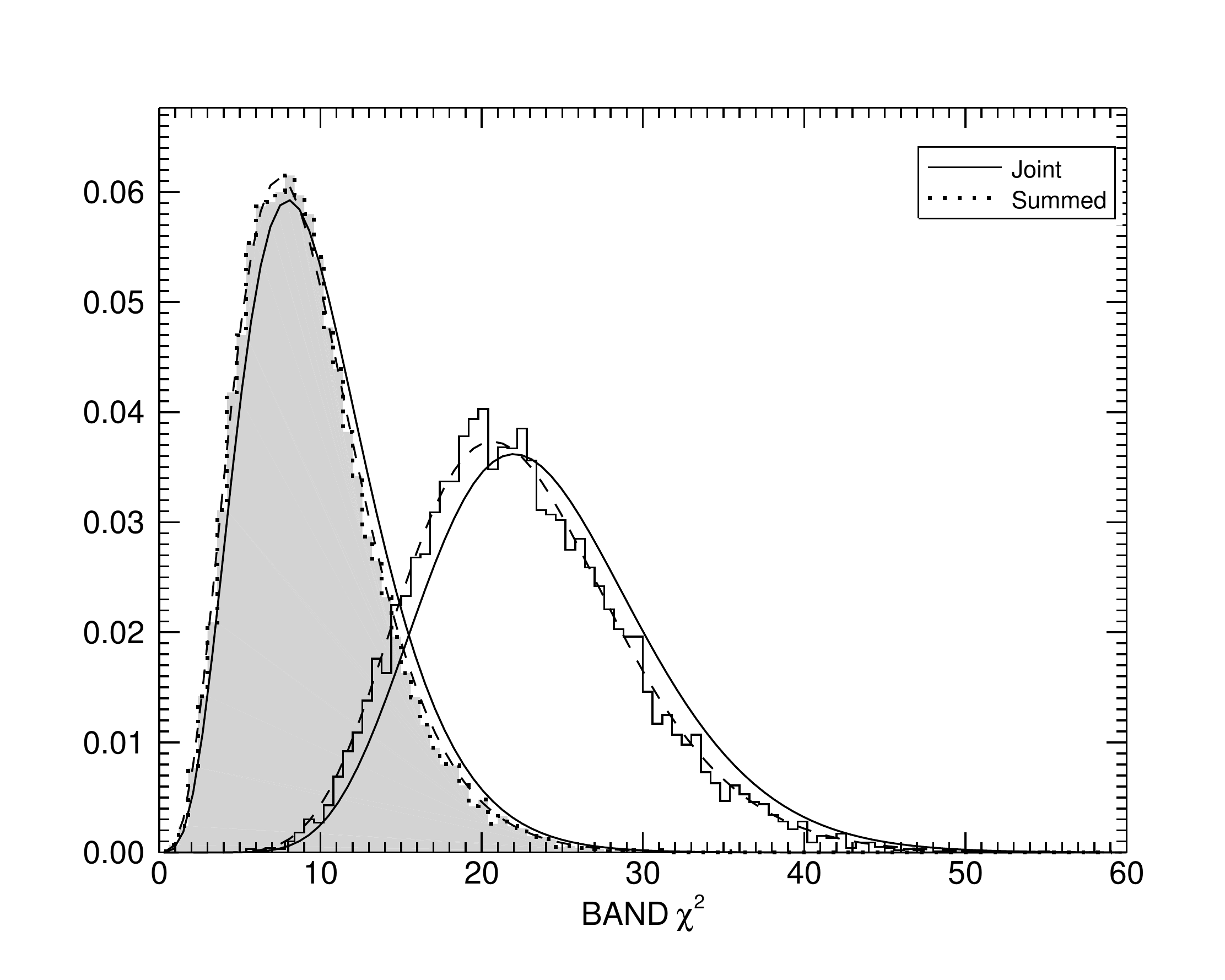}}\\
		\subfigure[]{\label{comparesbpl}\includegraphics[scale=0.25]{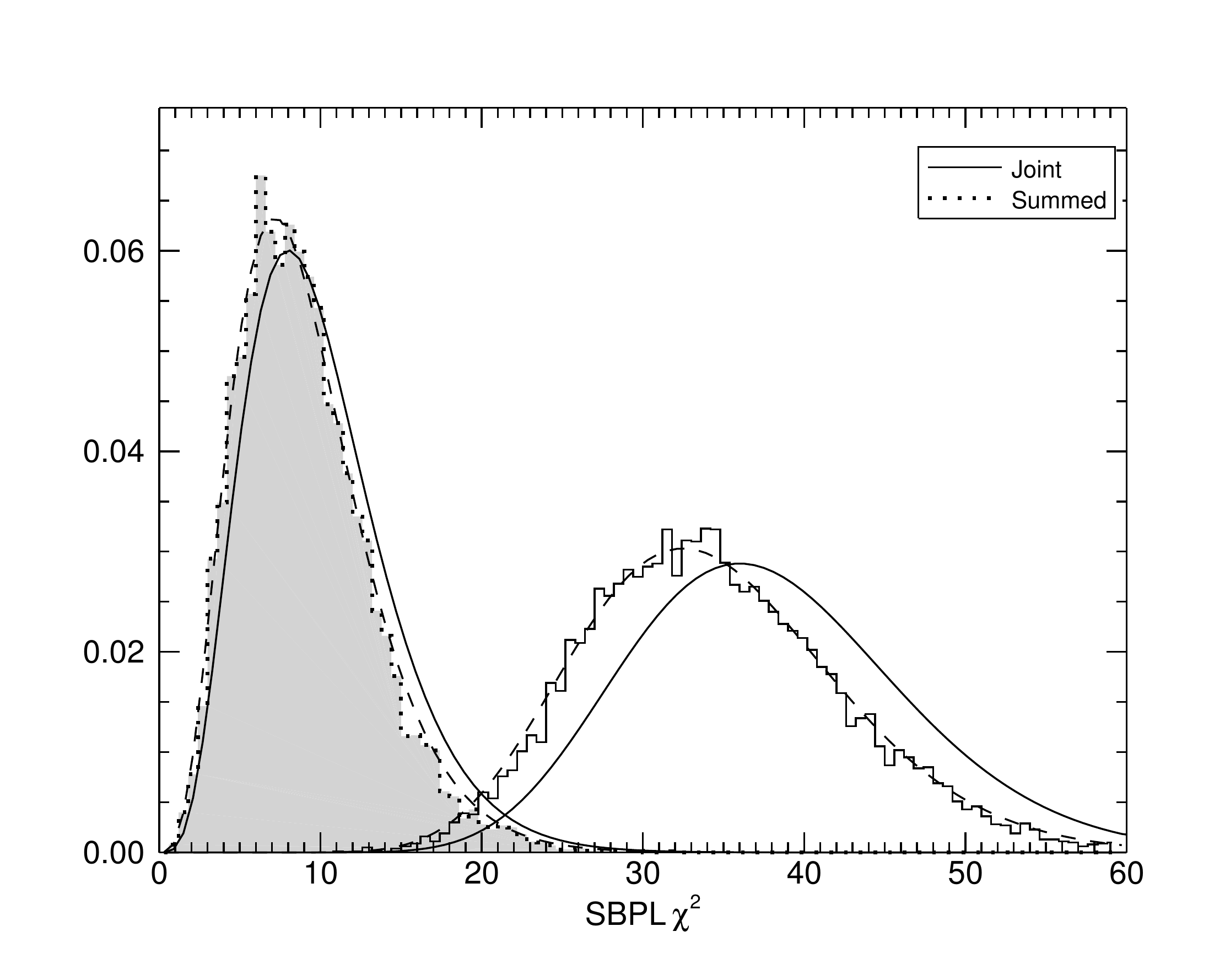}}
	\end{center}
\caption{Comparisons of the $\chi^2$ distributions between the joint and summed simulations.  The solid curve represents the $
\chi^2$ distribution for the defined number of degrees of freedom.  The dashed curve is a best fit $\chi^2$ distribution showing 
that the distribution of $\chi^2$ values from both simulations are shifted from what is expected, although by no more than 10\% in 
any one case.  Both the summed and joint distributions are shifted by roughly the same amount, therefore, the shift is not due to 
summing detectors.  \label{comparechisq}}
\end{figure}

\begin{figure}
	\begin{center}
		\subfigure[]{\label{redchipl}\includegraphics[scale=0.25]{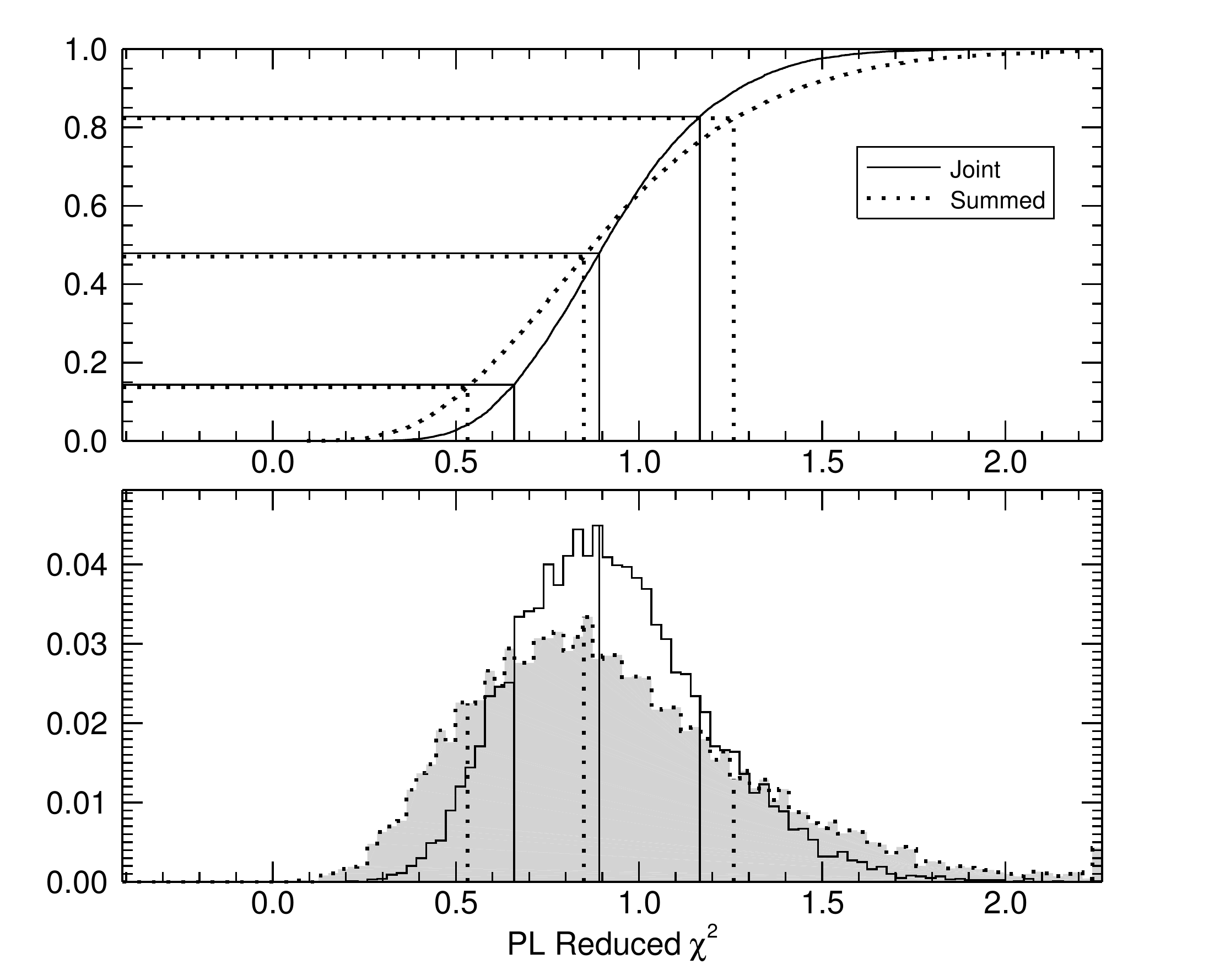}}
		\subfigure[]{\label{redchiglog}\includegraphics[scale=0.25]{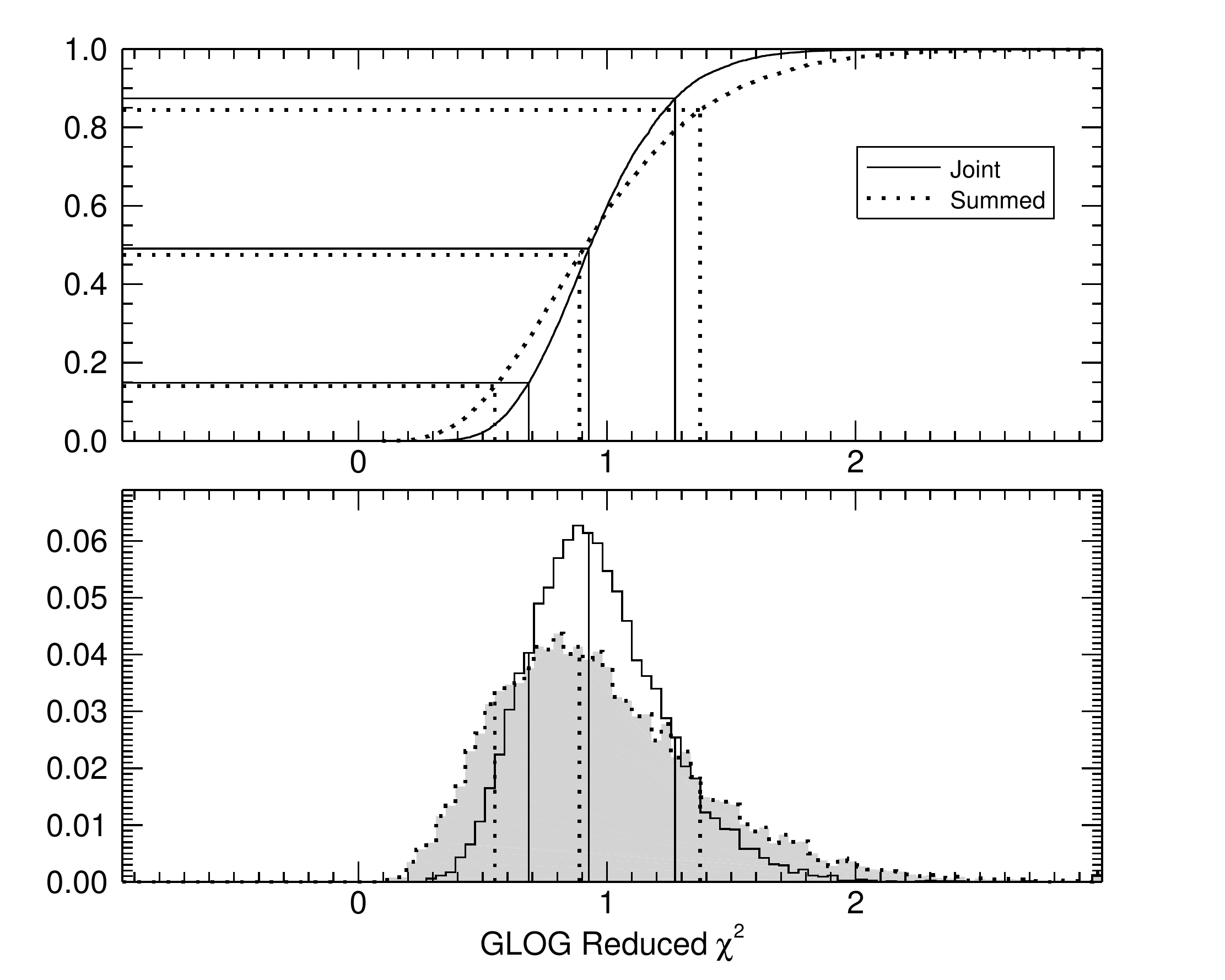}}\\
		\subfigure[]{\label{redchicomp}\includegraphics[scale=0.25]{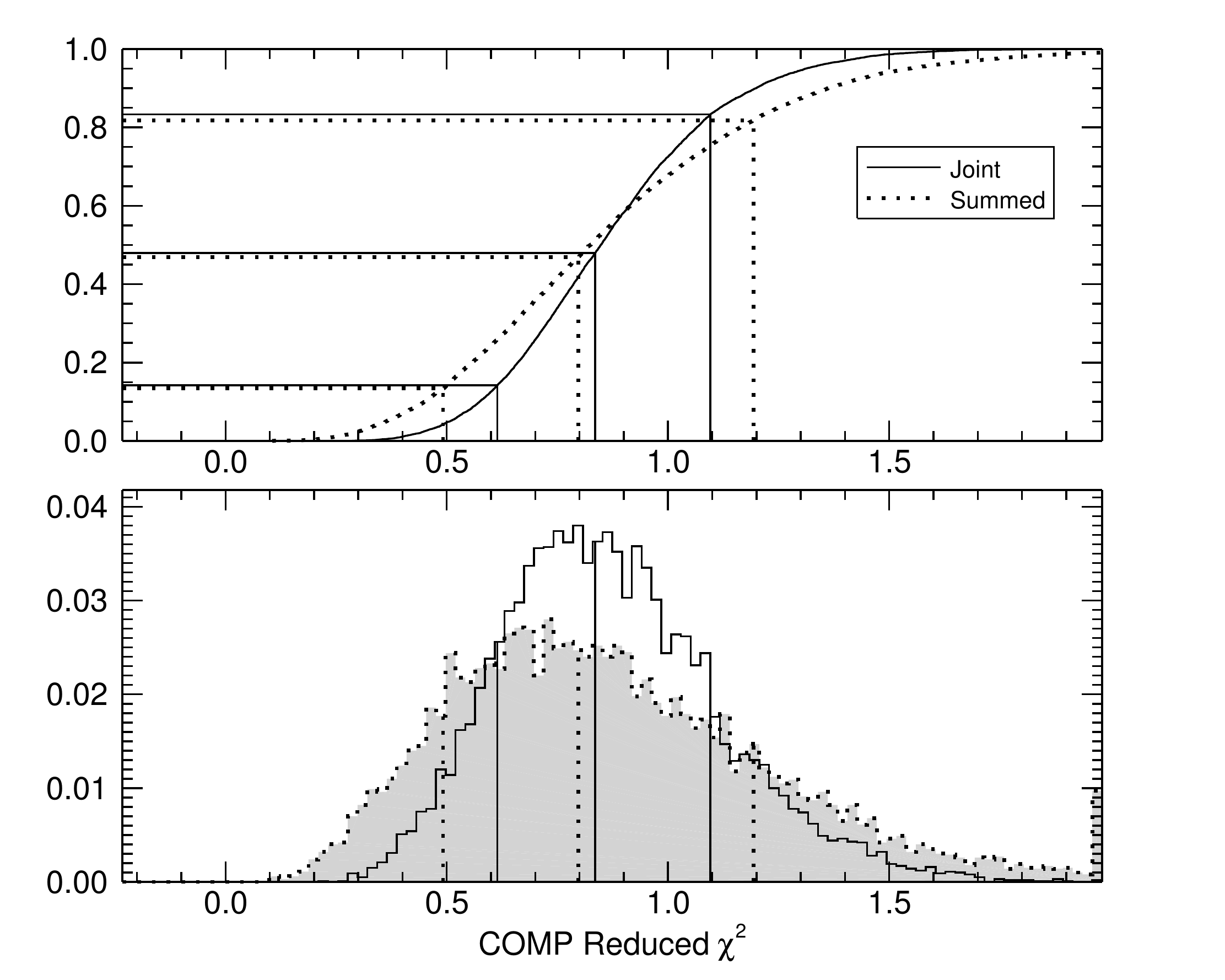}}
		\subfigure[]{\label{redchiband}\includegraphics[scale=0.25]{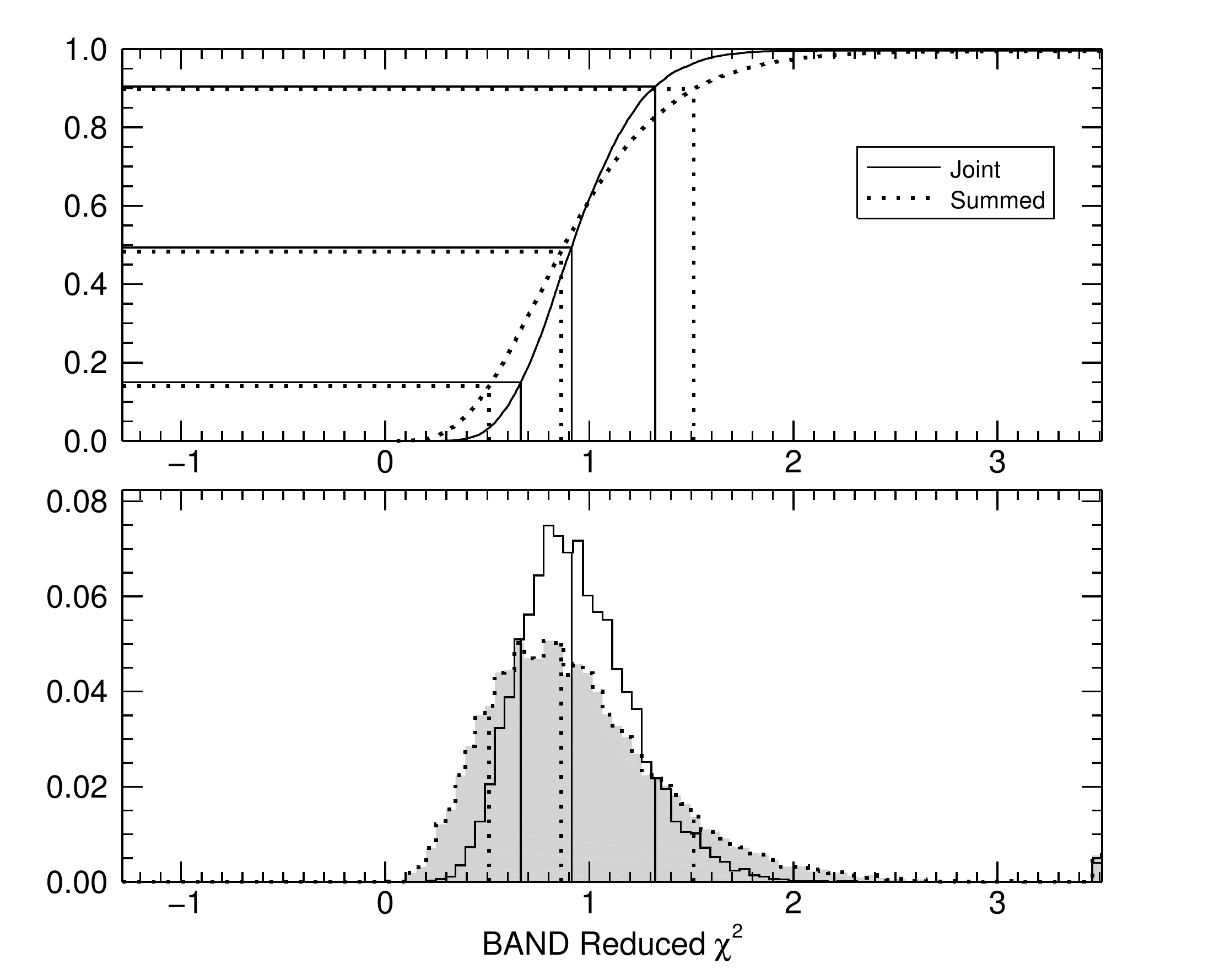}}\\
		\subfigure[]{\label{redchisbpl}\includegraphics[scale=0.25]{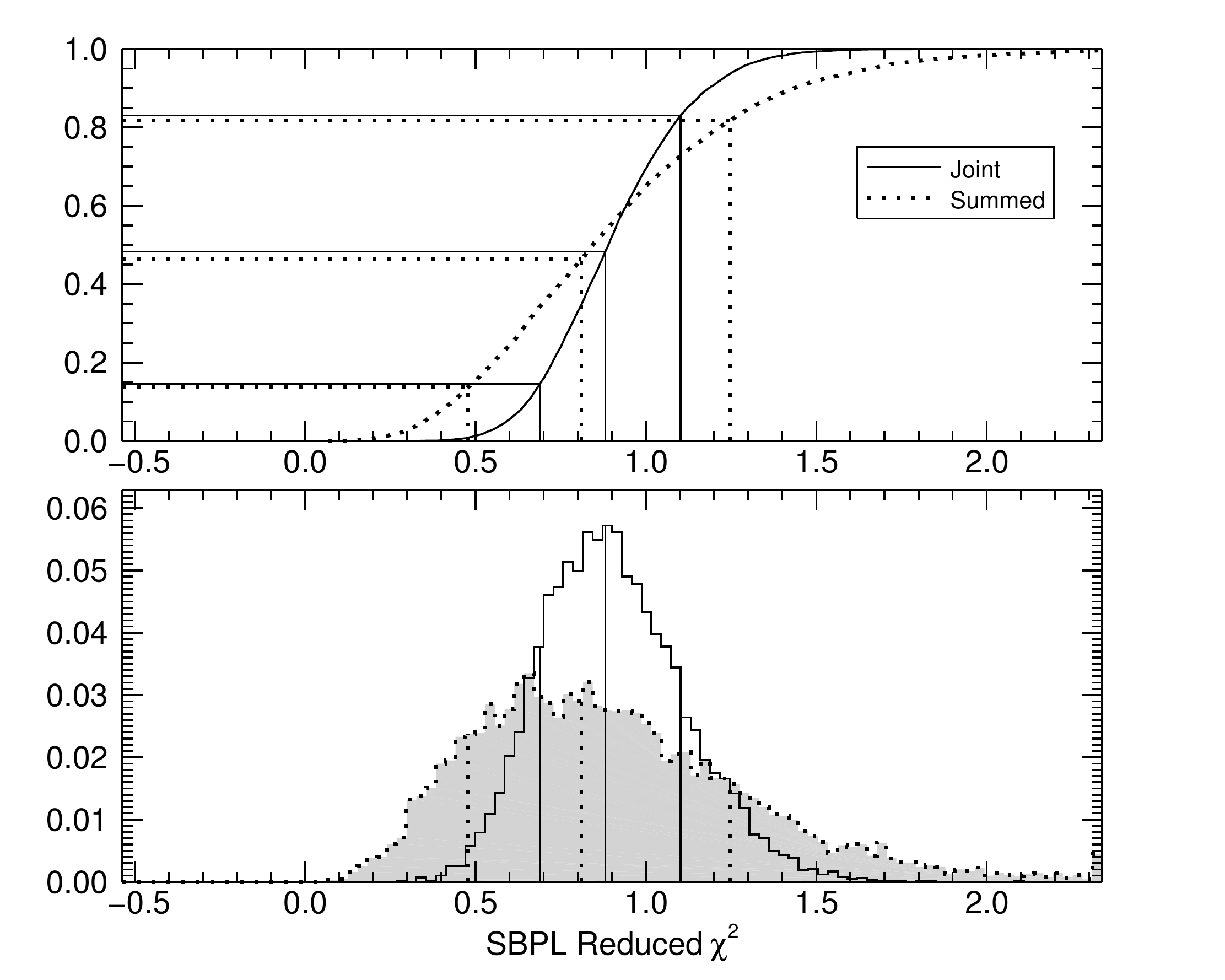}}
	\end{center}
\caption{Comparisons of the reduced $\chi^2$ distributions between the joint and summed simulations.  The upper plots show 
the cumulative distributions and reveal the large differences in the variance of the distributions shown in the lower plots.  The 
three vertical lines in each plot represent the sample means (middle line), and the bounds of the sample standard deviations.  In 
each case the means between the summed and joint simulations are approximately the same, yet the variance for the summed 
simulations are considerably larger than that for the joint simulations.  \label{compareredchisq}}
\end{figure}

Another potential consequence of summing detectors is that the resulting fit statistics could be skewed.  To measure the amount 
by which the $\chi^2$ distributions are affected, we compare the $\chi^2$ distributions of the simulated joint fits and the 
simulated summed fits.  Since a distribution of $\chi^2$ values from fitting a model to the data should be a $\chi^2$ distribution, we 
expect that when a BAND function is fit to the simulated data produced from a BAND function, the resulting distribution of $
\chi^2$ should also be a $\chi^2$ distribution defined by the number of degrees of freedom of the fit.  However, the figure of merit 
we are minimizing is not $\chi^2$, but C-stat, therefore the resulting distribution is not necessarily expected to represent the ideal 
case.  In Figure \ref{comparechisq} we show that indeed both the joint and summed $\chi^2$ distributions deviate from the 
distribution that is expected in the ideal case.  By fitting a $\chi^2$ distribution to the distribution of $\chi^2$, we can estimate the 
amount of deviation between what is expected and what is produced through the simulations.  In this case, we find that both the 
joint and summed $\chi^2$ distributions deviate by the same amount, therefore we conclude that this shift is due to the 
minimization of C-stat (instead of $\chi^2$) and not caused by summing the detectors.  Additionally, we can inspect the reduced 
$\chi^2$, which can be used as a measure of the goodness-of-fit and also represents a comparison between the data and model 
variances.  Ideally, a reduced $\chi^2$ distribution will peak close to a value of unity, denoting that the model and data variances 
are as expected.  A value less than 1 implies that the model variances are overestimated, and a value greater than 1 implies that 
the model variances are underestimated.  As shown in Figure \ref{compareredchisq},  the joint and summed reduced $\chi^2$ 
distributions in many cases are not the same.  In most cases, the medians of the distributions are the same, but the summed 
distribution is much broader.  This indicates that summing the detectors tends to change the model variance.  

\begin{figure}
	\begin{center}
		\subfigure[]{\label{deltachibandpl}\includegraphics[scale=0.35]{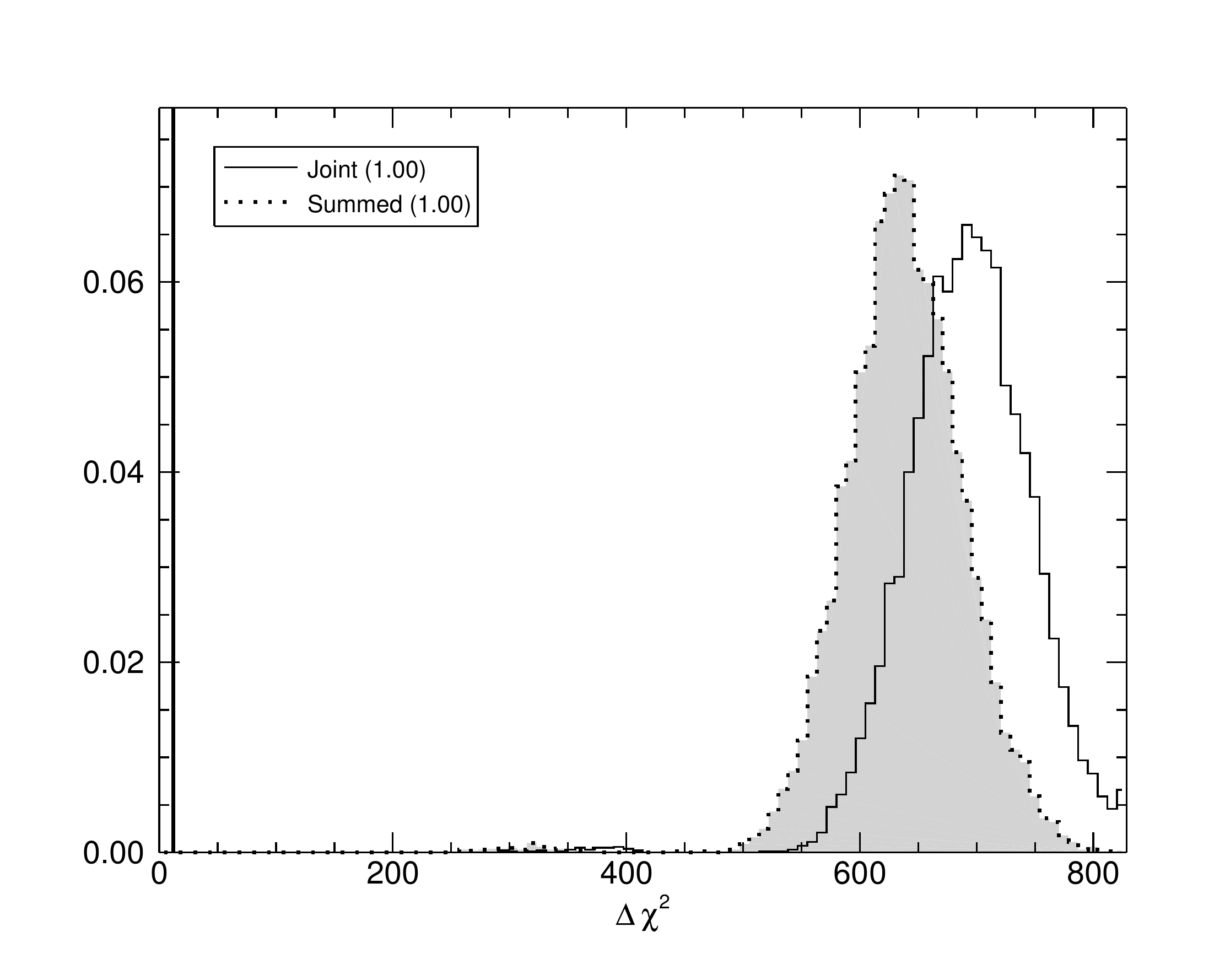}}
		\subfigure[]{\label{deltachibandglog}\includegraphics[scale=0.35]{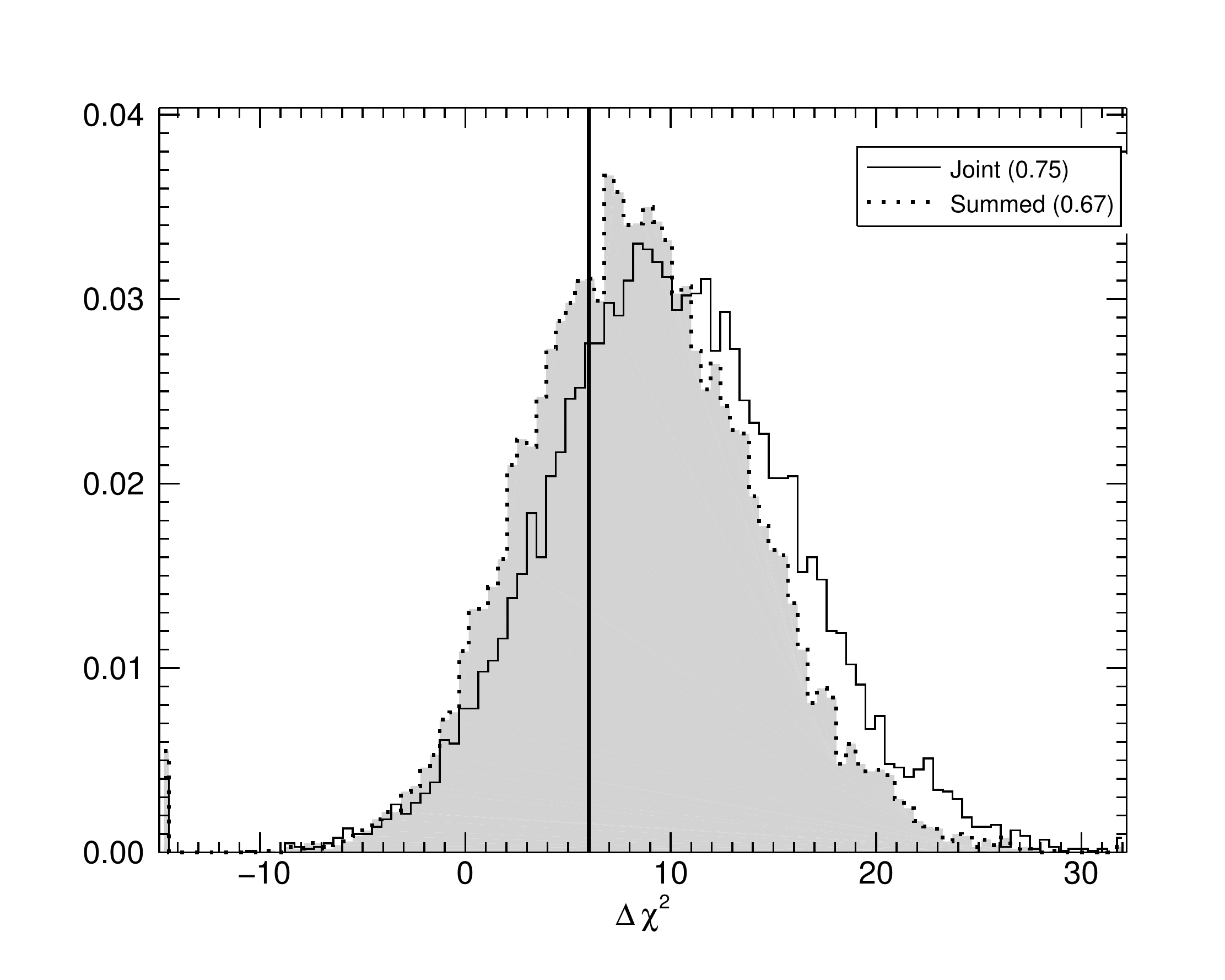}}\\
		\subfigure[]{\label{deltachibandcomp}\includegraphics[scale=0.35]{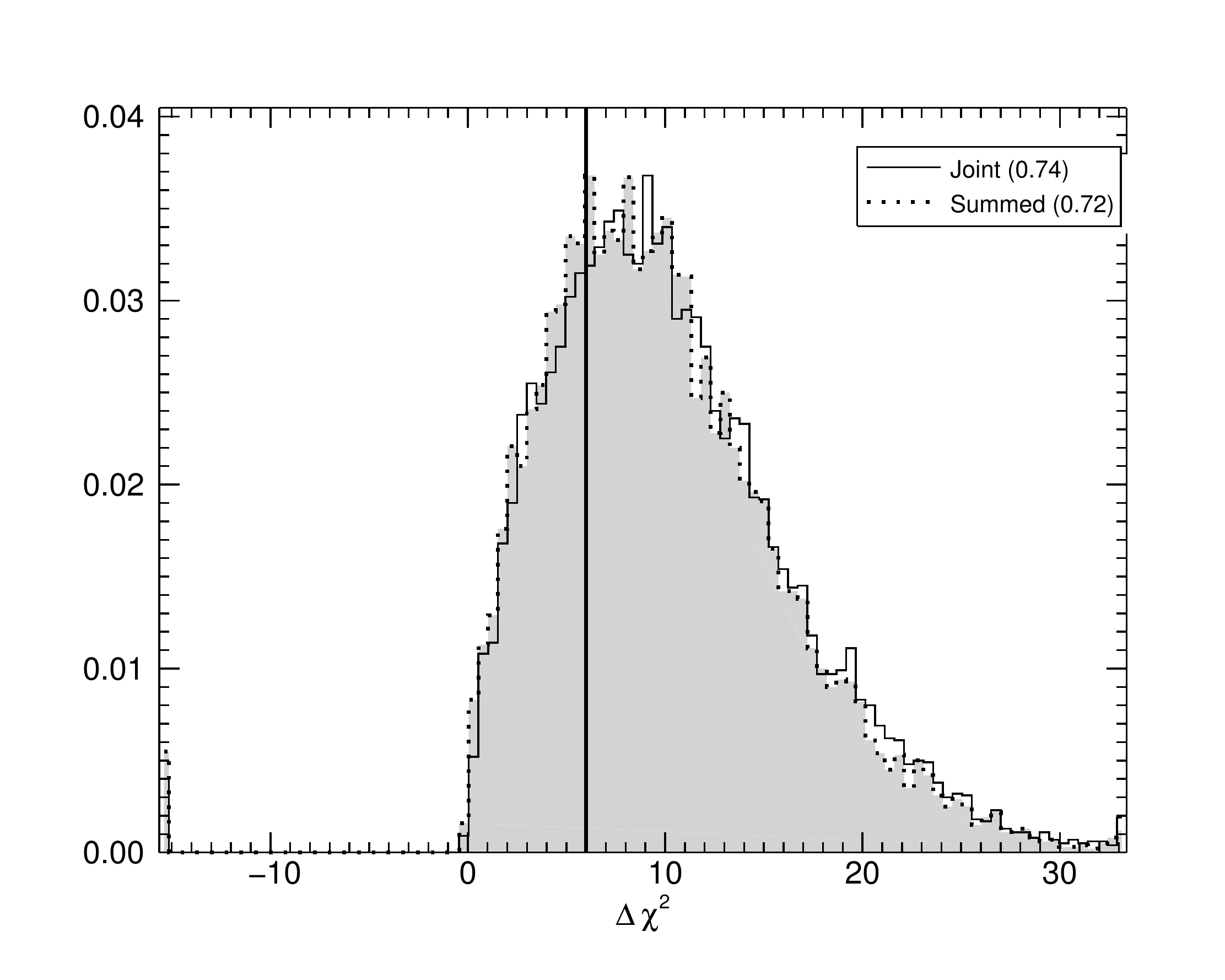}}
		\subfigure[]{\label{deltachibandsbpl}\includegraphics[scale=0.35]{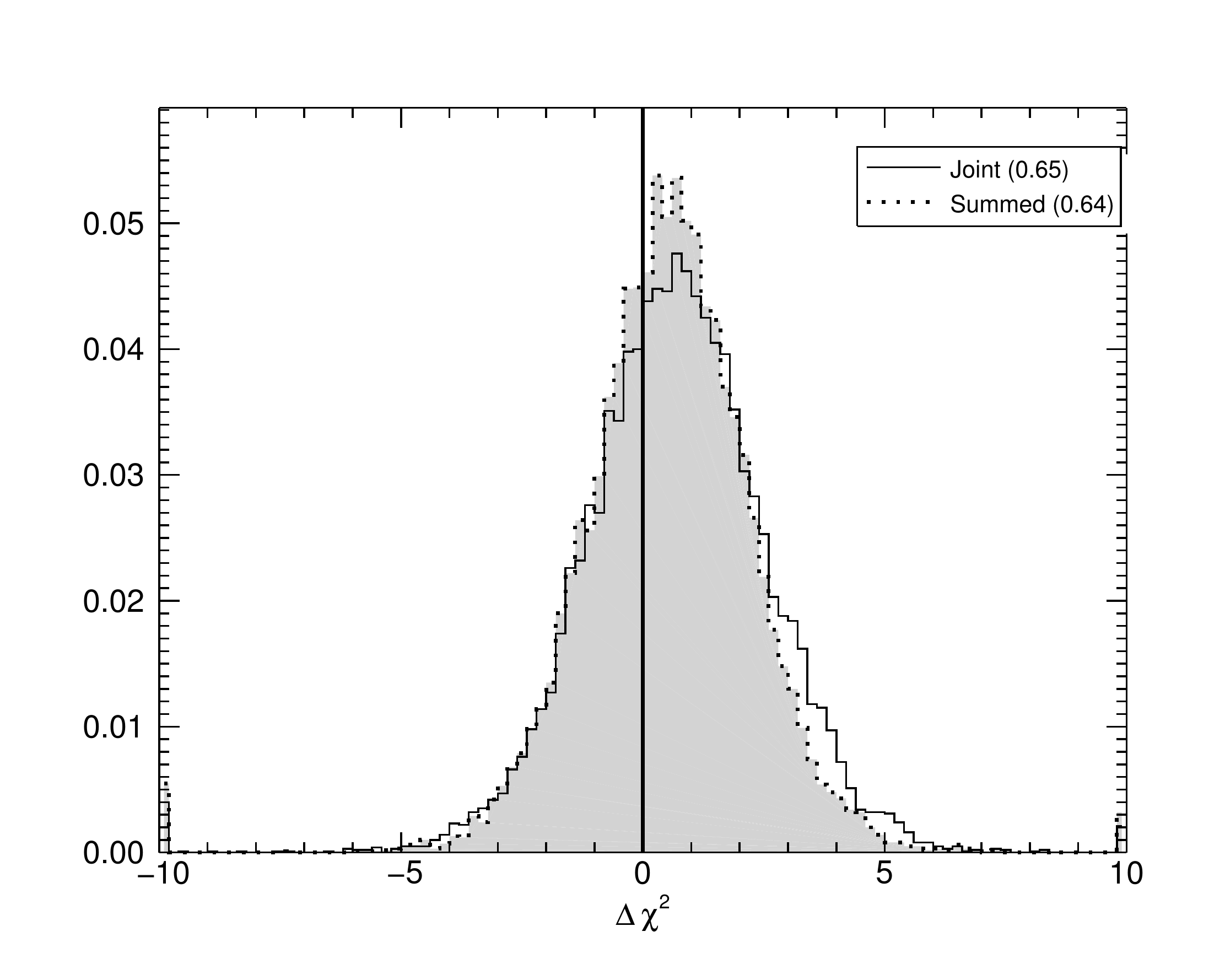}}
	\end{center}
\caption{Comparisons of the $\Delta \chi^2$ distributions between the joint and summed simulations.  10,000 simulated BAND 
spectra were created and the other four models were fit to the simulated data.  The vertical lines in each plot represent the cutoffs 
specified by our model selection criteria.  To determine if the model selections change between the summed and joint data, we 
measure how much of each distribution exists above the cutoff.  This indicates a probability that our cutoff will prefer a BAND 
model rather than the other model.  In all cases the distributions are different, but approximately the same amount of each 
distribution exists above the cutoff (percentages in parentheses).  \label{deltachiband}}
\end{figure}

For the purpose of model comparison, $\chi^2$ is an important statistic, and we need to determine if $\Delta \chi^2$ changes 
when summing detectors.  For each of the bursts on which we performed simulations, we also fit the other models to determine 
how accurate the summed detectors method is at model selection.  For this test, we calculate the change in $\chi^2$ and plot it 
as a distribution for each simulation.  We then determine the change in the degrees of freedom when fitting each model and the 
cutoff value we have used for this catalog.  Because of the additive property of $\chi^2$, the $\Delta \chi^2$ values are 
associated with a $\chi^2$ distribution represented by the difference in the degrees of freedom between the two models, 
therefore our criteria of 6 units of $\chi^2$ per difference in degrees of freedom represents a cutoff in a $\chi^2$ distribution.  This 
cutoff value admits a probability in achieving our predetermined change in $\chi^2$ by chance, so for the summed $\Delta 
\chi^2$ distributions to be acceptable, the percentage of the distribution existing above the cutoff must be approximately the 
same as that from the joint distribution. This would indicate that both methods produce the same model choice at the same 
probability.  Figure \ref{deltachiband} shows an example of the comparison of $\Delta \chi^2$ between joint and summed fits.  
Even though the distributions of $\Delta \chi^2$ are different in many cases, approximately the same amount of the summed 
distribution exists above the cutoff when compared to the joint distribution.

The culmination of the simulation tests between joint fitting of detectors and summing detectors indicates a relatively small loss of 
information by adding detector count rates and responses prior to fitting.  The details in the response are lost through the 
summation process, although since the 16-channel CONT data does not provide high spectral resolution, this effect is minimal.  
The parameter values from the fits vary only negligibly, and the parameter errors are only slightly larger when summing the 
detectors.  The $\chi^2$ values that are produced from summed fits slightly deviate from a $\chi^2$ distribution of the required 
number of degrees of freedom, but they deviate no more than the joint fit $\chi^2$, therefore they are unaffected by the summation.  
The reduced $\chi^2$, which measures the goodness-of-fit, can however, show significant deviation.  The means of the reduced $
\chi^2$ distributions are typically similar, although their distance from unity can be larger than expected.  In particular, if a fit is 
required to be within 1$\sigma$ of the mean ($\sim$1), the result from the summed detectors will return a small percentage of fits 
that would have been rejected if fitted with the joint detectors.  Finally, the $\Delta \chi^2$ found when comparing models using the 
summed detectors will succeed in choosing the same model almost all of the time when compared to fitting joint detectors.  From 
these results, we show that summing detectors for this catalog has little adverse effect on the spectral results.

\clearpage

\section{Appendix E \\
Spectral Table Formats}
Please note that the spectral tables list the raw results of each spectral fit to each GRB.  In cases where the spectral fit failed, the 
values reported are those that initialized the spectral fit.  If the uncertainty on the spectral parameters is reported as zero (no 
uncertainty), then the fit failed.  In a few cases throughout these tables, the uncertainties for certain spectral parameters may be 
reported as `9999.99' which indicates that the uncertainty on that parameter is completely unconstrained.  An example of this is when 
the spectral data from a burst is fitted with a BAND function but is unable to constrain the high-energy index.  In this case, the best fit 
centroid value of the high-energy index parameter is reported, and the `9999.99' is reported for the uncertainty.

\begin{deluxetable}{lll}
\tablecolumns{3}
\tablewidth{0pt}
\tabletypesize{\scriptsize}
\tablecaption{BAND Fluence \& Peak Flux Parameters Table Format\label{BANDparms}}
\startdata
\hline
Column & Format & Description\\ \hline
1 & I4 & Trigger number\\
2 & A4 & Detectors used\\
3 & A4 & Datatype\\
4 & F6.2 & Total integrated time\\
5 & E8.2 & Amplitude\\
6 & E8.2 & Uncertainty in Amplitude\\
7 & F7.2 & Low-Energy Spectral Index\\
8 & F7.2 & Uncertainty in Low-Energy Spectral Index\\
9 & F7.2 & High-Energy Spectral Index\\
10 & F7.2 & Uncertainty in High-Energy Spectral Index\\
11 & E8.2 & $E_{peak}$\\
12 & E8.2 & Uncertainty in $E_{peak}$\\
13 & F7.2 & Photon Flux\\
14 & F7.2 & Uncertainty in Photon Flux\\
15 & F7.2 & Photon Fluence\\
16 & F7.2 & Uncertainty in Photon Fluence\\
17 & E9.2 & Energy Flux\\
18 & E9.2 & Uncertainty in Energy Flux\\
19 & E9.2 & Energy Fluence\\
20 & E9.2 & Uncertainty in Energy Fluence\\
21 & F8.2 & $\chi^2$ Goodness-of-Fit Statistic\\
22 & I2 & Degrees of Freedom\\
\enddata
\end{deluxetable}

\begin{deluxetable}{lll}
\tablecolumns{3}
\tablewidth{0pt}
\tabletypesize{\scriptsize}
\tablecaption{COMP Fluence \& Peak Flux Parameters Table Format\label{COMPparms}}
\startdata
\hline
Column & Format & Description\\ \hline
1 & I4 & Trigger number\\
2 & A4 & Detectors used\\
3 & A4 & Datatype\\
4 & F6.2 & Total integrated time\\
5 & E8.2 & Amplitude\\
6 & E8.2 & Uncertainty in Amplitude\\
7 & F7.2 & Spectral Index\\
8 & F7.2 & Uncertainty in Spectral Index\\
9 & E8.2 & $E_{peak}$\\
10 & E8.2 & Uncertainty in $E_{peak}$\\
11 & F7.2 & Photon Flux\\
12 & F7.2 & Uncertainty in Photon Flux\\
13 & F7.2 & Photon Fluence\\
14 & F7.2 & Uncertainty in Photon Fluence\\
15 & E9.2 & Energy Flux\\
16 & E9.2 & Uncertainty in Energy Flux\\
17 & E9.2 & Energy Fluence\\
18 & E9.2 & Uncertainty in Energy Fluence\\
19 & F8.2 & $\chi^2$ Goodness-of-Fit Statistic\\
20 & I2 & Degrees of Freedom\\
\enddata
\end{deluxetable}

\begin{deluxetable}{lll}
\tablecolumns{3}
\tablewidth{0pt}
\tabletypesize{\scriptsize}
\tablecaption{GLOGE Fluence \& Peak Flux Parameters Table Format\label{GLOGEparms}}
\startdata
\hline
Column & Format & Description\\ \hline
1 & I4 & Trigger number\\
2 & A4 & Detectors used\\
3 & A4 & Datatype\\
4 & F6.2 & Total integrated time\\
5 & E8.2 & Amplitude\\
6 & E8.2 & Uncertainty in Amplitude\\
7 & F6.2 & Full-Width at Half Maximum\\
8 & F6.2 & Uncertainty in Full-Width Half Maximum\\
9 & E8.2 & Centroid Energy\\
10 & E8.2 & Uncertainty in Centroid Energy \\
11 & E8.2 & $E_{peak}$\\
12 & E8.2 & Uncertainty in $E_{peak}$\\
13 & F7.2 & Photon Flux\\
14 & F7.2 & Uncertainty in Photon Flux\\
15 & F7.2 & Photon Fluence\\
16 & F7.2 & Uncertainty in Photon Fluence\\
17 & E9.2 & Energy Flux\\
18 & E9.2 & Uncertainty in Energy Flux\\
19 & E9.2 & Energy Fluence\\
20 & E9.2 & Uncertainty in Energy Fluence\\
21 & F8.2 & $\chi^2$ Goodness-of-Fit Statistic\\
22 & I2 & Degrees of Freedom\\
\enddata
\end{deluxetable}

\begin{deluxetable}{lll}
\tablecolumns{3}
\tablewidth{0pt}
\tabletypesize{\scriptsize}
\tablecaption{PL Fluence \& Peak Flux Parameters Table Format\label{PLparms}}
\startdata
\hline
Column & Format & Description\\ \hline
1 & I4 & Trigger number\\
2 & A4 & Detectors used\\
3 & A4 & Datatype\\
4 & F6.2 & Total integrated time\\
5 & E8.2 & Amplitude\\
6 & E8.2 & Uncertainty in Amplitude\\
7 & F7.2 & Spectral Index\\
8 & F7.2 & Uncertainty in Spectral Index\\
9 & F7.2 & Photon Flux\\
10 & F7.2 & Uncertainty in Photon Flux\\
11 & F7.2 & Photon Fluence\\
12 & F7.2 & Uncertainty in Photon Fluence\\
13 & E9.2 & Energy Flux\\
14 & E9.2 & Uncertainty in Energy Flux\\
15 & E9.2 & Energy Fluence\\
16 & E9.2 & Uncertainty in Energy Fluence\\
17 & F8.2 & $\chi^2$ Goodness-of-Fit Statistic\\
18 & I2 & Degrees of Freedom\\
\enddata
\end{deluxetable}

\begin{deluxetable}{lll}
\tablecolumns{3}
\tablewidth{0pt}
\tabletypesize{\scriptsize}
\tablecaption{SBPL Fluence \& Peak Flux Parameters Table Format\label{SBPLparms}}
\startdata
\hline
Column & Format & Description\\ \hline
1 & I4 & Trigger number\\
2 & A4 & Detectors used\\
3 & A4 & Datatype\\
4 & F6.2 & Total integrated time\\
5 & E8.2 & Amplitude\\
6 & E8.2 & Uncertainty in Amplitude\\
7 & F7.2 & Low-Energy Spectral Index\\
8 & F7.2 & Uncertainty in Low-Energy Spectral Index\\
9 & F7.2 & High-Energy Spectral Index\\
10 & F7.2 & Uncertainty in High-Energy Spectral Index\\
11 & E8.2 & Spectral Break Energy\\
12 & E8.2 & Uncertainty in Spectral Break Energy\\
13 & E8.2 & $E_{peak}$\\
14 & E8.2 & Uncertainty in $E_{peak}$\\
15 & F7.2 & Photon Flux\\
16 & F7.2 & Uncertainty in Photon Flux\\
17 & F7.2 & Photon Fluence\\
18 & F7.2 & Uncertainty in Photon Fluence\\
19 & E9.2 & Energy Flux\\
20 & E9.2 & Uncertainty in Energy Flux\\
21 & E9.2 & Energy Fluence\\
22 & E9.2 & Uncertainty in Energy Fluence\\
23 & F8.2 & $\chi^2$ Goodness-of-Fit Statistic\\
24 & I2 & Degrees of Freedom\\
\enddata
\end{deluxetable}


\clearpage

\clearpage

\clearpage

\clearpage

\begin{deluxetable}{ c  c  c }
\tablecolumns{5}
\tablewidth{0pt}
\tabletypesize{\scriptsize}
\tablecaption{K-S Test between Fluence and Peak Flux Parameters \label{FluPFluxCompare}}
\startdata
	\hline
	Parameter & K-S Stat & Probability \\ \hline
	PL Index & 0.14 & $2.3\times 10^{-19}$ \\
	SBPL Alpha & 0.20 & $3.1\times 10^{-31}$ \\
	BAND Alpha & 0.23 & $1.0\times 10^{-27}$ \\
	COMP Index & 0.24 & $5.9\times 10^{-44}$ \\ \hline
	SBPL Beta & 0.03 & 0.68 \\
	BAND Beta & 0.05 & 0.15 \\ \hline
	SBPL $E_{peak}$ & 0.09 & $5.5\times10^{-5}$ \\
	BAND $E_{peak}$ & 0.09 & $9.3\times10^{-6}$ \\
	COMP $E_{peak}$ & 0.09 & $1.1\times10^{-6}$ \\
	GLOG $E_{peak}$ & 0.16 & $3.1\times10^{-7}$ \\ \hline
\enddata
\end{deluxetable}

\begin{deluxetable}{ c  c  c  c  c }
\tablecolumns{5}
\tablewidth{0pt}
\tabletypesize{\scriptsize}
\tablecaption{BEST GRB models \label{BestTable}}
\startdata
	\hline
	\bf PL & \bf SBPL & \bf BAND & \bf COMP & \bf GLOGE \\ \hline

	\multicolumn{5}{ c }{\bf Fluence Spectra} \\ \hline 
	
	506 (23\%) & 124 (6\%) & 77 (4\%) & 903 (42\%) & 535 (25\%) \\ \hline
	
	\multicolumn{5}{ c }{\bf Peak Flux Spectra} \\ \hline
	
	454 (21\%) & 150 (7\%) & 65 (3\%) & 847 (40\%) & 629 (29\%) \\ \hline	

\enddata
\end{deluxetable}

\begin{deluxetable}{ c  c  c  c  c  c  c  c  c }
\tablecolumns{9}
\tablewidth{0pt}
\tabletypesize{\scriptsize}
\setlength{\tabcolsep}{0.02in}
\tablecaption{Sample mean and standard deviation of the parameter distributions \label{ParamTable}}
\startdata
	\hline
	\multirow{2}{*}{\bf Model} & \bf Low-Energy & \bf High-Energy & \multirow{2}{*}{$\bf E_{peak} (keV)$} & \multirow{2}{*}{$\bf
	 E_{break} (keV)$} & \bf $\bf E_{cent}$ (keV) & \bf FWHM & \bf Photon Flux & \bf Energy Flux \\
 	& \bf Index & \bf Index & & & & \bf ($\bf Log_{10}$ keV) & \bf (ph $\bf s^{-1} \ cm^{-2}$) & \bf ($\bf 10^{-7} \ erg \ s^{-1} \ cm^
	{-2}$) \\ \hline

	\multicolumn{9}{ c }{\bf Fluence Spectra} \\ \hline 
	
	PL & $-1.79^{+0.35}_{-0.54}$ & - & - & - & - & - & $0.53^{+0.41}_{-0.22}$ & $1.16^{+0.99}_{-0.46}$ \\

	COMP & $-1.12^{+0.55}_{-0.43}$ & - & $195.11^{+256.43}_{-98.74}$ & - & - & - & $0.68^{+0.80}_{-0.33}$ & $1.22^
	{+1.96}_{-0.62}$ \\

	SBPL & $-1.36^{+0.47}_{-0.52}$ & $-2.62^{+0.56}_{-0.71}$ & $190.77^{+186.24}_{-92.77}$ & $173.18^{+150.14}_
	{-88.64}$ & - & - & $0.71^{+0.93}_{-0.35}$ & $1.50^{+2.47}_{-0.76}$ \\

	BAND & $-0.94^{+0.52}_{-0.38}$ & $-2.43^{+0.49}_{-0.54}$ & $167.08^{+163.86}_{-75.67}$ & $228.61^{+185.31}_	
	{-97.72}$ & - & - & $0.74^{+0.97}_{-0.37}$ & $1.54^{+2.37}_{-0.79}$ \\
	
	GLOGE & - & - & $131.53^{+161.72}_{-59.93}$ & - & $12.34^{+24.82}_{-7.25}$ & $1.22^{+0.30}_{-0.27}$ & $0.60^{+0.67}_
	{-0.29}$ & $1.14^{+1.68}_{-0.56}$ \\
	
	BEST & $-1.29^{+0.60}_{-0.63}$ & $-2.53^{+0.57}_{-0.71}$ & $195.94^{+267.66}_{-104.32}$ & $195.78^{+191.39}_	
	{-120.56}$ & $13.99^{+22.07}_{-7.15}$ & $1.10^{+0.24}_{-0.24}$ & $0.75^{+1.09}_{-0.37}$ & $1.40^{+2.49}_{-0.71}$\\ 
	\hline

	\multicolumn{9}{ c }{\bf Peak Flux Spectra} \\ \hline 
	
	PL & $-1.72^{+0.32}_{-0.39}$ & - & - & - & - & - & $0.80^{+0.78}_{-0.40}$ & $1.91^{+1.48}_{-0.85}$\\

	COMP & $-0.87^{+0.58}_{-0.47}$ & - & $215.06^{+252.45}_{-107.80}$ & - & - & - & $1.30^{+2.18}_{-0.80}$ & $2.65^
	{+5.20}_{-1.57}$\\

	SBPL & $-1.16^{+0.47}_{-0.57}$ & $-2.61^{+0.54}_{-0.69}$ & $214.48^{+184.74}_{-95.06}$ & $195.11^{+132.61}_	
	{-84.69}$ & - & - & $1.44^{+3.15}_{-0.90}$ & $3.69^{+9.50}_{-2.20}$\\
	
	BAND & $-0.73^{+0.57}_{-0.41}$ & $-2.47^{+0.52}_{-1.02}$ &$186.84^{+175.24}_{-86.56}$ & $263.36^{+219.75}_	
	{-115.17}$ & - & - & $1.56^{+3.30}_{-0.99}$ & $3.95^{+9.30}_{-2.31}$ \\

	GLOGE & - & - & $156.18^{+189.15}_{-73.65}$ & - & $20.68^{+29.45}_{-12.72}$ & $1.18^{+0.33}_{-0.28}$ & $1.19^{+1.96}
	_{-0.73}$  & $2.62^{+4.80}_{-1.53}$ \\

	BEST & $-1.10^{+0.62}_{-0.70}$ & $-2.69^{+0.46}_{-0.73}$ & $228.25^{+271.74}_{-119.73}$ & $273.38^{+196.39}_	
	{-128.46}$ & $21.92^{+27.75}_{-12.53}$ & $1.09^{+0.28}_{-0.24}$ &$1.61^{+4.32}_{-1.02}$ & $3.44^{+11.5}_{-2.09}$ \\ 
	\hline
	
\enddata
\end{deluxetable}

\begin{deluxetable}{ c  c  c  c  c  c  c }
\tablecolumns{7}
\tablewidth{0pt}
\tabletypesize{\scriptsize}
\tablecaption{Comparison of the sample mean and standard deviation from different catalogs \label{BestComparison}}
\startdata
	\hline
	\multirow{2}{*}{\bf Dataset} & \bf Low-Energy & \bf High-Energy & \multirow{2}{*}{$\bf E_{peak}$} & \multirow{2}{*}{$\bf
	 E_{break}$} & \bf Photon Flux & \bf Energy Flux \\
 	& \bf Index & \bf Index & \bf (keV) & \bf (keV) & \bf (ph $\bf s^{-1} \ cm^{-2}$) & \bf ($\bf 10^{-7} \ erg \ s^{-1} \ cm^{-2}$) \\ 
	\hline

	\multicolumn{7}{ c }{\bf Fluence} \\ \hline
	This Catalog BEST & $-1.29^{+0.60}_{-0.63}$ & $-2.53^{+0.57}_{-0.71}$ & $196^{+268}_{-104}$ & $196^{+191}_{-121}$ 
	& $0.75^{+1.09}_{-0.37}$ & $1.40^{+2.49}_{-0.71}$\\
	
	\citet{GBMSpecCat} & $-1.05^{+0.44}_{-0.45}$ & $-2.25^{+0.34}_{-0.73}$ & $205^{+359}_{-121}$ & $123^{+240}_{-80.4}$ 
	& $2.92^{+3.96}_{-1.31}$ & $4.03^{+9.38}_{-2.13}$\\
	
	\citet{Kaneko06} & $-1.07^{+0.42}_{-0.36}$ & $-2.43^{+0.38}_{-0.59}$ & $260^{+233}_{-116}$ & $203^{+129}_{-80.0}$ & 
	$3.32^{+6.01}_{-2.04}$ & $8.56^{+16.0}_{-5.47}$\\ \hline

	\multicolumn{7}{ c }{\bf Peak Flux Spectra} \\ \hline 
	This Catalog BEST & $-1.10^{+0.62}_{-0.70}$ & $-2.69^{+0.46}_{-0.73}$ & $228^{+271}_{-120}$ & 
	$273^{+196}_{-128}$ & $1.61^{+4.32}_{-1.02}$ & $3.44^{+11.5}_{-2.09}$ \\
	\citet{GBMSpecCat} & $-1.12^{+0.61}_{-0.50}$ & $-2.27^{+0.44}_{-0.50}$ & $223^{+352}_{-126}$ & 
	$172^{+254}_{-100}$ & $5.39^{+10.18}_{-2.87}$ & $8.35^{+22.61}_{-4.98}$ \\
	\hline 

\enddata
\end{deluxetable}

\begin{deluxetable}{ c  c  c  c }
\tablecolumns{4}
\tablewidth{0pt}
\tabletypesize{\scriptsize}
\tablecaption{Mean and Standard Deviation of Simulated Parameters \label{SimTable}}
\startdata
	\hline
	\bf Parameter & \bf Catalog & \bf Joint Sim. & \bf Summed Sim. \\ \hline

	\multicolumn{4}{ c }{\bf GRB 910425 - PL} \\ \hline 
	
	Index & $-1.678 \pm 0.043$ & $-1.677^{+0.040}_{-0.044}$ & $-1.678^{+0.042}_{-0.045}$ \\ \hline
	
	\multicolumn{4}{ c }{\bf GRB 910521 - GLOGE} \\ \hline
	
	$E_{cen}$ & $12.094 \pm 4.758$ & $12.428^{+4.149}_{-3.402}$ & $12.690^{+4.310}_{-3.482}$ \\
	FWHM & $0.904 \pm 0.096$ & $0.898^{+0.075}_{-0.076}$ & $0.893^{+0.074}_{-0.077}$  \\ \hline
	
	\multicolumn{4}{ c }{\bf GRB 910426 - COMP} \\ \hline
	
	$E_{peak}$ & $93.182 \pm 2.956$ & $93.163^{+2.842}_{-3.030}$ & $92.129^{+2.878}_{-3.089}$ \\
	$\alpha$ & $-1.095 \pm 0.109$ & $-1.093^{+0.110}_{-0.106}$ & $-1.096^{+0.113}_{-0.107}$  \\ \hline

	\multicolumn{4}{ c }{\bf GRB 910421 - BAND} \\ \hline
	
	$E_{peak}$ & $127.788 \pm 3.334$ & $127.692^{+3.253}_{-5.700}$ & $127.690^{+3.354}_{-6.562}$ \\
	$\alpha$ & $-0.996 \pm 0.060$ & $-0.995^{+0.076}_{-0.054}$ & $-0.995^{+0.090}_{-0.057}$  \\
	$\beta$ & $-2.939 \pm 0.171$ & $-2.938^{+0.158}_{-0.388}$ & $-2.940^{+0.168}_{-0.377}$  \\ \hline

	\multicolumn{4}{ c }{\bf GRB 910522 - SBPL} \\ \hline
	
	$E_{break}$ & $224.585 \pm 14.324$ & $224.826^{+14.734}_{-13.630}$ & $224.685^{+14.585}_{-13.651}$ \\
	$\alpha$ & $-1.343 \pm 0.019$ & $-1.343^{+0.020}_{-0.018}$ & $-1.343^{+0.020}_{-0.018}$  \\
	$\beta$ & $-2.461 \pm 0.058$ & $-2.462^{+0.052}_{-0.062}$ & $-2.460^{+0.052}_{-0.063}$  \\ \hline

\enddata
\end{deluxetable}

\end{document}